\documentstyle{article}
\def\emline#1#2#3#4#5#6{%
       \put(#1,#2){\special{em:moveto}}%
       \put(#4,#5){\special{em:lineto}}}
\def\newpic#1{}

\catcode`\@=11
\def\marginnote#1{}

\newcount\hour
\newcount\minute
\newtoks\amorpm
\hour=\time\divide\hour by60
\minute=\time{\multiply\hour by60 \global\advance\minute by-\hour}
\edef\standardtime{{\ifnum\hour<12 \global\amorpm={am}%
        \else\global\amorpm={pm}\advance\hour by-12 \fi
        \ifnum\hour=0 \hour=12 \fi
        \number\hour:\ifnum\minute<10 0\fi\number\minute\the\amorpm}}
\edef\militarytime{\number\hour:\ifnum\minute<10 0\fi\number\minute}

\def\draftlabel#1{{\@bsphack\if@filesw {\let\thepage\relax
   \xdef\@gtempa{\write\@auxout{\string
      \newlabel{#1}{{\@currentlabel}{\thepage}}}}}\@gtempa
   \if@nobreak \ifvmode\nobreak\fi\fi\fi\@esphack}
        \gdef\@eqnlabel{#1}}
\def\@eqnlabel{}
\def\@vacuum{}
\def\draftmarginnote#1{\marginpar{\raggedright\scriptsize\tt#1}}
\def\half{\frac{1}{2}}
\def\draft{\oddsidemargin -.5truein
        \def\@oddfoot{\sl preliminary draft \hfil
        \rm\thepage\hfil\sl\today\quad\militarytime}
        \let\@evenfoot\@oddfoot \overfullrule 3pt
        \let\label=\draftlabel
        \let\marginnote=\draftmarginnote
   \def\@eqnnum{(\theequation)\rlap{\kern\marginparsep\tt\@eqnlabel}%
\global\let\@eqnlabel\@vacuum}  }
\begingroup\ifx\undefined\newsymbol \else\def\input#1 {\endgroup}\fi
\input amssym.def \relax
\input amssym
\newfont{\hr}{msbm10}
\newfont{\ams}{msam10}
\def\W{Weierstrass\ }
\def\RS{Ruijsenaars-Schneider\ }
\def\XXX{{\bf XXX\ }}
\def\XXZ{{\bf XXZ\ }}
\def\XYZ{{\bf XYZ\ }}

\def \a {\alpha}
\def \b {\beta}

\def \tr{{\rm tr}}

\def \det{{\rm det}}

\def \log {{\rm log}}

\def \sn{{\rm sn}}
\def \cn{{\rm cn}}
\def \dn{{\rm dn}}

\def\tilde{\widetilde}
\def\bar{\overline}
\def\hat{\widehat}
\def\*{\star}
\def\({\left(}      
\def\){\right)}     
\def\[{\left[}      
\def\]{\right]}

\def\frac#1#2{{#1 \over #2}}

\def\2pi{\hbox{$2\pi i$}}

\def\dsl{\raise.15ex\hbox{/}\kern-.57em\partial}
\def\Dsl{\,\raise.15ex\hbox{/}\mkern-.13.5mu D}
\def\la{\lambda}    
     \def\De{\Delta}
     
\def\sig{\sigma}    

   \def\CE{{\cal E}}   \def\CF{{\cal F}}
      
      \def\CL{{\cal L}}
\def\CM{{\cal M}}   \def\CN{{\cal N}}   
\def\CP{{\cal P}}   \def\CQ{{\cal Q}}   
   \def\CT{{\cal T}}   
   \def\CW{{\cal W}}   
   
\font\numbers=cmss12
\font\upright=cmu10 scaled\magstep1
\def\stroke{\vrule height8pt width0.4pt depth-0.1pt}
\def\topfleck{\vrule height8pt width0.5pt depth-5.9pt}
\def\botfleck{\vrule height2pt width0.5pt depth0.1pt}
\def\Zmath{\vcenter{\hbox{\numbers\rlap{\rlap{Z}\kern 0.8pt\topfleck}\kern
2.2pt
                   \rlap Z\kern 6pt\botfleck\kern 1pt}}}
\def\Qmath{\vcenter{\hbox{\upright\rlap{\rlap{Q}\kern
                   3.8pt\stroke}\phantom{Q}}}}
\def\Nmath{\vcenter{\hbox{\upright\rlap{I}\kern 1.7pt N}}}
\def\Cmath{\vcenter{\hbox{\upright\rlap{\rlap{C}\kern
                   3.8pt\stroke}\phantom{C}}}}
\def\Rmath{\vcenter{\hbox{\upright\rlap{I}\kern 1.7pt R}}}
\def\Z{\ifmmode\Zmath\else$\Zmath$\fi}
\def\Q{\ifmmode\Qmath\else$\Qmath$\fi}
\def\N{\ifmmode\Nmath\else$\Nmath$\fi}
\def\C{\ifmmode\Cmath\else$\Cmath$\fi}
\def\R{\ifmmode\Rmath\else$\Rmath$\fi}
\newcounter{app}

\def\app{\setcounter{equation}{0}
\def\theequation{\Alph{app}.\arabic{equation}}\par
   \addvspace{4ex}
   \@afterindentfalse
  \secdef\@app\@dapp}

\newcommand\@app{\@startsection {app}{1}{0ex}%
                                   {-3.5ex \@plus -1ex \@minus -.2ex}%
                                   {2.3ex \@plus.2ex}%
                                   {\normalfont\Large\bf}}
\def\@dapp#1{%
{\parindent \z@ \raggedright  \bf #1}\par\nobreak}
\def\l@app#1#2{\ifnum \c@tocdepth >\z@
    \addpenalty\@secpenalty
    \addvspace{1.0em \@plus\p@}%
    \setlength\@tempdima{8em}%
    \begingroup
      \parindent \z@ \rightskip \@pnumwidth
      \parfillskip -\@pnumwidth
      \leavevmode \bfseries
      \advance\leftskip\@tempdima
      \hskip -\leftskip
      #1\nobreak\hfil \nobreak\hb@xt@\@pnumwidth{\hss #2}\par
    \endgroup\fi}
\def\stackreb#1#2{\mathrel{\mathop{#2}\limits_{#1}}}
\def\Tr{{\rm Tr}}
\def\res{{\rm res}}
\def\Bf#1{\mbox{\boldmath $#1$}}
\def\bP{{\Bf P}}
\def\balpha{{\Bf\alpha}}

\def\bphi{{\Bf\phi}}

\def\Im{{\rm Im}}

\def\2{{1\over 2}}
\def\N2{${\cal N}=2$}
\def\n=1{$\CN=1$}
\def\pf{{\rm Pf}}

\def\d{\partial}
\def\pdv{\partial_{\varphi}}
\def\intt {\int_{S^{1}}}
\def\ddd {\cal D}

\makeatletter
\newdimen\normalarrayskip              % skip between lines
\newdimen\minarrayskip                 % minimal skip between lines
\normalarrayskip\baselineskip
\minarrayskip\jot
\newif\ifold             \oldtrue            \def\new{\oldfalse}
\def\arraymode{\ifold\relax\else\displaystyle\fi} % mode of array entries
\def\eqnumphantom{\phantom{(\theequation)}}     % right phantom in eqnarray
\def\@arrayskip{\ifold\baselineskip\z@\lineskip\z@
     \else
     \baselineskip\minarrayskip\lineskip2\minarrayskip\fi}
\def\@arrayclassz{\ifcase \@lastchclass \@acolampacol \or
\@ampacol \or \or \or \@addamp \or
   \@acolampacol \or \@firstampfalse \@acol \fi
\edef\@preamble{\@preamble
  \ifcase \@chnum
     \hfil$\relax\arraymode\@sharp$\hfil
     \or $\relax\arraymode\@sharp$\hfil
     \or \hfil$\relax\arraymode\@sharp$\fi}}
\def\@array[#1]#2{\setbox\@arstrutbox=\hbox{\vrule
     height\arraystretch \ht\strutbox
     depth\arraystretch \dp\strutbox
     width\z@}\@mkpream{#2}\edef\@preamble{\halign
\noexpand\@halignto
\bgroup \tabskip\z@ \@arstrut \@preamble \tabskip\z@ \cr}%
\let\@startpbox\@@startpbox \let\@endpbox\@@endpbox
  \if #1t\vtop \else \if#1b\vbox \else \vcenter \fi\fi
  \bgroup \let\par\relax
  \let\@sharp##\let\protect\relax
  \@arrayskip\@preamble}
\def\eqnarray{\stepcounter{equation}%
              \let\@currentlabel=\theequation
              \global\@eqnswtrue
              \global\@eqcnt\z@
              \tabskip\@centering
              \let\\=\@eqncr
              $$%
 \halign to \displaywidth\bgroup
    \eqnumphantom\@eqnsel\hskip\@centering
    $\displaystyle \tabskip\z@ {##}$%
    \global\@eqcnt\@ne \hskip 2\arraycolsep
         $\displaystyle\arraymode{##}$\hfil
    \global\@eqcnt\tw@ \hskip 2\arraycolsep
         $\displaystyle\tabskip\z@{##}$\hfil
         \tabskip\@centering
    &{##}\tabskip\z@\cr}
\def\theequation{\thesection.\arabic{equation}}

\def\IB{\relax\hbox{$\inbar\kern-.3em{\rm B}$}}
\def\IC{\relax\hbox{$\inbar\kern-.3em{\rm C}$}}
\def\ID{\relax\hbox{$\inbar\kern-.3em{\rm D}$}}
\def\IE{\relax\hbox{$\inbar\kern-.3em{\rm E}$}}
\def\IF{\relax\hbox{$\inbar\kern-.3em{\rm F}$}}
\def\IG{\relax\hbox{$\inbar\kern-.3em{\rm G}$}}
\def\IGa{\relax\hbox{${\rm I}\kern-.18em\Gamma$}}
\def\IH{\relax{\rm I\kern-.18em H}}
\def\IK{\relax{\rm I\kern-.18em K}}
\def\IL{\relax{\rm I\kern-.18em L}}
\def\IP{\relax{\rm I\kern-.18em P}}
\def\IR{\relax{\rm I\kern-.18em R}}
\def\IZ{\relax\ifmmode\mathchoice
{\hbox{\cmss Z\kern-.4em Z}}{\hbox{\cmss Z\kern-.4em Z}}
{\lower.9pt\hbox{\cmsss Z\kern-.4em Z}}
{\lower1.2pt\hbox{\cmsss Z\kern-.4em Z}}\else{\cmss Z\kern-.4em Z}\fi}
\def\ocom{\relax{\raise1.2pt\hbox{$\otimes$}\kern-.5em\lower.1pt\hbox{,}}\,}
\def\Id{{\rm Id}}

\font\manual=manfnt \def\dbend{\lower3.5pt\hbox{\manual\char127}}

\def\half {{1\over 2}}

\def\inbar{\,\vrule height1.5ex width.4pt depth0pt}
\def\bea{\begin{eqnarray}}
\def\eea{\end{eqnarray}}
\def\nn{\nonumber}
\def\beq{\begin{equation}}
\def\eeq{\end{equation}}
\def\ba{\beq\new\begin{array}{c}}
\def\ea{\end{array}\eeq}
\def\be{\ba}
\def\ee{\ea}
\def\stackreb#1#2{\mathrel{\mathop{#2}\limits_{#1}}}
\def\f{1\over}
\def\SW{Seiberg-Witten theory\ }

\def\Q{\bar{Q}}

\def\LL{{\cal L}}
\def\ggg{\sf g}
\def\curve{\Sigma_{\tau}}
\def\intc{\int_{\Sigma_{\tau}}}
\def\haha{\theta_{\ast}}

\def\elg{{\hat {\sf g}}^{\Sigma_{\tau}}}

\def\ppz{\phi (z, {\bar z})}
\def\aaz{{\bar A} (z, {\bar z})}
\def\ggz{g (z, {\bar z}) }
\def\slnc{{\it sl}_{N} ({\bf C})}

\def\rw{\rightarrow}

\begin{document}
\begin{titlepage}
\setcounter{footnote}0
\def\thefootnote{\fnsymbol{footnote}}
\begin{center}
\hfill FIAN/TD-30/00\\
\hfill ITEP/TH-64/00\\
\hfill hep-th/0011197\\
\vspace{0.4in}
{\LARGE\bf Integrable Many-Body Systems and Gauge Theories}\\
\bigskip
\bigskip
\bigskip
{\Large A.Gorsky
\footnote{E-mail: gorsky@heron.itep.ru}}
\\
\bigskip
{ITEP, Moscow 117259, Russia}\\
\bigskip
and\\
\bigskip
{\Large A.Mironov
\footnote{E-mail:
mironov@lpi.ru, mironov@itep.ru}}\\
\bigskip
{Theory Department, Lebedev Physics
Institute, Moscow ~117924, Russia\\
and ITEP, Moscow 117259, Russia}
\end{center}
\bigskip \bigskip

\begin{abstract}
The review studies connections
between integrable many-body systems and gauge theories.
It is shown how the degrees of freedom in integrable systems
are related with topological degrees of freedom in
gauge theories. The relations between families of integrable systems
and \N2 supersymmetric gauge theories are described. It
is explained that the degrees of freedom in the many-body systems
can be identified with collective coordinates of D-branes,
solitons in string theory.
\end{abstract}

\end{titlepage}

\newpage
\tableofcontents
\newpage

\section{Introduction}
\setcounter{equation}{0}
\setcounter{footnote}{0}
\def\thefootnote{\arabic{footnote}}

This review is devoted to the issue of
hidden integrability in quantum field theory,
i.e. integrability in the systems that naively have not
enough many conserved integrals of motion. This situation
differs considerably from the picture familiar for the
experts in the theory of integrable systems. In the typical
integrable system,
one deals with a well-defined (field theory) Lagrangian,
with the fields given on a base space-time manifold. The
fields at some fixed moment of time, i.e. given on a space
slice of the base manifold, provide the dynamical degrees of freedom,
while integrable flows are
identified with the time direction in the base manifold.

On the contrary, in the systems we are interested in,
the phase space for the
integrable dynamics is provided by a moduli space essential
in the problem. The collective coordinates or moduli arise in the course of
consideration of various issues and have essentially
different origins. In particular, they can describe effective degrees of
freedom important in one or another regime of the field theory
not completely integrable by itself. The proper regime can be determined by
the kinematics of the process or by the choice of specific
topologically nontrivial data. In the cases when there are
several collective degrees of freedom, a hidden
integrability in the theory emerges as a rule. Its origin lies in
symmetry properties of the moduli space.

As a particular example, it appears that a hidden integrability
governs the low-energy effective actions of \N2 supersymmetric gauge theories,
which in four dimensions
were described by Seiberg and Witten \cite{SW1,SW2}.

The role of this integrability is twofold; it allows one to
uncover the hidden symmetries and to define proper degrees of freedom.
In what follows, we shall show how the low-energy sector of
supersymmetric gauge theories is described via finite dimensional
integrable systems.

Restricted frames of the review does not allow us to provide the reader
with all necessary information from the theory of integrable systems,
supersymmetric gauge theories and nonperturbative string theory
which might be relevant to the issue.
This kind of problem arises since the integrability appears at the
intersection of different branches of theoretical physics which
use different tools and approaches. Therefore,
we just recommend several (rather accidental) review papers, where the issues
used in what follows as well as the proper list of references can be found.
Other references can be found in the proper parts of the present review.

The finite gap integration is perfectly presented in \cite{konzon},
Whitham hierarchies are discussed in  \cite{KriW1,whitham}, while
background of the Seiberg-Witten approach is described in \cite{bilal}.
Nonperturbative string theory is reviewed in \cite{giveonkut}.
Necessary
background from algebraic geometry used in the brane approach
is described in  \cite{klemm},
while a generic point of view on integrability
in (nonperturbative) field/string theory
can be found in  \cite{morozov}. An intensive
discussion of integrability in \SW and related structures (Whitham hierarchies
and WDVV equations) can be found in the book \cite{BK} that contains
recent comprehensive reviews.
Note also that the very detailed review \cite{DPhongrev} may serve a good
complimentary text to the present paper.

\subsection{Moduli and nonperturbative degrees of freedom}

Description of the strong coupling regime remains one of the most
important problems of quantum field theory. Since the perturbation theory
does not work in this region, one can mostly reply on constructing
some effective degrees of freedom which would allow one to develop a new
perturbation theory with a small parameter distinct from the
coupling constant of the initial theory. A nice example
of the existence of different descriptions at different energy scales
is quantum chromodynamics (QCD). At high energies,
quarks and gluons giving the proper degrees of freedom are described by the
Yang-Mills Lagrangian coupled to fermions in the fundamental representation,
while at low energies it is a chiral ($\sigma$-model) Lagrangian that
describes the low-energy effective action, with the proper degrees
of freedom given by the
colorless mesons and baryons. Note that
the parameters of the chiral Lagrangian can not be calculated
from the first
principles (at the moment) and are fixed phenomenologically.

The general structure of low-energy actions is fixed by the symmetries.
It is expected that the effective actions enjoy a complete set
of the Ward identities of the initial field theory. Some of the Ward
identities may be anomalous. However, since the anomalies are
related to the level crossing phenomena, not renormalized
and, therefore,  can be equally well described in the UV and IR
terms \cite{shifmanan},
the anomalous Ward identities could still impose restrictions
on the structure of effective theories. For instance, the chiral anomalous
Ward identities fix the Lagrangian of mesons, while the
conformal Ward identities
yield the effective Lagrangian for the dilaton in ordinary as well
as in $\CN=1$ supersymmetric QCD \cite{shifmanmig}.

Another phenomenon that has much to do with the issues discussed is
universality. Effective actions typically enjoy this feature which
implies that several theories different in the UV regime can flow to the same
theory in the infrared region. The reason for such an universality to occur
is the rigidity of effective actions due to the symmetries.
Moreover, the symmetry in the IR regime
may be even higher then that in the initial UV theory.

The symmetry
nature of effective actions results in nontrivial
consequences. Presence of an underlying integrable systems
seems to be the most important of all. As a rule, the partition
function calculated in the effective theory appears to be the
$\tau$-function of some integrable system. On the other hand, the
$\tau$-function is a generating
function for the conservation laws in integrable system and
simultaneously the generating function of
correlators in some quantum field theory \cite{morozov,mironov}.

Identification of the variables in the integrable system
describing the effective action is a difficult problem. An invariant
way to introduce these variables in non-topological theories
is still missing. However, there are several examples of two-dimensional
theories when these variables
can be identified with the transition amplitudes between vacuum states
induced by nonperturbative configurations. The
couplings and sources serve as the ``space-time variables" in
integrable systems. Examples of the systems related to the $2d$ Toda
lattice and similar hierarchies (KP, KdV etc) can be borrowed from $2d$
$\sigma$-models \cite{vafacec} and
$2d$ gravity \cite{douglasstr,kawai,GMMMO,GKM,versus,morozov}.
Selecting the concrete solution to the nonlinear integrable
equations is a separate problem.
It is typically done by an additional
equation, string equation
 \cite{douglasstr}\footnote{The string equation does not
fix the possible discrete ambiguity of solutions \cite{David}.}
which is one of
the Ward identities imposed on the theory \cite{kawai,MM}.

Although one can fix the general structure of effective action
from the symmetry arguments, it is also interesting to evaluate it by the
direct
integration over heavy modes in the initial field theory. It is implied
that there is some energy scale and all modes above this scale are integrated
out. This integration should be done over the perturbative
configurations and the scale serves a cutoff in the
corresponding Feynman loop diagrams. It is also necessary to sum over the
nonperturbative configurations,
their typical sizes being restricted by this scale. The
nonperturbative configurations important in different dimensions are:
instantons, monopoles, vortexes as well as solitons
of different codimension in string theory.

The important feature of nonperturbative effects is that they
give rise to the moduli spaces of the corresponding solutions. The very
reason for the moduli space to emerge is the high symmetry of
the theory. Therefore, the moduli space itself
encodes symmetry properties. For instance, the
dimension of the instanton moduli space is fixed by the
space-time symmetries -- translations and rotations as well as by the gauge
symmetries \cite{adhm,douglasinst}. There are many other important
moduli spaces, for instance, the moduli spaces: of the complex structures
of Riemann
surfaces, of the flat connections on a fixed surface, of the
holomorphic vector bundles etc.
Each of these appears when considering some
nonperturbative configurations and can be derived via the reduction
procedure from the instanton moduli space. Thus, the universal instanton
moduli space comes as a most general object arising in field theories
in four or less dimensions.

As we explained,
the integration over moduli spaces involved into the derivation
of effective actions should ultimately
give rise to integrable theories. The same
kind of moduli spaces have much to do with integrable systems
also from another point of view. Namely,
the phase spaces of integrable systems
typically coincide with moduli spaces or cotangent bundles to the moduli
spaces of the same type.
Simple examples are given by solutions of the KdV hierarchy
related to the moduli space
of the complex structures or by solutions of the $2d$ Toda lattice hierarchy
related to the moduli of the flat
connections. Another important example is given by the Chern-Simons theory
defined on a surface, whose phase space is the
moduli space of the flat connections on this surface \cite{wittenkon},
 while any gauge invariant observable can serve as
a Hamiltonian.  In terms of integrable systems, the problem of calculation
of the nonperturbative contributions into effective actions is nothing
but the calculations of some vacuum expectation values of observables on
moduli spaces (see \cite{wittenkon} as an example).

A large class of theories related with the  moduli spaces is known
as topological theories. They were introduced into quantum field theory
by Witten \cite{wittentop} to obtain invariants of the vacuum manifolds in
field theories. The first examples of topological theories have been
elaborated in four dimensions for $\sigma$-models and gauge theories. In the
$\sigma$-model case, the topological theory provides invariants of the
holomorphic maps,
while in the gauge theory it gives invariants of the instanton
moduli space \cite{donaldson}. At the moment, there are a lot of examples of
topological theories in different dimensions, e.g., $\sigma$-models
in $d=2$, Chern-Simons theory in $d=3$, topological Lagrangians
 in $d=4$ etc.

In all the cases, the topological theories can be obtained from the
nontopological
 \N2 SUSY theories via the twisting procedure \cite{Li}. The partition
function of the topological theories appears to be some invariant of the
corresponding
moduli space.

There are also some examples of field theories
with the topological subsectors related solely with the
vacuum states. An example is given by
the ${\cal N}=1$ supersymmetric gauge theory,
where the perturbative $\beta$ function \cite{shifmanbeta} as well as
some correlators can be calculated exactly \cite{shifmantop}. At last,
let us mention that the associativity (WDVV) equations which typically
 emerge in topological theories
also appear in \N2 SUSY theories. This implies some hidden topological
structures (subsectors) in these theories as well \cite{mmm}.

Since the moduli spaces arising in different applications
are typically finite dimensional, the topological theories
that provide finite dimensional moduli spaces are of the prime
interest. It is these moduli spaces where the dynamics of
integrable many-body systems with nontrivial conservation laws
is developing. Gauge invariant observables with appropriate couplings
serve the Hamiltonians.

Note that the partition function for the perturbed
topological theory, i.e. the generating function for the correlators
is the most interesting and informative object.
It is simultaneously the generating function
for topological invariants of the moduli space manifold \cite{wittentop}
and the $\tau$-function of an integrable hierarchy \cite{GKM,kon}.

\subsection{Seiberg-Witten solution}

New possibilities to deal with nonperturbative physics at
strong coupling come from the exact derivation of the low energy effective
action for the \N2 SYM theory \cite{SW1,SW2}. In principle, to get the exact
answer one needs to perform the summation of
the infinite instanton series. Therefore, the result \cite{SW1,SW2}
due to N.Seiberg and E.Witten was the first example of the exact derivation of
the
total instanton sum.

However, the authors did not calculate instantonic series,
using
instead indirect and quite sophisticated methods.
The calculation of the exact mass spectrum of
stable particles was their second essentially new result.
What is important, this theory is non-trivial in one loop and
belongs to the class of asymptotically free theories.
In fact, the situation is a little bit unusual, since, despite the fact that
the direct instanton summation has not been performed yet,
there are no doubts in the results derived which
survives under different tests.

In order to get
the effective action unambiguously, the authors of \cite{SW1,SW2} used the
three ideas: holomorphy \cite{shifmanvai},
duality \cite{montonen} and their compatibility with the renormalization group
flows.
Holomorphy implies that the low energy effective
action can be only of the form $\int{\Im \cal{F}}(\Psi)$ depending
on a single
holomorphic function $\cal{F}$ called prepotential. The duality
principle which provides the relation between the strong and week coupling
regimes
is natural in the UV finite\footnote{In fact, one should require more:
not renormalized and, therefore, non-running coupling constants.},
say ${\cal N}=4$ SUSY \cite{wittenolive}, theory, where the evident
modular parameter built from the coupling constant and the $\theta$-term

\be
\tau=\frac{4\pi i}{g^{2}}+{\theta\over 2\pi},
\ee
is not renormalized  and coincides with its asymptotic value.
To construct duality transformation in asymptotically free theory explicitly,
one needs an artificial trick, to introduce into the game
an additional object, a higher genus Riemann surface
whose period matrix coincides with the matrix of couplings
depending on the values of condensate in a given vacuum.

In the \N2 SYM theory this scheme is roughly
realized as follows \cite{SW1,SW2}.
First of all, in the Lagrangian there is a
potential term for the scalar fields of the form

\be
V(\phi)=\Tr[\phi,\phi^{+}]^{2},
\ee
where the trace is taken over adjoint representation of the gauge group
(say, $SU(N_c)$). This potential gives rise to the valleys in the theory, when
$[\phi,\phi^+]=0$.
The vacuum energy vanishes along the valleys, hence,
the supersymmetry remains unbroken. One may always choose the v.e.v.'s of the
scalar field to
lie in the Cartan subalgebra $\phi=diag(a_{1},...,a_{n})$. These parameters
$a_i$ can not serve, however, good order parameters, since there is still
a residual Weyl symmetry which changes $a_i$ but leave the same vacuum state.
Hence, one should consider the set of the gauge invariant
order parameters $u_{k}=<\Tr\phi^{k}>$ that fix the vacuum state unambiguously.
Thus, we obtain a moduli space parametrized
by the vacuum expectation values of the scalars, which is known
as the Coulomb branch of the whole moduli space of the theory.
The choice of the point on the Coulomb branch is
equivalent to the  choice of the vacuum state and simultaneously
yields the scale which the coupling is frozen on. At the generic point of
the moduli space, the theory becomes effectively Abelian after the
condensation of the scalar.

As soon as the scalar field acquires the vacuum expectation value, the
standard
Higgs mechanism works and there emerge heavy gauge bosons at large values
of the vacuum condensate. To derive the effective low energy action one has
to sum up the loop corrections as well as the multi-instanton
contributions. The loop corrections are trivial, since they are vanishing
beyond the one loop due to the supersymmetry.
The explicit summation over instantons has not been done yet,
however, additional arguments allow one to define the effective action
indirectly.

The initial action of \N2 theory written in \N2 superfields has the simple
structure $S(\Psi )=\Im\tau \int \Tr\Psi^{2}$ with
$\tau=\frac{4\pi i}{g^{2}} + {\theta\over 2\pi}$. Now the \N2 supersymmetry
implies
that the low energy effective action gets renormalized only by holomorphic
contributions so that it is ultimately given by a single function
known as prepotential $S_{eff}(\Psi )=\Im\int{\cal{F}}(\Psi)$. The
prepotential
is a holomorphic function of moduli $u_k$ (or v.e.v.'s)
except for possible singular points at the values of moduli where additional
massless states can appear disturbing the low energy behaviour.

Thus, the problem effectively reduces to the determination of one
holomorphic function. If one manage to fix its behaviour nearby singularities,
the function can be unambiguously restored. One of the singularities,
corresponding
to large values of v.e.v.'s, i.e. to the perturbative limit is under control
(since the theory is asymptotically free). All other singularities are treated
with the use of duality and of the non-renormalization theorems
for the central charges of the SUSY algebra. A combination of these two ideas
allows one to predict the spectrum of the stable BPS states which become
massless
in the deep nonperturbative region and are in charge of all other
singularities.

The duality transformation can be easily defined
in the finite ${\cal N}=4$ SUSY theory just as the modular transformations
generated by $\tau \to \tau^{-1}$ and $\tau \to \tau +1$.
This makes a strong hint that
the duality can be related with a modular space of some Riemann surfaces,
where
the modular group acts.
In fact, the naive application of duality meets some serious difficulties.
The reason is that, in the asymptotically free
theory, one has to match the duality with the renormalization
group\footnote{In
fact,
there are more problems with duality even in \N2 SUSY theories. In particular,
it unifies into one duality multiplet the monopole and gauge boson
supermultiplets,
while they have different spins, see, e.g., \cite{SW1,bilal}.}. This is
non-trivial,
since now
$\tau$ depends on the scale which is supposedly involved into the duality
transformation. The solution of this problem is that one is still able
to connect the duality and
modular transformations if considering the theory at different vacua
connected by the duality. Then, the duality acts on the moduli space
of vacua and this moduli space is associated with the moduli space of
the auxiliary Riemann surface, where the modular transformations act.

At the next step, one has to find out proper variables whose modular
properties fit the field theory interpretation. These variables are
the integrals\footnote{
We define the symbols $\oint$ and $res$ with
additional factors $(2\pi i)^{-1}$ so that
$$
{\res}_0 \frac{d\xi}{\xi}
= -{\res}_\infty \frac{d\xi}{\xi} =
\oint \frac{d\xi}{\xi} = 1
$$
This explains the appearance of $2\pi i$ factors in the
Riemann identities and in perturbative prepotentials in sections 5 and 6.
Accordingly, the theta-functions are periodic
with period $2\pi i$, and
$$
\frac{\partial\theta(\vec\xi|{\cal T})}{\partial{\cal T}_{ij}}
= i\pi\partial^2_{ij}\theta(\vec\xi|{\cal T})
$$
since periods of
the Jacobi transformation $\xi_i \equiv \int^\xi d\omega_i$
belong to $2\pi i \left( \Z + {\cal T}\Z\right)$.} of a meromorphic 1-form $dS$
over the cycles on the Riemann surface,
$a_i$ and $a^D_i$

\be\label{aad}
a_{i}=\oint_{A_{i}}dS, \\
a_{D}=\oint_{B_{i}}dS,
\ee
(where $i,j=1,....,N_{c}-1$ for the gauge group $SU(N_{c})$).

These integrals play the two-fold role in the Seiberg-Witten approach.
First of all, one may calculate the prepotential ${\cal F}$ and, therefore,
the low
energy effective action through the identification of
$a_D$ and $\partial \CF/\partial a$ with
$a$ defined as a function of moduli (values of condensate) by formula
(\ref{aad}).
Then, using the property of the differential $dS$
that its variations w.r.t. moduli are holomorphic one may also calculate the
matrix of coupling constants

\be\label{Tij}
T_{ij}(u)=\frac{\partial^{2}{\cal{F}}}{\partial a_{i} \partial a_{j}},
\ee

The second role of formula (\ref{aad}) is that, as
was shown these integrals define the spectrum of the stable states
in the theory which saturate the Bogomolny-Prasad-Sommerfeld (BPS)
limit. For instance, the formula for the BPS spectrum in the $SU(2)$
theory reads as

\be\label{BPS}
M_{n,m}=\left|na(u) +ma_{D}(u)\right|,
\ee
where the quantum numbers $n,m$ correspond to the ``electric" and ``magnetic"
states. The reader should not be confused with the spectrum derived
from the low energy behaviour which fixes arbitrarily heavy
BPS states. The point is that the BPS spectrum is related to the central
charge
of the extended SUSY algebra $Z_{\{m\},\{n\}}=a_in^i+a^D_im^i$
\cite{3a} and, therefore, has an anomaly origin.
On the other hand, anomalies are not renormalized by the quantum
corrections and can be evaluated in either of the UV and IR regions.

Note that the column $(a_i,a^D_i)$ transforms under
the action of the modular group $SL(2,\Z)$ as a section of the linear bundle.
Its global behaviour, in particular, the structure of the singularities
is uniquely determined by the monodromy data. As we discussed earlier,
the duality transformation connects different singular points. In particular,
it interchanges ``electric", $a_i$ and ``magnetic", $a^D_i$
variables which describe the perturbative degrees of freedom at the
strong and weak coupling regimes (of the theory at different vacua)
respectively.
Manifest calculations with the Riemann surface allow one to analyze
the monodromy properties of dual variables when
moving in the space of the order parameters. For instance,
in the simplest $SU(2)$ case, on the $u$-plane of the single order parameter
there are three singular points, and the magnetic and
electric variables mix when encircling these points. Physically,
in the theory with non-vanishing $\theta$-term (not pure imaginary $\tau$),
the monopole acquires the electric
charge, while the polarization of the instanton medium yields the
induced dyons.

\subsection{\SW and theory of prepotential}

Although the both auxiliary objects, the Riemann surface and the differential
$dS$ have come artificially, an attempt to recognize them
in SUSY gauge theories results into discovery of the integrable
structures responsible for the Seiberg-Witten solutions.

Now let us briefly formulate the structures underlying Seiberg-Witten
theory. Later on, we often
refer to \SW as to the following set of data:
\begin{itemize}
\item
Riemann surface ${\cal C}$
\item
moduli space ${\cal M}$ (of the curves ${\cal C}$), the moduli space
of vacua of the gauge theory
\item
meromorphic 1-form $dS$ on ${\cal C}$
\end{itemize}
How it was pointed out in \cite{GKMMM},
this input can be naturally described
in the framework of some underlying integrable system.

To this end, first, we introduce bare spectral curve $E$ that is torus
$y^2=x^3+g_2x^2+g_3$ for the UV-finite\footnote{The situation is still
unclear in application to the case of fundamental matter with the number
of matter hypermultiplets $N_f$ equal to $2N_c$.
In existing formulation for spin chains the bare coupling constant
appears rather as a twist in gluing the ends of the chain together
\cite{GGM1} (this parameter occurs only when $N_f = 2N_c$) and is not
immediately identified as a modulus of a {\it bare} elliptic curve. This
problem is a fragment of a more general puzzle:  spin chains have not been
described as Hitchin systems; only the ``$2\times 2$'' Lax representation is
known for them, while its ``dual'' $N_c\times N_c$ one is not yet available.}
gauge theories with the associated holomorphic 1-form
$d\omega=dx/y$. This bare spectral curve degenerates into the
double-punctured sphere (annulus) for the asymptotically free theories
(where dimensional transmutation occurs): $x\to
w+1/w$, $y\to w-1/w$, $d\omega=dw/w$.
On this bare curve, there are given a
matrix-valued Lax operator $L(x,y)$. The corresponding dressed spectral
curve ${\cal C}$ is defined from the formula $\det(L-\lambda)=0$.

This spectral curve is a
ramified covering of $E$ given by the equation

\be\label{SWcurve}
{\cal P}(\lambda;x,y)=0
\ee
In the case of the gauge group  $G=SU(N_c)$, the function ${\cal P}$ is a
polynomial of degree $N_c$ in $\lambda$.

Thus, we have the spectral curve ${\cal C}$, the moduli space ${\cal M}$ of the
spectral curve being given just  by
coefficients of ${\cal P}$.
The third important ingredient of the construction is the
generating 1-form $dS \cong \lambda d\omega$ meromorphic on
${\cal C}$ (``$\cong$" denotes the equality modulo total derivatives).
{}From the point of view of
the integrable system, it is just the shortened action "$pdq$" along the
non-contractable contours on the Hamiltonian tori. This means that the
variables $a_i$ in (\ref{aad}) are nothing but the action variables in the
integrable system. The defining property of $dS$ is
that its derivatives with respect to the moduli (ramification points)
are holomorphic differentials on the spectral curve.  This, in particular,
means
that

\be
{\partial dS\over\partial a_i}=d\omega_i
\ee
where $d\omega_i$ are the canonical holomorphic differentials\footnote{I.e.
satisfying the conditions
$$
\oint_{A_i}d\omega_j=\delta_{ij},\ \ \ \ \oint_{B_i}d\omega_j=T_{ij}
$$
}.
Integrating this formula over $B$-cycles and using that
$a_D=\partial \CF/\partial a$, one immediately obtains (\ref{Tij}).

So far we reckoned without matter hypermultiplets.
In order to include them, one just needs to consider the
surface ${\cal C}$ with punctures. Then, the masses are proportional to the
residues of $dS$ at the punctures, and the moduli space has to be extended to
include these mass moduli. All other formulas remain in essence the same
(see \cite{WDVVlong} for more details).

Note that it is still unknown whether the set of data above is identically
equivalent to an
integrable system. Inversely, it is associated only with specific integrable
systems, which can be called analytic integrable systems. Although being
wide enough, this class of integrable systems far does not cover the systems
integrable in the Liouville sense.

The prepotential ${\cal F}$
and other ``physical" quantities are defined in terms of the
cohomology class of $dS$, formula (\ref{aad}). Note that
formula (\ref{Tij}) allows one to identify the prepotential with logarithm of
the
$\tau$-function of the Whitham hierarchy \cite{typeB}:
${\cal F}=\log{\CT}_{\rm Whitham}$.
The other important property of the prepotential related to
 the feature of the Seiberg-Witten fra\-me\-work
which merits the adjective "topological"  has
much to do with the associative algebras \cite{mmm}.
Namely, it turns out that the
prepotential of \SW satisfies a set of the
Witten-Dijkgraaf-Verlinde-Verlinde (WDVV) equations. These equations have
been originally presented in \cite{wdvv} (in a different form, see below)

\be\label{wdvv}
F_iF_j^{-1}F_k=F_kF_j^{-1}F_i
\ee
where $F_i$'s are matrices with the matrix elements that are the third
derivatives of a unique function $F$ of many variables $a_i$'s
(prepotential in the Seiberg-Witten theory) parameterizing a moduli space:

\be
\left(F_i\right)_{jk}\equiv {\partial^3 F\over \d a_i\d a_j\d a_k},
\ \ \ \ i,j,k=1,...,n
\ee
Although generally there are a lot of solutions to the matrix equations
(\ref{wdvv}), it is extremely non-trivial task to express all the matrix
elements through the only function $F$. In fact, there have been only known
the two different classes of non-trivial solutions to the WDVV equations,
both being intimately related to the two-dimensional topological theories of
type A (quantum cohomologies \cite{typeA}) and of type B (\N2 SUSY
Ginzburg-Landau theories that were investigated in a variety of papers,
see, for example, \cite{typeB} and references therein). Thus, the existence
of a new class of solutions connected with the four-dimensional (and
higher dimensional) theories looks quite striking.

\subsection{Duality, branes and gauge theories}

The progress in the description of the low energy SUSY gauge theories
initiated the investigation of duality in string theory. It appeared
that, in order to formulate the duality transformation in string
theory, one has to
introduce new objects called D-branes \cite{polchinski1}.
Their main property is that the open strings can end on the D-brane.
Therefore, the Dirichlet boundary condition
can be imposed.  Moreover, the charge in the RR sector of the theory is
attributed to the D-brane. Using D-branes, a stringy duality can be
self-consistently described,
and actually three different types of dualities can be
introduced \cite{duality1}.

Since the superstring theory is anomaly free only in ten dimensions,
the question about the role of remaining dimensions naturally arises.
The results of the last years definitely indicate that these
dimensions should be identified with the zero modes of the
scalar fields \cite{witbran} presented in $4d$ field theories.  In the most
symmetric situation in $d=4$, which is the ${\cal N}=4$ SUSY gauge
theory there are 3 complex scalars whose vacuum expectation values
should be considered as the six compactified dimensions.
Generically the structure of the field space is very complicated
and its metric is the subject of numerous investigations.

{}From the field theory point of view, ${\cal N}=4$ is maximal admissible
supersymmetry, since it is the case when
the gravitational degrees of freedom are still decoupled \cite{5a} from
the gauge sector of the theory, the case of our prime interest.

Now, in order to analyze a generic $4d$ gauge theory, one has to introduce
an arbitrary gauge group, to ask for the proper SUSY and to define all
the parameters in the theory such as masses and couplings. It appears
that the account of the brane degrees of freedom allows one
to fulfill this program at full. The important
properties of branes which allow one to obtain these results
are the presence of an $U(1)$ gauge field on their worldvolumes
as well the existence of brane bound states \cite{witbran}.

Let us briefly describe how the $4d$ field theory can be formulated
as a theory on the
brane world volume of some dimension. At the first step, we start with
the superstring theory
in ten dimensions of type II, and branes are the BPS states
breaking half of the supersymmetries (this gives us ${\cal N}=4$ SUSY in $4d$).
In one approach, the initial theory is
considered in the flat space
and SUSY is broken by the background brane configuration \cite{HW}. In the
other approach, known as ``geometrical engineering" \cite{eng} the
supersymmetry is broken by a special choice of metric of
the six dimensional compactification manifold. For instance, the metric of
the $K3$ manifold gives rise to breaking of the half of supersymmetries.

As a corollary of definition, the strings ending on the D-brane induce an
$U(1)$ gauge field living on its worldvolume.
At the next step, one should produce the general
gauge group, say $SU(N_c)$, which is absent in the original string theory.
Once again two approaches are possible; in the first approach, the
gauge group arises from the bound state of branes \cite{witbran},
while in the second approach it originates from choice of the compactification
manifold with the singularity structure related to the gauge group
root system \cite{OV}. The higher rank gauge group obtained
from the brane bound state can be explained as follows. Each of the branes
carries the $U(1)$ index while the open strings can stretch between them.
Therefore, having the stack of parallel $N_c$ branes, each string can be
associated with
an $U(N_c)$ matrix realizing the
$U(N_c)$ gauge theory\footnote{It can be shown that
the total $U(1)$ factor decouples.}. If one now splits the branes,
there are massive
modes of the gauge fields in the spectrum of open strings which correspond to
the strings stretched between different branes
and the corresponding masses are proportional to the
distance between branes in the embedding space. This is a realization of
the Higgs mechanism.
When the distance between branes vanishes, rank of the corresponding
``unbroken" gauge group increases. In the ``geometrical engineering"
approach the gauge
group comes from the wrapping of branes over the manifold with the
corresponding singularity structure \cite{vafan}.

Note that the scalars in field theory play the role of coordinates of the brane
or
the bound state of branes in the $6d$ space.

The next step of this rough (rather pictorial) construction
 of field theory is to introduce massive hypermultiplets and
couplings. To this end, one should further break the
supersymmetry\footnote{Note
 that, due to the conformal invariance, the ${\cal N}=4$ SUSY theory depends on
the coupling constant trivially and can not contain massive hypermultiplets.}.
It can be done by adding more, background branes, of different codimensions
and non-parallel to the stack of $N_c$ branes.

Say, one can introduce hypermultiplets by adding branes of
other codimensions so that the coordinates of these branes in $6d$
compactification
manifold correspond to the values of their masses \cite{HW}. Couplings are
introduced
similarly and their values are identified with the distance between the
background branes in one picture or by the size of the compact manifold in
the ``geometrical engineering" approach \cite{eng}.

\subsection{Seiberg-Witten solution and branes}

Realization of the Seiberg-Witten solution of the \N2 SUSY gauge theory
in brane terms seems to be the most impressive realization of the
approach described.
In the type IIA framework, the theory is solved along the following line
\cite{W}. The $SU(N_c)$ gauge theory arises on the worldvolume of $N_c$
parallel
D4 branes embedded into the flat $10d$ manifold. The worldvolume of the
branes is five dimensional and stretches
along, say, $x_0,x_1,x_2,x_3,x_6$-directions,
the branes being of finite size in the $x_4$-direction, with the first
four coordinates identified with the coordinates in our (infinite) $4d$
world.

Now one has to describe the compact six dimensions in field theory terms.
To this end,
one first assumes that all the branes are placed at the origin
with respect to three
coordinates\footnote{This is necessary to preserve, at least, \N2 SUSY.},
say, $x_7,x_8,x_9$. Now, the two remaining directions not stretching along
the brane, $x_4$ and $x_5$ form
the holomorphic variable, $x_{4}+ix_{5}$ which is associated with
the vacuum expectation value of the scalar in
the \N2 SUSY gauge theory, that is
to say, the coordinate on the Coulomb branch of the moduli space.
At last, the compact coordinate of brane, $x_6$ is just the coordinate
on (real section of) the auxiliary Riemann surface.

Since the D4 branes are restricted along the $x_6$-direction, one needs
some other, background branes where the D4 branes should end on. It is to
be two additional branes of the NS type.
It is just this configuration which breaks the ${\cal N}=8$ SUSY in ten
dimensions (32 generators) down to the
\N2 SUSY in four dimensions (8 generators).
The distance between the background branes along the $x_6$-direction
is identified with the coupling constant in the gauge theory
$\delta x_{6}=\frac{1}{g^{2}}$,  while the back reaction to the background
brane can be attributed to the perturbative renormalization of the coupling.
In order to incorporate the fundamental matter hypermultiplets
into the theory, one has
to introduce additional $N_f$ D6 branes whose coordinates\footnote{More
precisely,
the distance to the D4 branes.} along
$x_{4}+ix_{5}$ are related with masses of the hypermultiplets.
On the other hand, in order to add the adjoint matter,
the (complexified) coordinate $x_6$  should be made periodic.

The configuration considered is singular, since it is impossible to
describe the intersection of the D4 and background NS5 branes smoothly.
To resolve the singularities, Witten has suggested to lift the
configuration into $d=11$
manifold which plays the same role for membrane theory as the $d=10$ space
for superstrings. The theory defined in $d=11$ is known as M-theory and it
has the $d=11$ supergravity as its low energy
limit \cite{townsend1}.
It is important that there are solitons in M-theory, that is, M5 branes and
membranes (M2 branes). In the non-singular M-theory picture of \N2 SUSY
gauge theories, the eleventh dimension is assumed to be compact and there is
a single M5 brane wrapped around it. This M5 brane is assumed to be
topologically
non-trivial and to have 2 compact and 4 infinite dimensions, these latter being
associated with our $4d$ world. With three coordinates ($x_7,x_8,x_9$) being
set
to zero, the two compact dimensions of the M5 brane are
embedded into the
four-dimensional space (of $x_4,x_5,x_6,x_{10}$ coordinates)
in a tricky way. This
embedding is, however, holomorphic and describes the auxiliary Riemann
surface. Being projected onto ten dimensions, different pieces of this
two-dimensional surface are associated with the D4 and NS5 branes.

Note now that a similar picture
has been also developed in the M-theory framework for ${\cal N}=1$ gauge
theories
\cite{N1} which manifest confinement due to the monopole
condensation. This gives some hope for relevance of the brane picture also
for the ordinary, non-supersymmetric QCD.

In the other IIB approach \cite{V1} related to the
previous consideration via the T duality transformation \cite{ab1}, the
embedding manifold is topologically non-trivial and branes are assumed to be
wrapped around some cycles on the surfaces, for instance, on tori
(elliptic
curves) embedded into K3 manifold \cite{bsv}. This approach is less
transparent, although the topological theories on the curves which
are associated with the
integrable many-body systems arise within this framework automatically.
Similarly to the way of resolving the singular IIA configuration by lifting it
into higher dimensional M-theory, the IIB  configuration has to be lifted to
the $d=12$ theory known as F-theory \cite{vafaF}.
The additional two dimensions are assumed to be compactified onto the torus
whose modular parameter is identified with
the (imaginary and real parts of)
UV values of the coupling constant and the $\theta$-term in the gauge
theory.

\section{Many-body systems and gauge theories in two and three
dimensions}
\setcounter{equation}{0}

In this section we demonstrate how the many-body systems
naturally arise in two and three dimensional gauge theories
and the phase space of the system of particles is interpreted
in terms of the topological gauge degrees of freedom. From what follows
it becomes clear that the $2d$ YM theory is closely related to the
Calogero-Moser system  \cite{calogero1,calogero3,olper},
while the Chern-Simons theory to its relativistic counterpart
\cite{rru2,rud}, the Ruijsenaars-Schneider system.

\subsection{Calogero many-body system and Yang-Mills theory
in two dimensions}

Let us show how the Lagrangian formulation of the Calogero systems
can be related to the YM theory on a cylinder
\cite{g8}.
Remind that the Calogero system is defined as an $N$
particle system with the pairwise
interaction  \cite{calogero1,calogero3}:

\beq
V(x_{1},\ldots,x_{N}) = \sum_{i\neq j}
\frac{\nu^{2}}{(x_{i}-x_{j})^{2}},
\eeq
$\nu^{2}$ is the coupling constant.
As we already mentioned, the phase spaces of integrable systems
are typically related with group manifolds. Therefore, it should not be
surprising that the Lagrangians under consideration have a group origin.

Let us start with the central extension $\hat \ggg$ of the loop algebra
${\LL}{\ggg}$, where $\ggg$ is a semi-simple Lie algebra.
It is defined as the space of pairs $(\phi,c)$, where
$\phi$ is the $\ggg$-valued scalar field given on the circle and
 $c$ is the central element, just a real number. The Lie algebra structure
is introduced by formula

\be
[({\phi}_{1}({\varphi}),c_{1}),
({\phi}_{2}({\varphi}),c_{2})] =
\left([{\phi}_{1}({\varphi}),{\phi}_{2}({\varphi})],
\int_{S^{1}} <{\phi}_{1}, {\partial}_{\varphi} {\phi}_{2}>\right),
\ee
where $<.,.>$ denotes the Killing form on the algebra $\ggg$.

Now we set the cotangent bundle to this algebra, $T^{\ast}\hat g$
to be a ``bare" phase space. $T^{\ast}\hat \ggg$ is defined to be
a direct sum of $\hat \ggg$ and ${\hat \ggg}^{\ast}$, i.e. the space
of sets $(A,k;\phi,c)$, where
$A$ is the gauge field on the circle and $k$ is dual to $c$. The pairing
between $\hat \ggg$ and ${\hat \ggg}^{*}$ is given by

\be
<(A, {\kappa}) ; (\phi , c)> = \intt <\phi, A> + c {\kappa},
\ee
On this cotangent bundle, one can define the natural symplectic structure

\beq
\Omega = \intt Tr ( \delta \phi \wedge \delta A) + \delta c
\wedge \delta {\kappa}
\label{sympl}
\eeq
The adjoint and coadjoint actions of the loop group ${\cal L}{\sf G}$
on $\hat \ggg$ and ${\hat \ggg}^{*}$ are given by formulas

\beq
(\phi(\varphi) , c) \rw \left(g(\varphi) \phi(\varphi) g(\varphi)^{-1},
\intt \Tr (- \phi g^{-1}{\partial_{\varphi}}g) + c \right)
\label{adj}
\eeq
\beq
(A, {\kappa}) \rw \left(g A g^{-1} + {\kappa} g {\partial_{\varphi}}
g^{-1},
{\kappa}\right)
\label{coadj}
\eeq
This action clearly preserves the symplectic structure and thus defines a
moment map

\be
\mu : T^{*}{\hat \ggg} \rw {\hat \ggg}^{*},
\ee
which sends
$(\phi, c; A, {\kappa})$ to $( {\kappa} \partial_{\varphi} \phi + [A, \phi] ,
0)$.

Note that the choice of level of
the moment map has a simple physical meaning:
in the theory of the angular momentum it just fixes the sector of the
Hilbert space. In the gauge theory, one selects the gauge invariant states.

Let us now discuss how to choose the proper level of the moment map
in order to get further the Hamiltonian reduction w.r.t. this level.
Let us remind that in the case of finite-dimensional group ${\sf G}$ one
takes an  element $J$ from $\ggg^{*}$,
which has the maximal stabilizer different from the whole
${\sf G}$. It is easy to show that the representative of the
coadjoint orbit of this element has the following form

\beq
J_{\nu} = {\nu} \sum_{\alpha \in \Delta_{+}}(
e_{\alpha} + e_{-\alpha}),
\label{mom}
\eeq
where  $\nu$ is some real number,
$e_{\pm \alpha}$  are the elements of the nilpotent subalgebras,
${\sf n}_{\pm} \subset
\ggg$ corresponding to the root $\alpha$, and ${\Delta}_{+}$ is
the set of positive roots.
Let us denote the coadjoint orbit of $J_{\nu}$ as ${\cal O}_{\nu}$ and
the representation of ${\sf G}$ arising upon the quantization
of ${\cal O}_{\nu}$ as $R_{\nu}$.
For a generic $J \in \ggg^{*}$, let ${\sf G}_{J}$ be the stabilizer of $J$ in
the coadjoint orbit ${\cal O}_{J}$, thus ${\cal O}_{J} = {\sf G} / {\sf
G}_{J}$.

Now in order to deal with the affine case, one just needs to substitutes
the roots by their affine counterparts. There is also a more geometrical
standpoint: because of the vanishing level $k=_{\mu}$ of the moment map,
the coadjoint orbit of its generic value is huge, while the stabilizer
of the element $(J(\varphi);0)\in {\hat \ggg}^{\ast}$ is ``a continuous
product" $\prod_{\varphi\in S^1}{\sf G}_{J(\varphi)}$ which is very small as
compared with the whole loop group $\LL {\sf G}$ unless the group
${\sf G}_{J(\varphi)}$
coincides with ${\sf G}$ for almost all $\phi\in S^1$.

All this implies that one can reasonably choose the value
of the moment map to be

\beq
\mu = ({\cal J}[{\mu}],0) :\ {\cal
J}[{\mu}]({\varphi}) = \delta({\varphi}) J_{\nu} \label{affmom}
\eeq
The coadjoint orbit of $\mu$ is nothing but the
finite dimensional orbit ${\cal O}_{\nu}$.

Now in order to perform the Hamiltonian
reduction, one has to resolve the moment map\footnote{This is
nothing but the Gauss law.}

\beq
{\kappa} \partial_{\varphi} \phi + [A, \phi] =  J_{\mu}({\varphi}) =
\delta({\varphi}) J_{\nu}.
\label{redeq}
\eeq
It can be done in the following way.
First, we use the generic gauge transformation $g({\varphi})$ to
make
$A$ to be a Cartan subalgebra valued constant one-form $D$.
However,
we are still left with the freedom to use the constant gauge transformations
which do not touch
$D$.  Actually the choice of $D$ is not unique and is parametrized by the
conjugacy classes of the monodromy
$\exp (
\frac{2\pi}{\kappa} D) \in {\bf T} \subset {\sf G}$.
Let us fix a conjugacy class
and denote $i x_{i}$ the elements of the matrix $D =
iQ$.  Decompose $\ggg$-valued function $\phi$ on the $S^{1}$ on the
Cartan-valued $P(\varphi) \in {\sf t}$ and nilpotent-valued
${\phi}_{\pm}({\varphi}) \in
{\sf n}_{\pm}$ components. Let ${\phi}_{\alpha} = <{\phi}, e_{\alpha}>$. Then
(\ref{redeq}) reads:

\beq
{\kappa}\partial_{\varphi} P = \delta({\varphi}) [ J_{\nu}^{g}
]_{\gamma}
\label{careq}
\eeq

\beq
{\kappa}\partial_{\varphi}{\phi}_{\alpha} + <D, {\phi}_{\alpha}> =
\delta({\varphi}) {[ J_{\nu}^{g} ]}_{\alpha},\\
\label{rooteq}
\eeq
where $J_{\nu}^{\tilde g}$ is $Ad_{{\tilde
g}(0)}^{*}(J_{\nu})$,
$[J]_{\gamma}$ denotes the Cartan part of $J$ and
$[J]_{\alpha} = <J,e_{\alpha}>$.

{}From (\ref{careq}) one gets $D = const$ and
$[J_{\nu}^{g}]_{\gamma} = 0$.
Therefore, one  can twist back
$J^{g}_{\nu}$ to $J_{\nu}$.
Moreover, (\ref{rooteq}) implies that (if $\varphi \neq 0$ )
${\phi}_{\alpha}({\varphi})$ can be presented as

\beq
{\phi}_{\alpha}({\varphi}) =
\exp \left( - \frac{\varphi}{\kappa} <D, {\alpha}> \right) \times M_{\alpha},
\label{rootres}
\eeq
where $M_{\alpha}$ is a locally constant element in
$\ggg$. From (\ref{rooteq}) one sees that $M_{\alpha}$ jumps,
when ${\varphi}$ passes through
$0$. The jump is equal to

\beq
\left[\exp \left( - \frac{2\pi}{\kappa} <D, {\alpha}> \right)
- 1\right] \times M_{\alpha} = [J^{g}_{\nu}]_{\alpha}.  \label{jump}
\eeq

The final conclusion is that the physical degrees of freedom
are contained
in $\exp(-\frac{2\pi i}{\kappa}Q)$ and $P$ with the symplectic structure

\beq
\omega = \frac{1}{2 \pi i} \Tr ( \delta P \wedge \delta Q )
\label{redsym}
\eeq
and one has

$$
{\phi}_{\alpha}({\varphi}) = {\nu} \frac{\exp ( -
\frac{i\varphi}{\kappa} <Q, {\alpha}> )} {\exp ( - \frac{2\pi i}{\kappa} <Q,
{\alpha}> ) - 1}
$$
With this manifest expression, one easily restores the whole $\phi(\varphi)$
adding
the constant Cartan part. Note that, while the diagonal elements $q_i$ of $Q$
determine particle coordinates, the constant Cartan part which can not
be fixed from the moment map is
parameterized by particle momenta $p_i$. In these variables,
$\phi(\varphi)$ is nothing but
the Calogero Lax matrix and the particles are the Calogero ones.

Now if one takes a simple Hamiltonian system on
$T^{*}{\hat \ggg}$
with a Hamiltonian invariant under the action
(\ref{adj}),(\ref{coadj}), this
induces a relatively complicated system on the reduced
symplectic manifold, $T^{*}{\sf T}$. As a Hamiltonian on
$T^{*}{\hat \ggg}$, one can choose

\beq
{\cal H} = \int_{S^{1}} d{\varphi} P({\phi})
\label{anycas}
\eeq
where $P(\phi)$ is any invariant polynomial on the Lie algebra $\ggg$.
This is, indeed, the standard way to construct Hamiltonians from the Lax
matrix in integrable system.

Let us choose the Hamiltonian to be the quadratic Casimir

\beq
{\cal H}_{2} = \int_{S^{1}} d{\varphi} <{\phi},
{\phi}> ,
\label{2cas}
\eeq
and look at the action of this system. Taking into
account the constraint, i.e.
adding to the action the moment map with the Lagrangian multiplier,
one obtains

\beq
S= \int_{S^{1}\times Z} d{\varphi}dt( \Tr A \partial_{t} \phi -
<{\phi},{\phi}> -{1\over \kappa} \Tr A_{0}
({\kappa} \partial_{\varphi} \phi + [A, \phi] -
\delta({\varphi}) J_{\nu})).
\eeq
Denoting $A_1\equiv A$ and $g_{YM}\equiv 1/\kappa$, one can rewrite this action
in the form

\be
S=\int_{S^{1}\times Z} d^2 x \left[-\Tr\left(\phi F_{01}+\phi^2\right)+
g_{YM}\Tr \left(J_{\nu}\int_CA_{\mu}dx^{\mu}\right)\right]
\ee
where $F_{01}$ is the component of the gauge field strength tensor and $C$ is
the
contour passing through $\varphi =0$ along the time axis. This action of
induced
gauge theory, after integrating over the scalar field $\phi$ can be rewritten
as the action of the $2d$ YM theory with an additional Wilson line inserted

\be
S=\int_{S^{1}\times Z} d^2 x \left[-{1\over 4}\Tr F_{\mu\nu}^2 +
g_{YM}\Tr \left(J_{\nu}\int_CA_{\mu}dx^{\mu}\right)\right]
\ee

On the reduced manifold one obtains from the Hamiltonian
(\ref{2cas}) the trigonometric Calogero-Sutherland system
\cite{calogero3} with the Hamiltonian

\beq
H_{2} = \frac{1}{2} \Tr P^{2} +
\sum_{\alpha \in \Delta_{+}} \frac{{\nu}^{2}}
{\sin^{2}<Q,{\alpha}>} ,
\label{redham}
\eeq
For instance, for
${\sf G} = SU(N)$ one obtains the Hamiltonian
with the pairwise interaction

\be
V_{ij}^{A} = \frac{\nu^{2}}{\sin^{2}(q_{i}-q_{j})}
\ee
while for ${\sf G} = SO(2N)$ one
has

\be
V_{ij}^{D} = \nu_{2}^{2} \left[ \frac{1}{\sin^{2}(q_{i}-q_{j})} +
\frac{1}{\sin^{2}(q_{i}+q_{j})}\right] +
\nu_{1}^{2} \left[\frac{1}{\sin^{2}(q_{i})} + \frac{1}{\sin^{2}(q_{j})}\right]
\ee
where $\nu_{1}, \nu_{2}$  are the coupling constants.
In the general case, one has as many different coupling constants
as many orbits the Weyl group has in the root system.

Generalization of the consideration to the generic Wilson line
which amounts to the spin trigonometric Calogero system
was considered in \cite{Minahan}.

\subsection{Gauged ${\sf G}/{\sf G}$ $\sigma$-model and Ruijsenaars systems}

In this subsection we show that the Calogero system have
a natural relativization which also admits an interpretation in
field theory terms. The corresponding field theory (given on
the phase space before the reduction)
is nothing but the gauged ${\sf G}/{\sf G}$ sigma model which is equivalent
to the Chern-Simons Lagrangian   \cite{g13}, the central charge of
the Kac-Moody algebra being just the deformation parameter,
from the loop group point of view.

Note that this relativization arises in different physical
problems, for instance, in two dimensional systems
\cite{solitons3}  and in gauge theories in higher dimensions
\cite{nikita,nikita3,bmmm2}. Hence, the unified Lagrangian
description for these different physical problems is of evident importance.

As in the non-relativistic system, we perform the
Hamiltonian reduction procedure, however now the unreduced phase space is
the cotangent bundle to the central extended loop group $\hat {\sf G}$.
As a symplectic manifold, it is presented by the
following data:  $(g:S^{1} \rw {\sf G}, c \in
U(1) ; A \in \Omega^{1}(S^{1}) \otimes \ggg^{*}, {\kappa} \in {\bf R})$.

The group acts as

\be
g \rw h g h^{-1}, \; A \rw h A h^{-1} + {\kappa} h {\partial_{\varphi}}
h^{-1}\\
{\kappa} \rw {\kappa}, \;
c \rw c \times {\cal S}( g, h),
\ee
where ${\cal S}$ is constructed from the $U(1)$-valued 2-cocycle $\Gamma (g,
h)$
on the group ${\cal L}G$ ($\Gamma (g,g^{-1}$),

\be
{\cal S} (g, h) = \Gamma (h, g) \Gamma(hg, h^{-1}).
\ee
Then, the group invariant symplectic structure looks as follows

\be
\Omega = \int_{S^{1}} \Tr
\left[
A (g^{-1} \delta g)^{2}
+ \delta A \wedge
g^{-1} \delta g ] +
{\kappa}
{\pdv}g \dot g^{-1} (\delta g \dot g^{-1})^{2} -
\kappa \delta ({\pdv}g) \dot g^{-1} \delta g \dot g^{-1}\right]
+ c^{-1} \delta c \wedge \delta \kappa
\label{loopsympl}
\ee
The generalization of the Gauss law can be found from the group action

\beq
\mu (g,c; A,\kappa) = (g A g^{-1} + {\kappa} g {\partial_{\varphi}} g^{-1} - A,
0)
\label{momgru}
\eeq

At the next step, we fix the level of the moment map. It is
natural to assume that its color structure does not change keeping
the integrability intact

\beq
\mu (g,c; A,{\kappa}) = i \nu \left(\frac{1}{N}
\Id - e \otimes e^{+}\right) \delta (\varphi).
\label{momlev}
\eeq
As before, we make $A$ a diagonal matrix $D$
by a generic gauge transformation $h(\varphi)$
modulo the action of the affine Weyl group.

The moment map equation now looks

\beq
g D g^{-1} + {\kappa} g {\partial_{\varphi}} g^{-1} - D =
i \nu \left(\frac{1}{N}\Id - f \otimes f^{+}\right)\delta(\varphi)
\label{mastereq1}
\eeq
where $f = h(\varphi) e$
is the vector from ${\bf C}^{N}$ with the unit norm $<f,f>
= 1$. Using the residual gauge symmetry, one can make
$f \in {\bf R}^{N}$, and using (\ref{mastereq1}) one immediately gets

\be
g = \exp \left( \frac{\varphi}{\kappa} D\right) G(\varphi)
\exp\left( -\frac{\varphi}{\kappa} D\right);\ \ \ \
\partial_{\varphi} G = - \frac{J}{\kappa} G \delta(\varphi),
\ee
where $J = i \nu ( \frac{1}{N}\Id - f \otimes f^{+} )$.
It is useful to introduce the notation for the monodromy of connection
$D$: $Z = \exp ( - \frac{2\pi}{\kappa} D) =
diag(z_{1}, \dots, z_{N})$,  $\ \prod_{i} z_{i} = 1$, $\ z_{i} = \exp
(\frac{2{\pi}i q_{i}}{\kappa})$. It satisfies the boundary condition

\beq
 {\tilde G}^{-1} Z {\tilde G} = \exp\left(\frac{2\pi J}{\kappa}\right) Z
\label{commutant}
\eeq
where $\ {\tilde G} = G(+0)$.

Solving this equation is a separate
interesting issue, see \cite{g13}. The result
looks as follows. First, introduce the characteristic polynomial
$P(z)$ of the matrix  $Z$

\be
P(z)=\prod_{i} (z - z_{i}).
\ee
Define also

\be
Q^{\pm}(z) =
\frac{P({\lambda}^{\pm 1}z) - P(z) }{({\lambda}^{\pm N}-1)z P^{\prime}(z)},
\ee
where $\lambda = e^{\frac{2 \pi i\nu}{N \kappa}}$. At $\lambda \to 1$,
the rational functions $Q^{\pm}(z)$ tend to $\frac{1}{N}$.
In these notations, the matrix $\tilde G$ can be put in the form

\be\label{Lax}
{\tilde G}_{ij} = - \lambda^{-\frac{N-1}{2}}
\frac{{\lambda}^{-N}-1}{{\lambda}^{-1}z_{i}-z_{j}}
e^{ip_{i}}(Q^{+}(z_{i})Q^{-}(z_{j}))^{1/2}=\\
=e^{ip_{i}-\frac{\pi i}{\kappa} (q_{i}+q_{j}) }
\frac{\sin(\frac{\pi\nu}{\kappa})}{\sin(\frac{\pi(q_{ij}-
\frac{\nu}{N}}{\kappa})}
\prod_{k \neq i, l \neq j}
\frac{\sin(\frac{\pi{q_{ik} +
\frac{\nu}{N}}}{\kappa})}{\sin(\frac{{\pi}q_{ik}}{\kappa})}
\frac{\sin(\frac{{\pi}{q_{il} -\frac{\nu}{N}}}{\kappa})}{\sin(
\frac{{\pi}q_{il}}{\kappa})}
\ee
where $p_{i}$ are the quantities that can not be determined from
the moment map equation and, similarly to the Calogero case, are
identified with particle momenta, while $q_{i}$ are particle coordinates,
the symplectic structure on the reduced manifold being
$\sim \sum_{i} \delta p_{i} \wedge \delta q_{i}$.

In order to construct Hamiltonians, one may fix any $Ad$-invariant function
$\chi(g)$  and define

\be
H_{\chi} = \intt d{\varphi} {\chi}(g).
\ee
Such a Hamiltonian of the dynamical system on $T^{*}{\hat G}$ is evidently
invariant under the group action.

For instance, one may take $\chi_{\pm} (g) = \Tr (g {\pm} g^{-1})$ (where
the trace, $\Tr$ is taken over the fundamental representation of $SU(N)$)
obtaining

\beq
H_{\pm} = \sum_{i} (e^{i \vartheta_{i}} {\pm} e^{- i\vartheta_{i}})
\prod_{j \neq i} f(q_{ij})
\label{ruuham}
\eeq
where the function $f(q)$ is

\be
f^{2}(q) = [ 1 -
\frac{\sin^{2}
({\pi}{\nu}/{\kappa} N)}
{\sin^{2}({\pi}q/{\kappa})}],
\ee
One may recognize in this Hamiltonian the Hamiltonian of the
trigonometric \RS
model \cite{rru2,rud}, while the solution for $g(\varphi)$ constructed above
gives the
Ruijsenaars Lax matrix. Redefining $q_i\to \kappa q_i$ and
$\vartheta_i\to\vartheta_i/\kappa$,
which does not change the symplectic form, and taking the limit
$\kappa\to\infty$,
we return back to the Calogero-Sutherland system.

Now we come to the question what is the field theory our
system corresponds to. This time we choose the Hamiltonian to be zero
so that
the term $\int pdq$ serves as a free action and we also add to the action
the constraint, the moment map with the Lagrangian multiplier

\be
S( A, g)
= \int d{\varphi}dt
{\;\Tr} \left[A_{1} g^{-1} {\partial_{t}} g
+ \kappa {\partial_{t}}g \cdot g^{-1} \cdot {\pdv} g \cdot
g^{-1} \right.\\ \left.
- \kappa d^{-1}
({\pdv} g \cdot g^{-1} (dg \cdot
g^{-1})^{2}) -
{1\over \kappa} A_{0} \left( \kappa g^{-1}
{\pdv} g + g^{-1} A_{1} g -
A_{1} - J \delta(\varphi)
\right)\right].
\ee
where we have omitted the term $\kappa c^{-1}\partial_t c$.
This action is the action of the ${\sf G}/{\sf G}$ gauged $\sigma$-model
with the Wilson line inserted.

Now we present the equivalent representation of this theory
in terms of the Chern-Simons theory with some Hamiltonian. The
Chern-Simons theory is the gauge theory given on a three
manifold which we choose to be the product of an interval and a torus
$X = I \times T^{2}$,  with the action

\beq
S_{CS} = \frac{i {\kappa}}{ 4\pi}\int_{X} d^3x \Tr (A \wedge dA +
\frac{2}{3} A \wedge A \wedge A).
\label{CS}
\eeq
We are going to consider the theory with an inserted Wilson line.
The phase space of this theory is the moduli space of the flat connections
on the torus with a marked point with prescribed conjugacy class of the
monodromy of connection around the marked point. The path integral is of the
form

\be
\int {\ddd} A <v_{1}| \Tr(P \exp \int A ) |v_{2}> \exp( - S_{CS}(A) ).
\ee
The monodromy class $U$ around the marked point is fixed by the
highest weight $\goth h$ of the representation
$R_{\nu}$, $ U = \exp ( \frac{2\pi i}{{\kappa} + N} \mbox{diag} ( {\goth
h}_{i}))$.
The monodromy condition on the torus $\Sigma = T^{2}$ that relates
the monodromies along the cycles, $g_{A}, g_{B}$ and that around the
marked point $g_{C}$ is

\be
g_{A} g_{B} g_{A}^{-1} g_{B}^{-1} = g_{C}
\ee
With the condition on the conjugacy class of $g_C$ and taking into account
that for the representation $R_{\nu}$ the signature is like
$\nu\cdot diag(N,0,...,0)$,
one can get equation (\ref{commutant}) (see \cite{g13}
for detailed explanations).

Let us also point out the equivalence of the sector of
the Chern-Simons theory
on ${\Sigma} \times S^{1}$ with observables
depending on $A_{0}$, ${\partial}_{t}$
being tangent to $S^{1}$, and of the gauged ${\sf G}/{\sf G}$ theory.
This can be proved, say, by gauge fixing $A_0$ constant and diagonal.

The equivalence provides one with a tool for computation of the spectrum
of the \RS model which is contained in the spectrum of the
relativistic free particles on the circle. In the pure Chern-Simons theory,
the particles obey the Fermi statistics, since the phase space is
$({\bf T}^{N-1} \times {\bf T}^{N-1})/ {\cal S}_{N}$,  the moduli space of
the flat connections on the torus. Here
${\bf T}^{N-1}$ is the Cartan torus in
$SU(N)$, and ${\cal S}_{N}$  is the symmetric group. Since the wave function
vanishes at the diagonal in the $N-1$-dimensional torus, these particles
really behave like fermions.

The Shr\"odinger equation for the relativistic particle
is the difference equation, since
$\cos( \frac{2\pi i}{{\kappa} + N} {\partial}_{q})$
is a finite-difference operator.
Its eigenfunctions on the circle are
$e^{2 {\pi} in q}$
with the eigenvalues

\be
E_{n} = cos (\frac{2\pi n}{{\kappa} +N})
\ee
Then, the complete spectrum of the model reads

\be
\sum_{i} E_{n_{i}}
\ee
with additional conditions that follow
from the invariance
$E_{n} = E_{n + {\kappa} + N}$ and the symmetry group action

\beq
({\kappa} +
N ) > n_{N}> \dots > n_{i} > \dots > n_{1} \geq 0 \label{restr1}
\eeq
\beq
\sum_{i} n_{i} < ({\kappa} + N)
\label{restr2}
\eeq

\subsection{Elliptic Calogero model and gauge theories}

There is also another generalization of the trigonometric Calogero model,
the elliptic model. In next sections we
shall see that just the elliptic systems are responsible
for nonperturbative dynamics in the vacuum sector of  the softly broken
$\CN =4$ SUSY gauge theories, i.e. of the \N2 SUSY gauge theories with
the adjoint matter hypermultiplet.

In this subsection we show how the elliptic models are described
via Hamiltonian reduction of the holomorphic
theories in three dimensions  \cite{g10}. Let us consider a Hamiltonian
system whose phase space is the cotangent
bundle to the algebra of $SL(N,\C)$-valued functions on the elliptic curve
with modular parameter $\tau$, $T^{*} {\elg}$:

$$
( \phi, c ; {\kappa}{\bar \partial} + {\bar A}),
\ \  \phi :  \curve \to \slnc, \ \  {\bar A} \in
\Omega^{(0,1)}(\curve) \otimes \slnc, \ \  c, \kappa \in {\bf C}
$$
Let us also denote through $d\omega$
the holomorphic differential with integrals over the
$A$- and $B$- cycles equal to 1 and $\tau$, respectively.

On this phase space acts naturally the current group
${\it
SL}_{N}({\bf
C})^{\curve}$

\be
\ppz \to \ggz \ppz \ggz^{-1}
\\
\aaz d{\bar z} \to  \ggz \aaz \ggz^{-1} + {\kappa}
\ggz {\bar \partial} \ggz^{-1}
\\
\kappa \to \kappa, \ \ c \to c + \intc d\omega \wedge
\Tr (\phi g {\bar \partial} g
)
\label{graction}
\ee
preserving the holomorphic symplectic form
$\Omega$ on $T^{*}{\elg}$

\beq
\Omega = \delta c \wedge \delta \kappa +
\intc d\omega \; \Tr ( \delta \phi \wedge \delta {\bar A} )
\label{sympl2}
\eeq
Then, the moment map has the form

\beq
\mu = {\kappa} {\bar \partial}\phi + [{\bar A} , \phi ]
\label{moment}
\eeq

In order to get the system of interacting particles, one is to enlarge the
phase
space by adding a finite dimensional coadjoint orbit of $SL(N,\C)$.
A generic complex orbit
$SL(N,\C)/\left(SL(N-l,\C)\times \left(\C^*\right)^l\right)$
is given by a matrix $f_{ij}$ of rank $l$ which can be presented
as

\be
f_{ij}=\sum_{a=1}^l u_i^av_j^a
\ee
Then, the moment map $M$
on the enlarged manifold $X_{\nu}$ has the zero level and
reads as

\be\label{ellCalhr}
0 = {\kappa}{\bar \partial} {\phi} + [{\bar A}, {\phi} ] - J_{ij}
{\delta (z, {\bar
z}) dz \wedge d{\bar z}}{\omega}.
\ee
where $J_{ij}\equiv (\delta_{ij}-1)f_{ij}$.
Using gauge transformations, one can obtain
for the matrix elements of $\phi$

\beq
{\kappa} {\bar \partial}\phi_{ij} + q_{ij} \phi_{ij} =
J_{ij}{\delta}(z,{\bar z}),
\label{***}
\eeq
where $a_{ij} \equiv a_{i} - a_{j}$, and $a_{i} \in
{\bf C}$ are diagonal elements of $\bar A$.
 From this equation one obtains three consequences:
\begin{itemize}
\item $ J_{ii} = 0 $
\item $\phi_{ii} = p_{i} = const$, $i = 1, \dots, N $; $\sum_{i}
p_{i} = 0$
\item $\phi_{ij} = \frac{1}{\kappa}J_{ij}
\exp\left( \frac{a_{ij} (z -{\bar z})}{\kappa} \right)
\frac{{\haha}\left(z + \frac{\pi a_{ij}}{ \kappa \tau_{2}}\right)}{{\haha}(z)
{\haha}\left(\frac{{\pi}a_{ij}}{{\kappa}{\tau_{2}}}\right)}$, for $i \neq j$,
where $\tau_{2} = $\Im$\tau$,
$\haha$ is the $\theta$-function with odd characteristics\footnote{We also
use the notation $\theta_1$ for this $\theta$-function.}.
\end{itemize}
The first two statements merely follow from the diagonal components of
equation (\ref{***}), with $p_i$ being just arbitrary constants of integration
(restricted by $\sum_{i}p_{i} = 0$, since we consider the $SL(N,\C)$ group).
To prove the third statement, let us make the substitution

\be
\phi_{ij}(z,{\bar z}) = \exp \left( \frac{a_{ij} (z - {\bar z})}{\kappa}
\right)
\psi_{ij} (z, {\bar z}),
\ee
so that $\psi_{ij}$ has the nontrivial monodromy

\be
\psi_{ij} (z + 1, {\bar z} + 1) = \psi_{ij}(z, {\bar z})\\
\psi_{ij} (z + \tau, {\bar z} + {\bar \tau}) =
e^{-\frac{2\pi i\tau_{2}}{\kappa} a_{ij}}
\psi_{ij} (z, {\bar z})
\ee
According to (\ref{***}), we have ${\bar \partial} \psi_{ij} =
J_{ij} {\delta}(z, {\bar z})$, i.e. $\psi_{ij}$
is a meromorphic section of the holomorphic linear bundle with the pole at
the origin.

The matrix $\phi$ obtained is nothing but the Lax operator for the
elliptic spin Calogero system \cite{krispin,krispin2}. In order to get spinless
Calogero case, one needs to choose the simplest orbit $l=1$ (which describes
the
fundamental representation). In this case, the factor matrix
$J_{ij}$ in $\phi$ can be
trivially gauged away to an arbitrary scalar $-\kappa\nu$. One should
also introduce the new rescaled variables $q_i\equiv -\pi a_i /\kappa\tau_2$
identified with coordinates in the many-body system, $p_i$ being
the corresponding momenta,
and one obtains finally
the Lax operator for the elliptic Calogero system which
being calculated at $z \neq 0$, coincides with the Lax matrix found
by Krichever earlier \cite{KriF}.

Now invariants of the algebra yield Hamiltonians
of the integrable system. The simplest one
coincides, up to a constant,
with the second Hamiltonian of the elliptic Calogero-Moser system

\be\label{elcalham}
H=\int_{\Sigma_{\tau}} d^2z \frac{1}{2}\Tr \ppz^{2} =\int_{\Sigma_{\tau}} d^2z
\left\{\sum_{i} \frac{1}{2} p_{i}^{2} + {\nu}^{2}
\left[\sum_{i < j} {\wp} ( q_{ij}) -
{\wp}(z)\right]\right\}=\\=
\left\{\int_{\Sigma_{\tau}} d^2z \right\}\times\left\{
\sum_{i} \frac{1}{2} p_{i}^{2} + {\nu}^{2}
\sum_{i < j} {\wp} ( q_{ij})\right\} -
\frac{\nu^2N(N-1)}{2} \int_{\Sigma_{\tau}} d^2z {\wp}(z)
\ee
Let us remark that there is an essential difference
between elliptic model and its degeneration at the quantum level.
It appears that it is impossible to get the quantum
spectrum of the elliptic model analytically. The only
tool available is the Bethe anzatz approach which
yields the equation for the eigenvalues \cite{fv}.
Some quantum aspects of these systems within the representation theory
framework were discussed in \cite{etingof}.

One may want to obtain an elliptic deformation of the
two-dimensional YM theory. The action proposed in \cite{g10}
is of the form
\beq
S_{\tau} = \int_{\curve \times S^1} \omega \wedge \Tr
(\phi F_{t {\bar z}} - {\varepsilon} \phi^{2})
\label{ellaction}
\eeq
where $S^1$ is a time-like circle.
One should certainly also insert a (vertical) Wilson line
$P \exp \left(\nu \int A_{t}\right)$ in representation  $R_{\nu}$.

Let us emphasize that the
elliptic Calogero-Moser model is an example of the Hitchin systems given
on the moduli space of the holomorphic vector bundles
\cite{hitchin,markman,nikita2,er}
for the surfaces with marked points.
The elliptic Calogero-Moser system has a natural relativization -- the elliptic
\RS model which is generalization of the Hitchin systems.
A description of the elliptic \RS system via
Hamiltonian and Poissonian reductions can be found in \cite{Arutyunov}.

Recently the Hamiltonian reduction procedure for the
elliptic Calogero models
considered above for $A_n$ case has been extended to other classical groups in
\cite{hm}.

\section{Particle systems and supersymmetric gauge theories}
\setcounter{equation}{0}

After having presented a manifest construction of the integrable systems,
we are ready to show how these integrable many-body systems emerge
in the consideration of effective low-energy \N2
supersymmetric gauge field theories. As it will be clear later, the role
of the many-body systems is to provide the proper degrees of freedom
in the vacuum sector of the theory. The fact that the complete set of data
fixing the low-energy sector of the theory
is governed by the many-body system definitely implies that the
number of the effective degrees of freedom in the vacuum sector of the
supersymmetric gauge theory
is finite. In this section, we demonstrate how the corresponding
particle systems can be recovered, while later we comment on the role
of these degrees of freedom within the string theory context.

\subsection{\SW and integrable systems}

As explained in the Introduction,
the set of data a Riemann surface,
the corresponding moduli space and the differential $dS$ that gives the
physical quantities, in particular, prepotential is naturally associated
with an integrable system. In the case of the $SU(2)$ pure gauge theory, which
has been worked out in detail in \cite{SW1} this correspondence can be
explicitly checked \cite{GKMMM}.
In particular, one may manifestly check formulas (\ref{aad}),
since all necessary ingredients are obtained in \cite{SW1}. In this way, one
can
actually prove that this \SW corresponds to the integrable system (periodic
Toda chain with two particles, see below). However, to repeat
the whole Seiberg-Witten procedure for theories with more vacuum moduli
 is technically quite tedious if possible at all. Meanwhile, only in this way
one could unquestionably prove the correspondence in other cases.

In real situation, the identification between the \SW and
the integrable system
 comes via comparing the three characteristics
\begin{itemize}
\item number of the vacuum moduli and external parameters
\item perturbative prepotentials
\item deformations of the two theories
\end{itemize}
The number of vacuum moduli (i.e. the number of scalar fields that may
have non-zero v.e.v.`s) on the physical side should be compared with dimension
of
the moduli space of the spectral curves in integrable systems,
while the external parameters in
gauge theories (bare coupling constants,
hypermultiplet masses) should be also some external parameters
in integrable models (coupling constants, values of Casimir functions for
spin chains etc).

As for second item, we already pointed out the distinguished role of
the prepotential in Seiberg-Witten theory (which celebrates a lot of
essential properties which we shall discuss later).
In the prepotential, the contributions of particles and
solitons/mo\-no\-po\-les (dyons) sharing the same mass scale, are still
distinguishable, because of different dependencies on the bare coupling
constant, {\it i.e.} on the modulus $\tau$ of the bare coordinate elliptic
curve (in the UV-finite case) or on the $\Lambda_{QCD}$ parameter (emerging
after dimensional transmutation in UV-infinite cases).  In the limit $\tau
\rightarrow i\infty$ ($\Lambda_{QCD} \rightarrow 0$), the solitons/monopoles
do not contribute and the prepotential reduces to the ``perturbative'' one,
describing the spectrum of non-interacting {\it particles}.  It is
immediately given by the SUSY Coleman-Weinberg formula \cite{WDVVlong}:

\be\label{ppg}
{\cal F}_{pert}(a) \sim
\sum_{\hbox{\footnotesize reps}
\ R,i} (-)^F {\rm Tr}_R (a + M_i)^2\log (a + M_i)
\ee
Seiberg-Witten theory (actually, the identification of appropriate integrable
system) can be used to construct the non-perturbative prepotential,
describing the mass spectrum of all the ``light'' (non-stringy) excitations
(including solitons/monopoles).
Switching on Whitham times \cite{RG} presumably allows one to extract
some correlation functions in the ``light'' sector (see below).

The perturbative prepotential (\ref{ppg}) has to be, certainly, added with
the classical part,

\be
\CF_{class}=-\tau a^2
\ee
since the bare action is $S(\Psi )=\Im\tau \int \Tr\Psi^{2}$. $\tau$ here
is value of the coupling constant in the deeply UV region. It plays
the role of modular parameter of the bare torus in the UV finite theories.
On the other hand, in asymptotically free theories, where the perturbative
correction behaves like $a^2\log (a/\Lambda_{QCD})$, $\tau$ is just contained
in $\Lambda_{QCD}$, since can not be invariantly defined neither in integrable
nor in gauge theory.

As we already mentioned
the problem of calculation of the prepotential in physical
theory is simple only at the perturbative level, where it is given just by
the leading contribution,
since the $\beta$-function in ${\cal N}=2$ theories is non-trivial only in one
loop.
However, the calculation of all higher (instantonic) corrections in the gauge
theory can be hardly
done at the moment\footnote{In order to check the very ideology that
the integrable systems lead to the correct answers, there were
calculated first several corrections \cite{Instanton,Yung}.
The results proved to exactly
coincide with the predictions obtained within the integrable approach.}.
Therefore, the standard way of doing is to make an identification
of the \SW and an integrable system and then to rely on integrable
calculations.
This is why establishing the ``gauge theories $\leftrightarrow$
integrable theories" correspondence is of clear practical
(apart from theoretical) importance.

Now let us come to the third item in the above correspondence, namely, how one
can
extend the original Seiberg-Witten theory. There are basically three
different ways to extend the original \SW.

First of all, one may consider other gauge groups, from other simple
classical groups to those being a product of several simple factors.
The other possibility is to add some matter hypermultiplets in different
representations. The two main cases here are the matter in fundamental or
adjoint representations. At last, the third possible direction
to deform \SW is to consider 5- or 6-dimensional theories, compactified
respectively onto the circle of radius $R_5$ or torus with modulus $R_5/R_6$
(if the number of dimensions exceeds 6, the gravity becomes
obligatory coupled to the gauge theory).

Now we present a table of
known relations between gauge theories and integrable systems, Fig.1.

\begin{figure}[t]
\begin{center} \begin{tabular}{|c|c|c|c|c|}
\hline
{\bf SUSY}&{\bf Pure gauge}&{\bf SYM theory}&{\bf SYM theory}\\
{\bf physical}&{\bf SYM theory,}&{\bf with adj.}&{\bf with fund.}\\
{\bf theory}&{\bf gauge group ${\sf G}$}&{\bf matter}&{\bf matter}\\
\hline
        & inhomogeneous  & elliptic  & inhomogeneous        \\
         & periodic        & Calogero & periodic \\
{\bf 4d} & Toda chain & model & \XXX \\
& for the dual affine ${\hat {\sf g}}^{\vee}$  &({\it trigonometric}& spin
chain\\
        & ({\it non-periodic}  & {\it Calogero}& ({\it non-periodic}\\
        & {\it Toda chain})& {\it model})& {\it chain})\\
\hline
        & periodic  & elliptic & periodic   \\
        & relativistic  & Ruijsenaars&\XXZ\\
{\bf 5d} & Toda chain  & model & spin chain  \\
        & ({\it non-periodic} & ({\it trigonometric}& ({\it non-periodic}\\
        & {\it chain}) & {\it Ruijsenaars})&{\it chain})\\
\hline
    &periodic & Dell & periodic\\
    & ``Elliptic" & system & \XYZ (elliptic)\\
{\bf 6d} &  Toda chain & ({\it dual to elliptic} & spin chain\\
        & ({\it non-periodic}& {\it Ruijsenaars,}&({\it non-periodic} \\
& {\it chain})& {\it elliptic-trig.})&{\it chain})\\
\hline
\end{tabular}
\end{center}
\caption{SUSY gauge theories $\Longleftrightarrow$ integrable systems
correspondence. The perturbative limit is marked by the italic font (in
parenthesis).}\label{intvsYM}
\end{figure}
%\vspace{10pt}

Before coming into details, let us briefly describe different cells of the
table.

The original Seiberg-Witten model, which is the $4d$ pure gauge $SU(N)$ theory
(in fact, in their papers \cite{SW1,SW2}, the authors considered the $SU(2)$
case only, but the generalization made in \cite{3a} is quite
immediate), is the upper left square of the table. The
remaining part of the table contains possible deformations.
Here only two of the three possible ways to deform the original Seiberg-Witten
model are shown. Otherwise, the table would be three-dimensional.
In fact, the third direction related to changing the gauge group,
although being of an interest is slightly out of the main line. Therefore,
we only make several comments on it.

One direction in the table corresponds to matter hypermultiplets
added. The most interesting is to add matter in adjoint or fundamental
representations, although attempts to add antisymmetric and
symmetric matter hypermultiplets were also done (see \cite{asym} for
the construction of the curves and \cite{KPasym} for the corresponding
integrable systems). Adding matter in other representations in the basic
$SU(N)$ case leads to non-asymptotically free theories.

\paragraph{Columns: Matter in adjoint {\it vs.} fundamental
representations of the gauge group.}

Matter in adjoint representation can be described in terms of a
larger pure SYM model, either with higher SUSY or in higher
dimensional space-time.
Thus models with adjoint matter form a hierarchy, naturally
associated with the hierarchy of integrable models {\it Toda chain
$\hookrightarrow$ Calogero $\hookrightarrow$
Ruijsenaars $\hookrightarrow$ Dell}
\cite{GKMMM,dw,intadj,g14,IM1,IM2,bmmm2,bmmm3,MM1,MM2}. Similarly, the
models with fundamental matter \cite{SW2,4a}
are related to the hierarchy of spin chains
originated from the Toda chain: {\it Toda chain
$\hookrightarrow$ \XXX $\hookrightarrow$
\XXZ $\hookrightarrow$ \XYZ}
\cite{XXX,XYZ,GGM1,GGM2,MarMir}.

Note that, while coordinates in integrable systems describing pure gauge
theories and those with fundamental matter, live on the cylinder (i.e.
the dependence on coordinates is trigonometric), the coordinates in the
Calogero system (adjoint matter added)
live on a torus\footnote{Since these theories are
UV finite, they depend on an additional (UV-regularizing) parameter, which
is exactly the modulus $\tau$ of the torus.}. However, when one takes the
perturbative limit, the coordinate dependence becomes trigonometric.

\paragraph{Lines: Gauge theories in different dimensions.}

Integrable systems relevant
for the description of vacua of $d=4$ and $d=5$ models are
respectively the Calogero and Ruijsenaars ones (which possess the ordinary
Toda chain and ``relativistic Toda chain'' as Inozemtsev's limits
\cite{Ino}), while $d=6$ theories are described by the
double elliptic (Dell) systems. When we go from $4d$ (Toda, \XXX, Calogero)
theories to (compactified onto circle)
$5d$ (relativistic Toda, \XXZ, Ruijsenaars) theories
the momentum-dependence of the Hamiltonians
becomes trigonometric (the momenta live on the compactification circle)
instead of rational. Similarly, the (compactified onto torus) $6d$ theories
give rise to an elliptic momentum-dependence of the Hamiltonians, with
momenta living on the compactification torus. Since adding
the adjoint hypermultiplet elliptizes the coordinate dependence, the integrable
system corresponding to $6d$ theory with adjoint\footnote{Let us point out once
more that by adding the adjoint matter we always mean soft breaking of
higher supersymmetries.}
matter celebrates both
the coordinate- and momentum-dependencies elliptic. A candidate for ``the
elliptic Toda chain'' was proposed in \cite{Kriel}.

Further in this section we show how the Riemann surfaces and the meromorphic
differentials arise when considering a concrete integrable many-body system
and explain in more details different squares in the table of
correspondence between Seiberg-Witten theories and integrable systems.
However, of all $6d$ theories we describe here only the \XYZ chain case,
postponing the discussion of the most general adjoint matter system, Dell
system (the right bottom cell of the table) till section 7 where this
integrable model naturally emerges within the framework of duality.

\subsection{$4d$ pure gauge theory: Toda chain}

We start with the simplest case of the $4d$ pure gauge theory, which was
studied in details in the original paper \cite{SW1} for the $SU(2)$ gauge
group and straightforwardly generalized to the $SU(N_c)$ case in \cite{3a}.

This system is described by the periodical Toda chain
with period $N_c$ \cite{GKMMM,MW,6} whose equations of motion read

\be\label{Todaeq}
\frac{\partial q_i}{\partial t} = p_i \ \ \ \ \
\frac{\partial p_i}{\partial t} = e^{q_{i+1} -q_i}-
e^{q_i-q_{i-1}}
\ee
with the periodic boundary conditions $p_{i+N_c}=p_i$, $q_{i+N_c}=q_i$ imposed.
In physical system, there are $N_c$ moduli and there are no external
parameters ($\Lambda_{QCD}$ can be easily removed by rescaling).
In the Toda system there are exactly $N_c$ conservation laws.
These conservation laws can be constructed from the Lax operator
defined for any integrable system. The eigenvalues of the Lax operator
do not evolve, thus, any function of the eigenvalues is an
integral of motion.

There are two different Lax representation describing the periodic Toda chain.
In the first one, the Lax operator is represented by the $N_c\times N_c$
matrix depending on dynamical variables

\be\label{LaxTC}
{\cal L}^{TC}(w) =
\left(\begin{array}{ccccc}
p_1 & e^{{1\over 2}(q_2-q_1)} & 0 & & we^{{1\over 2}(q_1-q_{N_c})}\\
e^{{1\over 2}(q_2-q_1)} & p_2 & e^{{1\over 2}(q_3 - q_2)} & \ldots & 0\\
0 & e^{{1\over 2}(q_3-q_2)} & -p_3 & & 0 \\
 & & \ldots & & \\
\frac{1}{w}e^{{1\over 2}(q_1-q_{N_c})} & 0 & 0 & & p_{N_c}
\end{array} \right)
\ee
The characteristic equation for the Lax matrix

\be\label{SpeC}
{\cal P}(\lambda,w) = \det_{N_c\times
N_c}\left({\cal L}^{TC}(w) - \lambda\right) = 0
\ee
generates the conservation laws and
determines the spectral curve

\be\label{fsc-Toda}
w + \frac{1}{w} = 2P_{N_{c}}(\lambda )
\ee
where $P_{N_c}(\lambda)$ is a polynomial of degree $N_c$ whose
coefficients are integrals of motion. If one restores the dependence
on $\Lambda_{QCD}$ in this spectral curve, it takes the form

\be\label{fsc-Toda'}
w + \frac{\Lambda_{QCD}^{2N_c}}{w} = 2P_{N_{c}}(\lambda )
\ee
However, we typically re-scale it away, for exception of s.6.3, where
we discuss the manifest RG dynamics w.r.t. $\Lambda_{QCD}$.

The spectral curve (\ref{fsc-Toda}) is exactly the Riemann surface
introduced in the context of $SU(N_c)$ gauge theory. It can be also
put into the hyperelliptic form

\be\label{2}
2Y\equiv w-{1\over w},\ \ \ Y^2=P_{N_c}^2(\lambda)-1
\ee
The integrals of motion parameterize the moduli space of the
complex structures of the hyperelliptic surfaces of genus $N_c-1$, which is
the moduli space of vacua in physical theory.

If one considers the case of two particles ($SU(2)$ gauge theory), $P(\lambda)
=\lambda^2-u$, $u=p^2-\cosh q$, $p=p_1=-p_2$, $q=q_1-q_2$.
Thus we see that the order parameter
of the SUSY theory plays the role of Hamiltonian
in integrable system. In the perturbative regime of the
gauge theory, one of the exponentials in $\cosh q$ vanishes, and one
obtains the non-periodic Toda chain. In the equation (\ref{fsc-Toda}) the
perturbative limit implies vanishing the second term in the l.h.s., i.e.
the spectral curve becomes the sphere $w=2P_{N_c}(\lambda)$.

The alternative Lax representation appeals to introducing the
local operators at each site of the chain

\be\label{LTC}
L^{TC}_i(\lambda) =
\left(\begin{array}{cc} \lambda -p_i & e^{q_i} \\ -e^{-q_i} & 0
\end{array}\right), \ \ \ \ \ i = 1,\dots ,N_{c}
\ee
This Lax operator ``makes a shift" to the neighbour site so that the
auxiliary linear problem is

\be\label{lp2}
L_i(\lambda)\Psi_i(\lambda)=\Psi_{i+1}(\lambda)
\ee
where $\Psi_i(\lambda)$ is a two-component Baker-Akhiezer function.
The periodic boundary conditions are easily formulated in terms of
this Baker-Akhiezer function and read as

\be\label{TCbc}
\Psi_{i+N_c}(\lambda)=w\Psi_i(\lambda)
\ee
where $w$ is a free parameter (diagonal matrix). One can also introduce the
transfer matrix that shifts $i$ to $i+N_c$

\be\label{monomat}
T(\lambda)\equiv L_{N_c}(\lambda\ldots L_1(\lambda)
\ee
Then, the boundary conditions imply that $T(\lambda)\Psi_i(\lambda)=
w\Psi_i(\lambda)$, i.e.

\be\label{specTC0}
\det_{2\times 2}\left( T_{N_{c}}(\lambda )
- w\right) = 0
\ee
providing us with the same spectral curve

\be\label{specTC}
0=w^2 - w\Tr T_{N_{c}}(\lambda ) + \det
T_{N_{c}}(\lambda ) = w^2 - w\Tr T_{N_{c}}(\lambda ) + 1,\\ \hbox{ i.e.: }
\ \ \ \ \ {\cal P}(\lambda ,w) = \Tr T_{N_{c}}(\lambda) - w -
\frac{1}{w} = 2P_{N_{c}}(\lambda) - w - \frac{1}{w} = 0
\ee

The identification of the two representations can be also done
at the level of their linear problems. Indeed, let us consider the Lax operator
(\ref{LTC}) with the linear problem (\ref{lp2}) and the boundary conditions
(\ref{TCbc}). If we parameterize $\Psi_i=\left(
\begin{array}{c}\psi_i\\ \chi_i\end{array}\right)$, then the linear problem
(\ref{lp2}) can be rewritten as

\be
\psi_{i+1}+p_i\psi_i+e^{q_i-q_{i-1}}\psi_{i-1}=\lambda\psi_i,\ \ \
\chi_i=-e^{q_{i-1}}\psi_{i-1}
\ee
and, along with the periodic boundary condition (\ref{TCbc}) reduces to the
linear problem $\CL(w) \Phi=\lambda\Phi$ for the $N_c\times N_c$ Lax operator
(\ref{LaxTC}) with the $N_c$-component Baker-Akhiezer function $\Phi=
\left\{e^{-q_i/2}\psi_i\right\}$.

After having constructed the Riemann surface and the moduli space
describing the $4d$ pure gauge theory,
we turn to the third crucial ingredient of \SW
that comes from integrable systems, the generating differential $dS$.
The general construction was explained in the Introduction,
this differential is
in essence the ``shorten" action $pdq$. Indeed, in order to construct
action variables, $a_i$
one needs to integrate the differential ${\tilde {dS}}=\sum_ip_idq_i$
over $N_c-1$ non-contractable cycles in the Liouville torus which is nothing
but
the level submanifold of the phase space, i.e. the submanifold defined by
values of all $N_c-1$ integrals of motion fixed. On this submanifold, the
momenta $p_i$ are functions of the
coordinates, $q_i$. The Liouville torus in \SW
is just the Jacobian
corresponding to the spectral curve (\ref{SWcurve})
(or its factor over a finite subgroup) .

In the general case of a $g$-parameter family of complex curves
(Riemann surfaces) of genus $g$, the Seiberg-Witten differential
$dS$ is characterized by the property
$\delta dS = \sum_{i=1}^g \delta u_i dv_i$,
where $dv_i(z)$ are $g$ holomorphic 1-differentials
on the curves (on the fibers), while $\delta u_i$ are
variations of $g$ moduli (along the base).
In the associated integrable system, $u_i$ are integrals of motion
and $\pi_i$, some $g$ points on the curve are momenta.
The symplectic structure is

\be\label{dSjac}
\sum_{i=1}^g da_i\wedge dp_i^{Jac} = \sum_{i,k=1}^g
du_i\wedge dv_i(\pi_k)
\ee
The vector of the angle variables,

\be\label{pjac}
p_i^{Jac} = \sum_{k=1}^g \int^{\pi_k} d\omega_i
\ee
is a point of the Jacobian, and the Jacobi map identifies this with the
$g$-th power of the curve, $Jac\ \cong {\cal C}^{\otimes g}$.
Here $d\omega_i$ are {\it canonical} holomorphic differentials,
$dv_i = \sum_{j=1}^g d\omega_j \oint_{A_j} dv_i$.
Some details on the
symplectic form on the finite-gap solutions can be found in
\cite{konzon,KriPh}.

Technical calculation is, however, quite tedious. It is simple only in the
2-particle ($SU(2)$) case, when the Jacobian coincides with the curve itself.
In this case, the spectral curve is

\be\label{sc2}
w+{1\over w}=2(\lambda^2-u)
\ee
while $u=p^2-\cosh q$. Therefore, one can write for the action variable

\be\label{dSTC}
a=\oint pdq=\oint \sqrt{u-\cosh q}dq=\oint \lambda {dw\over w},\ \ \
dS=\lambda {dw\over w}
\ee
where we made the change of variable $w=e^q$ and used equation (\ref{sc2}).

Now one can naturally assume that this expression for the differential $dS$
is suitable for generic $N_c$. A long calculation (which can be borrowed, say,
from the book by M.Toda \cite{toda}) shows that this is really the
case. One can easily check that the derivatives of $dS$ w.r.t. to moduli are
holomorphic, up to total derivatives. Say, if one
parameterizes\footnote{The absence of the term $\lambda^{N_c-1}$ is due
to the $SU(N_c)$ group and corresponds to the total momentum equal to zero,
i.e. to the center mass frame. This is important for further counting of
holomorphic differentials.}
$P_{N_c}(\lambda)=-\lambda^{N_c}+s_{N_c-2}\lambda^{N_c-2}+...$ and note
that $dS=\lambda dw/w=\lambda dP_{N_c}(\lambda)/Y=P_{N_c}(\lambda) d\lambda/Y
+ \hbox{ total derivatives }$, then

\be
{\partial dS\over\partial s_k}={\lambda^kd\lambda\over Y}
\ee
and these differentials are holomorphic if $k\le N_c-2$.
Thus, $N_c-1$ moduli gives rise to $N_c-1$
holomorphic differentials which perfectly fits the genus of the curve (we
use here that there is no modulus $s_{N_c-1}$).

It turns out that the form of $dS=\lambda dw/w$ is quite general and
does not change even in more complicated cases of spin chains, which we discuss
in the next subsection.

To conclude this subsection, let us briefly consider
the theories with other gauge groups.
First of all, we present an invariant algebraic formulation
of the Toda Lax operator (\ref{LaxTC}). It can be obtained much along the line
discussed in the previous section, i.e. via the Hamiltonian reduction
\cite{olper}, or even obtained from the
elliptic Calogero model by a degeneration \cite{Ino} (we discuss it later).
Here we just write down the answer which looks like \cite{Bog,olper}

\be
\CL (w)=\sum_{i}^{N_c-1}\left[e^{{\bf \alpha}_i\cdot {\bf q}}\left(
{\bf e}_{\alpha_i} + {\bf e}_{\alpha_i}\right) + p_i {\bf h}_i\right] +
w{\bf e}_{\alpha_0}e^{{\bf \alpha}_0\cdot {\bf q}}+
{1\over w}{\bf e}_{-\alpha_0}e^{{\bf \alpha}_0\cdot {\bf q}}
\ee
where the sum runs over all simple roots $\alpha_i$ of the affine algebra
$A_{N_c-1}^{(1)}$, ${\bf e}_{\alpha_0}$ is the highest (long) root, ${\bf h}_i$
 span the Cartan subalgebra and the vector ${\bf q}$ has as its components
the particle coordinates. Note that the generators ${\bf e}_{\alpha_i}$
in (\ref{LaxTC}) are
taken in the simplest fundamental representation of the algebra.
However,
this requirement is not essential for the whole construction and can be
removed.

This expression is straightforwardly extended to any simple affine Lie
algebra\footnote{Sometimes one needs to add non-unit coefficients, depending on
the
root, in front of exponentials.}, $N_c-1$ being substituted by the rank of
the group. Now, in order to construct \SW with
a simple gauge group ${\sf G}$ one needs to consider the Toda chain for
the corresponding {\it dual} affine algebra ${\hat {\sf g}}^{\vee}$ \cite{MW}.
Say, the gauge theory with the group $Sp(2n)$ is described by the affine Toda
for $\left(C_n^{(1)}\right)^{\vee}$ etc.

Having the Lax operator, one may repeat the procedure of this subsection in
order
to get the spectral curve and the corresponding moduli space. As for the
generating differential $dS$, it preserves the form $dS=\lambda dw/w$ for all
groups.

Note that for any, at least, classical simple group there is also a $2\times 2$
Lax representation generalizing (\ref{LTC}). In this case, the Lax operator
is given on the Dynkin diagram as a substitute of the closed chain above
and
the definition of transfer matrix is to be extended to include the
reflection matrices \cite{SklRef}. Therefore, one can immediately
construct some \SW in this
way. We refer for the details to \cite{GorMir}.

\subsection{Adjoint hypermultiplet added: Calogero system}

As far as there exist two different Lax representations, there are, at least,
two possible generalizations of the periodic Toda chain (in fact, more as we
shall
see later). The first generalization is related to the $N_c\times N_c$
representation (\ref{LaxTC}).
On physical side, it corresponds to the simplest deformation
of the \N2 SUSY gauge theory by adding the adjoint hypermultiplet
\cite{dw,IM1,IM2,g14}. If the matter is
massless, the theory is finite and the coupling constant can be defined at
arbitrary scale. We mainly discuss the massive matter and find the
corresponding dynamical system with the additional parameter.
It appears that the deformation by the massive hypermultiplet can be
described in terms of the elliptic Calogero model. The hypermultiplet
mass plays the role of the coupling constant in the Calogero model, while the
UV value of the field theory coupling constant
defines the modulus of the torus where the Lax operator is defined.
When one sends the mass of the hypermultiplet to infinity, in field theory
the dimensional transmutation procedure is applied, which
has its counterpart in the many-body system and is equivalent to the
transition back from Calogero to the Toda system.

The matrix $N_c\times N_c$  Lax operator for the Calogero system was
found in section 2 and read as

\be
\label{LaxCal}
{\cal{L}}^{Cal}(\xi) =
\left(\sum_{i}^{N_c} p_i{\bf h}_i +
\sum_{\alpha\in\Delta}F({\bf q\balpha}|\xi){\bf e}_{\alpha}\right) =
\left(\begin{array}{cccc} p_1 & F(q_1-q_2|\xi) & \ldots & F(q_1 -
q_{N_c}|\xi)\\ F(q_2-q_1|\xi) & p_2 & \ldots & F(q_2-q_{N_c}|\xi)\\ & &
 \ldots  & \\ F(q_{N_c}-q_1|\xi) & F(q_{N_c}-q_2|\xi)& \ldots &p_{N_c}
\end{array} \right)
\ee
where the first sum runs over the Cartan subalgebra of $SU(N_c)$,
while the second one
runs over all roots (not only over the simple ones, in variance with the Toda
case).
The special functions that appear above are defined
\cite{KriF}

\be
F(q|\xi) = \frac{\sigma(q - \xi)}{\sigma(q)\sigma(\xi)},
\label{Ffun}
\ee
where $\sigma(\xi)$
denotes the (Weierstrass) $\sigma$-function \cite{BEWW}. One identity
used throughout the calculations with the Calogero (and Ruijsenaars
below) system is

\be
\frac{\sigma(u-v)\sigma(u+v)}{\sigma\sp2(u)\sigma\sp2(v)}=
\wp(v)-\wp(u).
\label{wpaddn}
\ee
where the Weierstrass $\wp$-function is defined as\footnote{In accordance
with the standard definition, \cite{BEWW} $\wp (\xi)$ is
the doubly periodic Weierstrass function with periods $2\omega$ and $2\omega'$,
and $\tau={\omega'\over\omega}$. We remark, however, that although the
Weierstrass function depends on two periods, the homogeneity relation
$$
\wp (tz|t\omega,t\omega')=t^{-2}\wp (z|\omega,\omega')
$$
enable us to arbitrary scale one of these. All our results here
are ultimately
independent of such scaling and this allows us to choose the real period to
be $\pi$ (that is $\omega=\pi/2$). We make this choice throughout.}

\be\label{ws}
\wp(\xi)\equiv\partial^{2}\log \sigma(\xi) =
\sum_{m} \frac{1}{\sinh^{2}(\xi+m\tau)}
\ee
This Lax operator (\ref{LaxCal}) can be just read off from the
Calogero Lax operator obtained in s.2.3 (after a trivial gauge
thransformation). The only external parameter
of the integrable system in this case, the coupling constant $\nu$
is proportional to the only external parameter of the physical theory,
 the hypermultiplet mass $M^2$.

The spectral curve $\Sigma^{Cal}$

\be\label{fscCal}
\det_{N_c\times N_c} \left({\cal L}^{Cal}(\xi) - \lambda\right) = 0
\ee
covers the ``bare" spectral curve, torus $E(\tau )$

\be\label{ell}
y^2 = (x -  e_1)(x - e_2)(x - e_3) \ \ \ \ \
\ee
manifestly parametrized by the variable $\xi$: $x=\wp (\xi)$, $y=-{1\over 2}
\wp '(\xi)$.

$\Sigma^{Cal}$ can be presented as a polynomial $\CP (x,y;\lambda)=0$ of
degree $N_c$ in $\lambda$, the manifest (quite involved) expression for it
can be found in \cite{dw,IM1,DPhong}. However, this manifest expression
can be always obtained by a degeneration from the Ruijsenaars spectral curve
discussed later. Amazingly, this latter curve can be written in a quite
compact form.

In the simplest case of two particles, the curve $\Sigma^{Cal}$ is

\be
\CP (x,y;\lambda)=\lambda^2 - u+\nu^2\wp (\xi)=0
\ee

The generating differential $dS$ on the Calogero curves has the form

\be
dS=\lambda d\omega,\ \ \ d\omega\equiv d\xi=-2{dx\over y}
\ee
Here $d\omega$ is the holomorphic 1-differential on the bare spectral torus and
one can check that the derivatives w.r.t. moduli of this differential are
holomorphic on $\Sigma^{Cal}$.

Now let us make the dimensional transmutation, i.e. reproduce the periodic Toda
chain from the Calogero system \cite{Ino,g14,IM1,IM2}. To this end,
first, we degenerate the bare spectral torus
$\tau\rightarrow i \infty$. Then, to provide the nearest neighbor
interaction, we introduce the homogeneous
coordinate ``lattice" with the large distance $\Delta$
between sites

\be
q_{j}=j\Delta+\phi_{j}
\ee
To see explicitly what kind of interaction emerges in this limit,
it suffices to look at the Weierstrass function giving
the Calogero potential, although the procedure can be easily repeated for the
Lax operator. The Calogero potential has form (see (\ref{elcalham}))

\be\label{potCal}
V(x_{ij})=\nu^2\sum_{i,j}\wp(q_{ij})
\ee
In the limit under consideration we introduce the renormalized coupling
constant $\nu=\nu_{0}\exp(\Delta)$ so that $nu_{0}$ will be ultimately nothing
but
$\Lambda_{QCD}$. Now, choosing $\Delta\sim\tau$ and taking the limit $\tau\to
i\infty$, we see that only $m=0$ term survives in the sum (\ref{ws})
in the potential (\ref{potCal}) so that the resulting potential reads as

\be
V_0(\phi_{i})=\nu_0^2\sum_{j=1}^{N_c-1}e^{\phi_{i+1}-\phi_i}
\ee
and describes the open (non-periodic) Toda chain.

In order to get the periodic Toda chain, one needs to fix $\Delta={\tau\over
n}$. In this case, the $m=-1$ term in (\ref{ws})
also contributes into the sum and one
finally obtains the potential

\be
V_{TC}(\phi_{i})=V_0+\nu_0^2e^{\phi_{1}-\phi_n}
\ee
describing the periodic Toda chain.

Now, the bare torus (\ref{ell}) under the procedure described degenerates
into a sphere with two punctures, at $w=0$ and $w=\infty$:
$\ y\to w-1/w$, $x\to w+1/w$, while the generating differential $dS$
turns into $\lambda dw/w$ ($d\omega\to dw/w$) coinciding with (\ref{dSTC}).

Returning to physical theory, making the limit to the pure gauge theory one
needs to send the hypermultiplet mass $m$ to infinity keeping finite
$\Lambda^{N_{c}} \sim M^{N_c}e^{2\pi i\tau}$. This is exactly the limit we just
discussed in integrable system.

To conclude our consideration, let us note that the Calogero Lax operator
considered above can be generalized to other groups \cite{DPhongE}, and
the corresponding \SW can be constructed \cite{DPhongESW}. However, since
the exact construction requires entering many technicalities we skip it here.

\subsection{Fundamental matter added (SQCD): \XXX spin chain}

The second possible generalization is related to the $2\times 2$
representation.
This generalization on physical side corresponds considering
theories with massive hypermultiplets in the fundamental representation.
To preserve the
asymptotic freedom one has to demand for
the number of matter hypermultiplets, $N_f$
 not exceed $2N_{c}$. In particular, the theory with zero $\beta$-function,
i.e.
UV finite one
corresponds to $N_f=2N_c$. On integrable side, this theory is described by
the inhomogeneous twisted \XXX
spin chain, while theories with $N_f< 2N_c$ are described by its
degenerations \cite{XXX,GGM1}.

The periodic inhomogeneous $sl(2)$ \XXX chain of length $N_c$ is given by the
$2\times 2$ Lax matrices

\be\label{LaxXXX}
L_i(\lambda) = (\lambda+\lambda_i)
\cdot {\bf 1} + \sum_{a=1}^3 S_{a,i}\cdot\sigma^a
\ee
$\sigma^a$ being the standard Pauli matrices and $\lambda_i$ being
the chain inhomogeneities, and periodic boundary conditions are imposed.
The linear problem in the spin
chain has the same form (\ref{lp2}) as in the Toda case.

One can also introduce the transfer matrix as in (\ref{monomat})
which provides the spectral curve equation (\ref{specTC0})
and generates a complete set of integrals of motion.

Integrability of the spin chain follows from the
{\it quadratic} r-matrix relations (see, e.g. \cite{FT})

\be\label{quadr-r}
\left\{L_i(\lambda)\stackrel{\otimes}{,}L_j(\lambda')\right\} =
\delta_{ij}
\left[ r(\lambda-\lambda'),\ L_i(\lambda)\otimes L_i(\lambda')\right]
\ee
with the rational $r$-matrix
\be\label{rat-r}
r(\lambda) = \frac{1}{\lambda}\sum_{a=1}^3 \sigma^a\otimes \sigma^a
\ee

The crucial property of this relation is that it
is multiplicative and any product like (\ref{monomat})
satisfies the same relation

\be\label{Tbr}
\left\{T(\lambda)\stackrel{\otimes}{,}T(\lambda')\right\} =
\left[ r(\lambda-\lambda'),\
T(\lambda)\otimes T(\lambda')\right]
\ee

The Poisson brackets of the dynamical variables $S_a$, $a=1,2,3$
(taking values in the algebra of functions)
are fixed by (\ref{quadr-r}) and are just

\be\label{Scomrel}
\{S_a,S_b\} = -i\epsilon_{abc} S_c
\ee
i.e. the vector
$\{S_a\}$ plays the role of angular momentum (``classical spin'')
giving the name ``spin-chains'' to the whole class of systems.
Algebra (\ref{Scomrel}) has an obvious Casimir function
(an invariant, which Poisson commutes with all the spins $S_a$),

\be\label{Cas}
K^2 = {\bf S}^2 = \sum_{a=1}^3 S_aS_a
\ee

The spectral curve (\ref{specTC0}) is explicitly now

\be\label{scsl2m}
w+{Q_{N_f}(\lambda)\over w}=2P_{N_c}(\lambda),\ \ \ \
2P_{N_c}(\lambda)\equiv\Tr T(\lambda),\ \
Q_{N_f}(\lambda)\equiv \det T(\lambda)
\ee
or in the hyperelliptic parameterization

\be\label{gm8}
Y^2=P_{N_c}^2(\lambda)-Q_{N_f}(\lambda)
\ee
Zeroes of $Q_{N_f}(\lambda)$ define the masses of the
hypermultiplets. Since

\be\label{detTxxx}
\det_{2\times 2} L_i(\lambda) = (\lambda+\lambda_i)^2 - K^2
\ee
one gets

\be
Q_{N_f}(\lambda)=\det_{2\times 2} T(\lambda) = \prod_{i=1}^{N_c}
\det_{2\times 2} L_i(\lambda) =
\prod_{i=1}^{N_c} \left((\lambda + \lambda _i)^2 - K_i^2\right) = \\
= \prod_{i=1}^{N_c}(\lambda - m_i^+)(\lambda - m_i^-)
\ee
where we assumed that the values of spin $K$ can be different at
different sites of the chain, and

\be
m_i^{\pm} = -\lambda_i \pm K_i.
\label{mpm}
\ee
Thus we obtain that $N_f$ is generally equal to $2N_c$.
Note that the hypermultiplet masses being external parameters in the
gauge theory are also such in the spin chain. Indeed, there are still
$N_c-1$ integrals of motion that parameterize the spectral curve and the
moduli space (i.e. parameterize the Coloumb branch of gauge theory), but
the system also depends on $2N_c$ additional external parameters, $N_c$
Casimir functions and $N_c$ inhomogeneities.

While the determinant of the monodromy matrix (\ref{detTxxx})
depends on dynamical variables
only through the Casimirs $K_i$ of the Poisson algebra, the dependence of
the trace $\Tr_{2\times 2}T(\lambda)$ is less trivial.
Still, it depends
on $S_a^{(i)}$ only through Hamiltonians of the spin chain (which are not
Casimirs but Poisson-commute with {\it each other}) -- see further details in
\cite{XXX}.

Let us note that we have some additional freedom in the definition of the
spin chain and the spectral curve. Namely, note that $r$-matrix (\ref{rat-r})
is proportional to the permutation operator $P(X\otimes Y)=Y\otimes X$.
Therefore, it commutes with any matrix of the form $U\otimes U$. Thus, one
can multiply Lax operator of the spin chain by arbitrary constant matrix
without changing the commutation relations and conservation laws.
Moreover,
one can also attach a constant (external magnetic field) matrix $V$ to the
end of the chain (to the $N_c$-th site). This is the same as to consider more
general boundary conditions -- those with arbitrary matrix $V^{-1}$. This is
why such a model is called twisted.

The described freedom allows one to fit easily the form of the spectral curve
proposed in \cite{SW2,AS}

\be\label{AS}
w+{Q(\lambda)\over w}=P(\lambda),\\ P(\lambda)=\prod_{i=1}^{N_c}
(\lambda-\phi_i),\ \ \ \ Q(\lambda)=h(h+1)\prod_{j=1}^{2N_c}
\left(\lambda-m_j-{2h\over n}\sum_i m_i\right),\\
h(\tau)={\theta_2^4\over \theta_4^4-\theta _2^4}
\ee
where $\tau$ is the bare curve modular parameter and $\theta_i$ are the
theta-constants.

It can be done, e.g., by
choosing the matrices $U$ and $V$ to be\footnote{To fit (\ref{AS}),
we also need to shift $\lambda_i\to\lambda_i-{2h\over n}\sum_i m_i$.}

\be\label{U}
U_i=\left(
\begin{array}{cc}
1&0\\
0&\alpha_i
\end{array}
\right)\ ,\ \ \ \ \
V=\left(
\begin{array}{cc}
1&h(h+1)\\
\prod_i\alpha_i&0
\end{array}
\right),
\\
\hbox{i.e.}\ \ \ \ \
V^{-1}={1\over\det V}\left(
\begin{array}{cc}
0&-h(h+1)\\
-\prod_i\alpha_i&1
\end{array}
\right)
\ee

Now let us briefly consider degenerating our UV finite system to an
$N_{f}<2N_c$ case. This can be done in the
standard way \cite{SW2} by sending $l$ masses $m_1,...,m_l$ to infinity
while keeping $\Lambda_{QCD}^l\equiv e^{i\pi\tau}m_1...m_l$ finite.
After this procedure the modular forms disappear from (\ref{AS}) so that
$\Lambda_{QCD}$ emerges instead.

Degenerations of the system can be studied
at a single site (for the sake of brevity, we omit the index of
the site). Let us consider the Lax operator $\tilde L=UL$ (see
(\ref{LaxXXX}), (\ref{U})) with spins satisfying the Poisson brackets
(\ref{Scomrel}). We are going to send $\alpha$ to zero. It still
reserves
two possibilities to get nontrivial Lax operator. The first possibility,
when the both masses (\ref{mpm})
disappear and one reaches the pure gauge theory, is described by
the periodic Toda chain. In order to get it, one needs to
redefine $S_+\to {1\over\alpha}S_+$, then to send
$\alpha$ to zero and to remove, after this, the inhomogeneity by the shift of
$S_0$. This brings us to the Lax operator of the form
(we introduce new notations $S_0=S_3$, $S_{\pm}=S_1\pm iS_2$)

\be
\left(
\begin{array}{cc}
\lambda+S_0&S_-\\
S_+&0
\end{array}
\right)
\ee
so that the Poisson brackets are

\be
\left\{S_{\pm},S_0\right\}=\pm S_{\pm},\ \ \ \ \left\{S_+,S_-\right\}=0
\ee
This algebra is realized in new (Heisenberg) variables $p$ and $q$

\be
S_{\pm}=\pm e^{\pm q},\ \ \ \ S_0=-p,\ \ \ \ \{p,q\}=1
\ee
This leads us finally to the Toda chain Lax operator (\ref{LTC})
and the Toda spectral curve (\ref{specTC}).

Now let us return to the second possibility of the asymmetric degeneration,
when one of the masses (\ref{mpm}) remains in the spectrum
while the second one goes to infinity. One can understand from (\ref{mpm})
that, in contrast to the Toda case, this degeneration requires a special fine
tuning of the Casimir function and inhomogeneity, so that both of them go to
infinity but their sum (difference) remains finite. In the Lax operator,
it can be done in the following way. Let us redefine $S_+\to {1\over\alpha}S_+$
{\it and} $S_0\to {1\over\alpha}S_0$. This means that the Poisson brackets
take the form

\be\label{specal}
\{S_+,S_-\}=-2S_0, \ \ \ \ \{S_{\pm},S_0\}=0
\ee
Now in order to preserve the finite Lax operator (\ref{LaxXXX}), one needs to
take care of its element $L_{11}(\lambda)$. This can be done by rescaling
$\lambda_i\to {1\over\alpha}\lambda_i$ and fixing
$\lambda_i+S_0$ to be $\alpha\cdot s_0$. This brings us to the
Lax operator

\be\label{Laxi}
L(\lambda)=\left(
\begin{array}{cc}
\lambda+ s_0& S_-\\
S_+& \lambda_i-S_0
\end{array}
\right)
=\left(
\begin{array}{cc}
\lambda+s_0& S_-\\
S_+& -2S_0
\end{array}
\right)
\ee
The determinant of this Lax operator is equal to $(\lambda -m)$ where $m$
is the finite mass
\be
m=2s_0S_0+S_+S_-
\ee
in perfect agreement with (\ref{mpm}).
Let us note that $2s_0S_0+S_+S_-$ is also the Casimir
function of the algebra (\ref{specal}), since $\{S_{\pm},s_0\}=\pm S_{\pm}$.

Let us note that another (equivalent) way to count all possible degenerations
\cite{Khar} is to consider $2\times 2$ Lax operator of the most general
form linear in $\lambda$, which satisfies the Poisson brackets
(\ref{quadr-r}) with the rational $r$-matrix (\ref{rat-r}) and to determine
all  Casimir functions with respect to this Poisson brackets. Then,
all possible degenerations are determined by
vanishing the Casimirs functions.

Let us describe another possible interpretation of formula (\ref{mpm}).
Consider a free particle in $d=4$ with the dispersion law

$$
E^{2}=\vec p^{2} +m^{2}
$$
and note that in the complex momentum space there is the level crossing
at  $\vec p=0$   in the massless case and on the surface
$p^{2}=-m^{2}$  in the massive situation. According
to general logic of the Berry phase phenomenon,
if one considers the momentum space as a parameter space
each level crossing point is associated with monopoles and each level
crossing surface with ``defects" of higher dimensions.

Therefore, the masses define the locations of singularities
in the momentum space, indeed. In section 8, we shall see
that at the points of the space of scalar field zero modes
 with coordinates equal to masses, there are branes
of different dimensions located. This fits the above arguments,
since, due to the gauge invariance, momenta and gauge fields
enter the equations of motion combined in covariant derivatives.

The monopole singularity at $\vec p=0$  in the massless case
results in an additional term in the rotation operator
in the momentum space, the phenomenon well-known
for the rotation operator in the monopole background.
Moreover, the monopole contribution to the rotation operator
allows one to define a chirality transformation
for the spin 0, 1 ,2 particles in free theory
as the rotation when carrying
over closed contour in the momentum space.
In the interacting theory, the chirality transformation
is generically anomalous and, therefore, such a simple reasoning fails.
However, the anomaly itself can be described
in terms of the Berry phase framework in the space of fields both for the
external and internal anomalies. Then, one has again a satisfactory
qualitative agreement, since the chiral invariance is broken either by mass
or by anomaly: in both cases the violation is related with
the appearance of singularities in the ``generalized momentum space".
An analogy with the Peierls model discussed in this context
in \cite{peierls} provides an additional support for this point of view.

To conclude this long subsection, let us mention two other deformations
of the original gauge theory. First of all, as in all other cases, one may
consider the theory with a different gauge group ${\sf G}$.
This case is treated
similarly to the Toda case by introducing reflection matrices and
considering the spin chain on the Dynkin diagram of
$\hat {\sf g}~{\vee}$ \cite{GorMir}.
Then, the spectral curves can be described by the general formula

\be
{\cal P}(\lambda, w) = 2P(\lambda) - w - \frac{Q(\lambda)}{w}
\label{curven}
\ee
Here $P(\lambda)$ is the characteristic polynomial of the
group ${\sf G}$ for all ${\sf G}\ne C_n$, i.e.

\be
P(\lambda) = \det(g - \lambda I) =
\prod_i (\lambda - \lambda_i)
\ee
where determinant is taken in the first fundamental representation
and $\lambda_i$'s are the eigenvalues of the algebraic element $g$.
For the pure gauge theories with the classical groups \cite{MW},
${Q}(\lambda)=\lambda^{2s}$ and\footnote{In the symplectic
case, the curve can be easily recast in the form with polynomial
$P(\lambda)$ and $s=0$.}

\be
A_{n-1}:\ \ \ P(\lambda) = \prod_{i=1}^{n}(\lambda - \lambda_i), \ \ \
\ s=0;\\
B_n:\ \ \ P(\lambda) = \lambda\prod_{i=1}^n(\lambda^2 - \lambda_i^2),
\ \ \ s=2;\\
C_n:\ \ \ P(\lambda) = \prod_{i=1}^n(\lambda^2 - \lambda_i^2)-
{2\over\lambda^2},
\ \ \ s=-2;\\
D_n:\ \ \ P(\lambda) = \prod_{i=1}^n(\lambda^2 - \lambda_i^2),
\ \ \ s=2
\label{charpo}
\ee
For exceptional groups, the curves arising from the characteristic
polynomials of the dual affine algebras do not acquire the hyperelliptic
form.

In order to include $N_F$ massive hypermultiplets in the
first fundamental representation one can just change
$\lambda^{2s}$ for ${Q}(\lambda) = \lambda^{2s}
\prod_{\iota = 1}^{N_F} (\lambda - m_\iota)$ if ${\sf G}=A_n$
and for ${Q}(\lambda) = \lambda^{2s} \prod_{\iota =
1}^{N_F}(\lambda^2 - m^2_\iota)$ if ${\sf G}
=B_n,C_n,D_n$ \cite{AS,fuma,GorMir}.

Another possibility of deformation, letting
the gauge groups be products of simple factors
and bi-fundamental matter be added,
was proposed in \cite{W}. This system
is described by a higher, $sl(p)$ \XXX magnet \cite{GGM1},
with the Lax operator given by the $p\times p$ matrix at each site

\be
L_i(\lambda)=K^{ab}S_{a,i}X_a+(\lambda+\lambda_i)\cdot{\bf 1}
\ee
where $S_{a,i}$ are dynamical variables, $X_a$ are generators of the $sl(p)$
algebra and $K^{ab}$ is its Killing form.

Thus, we described the families of spectral curves for theories with
fundamental matter. Note that, at any \XXX spin chain, the generating
differential has the same form as in the Toda theory (\ref{dSTC}),
$dS=\lambda dw/w$.

\subsection{$5d$ SQCD: twisted \XXZ chain}

Now we describe the integrable system behind the $5d$ theory with fundamental
matter hypermultiplets. The theory is considered with one dimension
compactified onto the
circle of radius $R_5$, i.e. given on the space ${\bf R}^4\times S^1$.
We start from the UV finite theory, i.e. that with $N_f=2N_c$ hypermultiplets.
The corresponding integrable theory is the inhomogeneous \XXZ spin chain
\cite{GGM2,MarMir}.

The Lax matrix for the \XXZ ($sl(2)$) spin magnet
has the form

\be\label{l-gen}
L(\mu)\,=\,\left(
\begin{array}{cc}
\mu e^{S_0}-\mu^{-1}e^{-S_0} & 2S_-\\
2S_+ & \mu e^{-S_0}-\mu^{-1}e^{S_0}
\end{array}\right)
\ee
Thus, this Lax operator is given on a cylinder.
In fact, from the point of view of integrable systems, some more natural
spectral parameter is $\zeta$: $\mu=e^{\zeta}$, and the Lax operator
(\ref{l-gen}) becomes clearly trigonometric.
This Lax operator is intertwined by the trigonometric $r$-matrix

\be\label{trigrmatrix}
r(\zeta)={i\over\sinh\pi\zeta}
\left(\sigma_1\otimes\sigma_1+\sigma_2\otimes\sigma_2+
\cosh\pi\zeta\sigma_3\otimes\sigma_3\right)
\ee
so that the Poisson bracket of the Lax operators (\ref{quadr-r})
gives rise to the Poisson brackets of the algebra of $S_i$'s:

\be\label{pois}
\{S_+,S_0\}=\pm S_{\pm};\ \ \ \{S_+,S_-\}=\sinh 2S_0
\ee
The second Casimir function of this algebra is

\be\label{Casimir}
C_2=\cosh 2S_0+2S_+S_-
\ee

The non-linear commutation relations (\ref{pois}) are ones from
the quantum deformed algebra $U_q(sl(2))$.
This generalizes the fact that the
\XXX magnet is described by the Poisson brackets that reproduce the
classical $sl(2)$ algebra.
In these Poisson brackets, the Plank constant $\hbar$ turns out to
be inessential and can be put equal to unity. In fact, it is proportional to
the radius $R_5$ of the space-time circle in the corresponding $5d$ SUSY
theory and can be easily restored with the replace of any generator
$S_i$ by $R_5S_i$. Hereafter, we omit $R_5$ from all the formulas.

Following the standard procedure, now we consider
the chain with $N_c$ sites with the Lax operators (\ref{l-gen}) associated with
each site and commuting with each other,
introduce the inhomogeneities $\zeta_i$
which depend on the site of chain by the replace $\zeta\to\zeta+\zeta_i$
and impose
the periodic boundary conditions. On this chain, the transfer matrix acts
and the periodic boundary conditions generate the spectral curve and
the conservation laws.

The manifest form of the curve in the \XXZ case can be derived
using explicit expression for the Lax matrix (\ref{l-gen}) and
reads\footnote{In fact, the coefficient in front of $P_{N_c}$, instead of 2,
is equal to $2\cosh\prod_i^{N_c}e^{S_{0,i}}$. It is, however, the integral
of motion that can be put equal to any number. We always fix
$\sum_i^{N_c}S_{0,i}$ to be zero.}

\be\label{hrena}
w+{Q_{2N_c}\left(e^{2\zeta}
\right)\over e^{2N_c\zeta+2\sum_i\zeta_i}w}=2e^{-N_c\zeta-\sum_i\zeta_i}P_{N_c}
\left(e^{2\zeta}\right)=2e^{-N_c\zeta-\sum_i\zeta_i}\left(e^{2N_c\zeta}+
\ldots+e^{2\sum\zeta_i}\right)
\ee
Changing the variables $e^{2\zeta}=\mu^2\equiv \lambda$, $w\to
e^{-N_c\zeta-\sum_i\zeta_i}w$, this curve can
be recast in the hyperelliptic form {\em in $\lambda$ variables}:

\be\label{sc22}
w+{Q_{2N_c}\left(\lambda\right)\over w}=2P_{N_c}\left(\lambda\right),\
\ \ Y^2=P_{N_c}^2(\lambda)-Q_{2N_c}(\lambda),\ \ \ Y\equiv
\2\left(w-{Q_{2N_c}\left(\lambda\right)\over w}\right)
\ee
while in terms of  the ``true" spectral parameter
$\zeta$ this curve looks considerably more tricky. However, one can work with
the variable $\lambda$ taking into account that it lives on a cylinder
or sphere with two marked points.
In equation (\ref{sc22}), the polynomial

\be\label{212}
Q_{2N_c}(\lambda)=\prod_i \left(\lambda^2
-2K_{i}e^{2\zeta_i}\lambda+e^{4\zeta_i}\right)\equiv
\prod_i\left(\lambda-e^{2m_i^{(+)}}\right)
\left(\lambda-e^{2m_i^{(-)}}\right)
\ee
defines the masses $m_i^{(\pm)}$ of matter hypermultiplets
(cf. with (\ref{mpm}))

\be\label{mpm5}
m_i^{(\pm)}=\zeta_i\pm{1\over 2}\log\left(K_{i}+\sqrt{K_{i}^2-1}\right)
\ee
Note that, being written in terms of
the variable $\lambda$, the curve (\ref{sc22})
is very similar to the curves
arising for $4d$ theories. However, the difference coming from different
generating 1-differentials $dS$ turns
out to be very crucial (see, for example, the
discussion of residue formula and perturbative prepotentials in section 5).

It is easy to compare the curves (\ref{sc22}) with other curves
appeared in literature \cite{theisen5d},
rewriting them in the form

\be\label{sinhsc}
w+{\prod_{\alpha}\sinh\left(\zeta-m_{\alpha}\right)\over w}=2\prod_i
\sinh\left(\zeta-a_i\right)
\ee
where we rescaled $w\to e^{N_c\zeta+\sum_i\zeta_i}w$,
denoted the roots of the polynomial $P_{N_c}(\lambda)$ (\ref{hrena})
$\lambda_i \equiv e^{2a_i}$ and made use of formula (\ref{mpm5}) and
the manifest form of the leading and the constant terms
in this polynomial.
Comparing (\ref{sinhsc}) with (\ref{hrena}), one finds, in particular, that

\be\label{strange}
\sum a_i=\sum\zeta_i=\2\sum m_{\alpha}
\ee
The condition (\ref{strange}) is not, however, absolutely necessary in the
context of \XXZ chains and $5d$ theories.
It emerges only
in the standard \XXZ chain. In the context of \SW one needs rather to
consider, similarly to the $4d$ case,
the {\em twisted} \XXZ model \cite{kundu,GGM2,MarMir}.
It is characterized by the Lax operator

\be\label{l-gentw}
L_i(\mu)\,=\,\left(
\begin{array}{cc}
\mu e^{S_{0,i}-\zeta_i}-\alpha_i\mu^{-1}e^{-S_{0,i}+\zeta_i} & 2S_{-,i}\\
2S_{+,i} & \mu e^{-S_{0,i}-\zeta_i}-\beta_i\mu^{-1}e^{S_{0,i}+\zeta_i}
\end{array}\right)
\ee
with generally non-unit constants $\alpha_i$, $\beta_i$.
These constants provide an arbitrary coefficient in front of the
product in the r.h.s. of (\ref{sinhsc}) (which can be also fixed by
the integral of motion $\sum_i^{N_c}S_{0,i}$) and, thus, break the condition
(\ref{strange}). They are also important for
a careful treatment of dimensional
transmutation (i.e. decreasing the number of massive hypermultiplets), see
below.

The maximally degenerated case is related to the pure gauge theory when all the
masses become infinite. In this case, one gets
the relativistic Toda chain, s.3.6.

There certainly exist all intermediate degenerations when all but $N_f$
masses are brought to infinity. This system, corresponding to the gauge theory
with $N_f<2N_c$ massive hypermultiplets, is described by the chain with some of
the sites degenerated and some of them not. The corresponding spectral curve
has the form

\be
w+{Q_{N_f}(\lambda)\over w}=2P_{N_c}(\lambda),\ \ \
Q_{N_f}(\lambda)=\prod_{\alpha}^{N_f}\left(\lambda-e^{2m_{\alpha}}\right),
\ \ \ P_{N_c}(\lambda)=\prod_i^{N_c}\left(\lambda-e^{2a_i}\right)
\ee
or
\be\label{scd5'}
w+{\prod_{\alpha}^{N_f}\sinh\left(\zeta-m_{\alpha}\right)\over w}=
2e^{(N_c-N_f/2)\zeta+\sum a_i-1/2\sum m_{\alpha}}\prod_i^{N_c}
\sinh\left(\zeta-a_i\right)
\ee
(where we again rescaled $w\to e^{\2(N_f\zeta+\sum m_{\alpha})}w$;
this curve also coincides with that in \cite{theisen5d}).
As we already noted, the particular coefficient
$e^{\sum a_i-1/2\sum m_{\alpha}}$ can be, in principle, replaced by any other
number, say put equal to unity, which literally corresponds to
\cite{theisen5d}.
It does not influence the result and we will discuss the reason
below.

Note that the form of the spectral curve (\ref{sinhsc})
is perfectly designed for taking the
$4d$ limit. Indeed, one can restore $R_5$-dependence in this formula
multiplying each $\zeta_i$ and mass parameter by $R_5$. In terms
of algebra (\ref{pois}) it means that each generator has to be
multiplied by $R_5$ and so does the spectral parameter. Then, one
immediately reproduces the results of the previous subsection.

As in the $4d$ case,
the description of gauge theory in terms of the spin chain
allows an extension to the group product case. It is described by
higher  $sl(p)$ \XXZ magnets given by
the Lax operator

\be\label{laxhg}
L(\mu)=\sum_{i,j} e_{ij}\otimes L_{ij}
\ee
and

\be
L_{ii}=\mu e^{S_{0,i}}-\mu^{-1} e^{-S_{0,i}},\ \ \
L_{ij}=2S_{ji},\ (i\ne j) .
\ee
Here $S_{0,i}$ are associated with the vectors $\epsilon_i$,
realizing the simple roots
$\alpha_i=\epsilon_i-\epsilon_{i+1}$, and
$e_{ij}$ has the only non-vanishing matrix element $(i,j)$. The generators
$S_i$ satisfy non-linear Poisson brackets that can be read off from the
commutation relations of the quantum algebra $U_q(sl(p))$ like it has been
done in (\ref{pois}). These Poisson brackets can be
certainly obtained from the quadratic relation (\ref{quadr-r}) with the
trigonometric $r$-matrix for the $sl(p)$ case \cite{FT}.
Note that, specially degenerating this system, one can easily reproduce
the $(p,q)$-Web constructed in \cite{ahk} that corresponds to different toric
varieties.

Now let us say some words on the generating differential $dS$ in
$5d$ theories with fundamental matter.

The general scheme presented for $4d$ theory can be almost
literally transferred to the
$5d$ \N2 SUSY gauge models
with one compactified dimension. It can be described through
involving trigonometric $r$-matrices and $L$-operators instead
of rational ones, i.e. coming from Yangian to affine algebras. It means
that now it is natural to consider the {\it both} parameters
$\zeta={1\over 2}\log\lambda$ and $\log w$ as coordinates on the cylinder.

Similar arguments imply that, instead of differential $dS=\lambda
d\log w$, one now has to consider the differential \cite{nikita,WDVVlong}

\be\label{dSRTC}
dS = \zeta{dw\over w} \sim \log\lambda {dw\over w}
\ee
so that, despite the similarity of the $5d$ spectral curves with $4d$
ones, periods of $dS^{(5)}$ are different from those of $dS^{(4)}$.
Note that the derivatives of this differential w.r.t. moduli again give
holomorphic differentials.

In fact, one can interpret the $5d$ theory as the $4d$ theory with infinite
number of (Kaluza-Klein) vector multiplets with masses $M_n=n/R_5$ and
infinite number of analogous (Kaluza-Klein) fundamental
hypermultiplets.
Then, one can equally consider either
the just described $5d$ picture, or the
$4d$ picture that involves Riemann surface presented as an infinite order
covering (see the spectral curve equation in terms of the
variable $\zeta$ (\ref{hrena}))
with infinitely many punctures. This latter picture can be effectively
encapsulated in the usual hyperelliptic Riemann surface (\ref{sc22}) with
finite number of punctures, but, as a memory of infinitely many multiplets,
the spectral parameter $\lambda$ now lives on the cylinder. Meanwhile, the
differential $dS^{(5)}$ now evidently should be of the form (\ref{dSRTC})
which ``remembers" of its $4d$ origin.

\subsection{$5d$ pure gauge theory: relativistic Toda chain}

We are going to get the pure gauge theory in $5d$ from the theory with
fundamental matter via the dimensional transmutation, sending masses of all
the hypermultiplets to infinity. Therefore, we study the maximal
degeneration of the \XXZ spin chain \cite{GGM2,MarMir}.
This degeneration is obtained from the
twisted \XXZ chain introduced above and is described by the
relativistic Toda chain \cite{nikita} (see also \cite{WDVVlong}).
The twisting
means just introducing some new parameters into the Lax operator, which are
central elements of the Poisson bracket algebra. A special fine tuning of
these parameters allows one to match smoothly different regimes and limiting
cases of the \XXZ spin chain.

As it has been explained in \cite{GGM1}, there are
two equivalent ways to twist the integrable system. One of them, which was
applied in the \XXX case, is to multiply the Lax
operator by an arbitrary constant matrix $U$. This is possible in the \XXX
case, since the corresponding (rational) $r$-matrix commutes with the tensor
product $U\otimes U$. The situation is much more restrictive in the
trigonometric case, when the $r$-matrix commutes only with the matrices $U$
of the very special form (say, it can be proportional to any one of the Pauli
matrices). Therefore, in the trigonometric case we apply the second way of
doing \cite{kundu,Khar}, that is, we consider the Lax matrix of the general
form but with some prescribed dependence of matrix elements on the
spectral parameter.  Then, the Poisson bracket (\ref{quadr-r}) dictates
Poisson brackets of matrix elements and, in particular, fixes some
coefficients to be centers of the Poisson bracket algebra.

More explicitly, we fix the Lax matrix to be of the form

\be\label{l-gener}
L(\mu)\,=\,\left(
\begin{array}{cc}
\mu A^{(+)}+\mu^{-1}A^{(-)} & B\\
C & \mu D^{(+)}+\mu^{-1}D^{(-)}
\end{array}\right)
\ee
and to be intertwined by the same $r$-matrix (\ref{trigrmatrix}). Then, up to
inessential total normalization, one can rewrite the Lax operator in the form

\be\label{l-general}
L(\mu)\,=\,\left(
\begin{array}{cc}
\mu e^{S_0}-\rho\mu^{-1}e^{-S_0} & 2S_-\\
2S_+ & \gamma\mu e^{-S_0}-\nu\mu^{-1}e^{S_0}
\end{array}\right)
\ee
where $\rho$, $\nu$ and $\gamma$ are
constants (centers of algebra).

In this case, the commutation relations are slightly changed to

\be\label{poisgen}
\{S_+,S_0\}=\pm S_{\pm};\ \ \ \{S_+,S_-\}={1\over 2}\left(
\nu e^{2S_0}-\rho \gamma e^{-2S_0}\right)
\ee
and the second Casimir function is equal to

\be
C_2={1\over 2}\left(\rho\gamma e^{-2S_0}+\nu e^{2S_0}\right)+2S_+S_-
\ee
thus, the determinant of the Lax operator (\ref{l-general}) is still the
quadratic polynomial with coefficient being the Casimir function.

Now we are ready to demonstrate how the limit to the relativistic Toda chain
can be done. To make the Lax operator (\ref{l-general}) looking more similar
to the relativistic Toda Lax operator, we multiply it by the function
$e^{S_0}$.
In the whole chain it results into the factor
$e^{\sum_i^{N_c}S_{0,i}}$ which is the integral of
motion and can be put zero. In particular, in the relativistic Toda case this
integral is equal to the full momentum of the system. Although being so
inessential, this redefining of the Lax operator still requires to modify
some expressions \cite{Khar}. For instance, the new Lax operator

\be\label{laxnew}
L(\mu)\,=\,\left(
\begin{array}{cc}
\mu e^{2S_0}-\rho\mu^{-1} & 2S_-e^{S_0}\\
2S_+e^{S_0} & \gamma\mu-\nu\mu^{-1}e^{2S_0}
\end{array}\right)
\ee
is intertwined by the new $r$-matrix (here $r$ is the $r$-matrix
from (\ref{trigrmatrix}))

\be
r^{(tw)}=r+{1\over 2}\left({\bf I}\otimes\sigma_3-
\sigma_3\otimes {\bf I}\right)
\ee
which is called twisted. This name comes from the quantum
generalization of this matrix which can be obtained from the standard quantum
trigonometric $R$-matrix by twisting \cite{Khar}

\be
R^{(tw)}(u)\,=\,F_{12}(\gamma)R(u)F^{-1}_{21}(\gamma)'\ \ \
F_{12}\;\equiv\;F^{-1}_{21}\,=\,
\exp\Big\{\frac{1}{2}\big(\sigma_3\otimes {\bf I}-
{\bf I}\otimes \sigma_3\big)\Big\}
\ee
An important property of the Lax operator (\ref{laxnew}) is that its
determinant is no longer dependent on only the Casimir function but also on
$e^{S_0}$. The product of this quantity over all the sites is still certainly
an integral of motion.

Now one can consider the particular reduction \cite{Khar}

\be
\gamma=\nu=0,\ \ \rho=1
\ee
i.e.

\be
L(\mu)\,=\,\left(
\begin{array}{cc}
\mu e^{2S_0}-\mu^{-1} & 2S_-e^{S_0}\\
2S_+e^{S_0} & 0
\end{array}\right)
\ee
Taking into account the commutation relations (\ref{poisgen}), that is,
$\{S_+,S_-\}=0$, one can realize the algebra as $S_{\pm}=e^{\pm q}$, $S_0=p$,
$\{p,q\}=1$ and reproduce the standard Lax operator for the relativistic
Toda chain \cite{relToda}.

The spectral curve corresponding to this degenerated case can be easily
written

\be
w+{1\over w}={1\over\lambda^{N_c/2}}P_{N_c}(\lambda)
\ee

Now one can also obtain from
the theory with $N_f=2N_c$
the theory with less number of massive multiplets
degenerating the general \XXZ chain at several sites. The curves for such
theories we already discussed in the previous subsection.

At last, let us note that the differential $dS$ constructed for the
\XXZ spin chain (\ref{dSRTC})
does not change when degenerating to the relativistic Toda chain.

\subsection{$5d$ adjoint matter: \RS system}

Now we are going to study the system emerging upon adding in $5d$ theory
the adjoint matter hypermultiplet. This UV finite system can
be described differently as follows. By starting with a five dimensional model
one may obtain four dimensional \N2 SUSY models (with fields only
in the adjoint representation of the gauge group)
by imposing non-trivial boundary conditions on half of the fields:

\be
\phi(x_5 +R_5) = e^{2i\epsilon}\phi(x_5).
\label{bc}
\ee
If $\epsilon = 0$ one obtains ${\cal N}=4$ SUSY in four dimensions,
but when $\epsilon \neq 2\pi n$ this is explicitly broken to \N2.
The low-energy mass spectrum of the  four dimensional theory contains
two towers of Kaluza-Klein modes:

\be
M_n = \frac {\pi n}{R_5}\ \ {\rm and} \ \
M_n = \frac{\epsilon + \pi n}{R_5}, \ \ \ n\in { Z}.
\label{spectrum}
\ee

According to the proposal of \cite{nikita},
this five dimensional theory may be
associated  with the elliptic \RS integrable model \cite{rru2,rud}.
In various double-scaling limits it reduces to systems we already studied:

(a) If $R_5\rightarrow 0$ (with finite $\epsilon$) the
(finite) mass spectrum (\ref{spectrum}) reduces to a single point
$M_n = 0$. This is the standard four dimensional \N2 SUSY pure gauge model
associated with the periodic Toda chain. As we already know,
in this situation \N2 SUSY in four dimensions is insufficient to ensure
UV-finiteness, thus $\tau \rightarrow i\infty$, but the phenomenon of
dimensional transmutation occurs whereupon one
substitutes the dimensionless $\tau$ by
the new dimensional parameter $\Lambda_{QCD}^{N_c} =
e^{2\pi i\tau}  (\epsilon/R_5)^{N_c}$.

(b) If $R_5 \rightarrow 0$ and $\epsilon \sim MR_5$ for finite $M$,
then UV finiteness is preserved. The mass spectrum (\ref{spectrum})
reduces to the two points $M_n = 0$ and $M_n = M$. This is the
four dimensional YM model with ${\cal N}=4$ SUSY softly broken to \N2.
The associated finite-dimensional integrable system
is, as we know, the elliptic Calogero model.
Case (a) is then obtained from (b) by the double scaling limit.

(c) If $R \neq 0$ but $\epsilon \rightarrow i\infty$ the
mass spectrum (\ref{spectrum}) reduces to a single
Kaluza-Klein tower, $M_n = \pi n/R_5$, $n\in Z$. This compactification of
the five dimensional model has ${\cal N}=1$ SUSY and is not
UV-finite. Here $\tau \rightarrow i\infty$ and
$\epsilon \rightarrow i\infty$, such that $2\pi\tau - N_c\epsilon$
remains finite. This system as we discussed in the previous subsection is
described by the relativistic Toda chain.

(d) Finally, when $R_5\neq 0$ and $\epsilon$ and $\tau$
are both finite one distinguished case still remains:
$\epsilon =\pi/2$.\footnote{The case $\epsilon = 0$
of fully unbroken five dimensional \N2 supersymmetry
is of course also distinguished, but trivial:
there is no evolution of effective couplings (renormalisation
group flows) and the integrable system is just that of $N_c$
non-interacting (free) particles.}
Here only periodic and anti-periodic boundary conditions
occur in the compact dimension. This is the
case analyzed in \cite{bmmm1}. It is clearly special
from both the point of view of Yang-Mills theory and integrable
systems.

Thus, the elliptic \RS model is a remarkable completely integrable system whose
various limits include the (finite dimensional)
Toda and Calogero-Moser models.
Here we shall review only a few of its salient features.
More comprehensive accounts of its structure
and applications may be found in \cite{ruijrev,ruijh}.
The $SU(N_c)$ model has an explicit Lax representation with
Lax operator \cite{rru2,rud,BC,RL}

\be
{\cal L}_{ij} = c(\xi|\epsilon)
e^{P_i} \frac{F(q_{ij}|\xi)}{F_(q_{ij}|\epsilon)}.
\label{RLa}
\ee
Here

\be
e^{P_i} = e^{p_i}
\prod_{l\neq i} \sqrt{\wp (q_{il})-\wp(\epsilon)}.
\label{RLb}
\ee
where the $p_i$ and $q_i$ are canonically conjugate momenta and
coordinates, $\{p_i, q_j\} = \delta_{ij}$.
The commuting Hamiltonians may be variously written

\be
 H_k =
\sum_{{J\subset \{1,\ldots n\}}\atop{ |J|=k}}
e\sp{\sum_{j\in J} p_j}
\prod_{{j\in J}\atop{ k\in \{1,\ldots n\}\backslash J}}
\frac{F(q_{ij}|\xi)}{F_(q_{ij}|\epsilon)}=
\sum_{1\leq i_1<\ldots<i_k\leq N_c} e^{P_{i_1}+\ldots+P_{i_k}}
\prod_{a<b}\frac{1}{\wp(q_{i_ai_b})-\wp(\epsilon)}.
\label{Hams}
\ee
(The final product is taken to be unity in the case $k=1$.)
These Hamiltonians arise in the description of the spectral curve.
Note that, after trigonometric degeneration, these Hamiltonians are proportional
to those obtained in s.2.2. We discuss this degeneration in more details
in section 5.

The integrability of the model depends on the functional
equations satisfied by $F$ \cite{rru2,rud}; the connection with functional
equations and integrable systems is part of a larger story \cite{funl}.

Some comment on the parameters $\xi$ and $\epsilon$ in these
formulae is called for. Here $\xi$ is precisely the spectral parameter
associated with the bare spectral curve, where the Lax operator
 is given. Further, the
additional parameter $\epsilon$ in (\ref{RLa}), (\ref{RLb}) above
is the  same parameter we introduced in (\ref{bc})
characterizing the boundary conditions.
Actually the integrability of  (\ref{Hams}) does not require
$\epsilon$ to be real, but such a choice guarantees the reality of the
the Hamiltonians. The identification of these two parameters has been
simplified by our choice of the real period of $\wp$ being $\pi$
so that both (\ref{bc}) and (\ref{RLa}), (\ref{RLb}) are
manifestly $\pi$ periodic.
The \lq\lq non-relativistic limit" $\epsilon \rightarrow 0$
which leads to the Calogero-Moser system means we can identify the
mass $M=\epsilon/R_5$ of the gauge multiplet with the Calogero-Moser
coupling constant.
The special point $\epsilon = \pi/2 $ ($=\omega$)
singled out earlier is now a half-period of the $\wp$-function,
and at this point $\wp'(\frac{\pi}{2}) = 0$.

Now inserting the Lax operator (\ref{RLa}) into the determinant (\ref{fscR})

\be\label{fscR}
\det_{N_c\times N_c} \left({\cal L}^{R}(\xi) - \lambda\right) = 0
\ee
one gets the spectral curve equation. Its coefficients are expressed through the
Hamiltonians (\ref{Hams}) and parameterize the corresponding moduli space.

To conclude constructing \SW we should determine the differential $dS$.
It is evident from our previous discussions that it is to be of the form

\be\label{dSR}
dS={1\over R_5}\log\lambda d\xi
\ee
Indeed, on one hand, it should be proportional to $d\xi$ since describes the
adjoint matter (compare with the Calogero case). On the other hand, the
proportionality coefficient is to be $R_5^{-1}\log\lambda$, since the \RS
system describes a $5d$ theory.

Now let us discuss more the spectral curve arising.
{}From the point of view of classical integrability the
overall normalization factor $c(\xi|\epsilon)$ of the Lax operator
does not change the integrals of motion apart from scaling.
This normalization factor does however lead to the rescaling of
$\lambda$ and this effects the explicit form of
the 1-form $dS$.\footnote{
Note that although the choice of  normalization $c(\xi|\epsilon)$
(or, equivalently, rescaling $\lambda$)
can change the manifest expression
for $dS$, it affects neither the symplectic 2-form
${d\lambda\over\lambda}\wedge{dw\over w}$,
nor the period matrix, which is the second derivative of the
prepotential.
}
 We shall see below
that significant simplifications occur with the choice

\be
c(\xi|\epsilon) =
c_0(\xi|\epsilon) \equiv
 \frac{1}{\sqrt{ \wp(\xi)-\wp (\epsilon)}}=
\frac{\sigma(\xi)\sigma(\epsilon)}{\sqrt{\sigma(\epsilon-\xi)
\sigma(\epsilon+\xi)}} = \sqrt{c_+(\xi|\epsilon)c_-(\xi|\epsilon)}
\label{cfun}
\ee
where

\be
c_{\pm}(\xi|\epsilon)\equiv \frac{\sigma(\xi)\sigma(\epsilon)}
{\sigma(\epsilon\pm\xi)}=\mp\frac{1}{F(\epsilon|\mp\xi)}.
\ee
Similar issues of normalization enter into discussion of
separation of variables \cite{separation}.

Before generalizing to the higher rank situation it is instructive
to first consider the case of gauge group $SU(2)$.
In the formulas of the previous section this corresponds to
setting \N2 and working in the center-of-mass frame $p_1+p_2=0$.
We define $p\equiv p_1$ and $q\equiv q_1-q_2$.

Using the explicit form of the Lax operator the
spectral curve (\ref{fscR}) is found to be

\be
\lambda^2 - cu\lambda + c^2(\wp(\xi) - \wp(\epsilon)) =
\lambda^2 - cu\lambda + \frac{c^2}{c_0^2} = 0,
\label{SU2c}
\ee
where

\be
u\equiv  H_1(p,q)=2\cosh p\sqrt{\wp (q)-\wp (\epsilon)}
\ee
is the  Hamiltonian (\ref{Hams}).
For the choice   $c(\xi|\mu) = c_0(\xi|\mu)$ this simplifies to yield

\be
\lambda^2 - uc_0 \lambda + 1 = 0,
\ee
or simply

\be
c_0^{-1}(\xi) = \frac{u\lambda}{\lambda^2 + 1} =
\frac{u}{\lambda + \lambda^{-1}}.
\label{SU2c1}
\ee
Observe that with our choice (\ref{cfun}) this equation,
which describes the Seiberg-Witten spectral curve, may be expressed
in the form

\be
u= H(\log\lambda,\xi).
\ee
Comparison with (\ref{dSR}) shows that the generating differential
$R_5 dS=\log\lambda d\xi$ takes the form $pdq$ here.

We now consider the general $SU(N_c)$ model. The first step is to
evaluate the spectral curve (\ref{fscR}) for the Lax matrix (\ref{RLa}).
Expanding the determinant about the diagonal yields

\be
\sum_{k=0}^{N_c} (-\lambda)^{N_c-k}c^k\left\{
\sum_{1\leq i_1<\ldots<i_k\leq N_c} e^{P_{i_1}+\ldots+P_{i_k}}
{\det}_{(ab)} \frac{F(q_{i_ai_b}|\xi)} {F(q_{i_ai_b}|\epsilon)}
\right\} = 0,
\label{Rsc1}
\ee
and Ruijsenaars \cite{rru2,rud} expressed the determinants appearing here
in terms of the Hamiltonians (\ref{Hams}) by means of a generalized
Cauchy formula.
This expansion was re-obtained in \cite{bmmm2} using a simple fact that
generalized Cauchy formulae can be derived in terms of
free-fermion correlators via Wick's theorem for fermions.
The \RS model (and its spin
generalizations) may be understood \cite{RL,R+T} in terms of a reduction
of the Toda lattice hierarchy. The Hirota bilinear identities
of that hierarchy may be expressed in terms of free-fermion
correlators \cite{DJKM,versus},
and are the origin of those here. We believe that
the free field expansions will ultimately lead to a better
field theoretic understanding of the appearance of these integrable
systems. The machinery of free-fermion correlators has already found
use, for example, in calculating within the context of the
Whitham hierarchy for these integrable
systems \cite{RG}.

The formulas for the determinants are

\be
\tilde D_{I_k} \equiv
{\det}_{(ab)} \frac{F(q_{i_ai_b}|\xi)} {F(q_{i_ai_b}|\epsilon)}
= (-)^{k(k-1)\over 2}
\frac{\sigma^{k-1}(\xi - \epsilon) \sigma(\xi + (k-1)\epsilon)}
{\sigma^k(\xi) \sigma^{k(k-1)}(\epsilon)}
\prod_{a<b} \frac{\sigma^2(q_{i_ai_b})\sigma^2(\epsilon)}
{\sigma(\epsilon+q_{i_ai_b}) \sigma(\epsilon-q_{i_ai_b})},
\label{tDk}
\ee
where $I_k = \{i_1,\ldots,i_k\}$ denotes the set of $k$ indices
and $a,b=1,...,k$.
The right hand side of this expression is in fact a doubly
periodic function of all the arguments ($q_i$, $\xi$ and $\epsilon$)
and so may rewritten in terms of \W functions.
In particular, upon using (\ref{wpaddn}), we find that
\be
\prod_{a<b} \frac{\sigma^2(q_{i_ai_b})\sigma^2(\epsilon)}
{\sigma(\epsilon +q_{i_ai_b}) \sigma(\epsilon -q_{i_ai_b})}
= \prod_{a<b} \frac{1}{\wp (q_{i_ai_b})-\wp (\epsilon))},
\ee
thus recovering the Hamiltonians (\ref{Hams}) noted earlier.

Now we are ready to obtain the final formula for the spectral curve in the
\RS system.
Upon substitution of (\ref{tDk}) into (\ref{Rsc1})
the curve takes the form

\be
\sum_{k=0}^{N_c} (-\lambda)^{N_c-k}c^k\left\{
\sum_{I_k} e^{P_{i_1}+\ldots+P_{i_k}}\tilde D_{I_k}\right\} =
\sum_{k=0}^N (-\lambda)^{N_c-k} c^kD_k(\xi|\epsilon)\  H_k
=0.
\label{Rsc}
\ee
Here we have collected the $q$-independent factors in (\ref{tDk})
into $D_k$ where

\be
D_k(\xi|\epsilon) =
(-)^{k(k-1)\over 2}
\frac{\sigma^{k-1}(\xi - \epsilon) \sigma(\xi + (k-1)\epsilon)}
{\sigma^k(\xi) \sigma^{k(k-1)}(\epsilon)}
 =\frac{1}{c_-\sp{k}}. \frac{\sigma(\xi + (k-1)\epsilon)}{
\sigma(\xi - \epsilon)}.\frac{(-1)\sp{k(k-1)\over 2}}
{\sigma^{k(k-2)}(\epsilon)}.
\label{Dfun}
\ee
Again this is doubly periodic function in both $\xi$ and $\epsilon$
and so is expressible in terms of the \W function and
its derivative, though the explicit formulae are rather complicated.
Although it may appear from the last expression in
(\ref{Dfun}) that (\ref{Rsc}) simplifies when $c=c_-$, such a choice
would break the double periodicity of our spectral curve and so
is inappropriate for the fully elliptic model\footnote{However, in
the perturbative limit when one of the periods becomes infinite
this choice is then available.}.

Part of the difficulty in dealing with the elliptic \RS
and Calogero-Moser models is the complicated nature of these
spectral curves.
For example, in the case of $SU(3)$ the spectral curve (\ref{fscR}) is

\be
\lambda^3 - cu\lambda^2 + c^2v\lambda(\wp(\xi) - \wp(\epsilon)) +\\
+ c^3\left(\frac{1}{2}\wp'(\xi)\wp'(\epsilon)  - 3\wp(\xi)\wp^2(\epsilon)
+ \wp^3(\epsilon) + 2\alpha \wp(\xi)\wp(\epsilon) + 2\beta
(\wp(\xi) + \wp(\epsilon)) + 2\gamma\right)
= 0\\
u\equiv H_1=H_+,
v\equiv H_2=H_-
\label{SU3cc}
\ee
where

\be
H_{\pm}=e^{\pm p_1}\sqrt{\wp(q_{12})-\wp(\epsilon)}
\sqrt{\wp(q_{13})-\wp(\epsilon)} +
e^{\pm p_2}\sqrt{\wp(q_{12})-\wp(\epsilon)}
\sqrt{\wp(q_{23})-\wp(\epsilon)}+\\
+e^{\pm p_3}\sqrt{\wp(q_{13})-\wp(\epsilon)}
\sqrt{\wp(q_{23})-\wp(\epsilon)},\ \ \ \ p_1+p_2+p_3=0.
\label{SU3c}
\ee
Again the issue is whether some choice of the normalization $c$
might simplify matters. With our choice of (\ref{cfun})
for example, the third term in (\ref{SU3cc}) turns into just
$v\lambda$, but no drastic simplification occurs in the constant
term unless at the special point $\epsilon = \frac{\pi}{2}$
(where $\wp'(\epsilon =\pi/2) = 0$) when -- modulo $c$ -- the whole
equation becomes linear in $x = \wp(\xi)$.

To conclude the discussion of the \RS system,
we  separately consider the special point where $\epsilon$
equals the real half-period $\omega=\pi/2$.
Using the fact that
$\sigma(\xi+2\omega)=-\sigma(\xi)e^{2\eta_1(\xi+\omega)}$
with $\eta_1 = \xi(\omega)$ \cite{BEWW}, we find that

\be
\frac{\sigma(\xi)\sigma(\omega)}{\sigma(\xi + \omega)}\equiv
c_+(\xi|\omega) = \alpha c_0(\xi|\omega) = \alpha^2 c_-(\xi|\omega)
\ee
with $\alpha = e^{-\xi\eta_1}$.
Then

\be
D_k(\xi|\epsilon)=\Bigg\{
\begin{array}{ll}
c_0\sp{-k} {\cal H}_1\sp{-k(k-2)/2}&k\quad{\rm even},\\
c_0\sp{-k} {\cal H}_1\sp{-k(k-2)/2} c_0 {\cal H}_1\sp{-1/2}&
k\quad{\rm odd}.
\end{array}
\ee
Here ${\cal H}_1=2 e_1\sp2 +e_2 e_3$ is independent of $\xi$
and we note that in the trigonometric limit ${\cal H}_1=1$.
The choice $c\sim c_0$ now essentially removes all $\xi$-dependence
in the spectral curve apart from a term linear
in $c_0$ that multiplies the odd  Hamiltonians. Similar to the $SU(3)$ case,
we choose $c=-ic_0$.
By absorbing
the (inessential) constant factors into the Hamiltonians,
$h_k= i^k{\cal H}_1\sp{-k(k-2)/2} H_k $
for even $k$ and
$h_k = i^{k-1}{\cal H}_1\sp{-1/2}{\cal H}_1\sp{-k(k-2)/2} H_k $
for odd $k$, one obtains the non-perturbative spectral curve
(\ref{Rsc}) at the special point $\epsilon = \omega$:

\be\label{spsc1}
ic_0^{-1}(\xi|\omega)=i\sqrt{\wp (\xi)-\wp (\omega)}=\frac
{\sum^N_{odd\ k}h_k(-\lambda)^{N-k}}
{\sum^N_{even\ k} h_k(-\lambda)^{N-k}}={P(\lambda)
-(-)^NP(-\lambda)\over P(\lambda)+(-)^NP(-\lambda)},
\ee
where

\be
P(\lambda)=\sum_{k}h_k(-\lambda)^{N-k}=\lambda^N+\ldots +1
\ee
These formulae may also be easily expressed in terms of the
Jacobi elliptic functions.  For example,
$\wp (\xi)-\wp(\omega)=(e_1-e_3){cn^2(u,k)\over sn^2(u,k)}$
and, therefore,
$c_0(\xi|\omega)=
{1\over \sqrt{e_1-e_3}}{sn(u,k)\over cn(u,k)}$; here $sn(u,k)$ and $cn(u,k)$
are the Jacobi functions \cite{BEWW},
$u=(e_1-e_3)^{1/2} \xi$, $k$ is elliptic modulus.

To conclude this subsection, note that one can immediately generalize the \RS model
to other groups, see \cite{ogRS}.

\subsection{Fundamental matter in $6d$: \XYZ chain}

At the last subsection of this long and detailed consideration of different
Seiberg-Witten theories, we make the first step towards $6d$ theories, that is
$6d$ theory with the fundamental matter. This is
the theory with {\em two} extra compactified dimensions,
of radii $R_5$ and $R_6$. The same argument as
we used for the transition from four to five dimensions, i.e. taking account
of the Kaluza-Klein modes, allows one to predict the structure of results
in the $6d$
case as well. Indeed, one expects that
coming from $4d$ ($5d$) to $6d$ means
one should replace the rational (trigonometric) expressions by the elliptic
ones, the (imaginary part of) modular
parameter being identified with the ratio of the compactification radii
$R_5/R_6$.

This would give rise to `` elliptization'' of an
integrable systems. In terms of $2\times 2$ Lax representation, the
``elliptization''
corresponds to replacing by elliptic $r$-matrices and $L$-operators
the rational (\ref{LaxXXX}) and trigonometric (\ref{l-gen}) ones. In
particular, it means
that now it is natural to consider the parameter $\zeta =
\2\log\lambda$
as co-ordinate on a torus, i.e. one of the spectral parameters
lives on a torus, while the other one remains on a cylinder.

In fact, we know the only system providing proper elliptization
of the \XXX spin chain, that is, the \XYZ Sklyanin chain \cite{Skl}.
It was propose as a candidate for the integrable system behind the $6d$ theory
in \cite{XYZ,GGM2,MarMir}.

Now we briefly describe the \XYZ chain.
All necessary facts on the elliptic
functions can be found in \cite{BEWW}. The Lax matrix
is now defined on the elliptic curve
$E(\tau)$ and is explicitly given by (see \cite{FT} and references therein):

\be\label{39}
L^{\rm Skl}(\zeta) = S^0{\bf 1} + i\sum_{a=1}^3 W_a(\zeta)S^a
\sigma_a
\ee
where

\be\label{sklyatheta}
W_a(\zeta) = \sqrt{e_a - \wp\left({\zeta}|\tau\right)} =
i\frac{\theta'_{\ast}(0)\theta_{a+1}\left({\zeta}\right)}{\theta_{a+1}(0)
\theta_*(\zeta)}
\ee
To keep the similarities with \cite{XYZ,GGM2,MarMir},
we redefine in this subsection the spectral
parameter $\zeta\to i{\zeta\over 2K}$, where $K\equiv\int_0^{{\pi\over
2}}{dt\over\sqrt{1-k^2\sin^2t}}={\pi\over 2}\theta_{1}^2(0)$,
$k^2\equiv{e_1-e_2\over e_1-e_3}$ so that $K\to{\pi\over 2}$ as $\tau\to
i\infty$. This factor results into the additional multiplier $\pi$ in
trigonometric functions in the limiting cases.

The Lax operator (\ref{39})
satisfies the Poisson relation (\ref{quadr-r}) with the
numerical {\it elliptic} $r$-matrix $r(\zeta)={-{i\over 2}}\sum_{a=1}^3
W_a(\zeta)\sigma_a\otimes\sigma_a$, which
implies that $S^0, S^a$ form the (classical)
Sklyanin algebra \cite{Skl}:

\be\label{sklyaln}
\left\{S^a, S^0\right\} = i
\left(e_b - e_c\right)S^bS^c
\\
\left\{S^a, S^b\right\} = -iS^0S^c
\ee
with the obvious notation: $abc$ is the triple $123$ or its cyclic
permutations.

One can distinguish three interesting
limits of the Sklyanin algebra: the rational, trigonometric and
double-scaling limits \cite{XYZ}. We are interested in only
trigonometric limit here, since it describes degeneration to the
$5d$ case. In this limit,
$\tau\rightarrow +i\infty$ and the Sklyanin algebra
(\ref{sklyaln}) transforms to

\be\label{trisklya}
\{ \hat S^3,\hat S^0\} = 0,\ \
\{\hat S^1,\hat S^0\} =i\hat S^2\hat S^3,\ \
\{\hat S^2,\hat S^0\} =-2i\hat S^3\hat S^1\\
\{\hat S^1,\hat S^2\} =-i\hat S^0\hat S^3,\ \
\{\hat S^1,\hat S^3\} = i\hat S^0\hat S^2,\ \
\{\hat S^2,\hat S^3\} = -i\hat S^0\hat S^1
\ee
The corresponding Lax matrix is

\be\label{laxxxz}
L_{XXZ} = \hat S^0{\bf 1}-{\f \sinh\pi\zeta}\left(\hat S^1\sigma_1+
\hat S^2\sigma_2+\cosh\pi\zeta\hat S^3\sigma_3\right)
\ee
and the $r$-matrix

\be
r(\zeta)={i\over\sinh\pi\zeta}
\left(\sigma_1\otimes\sigma_1+\sigma_2\otimes\sigma_2+
\cosh\pi\zeta\sigma_3\otimes\sigma_3\right)
\ee
Now one can note that the algebra (\ref{trisklya}) admits the identification
$\hat S_0=e^{S_0^{trig}}+e^{-S_0^{trig}}$,
$\hat S_3=e^{-S_0^{trig}}-e^{S_0^{trig}}$. With this identification and
up to normalization of the Lax operator (\ref{laxxxz}), we finally get the Lax
operator (\ref{l-gen}) of the \XXZ chain with the Poisson brackets
(\ref{pois}).

The determinant
$\det _{2\times 2} L^{\rm Skl}(\zeta)$ is equal to

\be
\det _{2\times 2} L^{\rm Skl}(\zeta) = \hat S_0^2 + \sum_{a=1}^3 e_a\hat S_a^2
- \wp(\zeta)\sum_{a=1}^3\hat S_a^2
= K - M^2\wp(\zeta) =  K - M^2 x
\ee
where

\be\label{casi}
K = \hat S_0^2 + \sum_{a=1}^3 e_a(\tau)\hat S_a^2 \ \ \
\ \ \ \
M^2 =  \sum_{a=1}^3 \hat S_a^2
\ee
are the Casimir functions of the Sklyanin algebra.

Now, in. order to construct the spectral curve, note that
the determinant of the monodromy matrix $T_{N_c}(\zeta)$ is

\be\label{54}
Q(\zeta) = \det _{2\times 2} T_{N_c}(\zeta) =
\prod_{i=1}^{N_c} \det _{2\times 2} L^{\rm Skl}(\zeta - \zeta_i) =
\prod_{i=1}^{N_c} \left( K_i - M^2_i\wp(\zeta - \zeta_i)\right) =
\\
= Q_0{\prod _{\alpha =1}^{N_f}\theta_{\ast}(\zeta - m_\alpha)\over
\prod _{i=1}^{N_c}\theta _{\ast}(\zeta - \zeta_i)^2},\ \ \
Q_0=\prod_i^{N_c}A_i
\ee
where we used the Appendix of \cite{MarMir}
in order to rewrite $K_i-M_i^2\wp(\zeta-\zeta_i)$ as
$A_i{\theta_*\left(\zeta-m_i^{(+)}\right)
\theta_*\left(\zeta-m_i^{(-)}\right)\over \theta_*^2(\zeta-\zeta_i)}$. Here

\be
A_i={M_i^2\left[\theta_*'(0)\right]^2\over \theta_*^2(\kappa_i)},\ \ \
m_i^{(\pm)}=\zeta_i\pm \kappa_i
\ee
and $\kappa_i$ is a solution to the equation

\be
{\theta_4^2(\kappa_i)\over\theta_*^2(\kappa_i)}=
{K_i\theta_4^2(0)\over M_i^2\left[\theta_*'(0)\right]^2}+
{\pi^2\theta_4^2(0)\left[\theta_2^4(0)+\theta_3^4(0)\right]\over 3
\left[\theta_*'(0)\right]^2}
\ee
Similarly to the $5d$ case,

\be
\sum _{\alpha =1}^{N_f}m_\alpha = 2\sum _{i=1}^{N_c}\zeta_i
\ee
As a particular example, let us consider
the case of the {\it homogeneous} chain \cite{XYZ} (all $\zeta_i = 0$
in (\ref{54})).
$\Tr T_{N_c}(\zeta)$ is the combination of
polynomials:

\be\label{478}
P(\zeta) = Pol^{(1)}_{\left[\frac{N_c}{2}\right]}(x) +
y Pol^{(2)}_{\left[\frac{N_c-3}{2}\right]}(x),
\ee
where $\left[\frac{N_c}{2}\right]$ is integral part
of ${N_c\over 2}$, and coefficients of $Pol^{(1)}$ and $Pol^{(2)}$
are Hamiltonians of the \XYZ model.
The spectral equation (\ref{specTC0}) for the \XYZ model
is:

\be\label{specXYZ}
w + \frac{Q(\zeta)} w = 2P(\zeta)
\ee
where for the homogeneous chain $P$ and $Q$
are polynomials in $x = \wp(\zeta)$
and $y =   \frac{1}{2}\wp'(\zeta)$.
Eq. (\ref{specXYZ}) describes the double covering of the elliptic
curve $E(\tau)$:
with generic point $\zeta \in E(\tau)$ one associates the
two points of $\Sigma^{XYZ}$, labeled by the two roots $w_\pm$
of equation (\ref{specXYZ}).
The ramification points correspond to
$w_+ =w_- = \pm\sqrt{Q}$, or
$Y = \frac{1}{2}\left(w - \frac{Q}w\right) =
\sqrt{P^2 - Q}= 0$. Note that, for the curve (\ref{specXYZ}),
$x = \infty$ is {\it not} a branch point,
therefore, the number of cuts on the both copies of $E(\tau)$ is $N_c$ and the
genus of the spectral curve is $N_c+1$.\footnote{Rewriting
analytically $\Sigma^{XYZ}$ as a system of equations
\be
y^2 = \prod_{a=1}^3 (x - e_a), \ \ \
Y^2 = P^2 - Q
\ee
the set of holomorphic 1-differentials on $\Sigma^{XYZ}$ can be chosen as
\be\label{hdb}
v = \frac{dx}{y},\ \ \
V_\alpha = \frac{x^\alpha dx}{yY} \ \ \
\alpha = 0,\ldots,
\left[\frac{N_c}{2}\right], \ \ \
\tilde V_\beta = \frac{x^\beta dx}{Y} \ \ \
\beta = 0, \ldots,
\left[\frac{N_c-3}{2}\right]
\ee
with the total number of holomorphic 1-differentials
$1 + \left(\left[\frac{N_c}{2}\right] + 1\right) +
\left(\left[\frac{N_c-3}{2}\right] + 1\right) = N_c+1$
being equal to the genus of $\Sigma^{XYZ}$.}

In the case of a generic {\em inhomogeneous} chain the trace of monodromy
matrix looks more complicated, being expression of the form

\be\label{trsklya}
\Tr T_{N_c}(\zeta ) = \Tr \prod _{N_c\ge i\ge 1}^{\curvearrowleft}
L^{\rm Skl}_i(\zeta - \zeta_i) =
\Tr\sum_{\{\alpha_i\}}\prod_{N_c\ge i\ge 1}^{\curvearrowleft}
\sigma^{\alpha_i}S^{\alpha_i}_i
W_{\alpha_i}(\zeta-\zeta_i) =
\\ =
\sum_{\{\alpha_i\}}\Tr\prod_{N_c\ge i\ge 1}^{\curvearrowleft}
\sigma^{\alpha_i}S^{\alpha_i}_i
W_{\alpha_i}(\zeta-\zeta_i)
\\
\alpha_i = 0,a_i;\ \ \ \ \ \ a_i=1,2,3;\ \ \ \ \ W_0(\zeta)=1
\ee
It is clear from (\ref{sklyatheta}) that each
term in the sum (\ref{trsklya}) (and thus the whole sum) has simple poles in
all inhomogeneities $\zeta = \zeta_i$. Thus, the trace of the monodromy matrix
can be represented as

\be
P(\zeta ) = {1\over 2}\Tr T_{N_c}(\zeta ) =
P_0{\prod_{i=1}^{N_c}\theta_{\ast}(\zeta - a_i)\over
\prod_{i=1}^{N_c}\theta_{\ast}(\zeta - \zeta_i)}
\ee
\be\label{strange6}
\sum _i a_i = \sum _i\zeta_i =\2\sum_{\alpha}m_{\alpha}
\ee
where $a_i$ should be identified with (shifted) gauge moduli.
The equation (\ref{specXYZ}) can be finally rewritten as

\be\label{specXYZ1}
w + Q_0{\prod _{\alpha =1}^{N_f}\theta_{\ast}(\zeta - m_\alpha)\over
w\prod _{i=1}^{N_c}\theta _{\ast}(\zeta - \zeta_i)^2} =
P_0{\prod_{i=1}^{N_c}\theta_{\ast}(\zeta - a_i)\over
\prod_{i=1}^{N_c}\theta_{\ast}(\zeta - \zeta_i)}
\ee
We see that in the $6d$ case it is necessary always to require $N_f=2N_c$ in
order to have a single-valued meromorphic functions on torus.
We also meet again the constraint (\ref{strange6}). In the
$5d$ case, it was removed, along with the condition $N_f=2N_c$
twisting the spin chain. Since no twisted \XYZ chain is known
we are forced to preserve both constraints\footnote{Formula
(\ref{strange6}) implies that,
while the sum of all the would-be gauge moduli $a_i$
still remains a constant, it is no longer zero. Therefore,
one would rather associate with the gauge
moduli the quantities $a_i$ shifted by the constant ${1\over 2N_c}
\sum m_{\alpha}$. We do not know the proper field theory interpretation
of this condition.}.

The curve (\ref{specXYZ1})
has no longer that simple expression in terms of polynomials of
$x$ and $y$ similar to (\ref{478}).
Thus, in the inhomogeneous case it is not convenient to work in terms
of ``elliptic'' variables $x$ and $y$ (analogous to $\lambda$ in the $5d$
case) instead we deal with the variable
$\zeta\in \C/\Gamma$.

Given the spectral curve and integrable system, we complete the description
with writing down the generating 1-differential $dS$.
It can be evidently chosen in the following way

\be\label{dsxyz}
dS^{XYZ} \cong \zeta {dw\over w}\cong - \log w d\zeta
\ee
since, as any spin chain, it is to be proportional to $dw/w$ and to $\zeta$
due to six dimensions (in 5 dimensions, $\zeta$ was just $\log\lambda$ and
in 6 dimensions $\lambda$ is replaced by the pair $(x,y)$ related with $\zeta$
via $\wp (\zeta)$ and $\wp '(\zeta)$).
Return to our oversimplified example of the homogeneous chain.
Now, under the variation of moduli (which are all contained in $P$,
while $Q$ is moduli independent),

\be
\delta(dS^{XYZ}) \cong \frac{\delta w}{w}d\zeta =
\frac{\delta P(\zeta)}{\sqrt{P(\zeta)^2-4Q(\zeta)}} d\zeta
= \frac{dx}{yY}\delta P
\ee
and, according to (\ref{hdb}), the r.h.s. is a {\it holomorphic}
1-differential on the spectral curve (\ref{specXYZ}).

\section{Integrability and $\CN=1$ supersymmetric YM theories}
\subsection{Superpotentials in $\CN=1$ theories from $\CN=2$ spectral curves}

In this section we shall discuss gauge theories with $\CN=1$ SUSY and
their relation with the \N2 SUSY theories. For the  detailed
review of $\CN=1$ theories we recommend  \cite{shifman}. Here we
explain some features of $\CN=1$ theories important for
treating them within the integrable framework.

The key difference between theories with \n=1 and \N2 SUSY is
the absence of vacuum valleys in the \n=1 case.
Therefore, the Coulomb branch of the moduli space is absent and
the number of the vacuum states is finite. The order
parameter which distinguishes between the vacua can be
identified with the gluino condensate. Due to  the
discreteness of vacua the spectrum of BPS states now
is different. Instead of the particles saturating
the BPS bound one has domain walls interpolating between
vacua as well as stringy objects which can be
identified with the strings themselves or with the
domain wall intersections. These objects turn out to be related
with the corresponding central
charges in the algebra of \n=1 supersymmetry \cite{ds,cc}.

The main target is to be an
object that presumably could
be determined exactly. Instead of the prepotential in
\N2 theory, the function which enjoys holomorphy is
the superpotential $\CW$; therefore, one could attempt to evaluate it.
There are generically perturbative and nonperturbative
contributions to the superpotential. As for the perturbative
contribution, the $\beta$-function  unlike the
\N2 case gets contributions from all loops. However, the higher
loops IR contributions can be calculated exactly \cite{nsvz}.
There is also a crucial difference between nonperturbative
contributions in \n=1 and \N2 theories.
Indeed, configurations of all possible
instanton  numbers contribute into the prepotential in the \N2 case, while
only one-instanton terms contribute into the \n=1 superpotential
due to the non-anomalous R-symmetry.

Despite all this difference one may expect links between the exact results
in \N2 theories and the superpotential. These expectations
are true indeed and below we sketch the relation following \cite{n1} and
especially \cite{giv}.
The starting point is the Seiberg-Witten curve $ P(x,y)=0$ in \N2 theory
with some matter content. Then, one should fix singular
points on the Coulomb branch of the moduli space which correspond
to the vacua in \n=1 theory. These singular points are in one-to-one
correspondence with vanishing of the
discriminant of the curve. Then, comparing the equations in \N2 theory
and the equation $\partial \CW =0$, where the derivative is taken w.r.t.
the appropriated composite field, one observes that these equations
coincide if the coordinate on the spectral curve is identified
with some composite operator.

Consider $\CN=1$ supersymmetric $SU(2)$ gauge theories in four dimensions,
with any possible content of matter superfields, such that the theory is
either one-loop asymptotic free or conformal. This allows one to introduce
$2N_f$ matter supermultiplets in the fundamental representation,
$Q_i^a$, $i=1,...,2N_f$ and
$N_A$ supermultiplets in the adjoint representation,
$\Phi_{\a}^{ab}$, $\a=1,...,N_A$,.
Here $a,b$ are fundamental representation indices, and $\Phi^{ab}=\Phi^{ba}$.
The numbers $N_f$, $N_A$  are limited by the condition to keep the
theory asymptotically free or conformal:

\beq
b_1=6-N_f-2N_A\geq 0,
\eeq
where $-b_1$ is the one-loop coefficient of the gauge coupling beta-function.

The effective
superpotential of an
$N=1$ supersymmetric $SU(2)$ gauge theory,
with $2N_f$ doublets and $N_A$ triplets  is\footnote{Throughout this section,
for the sake of brevity we omit the index $QCD$ from the notation of
$\Lambda_{QCD}$.}

\be\label{super}
\CW_{N_f,N_A}(M,X,Z) =
- (4-b_1)\Big\{\Lambda^{-b_1} \pf_{2N_f} X\Big[
{\det}_{N_A}(\Gamma_{\a\b})\Big]^2
\Big\}^{1/(4-b_1)}\\
+\tr_{N_A} \mu M +{1\over 2}\tr_{2N_f} mX
+{1\over\sqrt{2}}\tr_{2N_f} Y^{\a} Z_{\a} ,
\ee
where

\beq
\Gamma_{\a\b}(M,X,Z)=M_{\a\b}+\tr_{2N_f}(Z_{\a}X^{-1}Z_{\b}X^{-1}).
\eeq
The first term in (\ref{super}) is the exact
non-perturbative superpotential, while
the other terms give the tree-level superpotential.
$\Lambda$ here is a dynamically generated scale, while $\mu_{\a\b}$, $m_{ij}$
and $Y^{\a}_{ij}$ are the bare masses and Yukawa couplings, respectively
($\mu_{\a\b}=\mu_{\b\a}$, $m_{ij}=-m_{ji}$,
$Y^{\a}_{ij}=Y^{\a}_{ji}$). The gauge singlets, $X$, $M$, $Z$,  are
given in terms of the $\CN=1$ superfield doublets, $Q^a$, the triplets,
$\Phi^{ab}$  as follows:

\be
X_{ij}=Q_{ia} Q_j^a, a=1,2, ;i,j=1,...,2N_f, \\
M_{\a\b}=\Phi_{\a b}^a\Phi_{\b a}^b; \a ,\b=1,...,N_A; a,b=1,2, \\
Z_{ij}^{\a}=Q_{ia}\Phi_{\a b}^a Q_j^b
\ee
The gauge-invariant superfields $X_{ij}$ may be considered as a mixture of
$SU(2)$ ``mesons'' and ``baryons'', while the gauge-invariant superfields
$Z_{ij}^{\a}$ may be considered as a mixture of $SU(2)$ ``meson-like'' and
``baryon-like'' operators.

As an example, consider the theory with $N_f=1$, $N_A=1$.
%\end{center}
In this case, the superpotential  reads

\beq
\CW_{1,1}=-{\pf X\over \Lambda^3}\Gamma^2+\mu M +{1\over 2}\tr mX +
{1\over \sqrt{2}}\tr Y Z
\eeq
Here $m$, $X$ are antisymmetric $2\times 2$ matrices, $Y$, $Z$ are
symmetric $2\times 2$ matrices and

\beq
\Gamma=M+\tr (ZX^{-1})^2 .
\eeq

To find the
quantum vacua, we solve the equations: $\partial_M W =\partial_X W
=\partial_Z W =0$.
The equations $\partial W=0$ can be re-organized into the singularity
conditions of an elliptic curve:

\beq
y^2=x^3+ax^2+bx+c
\eeq
(and some other equations), where the coefficients $a,b,c$ are functions of
only the field $M$, the scale $\Lambda$, the bare quark masses, $m$, and Yukawa
couplings, $Y$. Explicitly,

\beq
a=-M, \qquad b={\Lambda^3\over 4}Pf m, \qquad c=-{\a\over 16} ,
\eeq
where

\beq
\a={\Lambda^6\over 4}\det Y .
\eeq
The parameter $x$ in the elliptic curve  is given in terms of the
composite field:

\beq
x\equiv {1\over 2}\Gamma .
\eeq
$\CW_{1,1}$ has $2+N_f=3$ vacua, namely, the  three singularities of the
elliptic curve. These are the three solutions,
$M(x)$, of the equations: $y^2=\partial y^2/\partial x=0$; the solutions
for $X$, $Z$ are given by the other equations of motion.

Similarly, if one considers a generic $SU(N_c)$ group,
the spectral curve for the generic $N_f$, $N_c$ can be put into the form
(\ref{gm8})

\beq\label{scn1}
y^2=P_{N_c}^2(\lambda) + Q_{N_f}(\lambda)
\eeq
with $\lambda=ZX^{-1}$ \cite{n1}.

 From this consideration, we can get an important lesson
for the integrability approach; that is, one can make an
identification  of one of the coordinates defining the spectral curve.
Indeed, the Lax operator of the spin chain (when acting on the eigenfunction,
Baker-Akhiezer function) can be identified with some
known composite operator.
To get the pure gauge theory
one has to decouple the fundamental matter sending its masses
to infinity. Therefore, the Lax operator in  the Toda system
can also be identified as a composite operator which involves
very heavy ``regulator" fermions in the fundamental representation.
The identification
of the second coordinate involved in the equation of the spectral curve
(\ref{scn1}) is more tricky.

Let us note that the relation between \n=1 superpotentials and \N2
curves can be established moving in the opposite direction as well \cite{gwy}.
To this aim one can consider the Affleck-Dine-Seiberg superpotential
supplemented by the contribution from the heavy adjoint
hypermultiplet
and use the Konishi anomaly relating the vacuum expectation values of
different operators

\be
\left\langle \frac{1}{2}\tr mX +
\frac{1}{\sqrt{2}}\tr YZ+
\frac12 \,\frac{{\Tr}\,W^2}{8\pi^2}
\right\rangle=0 \\
\left\langle 2\tr\mu M + \frac{1}{\sqrt{2}}\tr YZ
+2\,\frac{{\rm Tr}\,W^2}{8\pi^2}
\right\rangle=0
\label{konishirel}
\ee
where the vacuum expectation value of ${\Tr}\,W^2$ is the gluino
condensate.
The combination of the equation $\partial \CW=0$
and the Konishi relations amounts exactly to the condition of
vanishing discriminant of the Seiberg-Witten curve.

Generically one can consider more specific degenerations of the
Seiberg-Witten curves resulting into specific \n=1 theories. For
instance, one can consider  all $B$-cycles
on the \N2 curves vanishing simultaneously \cite{ds1}. In this
case, the very peculiar spectrum of strings emerges in
\n=1 theory \cite{hsz}. One more interesting possibility
comes from the phenomenon of collisions of vacua at
some points in the parameter space \cite{ad} which is expected
to correspond to nontrivial conformal field theories
in four dimensions \cite{hori}. It was recently shown
that, using the correspondence between \N2 and \n=1 theories,
one can identify the phase transition at these Argyres-Douglas points
as the deconfinement phase transition \cite{gwy}. The
analysis of the Argyres-Douglas points within the
integrability approach remains an interesting problem.

\subsection{Superpotentials from nonperturbative configurations}

We used the analysis above to recognize some meaning of the Lax operator.
Now let us make the next step
and identify the configuration space of the dynamical system
under consideration in terms of the field theory.
This can be done via calculating superpotentials
from nonperturbative configurations \cite{dorey,khoze,vafa}.
In the brief description below we follow the paper \cite{dorey}.
To have the  convenient starting point,
one compactifies the four-dimensional theory
to three dimensions on a circle of radius $R$. The
most obvious modification is that the theory
acquires new scalar degrees of freedom which come from the Wilson
lines of the four-dimensional gauge field swapped around the compact dimension.
Non-zero values for these scalars generically break the gauge group
down to its maximal Abelian subalgebra, $SU(N_c)\rightarrow U(1)^{N_c-1}$.
Hence, at the classical level, the compactified theory has a Coulomb branch.
By a gauge rotation, the Wilson line can be chosen to
lie in the Cartan subalgebra, $\phi=\phi^{a}H^{a}$.
Each scalar $\phi^{a}$ is related to the corresponding
Cartan component of the four-dimensional gauge field
as $\phi^{a}=\oint_{S^{1}} A^{a}\cdot dx$. As a consequence, $\phi^{a}$
can be shifted by an integer multiple of $2\pi$ by performing a
topologically non-trivial gauge transformation which is single-valued
in $SU(N_c)/Z_{N_c}$.  Thus we learn that each $\phi^{a}$
is a periodic variable with period $2\pi$.

In addition to the $N_c$ periodic scalars $\phi^{a}$, the low energy
theory on the Coulomb branch also includes $N_c$ massless photons $A^{a}_{i}$,
with $\sum_{a=1}^{N_c}A^{a}_{i}=0$,
where $i=1,2,3$ is a three-dimensional Lorenz index.
In three dimensions, a massless Abelian
gauge field can be eliminated in favour of a scalar by a duality
transformation.
Thus we can eliminate the gauge fields
$A^{a}_{i}$, in favour of $N_c$ scalars $\sigma^{a}$ subject to
$\sum_{i=1}^{N_c}\sigma^{a}=0$. In particular, the
fields $\sigma^{a}$ appear in the effective action as Lagrange multipliers
for the Bianchi identity:
${\cal L}_{\sigma}\sim \sigma\varepsilon^{ijk}\partial_{i}F_{jk}$.
Specifically, the normalization of $\sigma^{a}$ is chosen
so that the resulting term in the action is precisely
$-i\sigma^{a}k^{a}$ where $k^{a}$ is the magnetic charge in the
$U(1)$ subgroup corresponding to the Cartan generator $H^{a}$.
The Dirac quantization condition implies that each
magnetic charge is quantized in integer units:
$k^{a}\in Z$, and so the path integral is invariant under
shifts of the form
$\sigma^{a}\rightarrow \sigma^{a}+ 2\pi n^{a}$ with $n^{a}\in Z$.
Thus, like $\phi^{a}$, each $\sigma^{a}$ is a periodic variable with
period $2\pi$.

It is natural to combine the real scalars $\phi$ and $\sigma$ to form
a complex scalar $X=-i(\sigma+\tau\phi)$  \cite {Seiberg3d} where
$\tau=4\pi i/g^{2}+\theta/2\pi$ is the usual complexified coupling.
Let us now concentrate on the Euclidean
theory with gauge group $SU(2)$.
Gauge field configurations in the compactified theory are
labelled by
two distinct kinds of topological charge.
The first is the Pontryagin
number carried by instantons in four-dimensions,

\be
p=\frac{1}{8\pi^{2}}\int_{R^{3}\times S^{1}}{\rm Tr}
\left[ F\wedge F\right]\,
\ee
The $4d$ instanton number appears in the microscopic action as the term
$-ip\theta$.
An important feature of the compactified theory which differs
from the theory on $R^{4}$ is that $p$ is not quantized in integer units.
However, for each integer value of $p$, the theory has
solutions with action $S_{cl}=8\pi^{2} |p|/g^{2}-ip\theta$.
Thus we have $S_{cl}= 2\pi i p\tau$
for $p>0$ and $S_{cl}=-2\pi i p\bar{\tau}$ for $p<0$. When the instanton scale
size is much less than the compactification radius, these solutions are
close to their counterparts on $R^{4}$.

In addition to the usual four-dimensional instantons, the compactified theory
also has finite action configurations which carry three-dimensional
magnetic charge,

\be
k=\frac{1}{8\pi}\int_{R^{3}} \vec{\nabla}\cdot  \vec{B} \in Z
\ee
where $B_{i}=\varepsilon_{ijk}F^{jk}/2$ is the Abelian magnetic
field of the low-energy theory.
As these configurations do not depend either on the time or
on the coordinate
of the compactified spatial dimension, we will refer to them as
three-dimensional instantons.
In addition to the integer-valued magnetic charge $k$, these $3d$
instantons also carry fractional Pontryagin number $p=\phi/2\pi$ .
The magnetic charge appears in the action as the term
$-ik\sigma$ while, as above, $4d$ instanton number appears as the term
$-ip\theta$. Including these term the action of a $3d$ instanton can be
written as
$S_{3d}=-i\tau\phi|k|+ik\sigma$. Thus we have $S_{3d}=kX$ for $k>0$,
and $S_{3d}=k\bar{X}$ for $k<0$.

Putting everything together we can write out the most general possible
superpotential which can generated at leading semi-classical order,
\be
{\cal W}  =  m_{1}m_{2}m_{3}\sum_{p=1}^{\infty}a_{p}q^{p}+
 m_{1}m_{2}m_{3}\sum_{k=1}^{\infty}
 b_{k}\exp(-kX) + \\\ \ \ \ \ \ \ \  +
m_{1}m_{2}m_{3}\sum_{k=1}^{\infty}\,\sum_{n=1}^{\infty}
 c_{k,n}q^{kn}\exp(-kX) +
 d_{k,n}q^{kn}\exp(+kX)
 \label{expansion}
 \ee
The dependence of the superpotential on the masses of the adjoint
scalars $m_i$ is
fixed by  the zero modes counting.

The superpotential of the $SU(2)$ theory
is a holomorphic function on the chiral superfield
$X=-i(\sigma+\tau\phi)$. The weak coupling
arguments  indicate that this function
should respect the classical periodicity of the variables $\phi$ and $\sigma$.
Thus, ${\cal W}(X)$ is a holomorphic function defined on the complex torus
$E$. The superpotential should also be invariant under the Weyl group of
$SU(2)$ which acts as $X\rightarrow -X$. As ${\cal W}(X)$ is a
non-constant holomorphic
function on a compact domain it must have a singularity somewhere.
As usual such a singularity should correspond to a point at which
our low-energy description of the theory breaks down due to the
presence of extra light degrees of freedom. There is only one such point
on the classical moduli space: the origin $X=0$ where non-Abelian gauge
symmetry is restored. Thus we will assume that ${\cal W}$ has an isolated
singularity at this point and no other singularities.
Finally, as there is only one
 elliptic function of order two with a double pole at the origin
 we have the result \cite{dorey},

 \be
 {\cal W}(X)=m_{1}m_{2}m_{3}\left({\cal P}(X)+C(\tau) \right)
 \ee
Importantly, we have obtained a superpotential which is independent of $R$.

Starting from an ${\cal N}=2$
theory in four dimensions, the superpotential of the ${\cal N}=1$
theory on $R^{3}\times S^{1}$ obtained by introducing a mass for the
adjoint scalar in the ${\cal N}=2$ vector multiplet should coincide
with the complexified potential of the corresponding integrable system.
This connection should
be a general one which applies to all ${\cal N}=1$ theories obtained by
soft breaking of ${\cal N}=2$ supersymmetry.
The key point is  that the
Coulomb branch of an ${\cal N}=2$ theory on $R^{3}\times S^{1}$
coincides with the complexified phase space of the integrable Hamiltonian
system associated with the corresponding theory on $R^{4}$.
The dynamical variables of the integrable system, $\{X^{a},P^{a}\}$
yield holomorphic
coordinates on the Coulomb branch of the compactified theory.
The dimensional transmutation
provides then the corresponding superpotential for the Toda
case too. The superpotential for the Toda case has been also
derived from the brane picture for instantons as wrapped
branes \cite{vafa}.

\subsection{$\CN=1$ theories and integrability}

We complete this section with general comments concerning
the relation between integrable dynamics and \n=1 SYM theories.
First of all, one should expect that superpotentials in \n=1 theories are
associated with Hamiltonians of integrable systems\footnote{The idea of this
identification comes from the fact that supersymmetry is
basically broken by the $\tr \Phi^2$-term whose vacuum value is the
Hamiltonian of the corresponding \N2 integrable system (we discuss the $SU(2)$
case),
see also below.
}. Therefore,
the vacuum in $\CN=1$ theory correspond to a stationary
state of the dynamical system

\beq
\partial\CW / \partial \phi_i = \partial H / \partial x_i=0
\eeq
and the number of vacua in \n=1 theory coincides with the number of
stationary states. Note that
the locus corresponding to the stationary states of the
elliptic Calogero system plays the essential role
in the relation between the Calogero system and the KdV hierarchy.
Namely, the initial data in the KdV hierarchy upon this correspondence
should follow from the equilibrium locus in the Calogero system.

Let us now say how to come from the \N2 theory to the \n=1 one
in integrable terms.
First, the \N2 system has to be placed
at the singularities of the Coulomb branch. This means that
some integrals of motion in the integrable system have to be fixed.
If we are sitting at the point of maximal degeneracy, all the
integrals of motion have to be fixed and the finite-gap solution
degenerates to the solitonic one. This has been carefully
done in the paper \cite{bm}. At the next step, one has to give
a mass to the adjoint scalar that breaks \N2 SUSY down to \n=1.
The spectral curve is then ``rotated" within $\C\bP^3$ and
is given by two polynomial equations

\be
P(v,u,x)=0,\ \ \ \
Q(v,u,x)=0
\ee
Peculiar expressions for these polynomials for different \n=1
theories are presented in \cite{N1}.
These equations definitely mean that the integrable system is
to be modified.
However, the question of possible
interpretation of the spectral curve of in \n=1 theory as a
spectral curve of some integrable system still remains open.

We have discussed above the (very restricted) role
played by the \N2 integrable system
in the context of \n=1 theory. However, it seems that the
Whitham integrability (see section 6 below)
is much more prominent candidate for describing
the \n=1 theories, since it actually
governs the dependence of the effective action on
\N2 SUSY breaking parameters. In a sense,
the Whitham theory which is nothing but RG flows predicts a
very peculiar ``time''-dependence consistent with RG flows
\cite{Jose1}. All the ``times" but the discrete one are SUSY breaking
parameters and some predictions concerning even $\CN=0$ theories
can be made\footnote{Within the Whitham treatment, one can re-sort the argument
identifying superpotentials and Hamiltonians of integrable systems,
at least, in the $SU(2)$ case. Indeed, one just needs to note that
${\partial^2 \CF\over\partial(\log\Lambda)^2}={\partial h_2\over \partial\log
\Lambda}$ (see (\ref{an}), (\ref{an2})) is proportional to the gluino
condensate (as a consequence of low-energy theorems).
On the other hand, this condensate is obtained by taking the
derivative of the superpotential w.r.t. $\Lambda$ as it follows from
Ward identities.}.

\section{Perturbative prepotentials}
\setcounter{equation}{0}

After a short discussion of $\CN=1$ SUSY theories, we return to the main
subject
of this survey, \N2 theories.

In previous sections we established connections between Seiberg-Witten
theories and gauge theories. However, so far we studied only a kind of
general landscape in the theory, since the real, numerical results enter the
game only with the prepotential calculated. Note that even the correspondence
between gauge and integrable theories is supported by comparing the
perturbative
prepotentials. We discuss this issue in this section.

In fact, as we noted already,
calculating  the prepotential is quite a challenging task. It can be done,
mostly in the framework of integrable systems, order by order. However,
technically one needs to use different specific properties of prepotentials
to simplify these calculations.

The other essential point is that the prepotential celebrates many properties
that allow one to push forward the theory in many directions. This
makes the prepotential
an important object from both mathematical and physical
standing points.

\subsection{Perturbative prepotentials in gauge theories}

We start with the perturbative prepotentials, since, on one hand,
one is able
to compare the prepotentials calculated in physical theories and in integrable
systems only at perturbative level and, on the other hand, many important
properties of the prepotential are observed already at this level.
Therefore, now we
are going to give an explicit description of the perturbative prepotentials
for the Seiberg-Witten theories discussed in section 3.

The technical tool that allows one to proceed with effective
perturbative expansion of the prepotential is ``the residue formula'',
the variation of the period matrix, i.e. the third derivatives
of ${\cal F}(a)$ (see, e.g. \cite{WDVVlong}):

\be
\frac{\partial^3{\cal F}}{\partial a_i\partial a_j\partial a_k}
= \frac{\partial T_{ij}}{\partial a_k} =
{\rm res}_{d\xi = 0} \frac{d\omega_id\omega_jd\omega_k}{\delta dS},
\label{res}
\ee
where $\delta dS\equiv d\left({dS\over d\xi}\right)d\xi $, or, explicitly

\be\label{dSall}
\begin{array}{cll}
\delta dS&=d\lambda d\xi&\hbox{ for $4d$ models }\\
\delta dS&=R_5^{-1} d\log\lambda d\xi & \hbox{ for $5d$ models }\\
\delta dS&=d\zeta d\xi & \hbox{ for $6d$ models }
\end{array}
\ee
Here the differential $d\xi$ given on the bare torus
degenerates into $dw/w$ when torus degenerates into the punctured sphere
(Toda and spin chains).
We remark that although
$d\xi$ does not have zeroes on the {\it bare} spectral
curve when it is a torus or doubly punctured sphere, it does
in general however possess them on the covering ${\cal C}$.
Hereafter, we denote via $\xi$ the variable living on the bare spectral curve
(torus), while $\zeta$ is the variable living on the compactification torus
(in 6 dimensions) or on the compactification circle (in 5 dimensions).

In order to construct the perturbative prepotentials, one merely
can note that, at the leading order, the Riemann surface (spectral
curve of the integrable system) becomes rational. Therefore, the
residue formula allows one to obtain immediately the third derivatives
of the prepotential as simple residues on the sphere. In this way,
one can check
that the prepotential has actually the form (\ref{ppg}) \cite{WDVVlong}.
We discuss these calculations later,
but now see what the perturbative prepotentials are
expected from the gauge theory point of view.

As a concrete example, let us consider the $SU(N_c)$ gauge group. Then, say,
the perturbative prepotential for the pure gauge theory acquires the
form

\be
{\cal F}_{pert}^V={\f 8\pi i}\sum_{ij}
\left(a_i-a_j\right)^2\log\left(a_i-a_j\right)
\ee
This formula establishes that when v.e.v.'s
of the scalar fields in the gauge supermultiplet are non-vanishing
(perturbatively $a_r$ are eigenvalues of the vacuum
expectation matrix  $\langle\phi\rangle$), the fields in the gauge multiplet
acquire masses $m_{rr'} = a_r - a_{r'}$ (the pair of indices $(r,r')$ label
a field in the adjoint representation of $SU(N_c)$). The
eigenvalues are subject to the condition $\sum_ia_i=0$.
Analogous formula for the
adjoint matter contribution to the prepotential is

\be
{\cal F}_{pert}^A=-{\f 8\pi i}\sum_{ij}
\left(a_i-a_j+M\right)^2\log\left(a_i-a_j+M\right)
\ee
while the contribution of one fundamental matter hypermultiplet reads as

\be
{\cal F}_{pert}^F=-{\f 8\pi i}\sum_{i}
\left(a_i+m\right)^2\log\left(a_i+m\right)
\ee

Similar formulas can be obtained for the other groups. The
eigenvalues of $\langle\phi\rangle$
in the first
fundamental representation of the classical series of the Lie groups are

\be
B_n\ (SO(2n+1)):\ \ \ \ \ \{a_1,...,a_n,0,-a_1,...,-a_n\};\\
C_n\ (Sp(n)):\ \ \ \ \ \{a_1,...,a_n,-a_1,...,-a_n\};\\
D_n\ (SO(2n)):\ \ \ \ \ \{a_1,...,a_n,-a_1,...,-a_n\}
\ee
while the eigenvalues in the adjoint representation have the form

\be\label{adjnew}
B_n:\ \ \ \ \ \{\pm a_j;\pm a_j\pm a_k\};\ \ \ j<k\le n\\
C_n:\ \ \ \ \ \{\pm 2a_j;\pm a_j\pm a_k\};\ \ \ j<k\le n\\
D_n:\ \ \ \ \ \{\pm a_j\pm a_k\}, \ \ j<k\le n
\ee
Analogous formulas can be written for the exceptional groups too.  The
prepotential in the pure gauge theory can be
read off from the formula (\ref{adjnew}) and has the form

\be\label{adjvo}
B_n:\ \ \ \ \ {\cal F}_{pert}={1\over 8\pi i}\sum_{i,i} \left(\left(a_{i}-a_{j}
\right)^2\log\left(a_{i}-a_{j}\right)+\left(a_{i}+a_{j}
\right)^2\log\left(a_{i}+a_{j}\right)\right)+
{1\over 2}\sum_i a_i^2\log a_i;\\
C_n:\ \ \ \ \ {\cal F}_{pert}={1\over 8\pi i}\sum_{i,i} \left(\left(a_{i}-a_{j}
\right)^2\log\left(a_{i}-a_{j}\right)+\left(a_{i}+a_{j}
\right)^2\log\left(a_{i}+a_{j}\right)\right)
+2\sum_i a_i^2\log a_i;\\
D_n:\ \ \ \ \ {\cal F}_{pert}={1\over 8\pi i}\sum_{i,i} \left(\left(a_{i}-a_{j}
\right)^2\log\left(a_{i}-a_{j}\right)+\left(a_{i}+a_{j}
\right)^2\log\left(a_{i}+a_{j}\right)\right)
\ee
The perturbative prepotentials in $4d$ theories with fundamental matter
are discussed in detail in \cite{WDVVlong} (see also \cite{KDP1}).

All these formulas can be almost immediately extended to the
``relativistic"
 $5d$ \N2 SUSY gauge models
with one compactified dimension. One can understand the
reason for this ``relativization'' in the following way. Considering
four plus one compact dimensional theory one should take into account the
contribution of all Kaluza-Klein modes to each 4-dimensional field.
Roughly speaking it leads to the 1-loop contributions to the effective
charge of the form ($a_{ij}\equiv a_i-a_j$)

\be\label{relKK}
T_{ij} \sim \sum _{\rm masses}\log\hbox{ masses} \sim \sum _m\log
\left(a_{ij} +
{m\over R_5}\right) \sim \log\prod _m\left(R_5a_{ij} + m\right)
\sim\log\sinh R_5a_{ij}
\ee
i.e. coming from $4d$ to $5d$
one should make a substitution $a_{ij} \rightarrow
\sinh R_5a_{ij}$, at least, in the
formula for perturbative prepotentials.

The same general argument
can be equally applied to the $6d$ case,
or to the theory with {\em two} extra compactified dimensions,
of radii $R_5$ and $R_6$. Indeed, the account of
the Kaluza-Klein modes allows one to predict the perturbative form
of charges in the $6d$ case as well. Namely, one should expect them
to have the form

\be\label{6dKK}
T_{ij} \sim \sum _{\rm masses}\log\hbox{ masses} \sim \sum _{m,n}\log
\left(a_{ij} +
{m\over R_5} +{n\over R_6}
\right) \sim
\\
\sim\log\prod _{m,n}\left(R_5a_{ij} + m+n{R_5\over R_6}\right)
\sim\log\theta_*\left(R_5a_{ij}\left|i{R_5\over R_6}\right)\right.
\ee
i.e. coming from $4d$ ($5d$) to $6d$
one should replace the rational (trigonometric) expressions by the elliptic
ones, at least, in the
formulas for perturbative prepotentials.

\subsection{Perturbative prepotentials in integrable systems}

Now we are going to compare these results with calculations done for
integrable models. We start with the
integrable chains \cite{WDVVlong,GGM2,MarMir}.

As we saw in the previous section (ss.3.2,3.4,3.5,3.6,3.8),
the spectral curve in all cases of spin chains (fundamental matter)
can be put in the form

\be\label{scg}
w+{Q^{(d)}(\zeta)\over w}=2P^{(d)}(\zeta)
\ee
or
\be\label{scg'}
W+{1\over W}={2P^{(d)}(\zeta)\over\sqrt{Q^{(d)}(\zeta)}},
\ \ \ \ \ \ W \equiv {w\over\sqrt{Q^{(d)}(\zeta)}}
\ee
In the perturbative limit, only the first term in the r.h.s.
of these formulas survives so that the spectral curve becomes
rational.

The generating differential $dS$ is always of the form (to unify the
notations, we use the same letter $\zeta$ for $\lambda$ in $4d$  and
${1\over 2}\log\lambda$ in $5d$)

\be\label{dSr}
dS=\zeta d\log W
\ee
The concrete forms of the functions introduced here are:

\be
Q^{(4)}(\zeta)\sim\prod_\alpha^{N_f}(\zeta-m_\alpha),\ \ \
Q^{(5)}(\zeta)\sim\prod_\alpha^{N_f}\sinh(\zeta-m_\alpha),\ \ \
Q^{(6)}(\zeta)\sim\prod_\alpha^{N_f}{\theta_*(\zeta-m_\alpha)\over\theta_*^2
(\zeta-\zeta_i)}
\ee
\be
P^{(4)}\sim\prod_i^{N}(\zeta-a_i),\ \ \
P^{(5)}\sim\prod_i^{N}\sinh(\zeta-a_i),\ \ \
P^{(6)}\sim\prod_i^{N}{\theta_*(\zeta-a_i)\over\theta_*(\zeta-\zeta_i)}
\ee
(in $P^{(5)}(\zeta)$, there is also some exponential of $\zeta$ unless
$N_f=2N$,
see (\ref{scd5'})).

In order to come to the pure gauge theory in $4d$ and $5d$, one should just
put $Q^{(d)}(\zeta)$ equal to unity (it corresponds to sending hypermultiplet
masses to infinity), while in $6d$ it is still unclear how to
remove the restriction $N_f=2N_c$ which prevents constructing the pure
gauge theory.

The perturbative part of the prepotential can be calculated using
these manifest expressions and the residue formula.

\subsection{$4d$ theory: a simple example of calculation}

Let us start from the simplest
example of the Toda chain. In the perturbative limit
(${\Lambda_{QCD}\over\langle\bphi\rangle}\rightarrow 0$)
the second term in (\ref{scg}) vanishes, the curve acquires the form

\be\label{pertcurv}
W = 2P_{N_c}(\lambda),\ \ \ P_{N_c}(\lambda)=\prod (\lambda -\lambda_i)
\ee
(a rational curve with punctures
\footnote{These punctures emerge as a
degeneration of the handles of the hyperelliptic surface so that the
$A$-cycles encircle the punctures.}) and the generating differential
(\ref{dSTC}) turns into

\be\label{pertdS}
dS^{(4)}_{\rm pert} = \lambda d\log P_{N_c}(\lambda)
\ee

Now the set of the $N_c-1$ independent
canonical holomorphic differentials is

\be\label{difff}
d\omega_i=\left({1\over\lambda-\lambda
_i}-{1\over\lambda -\lambda_{N_c}}\right)d\lambda=
{\lambda_{iN_c}d\lambda\over
(\lambda -\lambda_i)(\lambda -\lambda_{N_c})},\ \ \ i=1,...,N_c-1,
\ \ \ \lambda_{ij}\equiv \lambda_i-\lambda_j
\ee
and the $A$-periods $a_i = \lambda_i$ (the independent ones are, say, with
$i=1,\dots,N_c-1$) coincide with the roots of polynomial $P_{N_c}(\lambda)$,
$\lambda_i$.
The prepotential can be now computed via the residue formula

\be
\CF_{pert,ijk}=\stackreb{d\log {P_{N_c}}=0}{\res}
{d\omega_id\omega_jd\omega_k\over d\log
{P_{N_c}(\lambda)}d\lambda}
\ee
which can be explicitly done. The only technical trick is that it
is easier to compute the residues at the poles of $d\omega$'s instead of
the zeroes of $d\log P$. This can be done immediately since there are no
contributions from the infinity $\lambda=\infty$.
The final results have the following form

\be
2\pi i\CF_{pert,iii}=\sum_{k\ne i}{1\over \lambda_{ik}}+{6\over
\lambda_{iN_c}}+
\sum_{k\ne N_c}{1\over \lambda_{kN_c}},
\\
2\pi i\CF_{pert,iij}= {3\over \lambda_{iN_c}}+{2\over
\lambda_{jN_c}}+\sum_{k\ne i,j,N_c}
{1\over \lambda_{kN_c}}-{\lambda_{jN_c}\over \lambda_{iN_c}\lambda_{ij}},
\ \ \ i\ne j,\\
2\pi i\CF_{pert,ijk}=2\sum_{l\ne N_c}{1\over \lambda_{lN_c}}-\sum_{l\ne
i,j,k,N_c}
{1\over \lambda_{lN_c}},
\ \ \ i\ne j\ne k;
\ee
giving rise to the prepotential
formula

\be\label{FA}
{\cal F}_{pert} = {1\over 8\pi i}\sum _{ij}f^{(4)}(\lambda_{ij})
\\
f^{(4)}(x) = x^2\log x^2
\ee

Now let us turn to the case of $4d$ \N2 QCD with massive hypermultiplets.
Then, $Q^{(4)}(\zeta)\ne 1$ and there
arise additional differentials corresponding to the derivatives of $dS$
with respect to masses. They are of the form

\be\label{massdiff}
d\omega_{\alpha}=-{1\over 2}{d\lambda\over \lambda -m_{\alpha}}
\ee
It is again straightforward
to use the residue formula\footnote{Let us note that, despite
the differentials
$d\omega_{\alpha}$ have the pole at infinity, this does not contribute
into the residue formula because of the quadratic pole of $d\lambda$ in the
denominator.\label{5}}

\be\label{residueQ}
\CF_{pert,IJK}=\stackreb{d\log {P\over\sqrt{Q}}=0}{\res}
{d\omega_Id\omega_Jd\omega_K\over d\log
{P(\lambda)\over\sqrt{Q(\lambda)}}d\lambda},
\ \ \ \ \
\left\{I,J,K,\dots\right\}=\left\{i,j,k,\ldots|\alpha,\beta,
\gamma,\ldots\right\}
\ee
and obtain for the prepotential\footnote{The term in the prepotential
that depends only on
masses is not essential for the standard \SW but is crucial for
the prepotential to enjoy its main properties, similar to the WDVV
equations. This term is unambiguously restored from the residue formula.}

\be\label{f4d}
\CF_{pert}={1\over 8\pi i}\sum_{i,j}f^{(4)}(a_{ij})-{1\over 8\pi
i}\sum_{i,\alpha}
f^{(4)}(a_i-m_{\alpha})+{1\over 32\pi i}\sum_{\alpha,\beta}
f^{(4)}(m_{\alpha}-m_{\beta})
\ee
upon the
identification $\lambda_i=a_i$.

\subsection{From $4d$ to higher dimensions}

Now let us turn to the $5d$ case, to the \XXZ chain.
In the perturbative limit, the hyperelliptic curve (\ref{sc2}) turns into the
rational curve

\be\label{pertcurve}
W=2{P_{N_c}(\lambda)\over\sqrt{Q_{N_f}(\lambda)}}\equiv
2{\prod_i^{N_c}\left(\lambda-\lambda_i\right)\over\sqrt{
\prod_{\alpha}^{N_f}\left(\lambda-\lambda_{\alpha}\right)}}
\ee
with $\lambda_i=e^{2a_i}$, $\sum_i^{N_c}a_i=0$, $\lambda_{\alpha}=
e^{2m_{\alpha}}$. First, let us construct the set of holomorphic differentials
$d\omega_I$.

The perturbative differential $dS$ is

\be\label{pertdS5}
dS =\2\log\lambda\  d\log\left(\frac{P(\lambda)}{\sqrt{Q(\lambda)}}\right)=
\2\log\lambda d\lambda
\left(\sum_{i=1}^{N_c}{1\over\lambda-\lambda_i}-\2\sum_{\alpha=1}
^{N_f}{1\over\lambda-\lambda_{\alpha}}\right)
\ee
or, equivalently,

\be\label{pertdS5xi}
dS=\zeta d\zeta\left(\sum_{i=1}^{N_c}\coth(\zeta-a_i)-\2\sum_{\alpha=1}^{N_f}
\coth(\zeta-m_{\alpha})+N_c-\2 N_f\right)
\ee
The variation of (\ref{pertdS5}) gives

\be\label{varpert}
2\delta dS = -\log\lambda d\left(\sum _{i=1}^{N_c}
{\delta\lambda _i\over\lambda - \lambda _i}  -\2\sum _{\alpha = 1}^{N_f}
{\delta\lambda _{\alpha}\over\lambda - \lambda _{\alpha}}\right) \cong
\nn \\
\cong {d\lambda\over\lambda} \sum _{i=1}^{N_c-1}
{\delta\lambda _i\over\lambda _i}\left( {\lambda _i\over\lambda - \lambda _i}-
{\lambda _{N_c}\over\lambda - \lambda _{N_c}}\right) -
{1\over 2}{d\lambda\over\lambda} \sum _{\alpha=1}^{N_f}
{\delta\lambda _{\alpha}\over\lambda _{\alpha}}\left(
{\lambda _{\alpha}\over\lambda - \lambda _{\alpha}}\right)
\ee
where it is implied that only $N_c -1$ parameters $\lambda _i$ are independent
so that ${\delta\lambda _{N_c}\over\lambda _{N_c}} = - \sum _{i=1}^{N_c-1}
{\delta\lambda _i\over\lambda _i}$. Now the set of the $N_c-1$ independent
canonical holomorphic differentials is

\be\label{difff5}
d\omega_i=
{d\lambda \over \lambda }\left( {\lambda _i\over\lambda - \lambda _i}-
{\lambda _{N_c}\over\lambda - \lambda _{N_c}}\right)
= {\lambda_{iN_c}d\lambda\over
(\lambda-\lambda_i)(\lambda-\lambda_{N_c})}\ \ \
i=1,...,N_c-1,
\ \ \ \lambda_{ij}\equiv \lambda_i-\lambda_j
\ee
and one can check that the $A$-periods of $dS$
are $\log\lambda _i = 2a_i$, while the condition of vanishing
the sum of all residues gives the relation (\ref{strange})
$\sum_i a_i=\2\sum m_{\alpha}$. In s.3.5, we discussed how this
strange constraint should be removed off by using the twisted \XXZ
spin chain. The analytical explanation of this is that one can easily
reproduce any lack of residues by putting some additional poles
at infinities in $\zeta$ or at $\lambda=0,\infty$. It is analogous to
the procedure in $4d$. We saw, however, that the same procedure
does not work in $d=6$.

The additional differentials corresponding to the derivatives of $dS$
(\ref{pertdS5}) with respect to masses have the form
\footnote{It can be instructive to
describe in detail why (\ref{difff}) arises not only in $4d$, but also
in $5d$ models. Strictly speaking, in the $5d$ case,
one should consider
differentials on annulus -- on sphere with 2 marked points. Therefore,
instead of
${d\lambda\over\lambda-\lambda_i}$, one rather needs to take

\be\label{diffdr}
\sum_{m=-\infty}^{+\infty}{d\zeta\over
\zeta-a_i+m}\sim\coth{(\zeta-a_i)}d\zeta
={\lambda+\lambda_i\over\lambda-\lambda_i}{d\lambda\over 2\lambda}
\ee
Taking now, instead of (\ref{difff}), the differentials (\ref{diffdr})

\be
d\omega_i={\lambda+\lambda_i\over\lambda-\lambda_i}
{d\lambda\over 2\lambda} -{\lambda+\lambda_{N_c}\over\lambda-\lambda_{N_c}}
{d\lambda\over
2\lambda}= {\lambda_{iN_c}d\lambda\over
(\lambda-\lambda_i)(\lambda-\lambda_{N_c})}
\ee
we obtain again the formula (\ref{difff}).
 On the other hand, the mass
differentials (\ref{massdiff5}) look different from (\ref{diffdr}). It turns
out, however, that the difference does not contribute into the results and,
therefore, one is free to choose any of these two possible mass
differentials.}

\be\label{massdiff5}
d\omega_{\alpha}=-{d\lambda\over 2\lambda}{\lambda_{\alpha}
\over \lambda -\lambda_{\alpha}}
\ee
The better choice for these differentials (see footnote \ref{5}) is, however,
\cite{MarMir}

\be\label{massdiff5'}
d\omega_{\alpha}=-{d\lambda\over 4\lambda}{\lambda+\lambda_{\alpha}\over
\lambda-\lambda_{\alpha}}=-{d\xi\over 2}\coth (\xi-m_\alpha)
\ee
Now one can insert these manifest expressions for differentials into
the residue formula in order to calculate prepotential.

Similar formulas are also correct for the \XYZ chain.
A generic \XYZ curve (\ref{specXYZ1}) in the perturbative limit
turns into

\be\label{pertXYZ}
W = h_0{\prod_{i=1}^{N_c}\theta_{\ast}(\zeta - a_i)\over
\sqrt{\prod _{\alpha =1}^{N_f}\theta_{\ast}(\zeta - m_\alpha)}}
\ee
where $h_0 = {2P_0\over\sqrt{Q_0}}$. Then,
the rescaled generating differential (\ref{dsxyz}) is

\be\label{pertdS6}
dS= \left(\sum\log\theta_{\ast}(\zeta-a_i) - \2\sum\log\theta_{\ast}(\zeta-m_a)
\right) d\zeta + \log h_0d\zeta\cong\\
\cong
\zeta d\zeta\left(\sum_{i=1}^{N_c} {\goth z} (\zeta - a_i) -
\2\sum_{\alpha=1}^{N_f}{\goth z} (\zeta - m_\alpha)\right)
\ee
where the linear piece disappears due to $\sum a_i=\2\sum m_\alpha$ and
${\goth z}$ denotes the \W $\zeta$-function \cite{BEWW}. This
expression should be compared with (\ref{pertdS5xi}) in the $5d$ case.

Note that the sum of residues vanishes only provided the constraint
(\ref{strange6}) is fulfilled. It can not be removed by adding an
additional pole at infinity, since the variable $\zeta$, in contrast to
the $4d$ and $5d$ cases, lives on the compact surface with out
boundaries (torus) and, therefore, is always finite.

Now one can take the
total variation of $dS$ (we neglect the trivial variation $\delta h_0$
giving rise to the holomorphic differential $d\zeta $ on the bare torus)

\be
\delta dS^{XYZ} \cong \sum_{i=1}^{N_c} {\bar {\goth z}}
(\zeta - a_i)\delta a_id\zeta -
\2\sum_{\alpha=1}^{N_f}{\bar {\goth z}} (\zeta - m_\alpha)\delta m_\alpha
d\zeta
\ee
where ${\bar {\goth z}} (\zeta|\tau)\equiv
{\goth z}(\zeta|\tau)-\eta (\tau) \zeta$.
Therefore, the differentials related to gauge moduli are

\be\label{diff6}
d\omega_i=\left({\bar {\goth z}} (\zeta-a_i)-
{\bar {\goth z}} (\zeta-a_{N_c})\right)d\zeta
\ee
while those related to masses are

\be\label{massdiff6}
d\omega_\alpha=-\2\left({\bar {\goth z}} (\zeta-m_\alpha)
-{\bar {\goth z}} (\zeta-a_{N_c})\right)d\zeta
\ee
The second term in this expression is due to the condition
(\ref{strange6}). Thus, one can see that, in the $6d$ case, mass moduli
are practically identical to the gauge ones.

\subsection{Perturbative prepotentials: fundamental matter}

The result of calculation for the prepotentials using all these manifest
expressions for the differentials and the residue formula
is always of the form

\be\label{gp}
{\cal F}_{pert}={1\over 8\pi i}\sum_{i,j}f^{(d)}(a_{ij})-
{1\over 8\pi i}\sum_{i,\alpha}
f^{(d)}(a_i-m_{\alpha})+{1\over 32\pi i}\sum_{\alpha,\beta}
f^{(d)}(m_{\alpha}-m_{\beta})+\\+{\delta_{d,5}\over 24\pi i}(2N_c-N_f)
\left(\sum_i a_i^3+{1\over 4}\sum_{\alpha}m_{\alpha}^3\right)
\ee
The explicit form of these functions is

\be
f^{(4)}(x)=x^2\log x,\ \ \
f^{(5)}(x)=\sum_{n}f^{(4)}\left(x+{n\over R_5}\right)=
{1\over 3}\left|x^3
\right|-{1\over 2}{\rm Li}_3\left(e^{-2|x|}\right),\\
f^{(6)}(x)=\sum_{m,n}f^{(4)}\left(x+\frac{n}{R_{5}}+\frac{m}{R_{6}}\right)=
\sum_n f^{(5)}\left(x+n{R_{5}\over R_6}\right)=
\\
= \left({1
\over 3}\left|x^3\right|
-{1\over 2}{\rm Li}_{3,q}\left(e^{-2|x|}\right)
+ {\rm quadratic\ \ terms}\right)
\ee
so that

\be
{f^{(4)}}''=\log x,\ \ \ {f^{(5)}}''(x)=\log\sinh x,\ \ \
{f^{(6)}}''(x)=\log\theta_*(x)
\ee
Note that, in the $6d$ case,
$N_f$ is always equal to $2N$, and, in $d=5,6$,
there is a restriction $\sum a_i=\sum\zeta_i=\2\sum m_\alpha$
which implies that the gauge moduli would be
rather associated with $a_i$ shifted by the constant
${1\over 2N}\sum m_\alpha$.
In these formulas, ${\rm Li}_{3}(x)$  is the tri-logarithm, while
${\rm Li}_{3,q}(x)$  is the elliptic tri-logarithm
\cite{mamont}. In order to get the prepotential for the pure gauge theories,
one just need to omit all the terms containing mass moduli $m_{\alpha}$.

These formulas perfectly match the perturbative prepotentials obtained in s.5.1
 in supersymmetric gauge theories. This gives another
support for their identification with integrable systems.

Let us discuss now the cubic (in moduli) terms
appearing in the perturbative prepotential
only in higher-dimensional theories.
First, take the $5d$ case, where these terms correspond to
the Chern-Simons term in the field theory Lagrangian
$\Tr (A\wedge F\wedge F)$. In our computation they appear due to the
constant piece $(N_c-\2 N_f)$ in the brackets in (\ref{pertdS5xi}).

In fact, let us
choose the special values of the second Casimir functions at $m$ sites so that
the polynomial $Q(\lambda)$ acquires the form
$\lambda^m\bar Q_{N_f}(\lambda)$, i.e. $m$ of $2N_c$
factors in (\ref{212}) turn into $\lambda^m$. In this case one gets
the constant piece $N_c-m-N_f/2$ in $dS^{(5)}$,
with an integer $m\le 2N_c-N_f$. Therefore, the coefficient in front of
cubic term for $d=5$
in (\ref{gp}) can be made equal to

\be\label{c}
c={1\over 6}(N_c-m-N_f/2)
\ee

Restoring the dependence on $R_5$ in $d=5$ in (\ref{gp}) ($a_i\to
a_iR_5$ and $m_{\alpha}\to m_{\alpha}R_5$) one can study the different
limits of the system.
The simplest limit corresponds to the $4d$ case and is given by
$R_5\to 0$. In this limit, $f^{(5)}(x)
\stackreb{x\sim 0}{\to}f^{(4)}(x)$, the cubic terms vanish
and we reproduce the perturbative $4d$ prepotential (\ref{f4d}).
At the level of
integrable system it is enough to replace $S_i\to R_5S_i$,
$\mu\to e^{R_5\mu}$ in the Lax operators (\ref{l-gen}), (\ref{l-gentw}) in
order to reproduce the Lax operators of the \XXX chain.

Another interesting limit is the limit of flat $5d$ space-time, i.e.
$R_5\to\infty$. In this limit, only cubic terms survive in the prepotential
(\ref{gp}) for $d=5$ (one should carefully fix the branch of $f^{(6)}(x)$ which
leading to appearing of the absolute value in
(\ref{seith})):

\be\label{seith}
2\pi i\CF_{pert}=
{1\over 12}\sum_{i,j}\left|a_{ij}\right|^3-{1\over 12}\sum_{i,\alpha}
\left|a_i+m_{\alpha}\right|^3+{1\over 48}\sum_{\alpha,\beta}
\left|m_{\alpha}-m_{\beta}\right|+\\
+{1\over 12}(2N_c-N_f)
\left(\sum_{i}a_i^3+{1\over 4}\sum_{\alpha}m_{\alpha}^3\right)
\ee
In fact, there are two different sources
of cubic terms \cite{Sei}.
The first one comes from the function $f^{(5)}(x)$.
Since this function can be obtained as the sum over the $4d$
Kaluza-Klein perturbative contributions,
these cubic terms have perturbative origin and come from the 1-loop (due to
the well-known effect of generation of the CS terms in odd-dimensional gauge
theories).

The second source of the cubic terms is due to the bare CS Lagrangian. As it
was shown in \cite{Sei}, one can consider these terms with some
coefficient $c_{cl}$:

\be
{c_{cl}\over 6}\sum_i a^3_i
\ee
restricted only to satisfy the quantization condition $c_{cl}+{N_f\over
2}\in \Z$ and the inequality $\left|c_{cl}\right|\ge N_c-{N_f\over 2}$.
In (\ref{c}), we easily reproduce these conditions so that
(\ref{seith}) coincides with \cite{Sei}, provided
the prepotential is
defined in a fixed Weyl chamber. We leave, however, the wall crossing jumps
out of the discussion.

In the $6d$ case the calculations are quite similar, and one would expect
the similar cubic terms correspond to $\Tr (F\wedge F\wedge F)$. However,
although all the difference
in calculations is due to slightly different mass differentials
(\ref{massdiff5'}) and (\ref{massdiff6}),
the bare cubic terms in $6d$ are absent. This occurs due to
the $N_f=2N_c$ condition (since the coefficient in front of cubic terms is
proportional to $N_c-\2 N_f$) and, therefore, due to the
cancellation of the constant term in the brackets of (\ref{pertdS6})
as compared with (\ref{pertdS5xi}).
This is one of the problems with $6d$ theories: one can not
deform (twist) the theory so that the constraints $N_f=2N_c$ and
(\ref{strange6}) look unavoidable.
As for the ``quantum generated" terms, i.e. those coming from the function
$f^{(6)}$, they are certainly presented, with the coefficient equal to
the $5d$ coefficient.

\subsection{$SU(2)$ adjoint matter in $5d$}

Now let us consider the adjoint matter case. We discuss here only the $5d$
theory, i.e. the \RS system, since the limit to the $4d$ case is done
straightforwardly, while many concrete calculations in
the Calogero-Moser system
look even more tedious.

Suppose we parameterize a bare curve that is elliptic by the
algebraic equation,

\be
\hat y^2
%= (\hat x - e_1(\tau)) (\hat x-e_2(\tau)) (\hat x- e_3(\tau))
=\hat x^3 - \alpha \hat x^2 - \beta \hat x - \gamma.
\label{elc}
\ee
Then we can use the standard parameterization via \W $\wp$-function and
its first derivative:  $\hat x = \wp(\xi)+\alpha/3$ and
$\hat y= -\frac{1}{2}\wp'(\xi)$ so that $d\xi =-2\frac{d\hat x}{\hat y}$.

The perturbative (weak-coupling) limit is then given by the imaginary period
$2\omega'$ becoming infinite, $\tau \rightarrow i\infty$.
In this weak coupling limit (without any double-scaling which we
did considering the degeneration to the periodic Toda chain) the
bare curve (\ref{elc}) becomes

\be
y^2 = x^2(x-1).
\label{perelc}
\ee
Then $\hat x\rightarrow x-1/3$ with $x = \frac{1}{\sin^2\xi}$,
$\hat y\rightarrow y = -\frac{\cos\xi}{\sin^3\xi}$,
$\alpha = 1$, $\beta = \gamma = 0$ and
$d\xi\rightarrow {dx\over{x\sqrt{x-1}}}$, i.e.
$dS=R_5^{-1}\log\lambda d\xi\rightarrow R_5^{-1}\log\lambda
{dx\over{x\sqrt{x-1}}}$. At the same limit,

\be\label{ctrig}
c_0(\xi|\epsilon)\rightarrow{1\over\sqrt{\sin^{-2}\xi-
\sin^{-2}\epsilon}}=
\frac{\sin\xi\sin\epsilon}{\sqrt{\sin(\epsilon-\xi)
\sin(\epsilon+\xi)}},\ \ \
c_{\pm}(\xi|\epsilon) \rightarrow
\frac{\sin\xi\sin\epsilon}
{\sin(\epsilon\pm\xi)}e^{\mp\epsilon\xi/3}\equiv
\bar c_\pm e^{\mp\epsilon\xi/3}.
\ee
The corresponding integrable system is  then nothing but the
trigonometric \RS model we already met in s.2.2.
Finally the special point $\epsilon = \omega = \frac{\pi}{2}$
sees

\be
\wp(\omega) = e_1 \rightarrow  {2\over 3}
\ee
and the particular combination (\ref{cfun}) becomes

\be
c_0(\xi|\epsilon =\pi/2) \rightarrow \tan \xi.
\label{cfunp}
\ee

Now let us look at the period matrix in the perturbative limit in the
simplest $SU(2)$ case, the corresponding spectral curve being (\ref{SU2c1}).
As usual, the derivatives of the 1-form $dS$ with respect to the
moduli are holomorphic differentials.  A particular choice of
coordinates for the moduli will lead to the canonical  holomorphic
1-differentials. For the example at hand there is only one modulus and

\be
dv = \frac{\partial dS}{\partial u} =
R_5^{-1}\frac{1}{\lambda}\frac{\partial\lambda}{\partial u} d\xi
= R_5^{-1}\frac{c_0d\xi}{2\lambda - c_0u} =
{1\over R_5}\frac{d\xi}{\sqrt{u^2 - 4/c_0^2}}
={1\over{ 2 R_5}}\frac{d\xi}{\sqrt{\frac{u^2}{4} +
\wp(\epsilon)-\wp(\xi)}}.
\ee
If we order the roots  $e_3\le e_2\le e_1$,  we may take
for the $A$-integral

\be
\oint_A dv = {1\over i\pi}\int_{e_3}\sp{ e_2} dv
=\frac{1}{i\pi R_5 \sqrt{( \frac{u^2}{4} +\wp(\epsilon)-e_2)(e_1-e_3)}}
K\left(\sqrt{\frac{
(\frac{u^2}{4} +\wp(\epsilon)-e_1)(e_2-e_3)
                  }
{(\frac{u^2}{4} +\wp(\epsilon)-e_2)(e_1-e_3)
                  }
             }
 \,\right).
\label{intAdvtorus}
\ee
Here $K(q)$ is the complete elliptic integral of the first kind.
Dividing $dv$ by the right hand side of this expression would then
give us the canonical holomorphic 1-differential $d\Omega$.
Similarly for the $B$-integral we have

\be
\oint_B dv = {1\over i\pi}\int_{e_2}\sp{ e_1} dv
=-\frac{1}{\pi R_5 \sqrt{( \frac{u^2}{4} +\wp(\epsilon)-e_2)(e_1-e_3)}}
K\left(\sqrt{\frac{
(\frac{u^2}{4} +\wp(\epsilon)-e_3)(e_1-e_2)
                  }
{(\frac{u^2}{4} +\wp(\epsilon)-e_2)(e_1-e_3)
                  }
             }
 \,\right),
\ee
and so the period matrix is

\be
T=\oint_B d\Omega=
-i K\left(\sqrt{\frac{
(\frac{u^2}{4} +\wp(\epsilon)-e_3)(e_1-e_2)
                  }
{(\frac{u^2}{4} +\wp(\epsilon)-e_2)(e_1-e_3)
                  }
             }
 \,\right)
\Bigg/
K\left(\sqrt{\frac{
(\frac{u^2}{4} +\wp(\epsilon)-e_1)(e_2-e_3)
                  }
{(\frac{u^2}{4} +\wp(\epsilon)-e_2)(e_1-e_3)
                  }
             }
 \,\right).
\label{intBdvtorus}
\ee
One can take the limit $\epsilon\to 0$ in this expression to
obtain $T=-\tau$. This agrees perfectly with formula $\CF_{class}\sim\tau a^2$
for the classical
part of the prepotential and justifies our identification of the gauge theory
coupling constant and the modulus of the bare spectral curve.

Let us now calculate the same quantities in the perturbative limit:

\be
dv \rightarrow \left(\frac{1}{2R_5}\right)
\frac{dx}{x\sqrt{(x-1)(U^2 - x)}},
\ee
with $U^2 = \frac{1}{\sin^2\epsilon} + \frac{u^2}{4}$.
We have $e_2=e_3=-1/3$ and $e_1=2/3$ in this limit and so
the $A$-period in this case shrinks to a contour around $x=0$. Now
$\oint_A dv = 1/(2iR_5U)$ and we may identify the
canonical differential $d\Omega=2iR_5Udv$.
This result also follows from (\ref{intAdvtorus}) upon using

$$K(q)\stackreb{q\rightarrow 0}{=}
\frac{\pi}{2}(1+\frac{q^2}{4}+\ldots)
$$
The $B$-period of $d\Omega$ again
gives the period matrix. The corresponding
integral now goes (twice) between $x=0$ and
$x=1$, and the integral $\oint_B dv$  diverges logarithmically
in the vicinity of $x= 0$. This divergence was to be expected
because the period matrix contains a term $\tau$ on the
right hand side (coming from the classical part of the prepotential
$\CF_{class}=-\tau a^2$) and the perturbative limit is given by
$\tau \rightarrow i\infty$.
Upon making the rational substitution $x=\frac{v}{1+v}$ we obtain

\be
\begin{array}{rl}
T_{pert}&= \oint_B d\Omega=
{U\over\pi}
\int_0^1 \frac{dx}{x\sqrt{(x-1)(U^2-x)}} = \lim_{\varepsilon\to 0}
\frac{U}{i\pi\sqrt{U^2-1}}\int_\varepsilon^\infty
\frac{dv}{v\sqrt{v+ \frac{U^2}{U^2-1}}}
\\
&=-
{1\over i\pi}\lim_{\varepsilon\to 0}
\left(\log {\varepsilon\over 4}\right)+
{1\over i\pi}\log\frac{U^2}{1-U^2}
\end{array}
\label{int2}
\ee
where $\varepsilon$ is a small-$x$ cut-off. Thus,
the $U$ dependent part of this integral is finite and can be
considered as the \lq\lq true" perturbative correction, while
the divergent part just renormalizes the bare  \lq\lq classical"
coupling constant $\tau$. Again the same result follows from
(\ref{intBdvtorus}) upon using

$$\lim_{q\rightarrow 1}
\bigg( K(q) -\frac{1}{2}\ln (\frac{16}{1-q\sp2})\bigg)=0.
$$

The final ingredient we wish are the $a$-variables,
i.e. the $A$-period of $dS$ itself. This will correspond to the
integral of (\ref{intAdvtorus}) with respect to $u$, which is a
rather complicated integral. For our purposes
the perturbative limit will suffice when there are several
simplifications. From the definition of (\ref{aad}) and that of
$dS$ we find

$$a=\frac{1}{ R_5}\oint\frac{\log \lambda dx}{x\sqrt{x-1}}
=\frac{\log\lambda|_{x=0}}{i R_5}
$$
Now at $x=0$ we have that $c_0 =i \sin\epsilon$ while at
the same time $\lambda + \lambda^{-1} = c_0u$. Together these yield

$$2\cos aR_5=i u\sin\epsilon$$
and

\be
U^2 = \frac{u^2}{4} + \frac{1}{\sin^2\epsilon} =
\frac{1}{\sin^2\epsilon} (1 - \cos^2 aR_5) =
\frac{\sin^2aR_5}{\sin^2\epsilon}
\ee
Substituting this expression for $U^2$ into (\ref{int2})
we obtain for the finite part of the period matrix

\be\label{su2pp}
T_{pert}^{\rm{finite}}
 = {1\over i\pi}\log \frac{\sin^2 aR_5}{\sin^2\epsilon - \sin^2 aR_5}=
{1\over i\pi}\log\frac{\sin^2aR_5}{\sin\left(\epsilon+aR_5\right)
\sin\left(\epsilon -aR_5\right)}.
\ee

\subsection{Perturbative prepotentials: adjoint matter}

It is quite clear that to repeat this calculation for more general
curve is a hard job. Instead, let us see what is happening with
the full spectral curves under the
perturbative degeneration\footnote{Similar degeneration for the Calogero curves
was
studied in \cite{van}.}. The results turns out to be so strikingly simple,
despite quite involved form of the full spectral curve that this allows
us to immediately calculate the perturbative prepotential for arbitrary
$SU(N_c)$.

We again start with
a warm-up example of $SU(2)$, with the curve (\ref{SU2c1})
at the special point $\epsilon =\pi/2$. Then,
upon using (\ref{cfunp}) we see that (\ref{SU2c1}) reduces in the
perturbative limit to

\be
\cot\xi = \frac{u\lambda}
{\lambda^2 + 1}
\label{SU2cot}
\ee
The rational spectral curve for this situation can be put in the form

\be
w = \frac{(\lambda - \lambda_1)(\lambda - \lambda_1^{-1})}
{(\lambda + \lambda_1)(\lambda + \lambda_1^{-1})}.
\label{SU2B}
\ee
upon setting
$w = -e^{2 i\xi}$ and $i u = (\lambda_1 + \lambda_1^{-1})$.

Similarly in the $SU(3)$ case,
in perturbative limit and at the special point
$\epsilon =\pi/2$ (\ref{SU3cc}) turns into

\be
\lambda^3 - v\lambda = i(1-u\lambda^2)\tan \xi
\label{SU3tan}
\ee
for the choice $c=-ic_0$. Equivalently,

\be
i\tan \xi = \frac{\lambda(\lambda^2 + v)}
{u\lambda^2 + 1}.
\label{SU3cot1}
\ee

This curve can be put in the form

\be
w = -\frac{(\lambda -  \lambda_1)(\lambda - \lambda_2)(\lambda
-  \lambda_3)}
{( \lambda + \lambda_1)( \lambda + \lambda_2)( \lambda+ \lambda_3)}.
\label{SU3B}
\ee
upon the
identification $w = -e^{2i\xi}$ and
with $u = \lambda_1 + \lambda_2 + \lambda_3$
and $v =  \lambda_1\lambda_2 + \lambda_2\lambda_3 + \lambda_3\lambda_1
= \lambda_1\lambda_2\lambda_3
\left(\lambda_1^{-1} + \lambda_2^{-1} + \lambda_3^{-1}\right)$,
$\lambda_1\lambda_2\lambda_3=1$.

Now let us see what is the curve at the special point where $\epsilon=\pi/2$
for the general $SU(N_c)$ case \cite{bmmm1}.

We saw already that substantial simplifications occur at the
special point $\epsilon = \omega$ before even considering the
perturbative limit. Now in the perturbative limit
$\sigma(z)\rightarrow \sin(z)e\sp{z\sp2 /6}$.
Using this we easily obtain that

\be
i\cot\xi={P(\lambda)-(-)^{N_c}P(-\lambda)\over P(\lambda)+
(-)^{N_c}P(-\lambda)},\qquad{\rm and}\qquad
w\equiv -e^{2i\xi}=(-)^{N_c}{P(\lambda)\over P(-\lambda)}.
\ee
with $dS\sim \log\lambda {dw\over w}$.

We finally turn to the case of generic $\epsilon$ in the
perturbative limit.  The system in this limit is described by the
trigonometric \RS model. We have already seen that in this limit
$\sigma$-functions are proportional to sines with
exponential factors. However, the periodicity of
our spectral curve means that these exponential factors
must cancel amongst themselves.
For example we find that in this limit

\be
D_k(\xi|\epsilon) =
 \frac{1}{\bar c_-\sp{k}}. \frac{\sin(\xi + (k-1)\epsilon)}{
\sin(\xi - \epsilon)}.\frac{(-1)\sp{k(k-1)\over 2}}
{\sin^{k(k-2)}(\epsilon)},
\label{Dfuntrig}
\ee
where $\bar c_-$ is given by (\ref{ctrig}).
Now the ratio

\be
\frac{\sin(\xi + (k-1)\epsilon)}{
\sin(\xi - \epsilon)}
=\cos k\epsilon + \sin k\epsilon \cot(\xi - \epsilon)
\label{fractocot}
\ee
is expressible in terms of the single function
$\cot(\xi - \epsilon)$.
This simple observation
enables us to  separate the variables $\xi$
and $\lambda$ in the equation for the spectral curve
upon choosing\footnote{
Note that this separation is not in general  possible
non-perturbatively where there is no analogue of eq.(\ref{fractocot}).
(This is because the ratios
$\frac{\sigma(\xi + (k-1)\epsilon)}{\sigma(\xi-\epsilon)}$ transform
differently under
$\xi-\epsilon \rightarrow \xi-\epsilon + 1$ and
$\xi-\epsilon \rightarrow \xi-\epsilon + \tau$ for different $k$.)
The two notable exceptions are:

(i) The case of $SU(2)$, when the variables  separated
for the standard choice $c = c_0$, see eq.(\ref{SU2c1});

(ii) The case of $\epsilon = \pi/2$ (for any $SU(N_c)$),
when separation of variables survives non-perturbatively,
see eq.(\ref{spsc1}).
}
  $c =-ie^{i\epsilon}\bar c_-$.
With $\bar h_k= {H_k}/ {\sin^{k(k-2)}(\epsilon)}$
we may simplify the spectral curve to give

\be
i\cot(\xi - \epsilon)=
-\frac{\sum_{odd\ k}^{N_c} {\bar h_k}
(-\lambda)^{N_c-k}\left(e^{2ik\epsilon}+1\right)}
{\sum_{even\ k}^{N_c} {\bar h_k}
(-\lambda)^{N_c-k}\left(e^{2ik\epsilon}-1\right)}=
\frac{P(\lambda)+e^{2iN_c\epsilon}P(\lambda
e^{-2i\epsilon})}{P(\lambda)-e^{2i\epsilon N_c}P(\lambda
e^{-2i\epsilon})}
\ee
where $P(\lambda)=\sum_{all\ k}^{N_c} \bar h_k(-\lambda)^{N_c-k}=
\prod_i^{N_c}(\lambda -e^{2ia_i})$ with some constants $a_i$,
$\sum_i^{N_c} a_i=0$. Introducing the variable $w=e^{2i(\xi-\epsilon)}$,
one finally arrives the spectral curve in the form

\be\label{scf}
w=e^{-2i\epsilon N_c}\frac{P(\lambda)}{P(\lambda e^{-2i\epsilon})}
\ee
with
\be\label{dSc}
dS\cong\log\lambda {dw\over w}.
\ee

Thus, we have shown that our system leads in the perturbative limit
to the rational spectral curve
(\ref{scf}) and the generating differential (\ref{dSc}).
One may now calculate the corresponding prepotential using
the residue formula. In fact, this calculation has been
already done for the spin chains in the previous subsections,
where the residue formula is applied to a general
rational function of the form

$$
w\sim \frac{\prod_i\sp{N_c}(\lambda-\lambda_i)}{\sqrt{\prod_\alpha
\sp{N_f}(\lambda-\lambda_\alpha)}},
$$
with $\lambda_i=e\sp{2 a_i}$, $\sum_i a_i =0$, $\lambda_\alpha=
e\sp{2 m_\alpha}$. By choosing $N_f=2 N_c$, with the
hypermultiplets  masses pairwise coinciding and
equal to $a_i+\epsilon$ we obtain our curve (\ref{scf}).
One then finds the prepotential from (\ref{gp})
gives the prepotential for the \RS system

\be\label{appR}
{\cal F}_{pert}^{R}={1\over 8\pi i}\sum_{i,j}f^{(5)}(R_5 a_{ij})-
{1\over 8\pi i}\sum_{i,j}f^{(5)}(R_5a_{ij}+\epsilon)
\ee
and we are done.
Further, upon setting $\epsilon=\pi/2$ we reproduce the results above.

Now in order to obtain the prepotential in
the $4d$ theory with adjoint matter
described by the Calogero-Moser system, one can just take the limit we
described above, $R_5\to 0$ and $\epsilon=M R_5$. Therefore, there is no
special point $\epsilon={\pi\over 2}$ in the Calogero-Moser system!
The perturbative prepotential for the Calogero-Moser system
\cite{WDVVlong,DPhongCP,bmmm2}
is obtain by degenerating (\ref{appR}) so that $f^{(5)}(a)\longrightarrow
f^{(4)}(a)$:

\be\label{appC}
{\cal F}_{pert}^{C}={1\over 8\pi i}\sum_{i,j}f^{(4)}(a_{ij})-
{1\over 8\pi i}\sum_{i,j}f^{(4)}(a_{ij}+M)
\ee
which can be also calculated from the spectral curve obtained by the
degeneration of (\ref{scf})

\be\label{scfC}
w=\frac{P(\lambda)}{P(\lambda - M)}
\ee
with the same differential (\ref{dSc}).

In section 7, we shall see that this form of the perturbative prepotential
persists in the Dell systems.

\section{Theory of prepotential}
\setcounter{equation}{0}

So far we discussed only the properties of \SW related to
underlying integrable systems. However, the whole construction turns out
to give far more. In fact, in previous section we just discussed how
{\it to construct} different physical data in terms of integrable systems, but
almost missed {\it meaning} of several key notions. Maybe the main object
that is to be discussed in far more details is the prepotential, the quantity
not properly interpreted yet in previous sections.

Indeed, looking from the integrable standing point, one should find
natural both emerging the Riemann surfaces with their moduli space as the
corresponding spectral curves, as well
as the proper coordinates on the moduli space given by the action integrals
(\ref{aad}). However, the notion of the prepotential which allows one to
make quantitative predictions and defines the low-energy effective action was
so far introduced by hands, just via formula (\ref{aad}). This formula
implies that not only $A$- but also $B$-periods of the generating differential
$dS$
should get a proper interpretation. This means that one should better
understand
meaning of $dS$ and its properties. One has particularly to recover
the reason for $dS$ to give rise
to holomorphic differentials, or, to put it differently, the reason for the
second derivative of the prepotential to be a period matrix.

In this section we are going to recover another kind of integrability, Whitham
integrability which stands behind the notion of the prepotential and
generating differential $dS$ and discuss
several very essential its features, in particular, the WDVV equations
(\ref{wdvv}).

\subsection{From \SW to Whitham hierarchies: RG dynamics}

Let us consider a generic recipe how the generating differential
is introduced in integrable systems \cite{KriW1,whitham,KriW2}.
It is essentially
based on the analogue of the renormalization group flows in field theory, where
one performs the integration over heavy modes giving rise to the effective
action for light modes. Within the framework of integrable many-body
systems, an
analogous procedure looks as follows. First, one should start with
a solution to the equations of motion with moduli.
At the next step, one assumes that the moduli adiabatically
depend on some new time variables.
In this way, one obtains {\it the Whitham dynamics} on the moduli space,
with the generating differential $dS$ playing the role of
action differential and the prepotential being logarithm
of the $\tau$-function. For instance, for the finite-gap solutions
that are in charge of supersymmetric gauge theories without adjoint matter,
i.e. those described by hyperelliptic curves,
it is natural to choose the corresponding
branching points as these slowly evolving moduli.

Therefore, we come to the conclusion \cite{GKMMM} that \SW
has much to do with {\it two} different
structures: with finite-gap solutions of integrable systems of the
Hitchin type or of spin chains (so that the integrals of motion are
fixed and related to the vacuum expectation values
(vev's) of scalar fields in supersymmetric theories)
{\it and} with the corresponding {\it Whitham} hierarchy constructed in the
vicinity of  these
solutions and describing the evolution of integrals of motion.

Since the vev's of scalar fields physically depend on the scale
$\Lambda_{QCD}$, one associates one of the Whitham evolutions with
the RG dynamics w.r.t. this scale (in fact, $\Lambda_{QCD}$ is identified
with the first Whitham time). Other slow (Whitham) times
describes vev-dependence on perturbations of the original SYM action by
different operators (from topological sector).
To put it
differently, while \SW describes the low-energy effective action,
the Whitham dynamics governs the behaviour of correlators at low energies.
This correlators are from the topological subsector of the theory, since
only such correlators survive at low energies. Note that this subsector is
manifestly described by the Donaldson theory, \cite{MorWit,LNS,MooreM}.

In fact, this is the prepotential that is the generating function of these
(connected) topological correlators, i.e. logarithm of the corresponding
partition function. Therefore, now the prepotential depends on both the
integrals
of motion and the Whitham times as on {\it independent} variables \cite{RG}.
Before defining such an object, let us briefly review the very framework of
Whitham hierarchies, from the very general point of view.

Following I.Krichever \cite{KriW1,KriW2}, we consider
initially a {\it local} system
of holomorphic functions $ \Omega _I$ on complex curve, i.e. functions of
some local parameter $\xi$ in a neighborhood of a point $P$.
One can introduce a set of parameters $t_I$ and define on the space
$(\xi, t_I)$ a $ 1$-form

\begin{equation}\label{1form}
\omega = \sum_I{\Omega _I(P,{\bf t}) \delta t_I}
\end{equation}
Considering its total external derivative
$\delta \omega = \sum_I\delta {\Omega _I(P,{\bf t}) \wedge\delta t_I}$, where
$\delta \omega
= \partial _{\xi} \Omega _I \delta\xi\wedge\delta t_I +
\partial _J \Omega _I\delta t_I \wedge\delta t_J$, one can define general
{\it Whitham equations} as

\begin{equation}\label{gwhi}
\delta \omega\wedge\delta \omega = 0
\end{equation}
so that it is necessary to check the independent vanishing of two
different terms: $ \delta t ^4$ and $ \delta t^3\delta\xi$.
The second term gives

\begin{equation}\label{3whi}
\sum{  \partial _\xi \Omega _{[I}\partial _J \Omega _{C]}} = 0
\end{equation}
where $[...]$ means antisymmetrization. Fixing now some $I=I_0$ and
introducing the ``Darboux'' co-ordinates

\be\label{t0}
t_{I_0}\equiv x \ \ \ \ \ \ \ \Omega _{I_0}(P,{\bf t})\equiv p
\ee
one can get from (\ref{3whi}) the Whitham equations in their standard
form \cite{KriW2}

\be\label{pwhi}
\partial _I \Omega _J - \partial _J \Omega _I + \left\{ \Omega _I ,
\Omega _J \right\} = 0
\\
\left\{ \Omega _I , \Omega _J \right\} \equiv {\partial \Omega _I \over
\partial x}{\partial \Omega _J \over \partial p} - {\partial \Omega
_J \over \partial x}{\partial \Omega _I \over \partial p}
\ee
The explicit form of the equations (\ref{pwhi}) strongly depends on choice
of the local co-ordinate $p$. Equations (\ref{gwhi}), (\ref{3whi}) and
(\ref{pwhi}) are defined only locally and have a huge amount of solutions.

Now let us note that the Whitham equations (\ref{pwhi}) can be considered as
compatibility conditions of the system

\be\label{lam-coo}
{\d\lambda\over\d t_I} = \{\lambda ,\Omega _I\}
\ee
The function $\lambda$ can be used itself in order to define a new local
parameter. In this co-ordinate,
the system (\ref{pwhi}) turns into \cite{KriW2}

\be\label{lam-whi}
\left.{\d\Omega _I(\lambda,t)\over\d t_J}\right|_{\lambda=const}
= \left.{\d\Omega _J(\lambda,t)\over\d t_I}\right|_{\lambda=const}
\ee
or for the differentials $d\Omega_I \equiv d_{\Sigma}\Omega_I =
{\d\Omega_I\over\d\lambda}d\lambda$

\be\label{lam-whid}
\left.{\d d\Omega _I(\lambda,t)\over\d t_J}\right|_{\lambda=const}
= \left.{\d d\Omega _J(\lambda,t)\over\d t_I}\right|_{\lambda=const}
\ee
The Whitham equations in the form (\ref{lam-whi}) imply the existence
of a potential $S$ such that

\be\label{potential}
\left.\Omega _I(\lambda,t)\right|_{\lambda=const}
= \left.{\d S(\lambda,t)\over\d t_I}\right|_{\lambda=const}
\ee
With this potential, the 1-form $\omega$ (\ref{1form}) can be rewritten as

\be
\omega=\delta S - {\d S(\lambda,t)\over\d\lambda}d\lambda
\ee
To fill all these formulas with a real content,
one can consider interesting and important
examples of solutions to the Whitham hierarchy arising when one
takes the basis functions $\Omega _I$ to be the ``quasi-energies'' of the
finite-gap solution;
these are the (globally multi-valued) functions
whose periods determine the ``phase'' of the quasi-periodic Baker-Akhiezer
function corresponding to this finite-gap solution.
The potential should be then identified with

\be\label{intds}
S = \int^P dS
\\
{\d S\over\d t_I} = \int^P {\d dS\over\d t_I} = \int^P d\Omega _I =
\Omega _I
\ee
where $dS$ is, as usual, the generating differential for the finite gap system.
Usually for KP/Toda hierarchy these functions are taken to be
``half''-multi-valued, i.e. their ${\bf A}$-periods are fixed to be zero:
$\oint_{\bf A}d\Omega_I = 0$. The equations (\ref{intds}) are obviously
satisfied in a trivial way in the case of the (finite-gap solutions to)
KP/Toda
theory, when $\{ t_I\}$ are taken to be {\em external} parameters and
$\{\Omega _I\}$ do not depend themselves on $\{ t_I\}$. The idea, however, is
to deform this trivial solution to (\ref{gwhi}), (\ref{3whi}) and
(\ref{pwhi}) into a nontrivial one by {\em a flow} in the moduli space.
Practically it means that the formulas (\ref{intds}) should be preserved even
when $\{\Omega _I\}$ {\em depend} on $\{ t_I\}$, but in a special way --
determined by the equations of Whitham hierarchy. In other words, the Whitham
equations correspond to the dynamics on moduli space of complex curves and
the moduli become depending on the Whitham times.

To realize all this technically, one choose $d\Omega_I$ to be a set of
(normalized) differentials on the complex curve holomorphic outside the
marked points where they have some fixed behavior.
These differentials depend on Whitham times only through moduli dependence.

In fact, for this
whole construction to work (i.e. for the Whitham hierarchy to have some
non-trivial solutions), one needs to choose both differentials (with their
normalization) and local parameter in a clever way. This is the thing
to be added to Seiberg-Witten theory. Indeed, \SW is given by the set of data
that
includes Riemann surfaces, their moduli space and the differential $dS$, while
the
Whitham set-up additionally includes a (non-invariant geometrically) proper
choice of the local parameter.

An important example of the $\lambda $ co-ordinate is the
{\em hyperelliptic} co-ordinate on the hyperelliptic curve
(corresponding to the KdV equation)

\be
y^2= R(\lambda) = \prod_{\beta}
^{2g+2} (\lambda - r_{\beta})
\ee
Properly choosing the differentials $d\Omega_I$ and
expanding the equations
(\ref{lam-whid}) near the point $\lambda = r_{\alpha}$ one gets
the Whitham equations over the finite-gap solutions
of the KdV equation

\be
{\d r_{\alpha}\over\d t_I} =
\left.{d\Omega_J\over d\Omega_I}\right|_{\lambda=r_{\alpha}}
{\d r_{\alpha}\over\d t_J} \equiv
v_{IJ}^{(\alpha )}({\bf r}){\d r_{\alpha}\over\d t_J}
\label{whirie}
\ee
The set of ramification points $\{ {\bf r}\} = \{ r_{\beta}\}$ is usually
called Riemann invariants for the Whitham equations.

The particular case of Whitham hierarchy associated with the finite-gap
solution of the KdV hierarchy described by the hyperelliptic curve

\be\label{KdVsc}
y^2= R(\lambda) = \prod_{\beta}
^{2g+1} (\lambda - r_{\beta})
\ee
was considered first in \cite{FM}.

Note also that in the framework of Whitham hierarchy
it becomes possible to introduce a counterpart of the tau-function,
again a generating function for the solutions, which can be symbolically
defined as \cite{KriW2}

\be\label{taukri}
\log{\cal T}_{\rm Whitham} = \int_{\Sigma}\bar dS\wedge dS
\ee
where the two-dimensional integral over $\Sigma$ is actually ``localized''
on the non-analyticities of $S$. In fact, this $\log{\cal T}_{\rm Whitham}$
is nothing but the prepotential.

Recently an analogue of
Whitham dynamics for deformations of the analytic curves
instead of the complex surfaces has been found \cite{zabrodin,curves}.
At the language of dynamical systems it means that the
dynamics in the space of real Hamiltonians instead of
holomorphic ones is considered.

\subsection{Whitham hierarchy for the Toda chain}

Now let us discuss how one can choose the differentials
$d\Omega_I$ in order to get non-trivial solutions to Whitham hierarchy.
They can be defined for the following set of data:
\begin{itemize}
\item  complex curve (Riemann surface) of genus $g$;
\item  a set of punctures $P_i$ (marked points);
\item co-ordinates $\xi_i$ in the vicinities of the punctures $P_i$;
\item pair of differentials, say, ($d\lambda$, $dz$) with fixed periods.
\end{itemize}
We start with considering the simplest case of a single
puncture, say at $\xi = \xi_0 = 0$. This situation is typical, e.g., for
the Whitham hierarchy in the vicinity of the finite-gap solutions to
the KP or KdV hierarchies. For instance, in the KdV case the
spectral curve has the form (\ref{KdVsc}) so that the marked point is at
$\lambda=\infty$ but, in contrast to the Toda curve (\ref{2}), there is only
one infinite point (since it is the ramification point).

Given this set of data, one can introduce meromorphic
differentials with the poles only at some point $P_0$
such that in some local co-ordinate $\xi$: $\xi (P_0)=\xi_0=0$

\be
d\Omega_n \stackreb{P\to P_0}{=} \left( \xi^{-n-1} + O(1) \right)d\xi,
\ \ \  n\geq 1
\label{canbas}
\ee
This condition defines $d\Omega_n$ up to arbitrary
linear combination of $g$ holomorphic differentials
$d\omega_i$, $i = 1,\ldots,g$ and there are two different
natural ways to fix this ambiguity.
The first way already mentioned
above is to require that $d\Omega_n$ have vanishing
$A$-periods,

\be
\oint_{A_i} d\Omega_n = 0    \ \ \ \forall i,n
\label{vanAp}
\ee
Their generating functional ($\zeta\equiv\xi (P')$, $\xi\equiv\xi (P)$)

\be
W(\xi,\zeta) = \sum_{n=1}^\infty n\zeta^{n-1} d\zeta d\Omega_n(\xi)
+ \dots
\ee
is well known in the theory of Riemann surfaces.
It can be expressed through the Prime form $E(P,P')$ (see Appendix A
of \cite{RG} for details):
\footnote{
For example, for genus $g=1$ in the co-ordinate $\xi \sim \xi+1 \sim
\xi+\tau$
the formulas acquire the form
$$
d\Omega_1 = (\wp (\xi) - const)d\xi ,
\ d\Omega_2 = -\frac{1}{2}\wp'( \xi )d \xi ,
\ \ldots, d\Omega_n = \frac{(-)^{n+1}}{n!}\partial^{n-1}\wp(\xi)d\xi
$$
and
$$
W(\xi,\xi') = \sum_{n=1}^\infty \frac{(-)^{n+1}\xi'^{n-1}}{(n-1)!}
\frac{\partial^{n-1}}{\partial \xi ^{n-1}}\wp( \xi )d \xi d\xi'
- const\cdot d \xi  d\xi' =
\wp(\xi  - \xi')d\xi  d\xi' =
\partial_\xi  \partial_{\xi'} \log \theta_*(\xi - \xi')
$$
where $*$ denotes the (on torus the only one) odd theta-characteristic.
For $g=1$ $\nu_*^2(\xi)
=\theta_{*,i}(0)d\omega_i$ is just $d\xi$. Let us also point out that
chosen in this way
co-ordinate $\xi$ is not convenient from the point of view of Whitham
hierarchy since its ``periods" ($\tau = \oint_B d\xi$) depend on moduli
of the curve.}

\be
W(P,P') = \partial_P \partial_{P'} \log E(P,P')
\label{WE}
\ee
Such  $W(P,P')$ has a second order pole on diagonal $P\to P'$,

\be
W(\xi,\zeta) \sim \frac{d\xi d\zeta}{(\xi - \zeta)^2} + O(1) =
\sum_{n=1}^\infty n\frac{d\xi}{\xi^{n+1}}\zeta^{n-1}d\zeta + O(1)
\label{Wexp}
\ee
It is
the differentials (\ref{canbas}) that should be related with the potential
differential $dS$ by

\be
d\Omega_n = {\d dS\over\d t_n}
\ee
The second way to normalize differentials is to impose the condition

\be
\frac{\partial d\hat\Omega_n}{\partial\ {\rm moduli}} =
{\rm holomorphic}
\label{derdS1}
\ee
so that these differentials becomes similar to the generating differential
$dS_{SW}$ of integrable system.

Now let us specialize this description to the simplest case of \SW
corresponding to the $4d$ pure gauge theory, i.e. to the Toda chain.
Therefore, we deal with the family of spectral curves
(\ref{fsc-Toda}). The first problem is that naively there are no
solutions to eq.(\ref{derdS1}) because the curves (\ref{fsc-Toda}) are
spectral curves of the {\it Toda-chain} hierarchy (not of KP/KdV type).  The
difference is that the adequate description in the Toda case requires {\it
two} punctures instead of one. As already mentioned in sect.3.3, the
curves (\ref{fsc-Toda}) have two marked points and
there exists a function $w$
with the $N_c$-degree pole and zero at two (marked) points $\lambda =
\infty_\pm$, where $\pm$ labels two sheets of the hyperelliptic
representation (\ref{fsc-Toda}), $w(\lambda = \infty_+) = \infty$, $w(\lambda =
\infty_-) = 0$.  Accordingly, there are two families of the differentials
$d\Omega_n$:  $d\Omega^+_n$ with the poles at $\infty_+$ and
$d\Omega^-_n$ with the poles at $\infty_-$.

However, there are {\it no} differentials $d\hat\Omega^\pm$,
only $d\hat\Omega_n = d\hat\Omega^+_n + d\hat\Omega^-_n$, i.e.
condition (\ref{derdS1}) requires $d\hat\Omega_n$ to have the
poles at both punctures. Moreover, the coefficients in front of
$w^{n/N_c}$ at $\infty_+$ and $w^{-n/N}$ at $\infty_-$
(\ref{canbas}) coincide (in Toda-hierarchy
language, this is the Toda-{\it chain} case with the same dependence upon
negative and positive times).

The differentials $d\hat\Omega_n$ for the family (\ref{fsc-Toda})
have the form \cite{RG}:

\be
d\hat\Omega_n = R_n(\lambda)\frac{dw}{w}
= P^{n/N}_+(\lambda)\frac{dw}{w}
\label{expldO}
\ee
The polynomials $R_n(\lambda)$ of degree $n$ in $\lambda$
are defined by the property that
$P'\delta R_n - R_n'\delta P$ is a polynomial of degree
less than $N_c-1$. Thus, $R_n(\lambda) = P^{n/N_c}_+(\lambda)$,
where $\left(\sum_{k=-\infty}^{+\infty} c_k\lambda^k\right)_+
= \sum_{k=0}^{+\infty} c_k\lambda^k$.
For example\footnote{In the case of the Toda chain, since there
are two punctures, one can consider the meromorphic differential
with the simple pole at each of them.}:

\be\label{Rn}
R_0=1,\\
R_1 = \lambda, \nn \\
R_2 = \lambda^2 - \frac{2}{N_c}u_2, \nn \\
R_3 = \lambda^3 - \frac{3}{N_c}u_2\lambda - \frac{3}{N_c}u_3, \nn \\
R_4 = \lambda^4 - \frac{4}{N_c}u_2\lambda^2 - \frac{4}{N_c}u_3\lambda -
\left(\frac{4}{N_c}u_4 + \frac{2(N_c-4)}{N_c^2}u_2^2\right), \nn \\
\ldots
\ee
These  differentials satisfy (\ref{derdS1})
provided the moduli-derivatives are
taken at constant $w$ (not $\lambda$!). Thus, the
formalism of the previous section is applicable
for the local parameter $\xi = w^{\mp 1/N_c}$.\footnote{Rule of thumb
generally is to choose as the local co-ordinate for the Whitham hierarchy the
parameter living on the bare spectral curve. For the Toda case,
one may choose as a bare
curve the $w$-cylinder
with the corresponding generating differential $dS=\lambda
d\log w$.}

Thus, among the data one needs for the definition of solution to
the Whitham
hierarchy associated with the Toda chain, there are the punctures at
$\lambda = \infty_\pm$ and the relevant co-ordinates in the vicinities of
these punctures $\xi \equiv w^{-1/N_c} \sim \lambda^{-1}$ at $\infty_+$ and
$\xi \equiv w^{+1/N_c} \sim \lambda^{-1}$ at $\infty_-$.  The parameterization
in terms of $w$, however, does not allow one to use the advantages of
hyperelliptic form (\ref{2}).

\subsection{Solution to Whitham hierarchy and prepotential}

Now we are going to extend the system of differentials $d\Omega_i$ introduced
in the previous section to include the holomorphic differentials. This will
allow us to construct the function that, as a function of
one set of variables, is the prepotential of the
Seiberg-Witten theory (as defined in sect.2) and in the other set of {\it
independent} variables is the (logarithm of) $\tau$-function
${\cal T}_{\rm Whitham}$ of the associated solution to
the Whitham hierarchy. On this way, we also get a generic solution to
the associated Whitham equations.

Thus, we extend the set of Whitham equations by the
equations involving the holomorphic differentials $d\omega_i$
and moduli $\alpha_i$:

\be\label{whd}
{\d d\Omega_n\over\d \alpha_i}={\d d\omega_i\over\d t_n},\ \ \ \
{\d d\omega_i\over\d \alpha_j}={\d d\omega_j\over\d \alpha_i}
\ee
This system is solved by the differential $dS$ that satisfies

\be\label{dSe}
\frac{\partial dS}{\partial \alpha_i} = d\omega_i, \ \ \
\frac{\partial dS}{\partial t_n} = d\Omega_n
\label{derfdS}
\ee
Then, the first equation in (\ref{dSe}) implies that $dS$ is to be
looked for as a linear combination of the differentials $d\hat\Omega_n$,
satisfying (\ref{derdS1}).
Let us, following \cite{IM1,IM2,RG} introduce a
generating functional for $d\hat\Omega_n$
with infinitely many auxiliary parameters $t_n$:

\be
dS = \sum_{n\geq 1} t_n d\hat\Omega_n =
\sum_{i=1}^g \alpha_i d\omega_i + \sum_{n\geq 1} t_nd\Omega_n
\label{dSexp}
\ee
The periods

\be
\alpha_i = \oint_{A_i} dS
\label{perdef}
\ee
can be considered as particular co-ordinates on the moduli space.
Note that these periods do not exactly coincide with (\ref{aad}),
eq.(\ref{perdef}) defines $\alpha_i$ as functions of $h_k$ and $t_n$,
or, alternatively, $h_k$ as functions of $\alpha_i$ and $t_n$
so that derivatives $\partial h_k/\partial t_n$ are non-trivial.
In what follows we shall consider $\alpha_i$ and $t_n$,

\be
t_n = {\res}_{\xi=0}\ \xi^n dS(\xi)
\label{Tdef}
\ee
as independent variables so that partial derivatives w.r.t.
$\alpha_i$ are taken at constant $t_n$ and partial derivatives
w.r.t. $t_n$ are taken at constant $\alpha_i$.

The differential $dS$ (\ref{dSexp}) determines a generic form of the
solution associated to Seiberg-Witten type Whitham hierarchy. The
Whitham dynamics itself for given $dS$ can be formulated in terms of
equations (\ref{dSe}).  Note that, if one restricts himself to the Whitham
hierarchy with several first times (generally, for the genus $g$ complex
curve there are $g+n-1$ independent times, with $n$ being the number of
punctures), all higher times in (\ref{dSexp}) play the role of constants
(parameters) of generic Whitham solution (see example of $SU(2)$ below).

Note that the Seiberg-Witten differential $dS_{SW}$ is
$dS_{SW} = d\hat\Omega_2$, i.e.

\be
\left.dS\right|_{t_{n}=\delta_{n,1}} = dS_{SW}, \ \ \
\left.\alpha^i\right|_{t_{n}=\delta_{n,1}} = a^i, \ \ \
\left.\alpha^D_i\right|_{t_{n}=\delta_{n,1}} = a^D_i
\ee
and $\alpha$-variables are naturally associated with the Seiberg-Witten
moduli, while $t$-variables -- with the corresponding Whitham times
(although $\alpha_i$'s can be also considered as variables of
Whitham dynamics, cf. with (\ref{whd})).

Now one can introduce the Whitham tau-function (\ref{taukri}) whose logarithm
is a {\it prepotential}
${\cal F}(\alpha_i,t_n)\equiv\log{\cal T}_{\rm Whitham}$
by an analog of conditions
(\ref{aad}):

\be
\frac{\partial {\cal F}}{\partial \alpha_i} = \oint_{B_i} dS, \ \ \
\frac{\partial {\cal F}}{\partial t_n} = \frac{1}{2\pi in}
 {\res}_0\ \xi^{-n}dS
\label{prepdef}
\ee
Their consistency follows from (\ref{derfdS}) and Riemann
identities. In particular,

\be
\frac{\partial^2{\cal F}}{\partial t_m\partial t_n} =
\frac{1}{2\pi in} {\res}_0\ \xi^{-n} \frac{\partial dS}{\partial t_m}
= \frac{1}{2\pi in}{\res}_0\ \xi^{-n} d\Omega_m =
\frac{1}{2\pi im} {\res}_0\ \xi^{-m} d\Omega_n
\label{secderF}
\ee
This calculation implies that the definition
(\ref{prepdef}) assumes the co-ordinate $\xi$ is not changed
under the variation of moduli. It means that this provides a
moduli-independent parameterization of the entire family --
like $w$ in the case of (\ref{fsc-Toda}). Since
moduli-independence of $\xi$ should be also consistent with
(\ref{derdS1}), the choice of $\xi$ is strongly restricted:
to $w^{\pm 1/N_c}$ in the case of (\ref{fsc-Toda}) (see the discussion in the
previous section).

The last relation in (\ref{secderF})
(symmetricity) is just an example of the Riemann relations and it
is proved by the standard argument:

\be
0 = \int d\Omega_m \wedge d\Omega_n =
\oint_{A_i} d\Omega_m \oint_{B_i} d\Omega_n -
\oint_{A_i} d\Omega_n \oint_{B_i} d\Omega_m  +
\frac{1}{2\pi i}{\res} \left(d\Omega_m d^{-1} d\Omega_n\right) =\\=
0 + 0 +  \frac{1}{2\pi in}{\res}_0\ \xi^{-n} d\Omega_m -
\frac{1}{2\pi im}{\res}_0\ \xi^{-m} d\Omega_n
\label{pro}
\ee
where (\ref{vanAp}) and (\ref{canbas}) are used at the final
stage. The factors like $n^{-1}$ arise since
$\xi^{-n-1}d\xi = -d(\xi^{-n}/n)$.  We shall also
use a slightly different normalization $d\Omega_n \sim
w^{n/N_c}\frac{dw}{w} = \frac{N_c}{n}dw^{n/N_c}$, accordingly the residues
in (\ref{pro}) and (\ref{prepdef}) and (\ref{secderF})
will be multiplied by $N_c/n$ instead of $1/n$.

By definition, the prepotential is a homogeneous function of
its arguments $a_i$ and $t_n$ of degree 2,

\be
2{\cal F} =
\alpha_i \frac{\partial{\cal F}}{\partial \alpha_i} +
t_n\frac{\partial{\cal F}}{\partial t_n} =
\alpha_i\alpha_j
\frac{\partial^2{\cal F}}{\partial \alpha_i\partial \alpha_j} +
2\alpha_i t_n
\frac{\partial^2{\cal F}}{\partial \alpha_i\partial t_n} +
t_mt_n \frac{\partial^2{\cal F}}{\partial t_m\partial t_n}
\label{hom}
\ee
Again, this condition can be
proved with the help of Riemann identities, starting from
(\ref{prepdef}), (\ref{perdef}) and (\ref{Tdef}).
At the same time, ${\cal F}$ is not just a quadratic
function of $a_i$ and $t_n$, a non-trivial dependence on
these variables arise through the dependence of $d\omega_i$
and $d\Omega_n$ on moduli (like $u_k$ or $h_k$)
which in their turn depend on $a_i$ and $t_n$. This
dependence is obtained, for example, by substitution of (\ref{dSexp})
into (\ref{derfdS}):

\be\label{whian}
d\hat\Omega_n + t_m\frac{\partial d\hat\Omega_m}{\partial u_l}
\frac{\partial u_l}{\partial t_n} = d\Omega_n,
\ee
i.e.

\be
\left(\sum_{m,l} t_m\frac{\partial u_l}{\partial t_n}\right)
\oint_{A_i}\frac{\partial d\hat\Omega_m}{\partial u_l}
= - \oint_{A_i} d\hat\Omega_n
\ee
The integral in the l.h.s. is expressed, according to (\ref{derdS1}),
through the integrals of holomorphic 1-differentials, while
the integral in the r.h.s. --  through the periods of $dS$.
If

\be
\oint_{A_i} \frac{\partial dS}{\partial u_l} =
\oint_{A_i} dV_l = \Sigma_{il}
\ee
then

\be
t_m\frac{\partial u_k}{\partial t_m}
= \Sigma^{-1}_{ki}\alpha_i
\label{Whitheq}
\ee
The relations (\ref{Whitheq}) can be thought of as the other form of the
Whitham hierarchy.

Thus, we have constructed the prepotential for the Whitham hierarchy
associated with Seiberg-Witten theory.
The Whitham equations themselves describe the dependence of
effective coupling constants given by the period matrix of the
Seiberg-Witten curve on bare coupling constants $t_n$. The
prepotential also depends on characteristics of the vacuum parametrized by
vev's $u_k$. Using this interpretation, one expects \cite{RG} that the
constructed Whitham prepotential has to be associated with the generating
function of topological correlators, the Whitham times being bare coupling
constants of different local operators.
It issues the problem of calculating derivatives of the
prepotential with their further comparison with the topological
correlators. Indeed, the first two derivatives have been calculated in
\cite{RG} and turned out to be

\be
\frac{\partial{\cal F}}{\partial t^n} = \frac{\beta}{2\pi i n}
\sum_m  mt_m {\cal H}_{m+1,n+1}
= \frac{\beta}{2\pi in}t_1{\cal H}_{n+1} + O(t_2,t_3,\ldots)
\label{1der}
\ee
\be
\frac{\partial^2{\cal F}}{\partial \alpha^i\partial t^n}
= \frac{\beta}{2\pi in} \frac{\partial {\cal H}_{n+1}}{\partial a^i}
\label{11der}
\ee
\be
\frac{\partial^2{\cal F}}{\partial t^m\partial t^n}
= -\frac{\beta}{2\pi i} \left({\cal H}_{m+1,n+1}
+ \frac{\beta}{mn}\frac{\partial {\cal H}_{m+1}}{\partial a^i}
\frac{\partial {\cal H}_{n+1}}{\partial a^j}
\partial^2_{ij} \log \theta_E(\vec 0|T)\right)
\label{2der}
\ee
etc. In these formulas parameter $\beta = 2N_c$, $m,n = 1,\ldots,N_c-1$ and
${\cal H}_{m,n}$ are certain homogeneous combinations of
$u_k$, defined in terms of
$P_N(\lambda)$:

\be
{\cal H}_{m+1,n+1} =
-\frac{N_c}{mn}
{\res}_\infty\left(P^{n/N_c}(\lambda)d P^{m/N_c}_+(\lambda)\right)
= {\cal H}_{n+1,m+1}
\ee
and

\be
{\cal H}_{n+1} \equiv {\cal H}_{n+1,2}
= -\frac{N_c}{n}{\res}_\infty P^{n/N_c}(\lambda)
d\lambda
\ee

The \SW itself allows one to evaluate derivatives (\ref{1der}),
(\ref{2der})
at $n=1$, since after appropriate rescaling

\be
h_k \rightarrow t_1^k h_k,\ \ \
{\cal H}_k \rightarrow t_1^k {\cal H}_k
\label{hH}
\ee
$T_1$ can be identified with $\Lambda_{QCD}$. Note that, for doing this, one
needs to use the Toda curve in the form (\ref{fsc-Toda'}).
Eqs. (\ref{1der}),
(\ref{2der}) at $n=1$ are
naturally interpreted in terms of the stress-tensor anomaly,

\be
\ldots +\vartheta^4\Theta^\mu_\mu =
\beta {\rm tr} \Phi^2  = \ldots + \vartheta^4 \beta {\rm tr}
\left(G_{\mu\nu}G^{\mu\nu} + iG_{\mu\nu}\tilde G^{\mu\nu}\right),
\ee
since for any operator ${\cal O}$

\be
\frac{\partial}{\partial \log\Lambda} \langle {\cal O}\rangle \ =
\ \langle \beta {\rm Tr} \Phi^2, {\cal O}\rangle
\label{an}
\ee
Analogous interpretation for $n\geq 2$ involves anomalies of
$W_{n+1}$-structures.

For $n=1$, eq.(\ref{1der}), i.e. eq.(\ref{an}) for
${\cal O}=I$, has been derived
in \cite{Ma} in the form

\be\label{an2}
\frac{\partial{\cal F}_{SW}}{\partial\log\Lambda_{QCD}} =
\frac{\beta}{2\pi i} (t_1^2h_2)
\ee
Eq.(\ref{an}) for ${\cal O} = h_m$, the analogue of (\ref{2der})
for $n=1$ and any $m = 1,\ldots,N_c-1$, is

\be
\frac{\partial h_m}{\partial\log\Lambda_{QCD}} =
-\beta\frac{\partial h_{2}}{\partial a^i}
\frac{\partial h_{m}}{\partial a^j}
\partial^2_{ij} \log \theta_E({\vec 0}|T)
\label{lns}
\ee
Formulas of such a type first appeared for the $SU(2)$ case in \cite{MorWit},
and
literally in this form for the $SU(2)$ case in \cite{LNS,MooreM} and for the
general $SU(N_c)$ case in \cite{LNS}. It was deduced from its modular
properties and also from
sophisticated reasoning in the framework of the Donaldson theory
(the topological correlator is an object
from the Donaldson theory, i.e. topological subsector of
the \N2 SUSY Yang-Mills theory, which can be investigated with
the help of the twisting procedure).
In (\ref{lns}), one can substitute $h_m$ by any homogeneous function
of $h$'s, e.g. by ${\cal H}_m$.

Eq. (\ref{2der}) can be also rewritten as:

\be
\frac{\partial^2}{\partial t^m\partial t^n}
\left({\cal F}(\alpha,t) -
\frac{\beta^2}{4\pi i}{\cal F}^{GKM}(\alpha,t)\right)
= -\frac{\beta^2}{2\pi imn}
\frac{\partial {\cal H}_{m+1}}{\partial a^i}
\frac{\partial {\cal H}_{n+1}}{\partial a^j}
\partial^2_{ij} \log \theta_E(\vec 0|T)
\ee
where the prepotential of the Generalized Kontsevich Model \cite{t+T},

\be
{\cal F}^{GKM}(\alpha|t) \equiv \frac{1}{2N} \sum_{m,n} t_mt_n
{\cal H}_{m+1,n+1}
\ee
implicitly depends on the moduli $\alpha^i$ through
the coefficients of the polynomial $P(\lambda)$.

Here we considered only the simplest example of the $4d$ pure gauge theory.
This approach was generalized to the theories with fundamental matter and with
other gauge groups in \cite{Jose1,Jose2,RGTak}). Further development on the
connections between Whitham hierarchies, $\tau$-functions of integrable
hierarchies
of the KP/Toda type and Donaldson topological theories can be found in
\cite{Jose3}. These latter, very interesting results are still waiting to be
re-interpreted in spirit of the approach of \cite{t+T} to the Generalized
Kontsevich
Model.

As far as we know, the adjoint matter case has not been done yet.

\subsection{Whitham hierarchy for $SU(2)$: an example}

To conclude our discussion of Whitham hierarchies, let us illustrate
the scheme developed above in the simplest case of 2-site periodic Toda
chain\footnote{We do not discuss here how to calculate the second derivative
of the prepotential in this case, since it is not too illuminating. The details
can be found in Appendix B of \cite{RG}.}.
One can restrict himself to the first two
differentials $d\Omega_n$ and the first two times $t_0$,
$t_1$.\footnote{It can be done since the higher flows are dependent on
the two first flows (or, which is the same, $u$ depends only on
two independent combinations of times). Generically,
for a complex curve of genus
$g$, there are $g+1$ independent flows.}
Then, the Whitham equations

\be
{\d d\Omega_i\over\d t_j} = {\d d\Omega_j\over\d t_i}
\label{whitham}
\ee
are reduced to the only equation

\be
{\d d\Omega_0\over\d t_1} = {\d d\Omega_1\over\d t_0}
\label{whithamsl2}
\ee

To write it down explicitly one should remember that
\begin{itemize}
\item
There are two independent differentials (see (\ref{expldO}) and
(\ref{Rn}))

\be
d\Omega_0 = dz + \gamma_0{d\lambda\over y} =
\left(1 +{\gamma_0\over 2\lambda} \right)dz
\\
d\Omega_1 = \lambda dz + \gamma_1{d\lambda\over y} =
\left(\lambda +{\gamma_1\over 2\lambda} \right)dz
\label{diffw}
\ee
and two corresponding Whitham times $t_0$ and $t_1$. Here we introduced the
notation $dz\equiv dw/w$.
\item
The coefficients $\gamma_i$ ($i=0,1$) are fixed, as usual, by
vanishing of the correspondent $A$-periods

\be
\oint_A d\Omega_i = 0
\ee
i.e.

\be
\gamma_1 = - {a\over\sigma},
\ \ \ \ \ \ \
\gamma_0 = -{1\over\sigma}
\label{alphann}
\ee
where

\be
a = \oint_A dS = \oint_A \lambda dz,
\ \ \ \ \ \
\sigma = \oint_A {d\lambda\over y} = {\d a\over\d u}
\label{asigma}
\ee
\item
The derivatives over moduli in (\ref{whitham}) are taken at fixed values
of $z$-variable, while the $\lambda$-variable depends on moduli and this
dependence is given by (\ref{sc2}). This rule leads to the relations

\be
{\d\lambda\over\d u} = {1\over 2\lambda},
\ \ \ \ \ \
{\d a\over\d u} = \oint_A {\d\lambda\over\d u}dz =
\oint_A{dz\over 2\lambda} = \sigma
\ee
\end{itemize}
Using these relations one easily gets from (\ref{whitham}) the following
Whitham equation on moduli

\be\label{whisl2}
{\d u\over\d t_1} = a(u){\d u\over\d t_0}
\ee
The function $a(u)$ is an elliptic integral. Its explicit expression depends
on choice of the cycles. With the choice used in \cite{RG}, i.e. when
the $\bf A$-cycles encircle the points
$\lambda=\sqrt{u-2\Lambda^2}\equiv r^-$ and
$\lambda=\sqrt{u+2\Lambda^2}\equiv r^+$,

\be\label{elint}
a={2\over\pi}r^+E(k),\ \ \ \ \sigma={1\over\pi r^+}K(k)
\ee
where $K(k)$ and $E(k)$ are complete elliptic integrals of the first and
the second kinds respectively and the elliptic modulus is $k={2\over r^+}$.

The Whitham equation (\ref{whisl2}) has the general solution:

\be\label{sol}
u=F\left(t_0+a(u)t_1\right)
\ee
where $F(x)$ is an arbitrary function. This solution can be also
rewritten in the form

\be\label{soli}
\Phi(u)=t_0+a(u)t_1
\ee
where $\Phi(u)$ is the function inverse to $F(x)$.

Thus, we constructed the solution to the Whitham hierarchy just
straightforwardly. Now we do the same, using the solution for
the differential $dS$ (\ref{dSexp}) found in the previous subsection.
We start first with only two first non-zero times so that $dS$ has a
particular form

\be\label{anzatz}
dS=t_0d\hat\Omega_0+t_1d\hat\Omega_1=t_0dz+t_1\lambda dz
\ee
Then,
using formulas of the end of the previous section for the
differentials $d\Omega_{1,2}$, one can easily obtain
from (\ref{whian}) for $n=0,1$ the following two equations:

\be\label{hrenazh}
t_1{\d u\over\d t_1}=-{a\over\sigma},\\
t_1{\d u\over\d t_0}=-{1\over\sigma}
\ee
Any solution to the equations (\ref{hrenazh}) evidently solves
simultaneously (\ref{whisl2}), but not {\it vice versa}.

Let us use the manifest form of functions $a(u)$ and $\sigma(u)$ from
(\ref{elint}):

\be
t_1{\d u\over\d t_1}=-{8\over k^2}{E(k)\over K(k)},\\
t_1{\d u\over\d t_0}=-{2\pi\over\ k}{1\over K(k)}
\ee
Then, the first equation of (\ref{hrenazh}) has the
solution \cite{HB}

\be
t_1=c(t_0){k\over E(k)}
\ee
with arbitrary function $c(t_0)$.
The solution to the both equations (\ref{hrenazh}) takes the form

\be\label{cf}
a(u)={const-t_0\over t_1}
\ee
This particular solution has to be compared with (\ref{soli}) with the
constant function $\Phi(u)$ (since it has no inverse, the second form
(\ref{sol}) does not exist for this concrete solution).

In order to get the general solution (\ref{soli})
to the Whitham hierarchy, one has to require for
all higher odd times in (\ref{dSexp}) (i.e. with $n>1$)
to be non-zero constants.
Then, these higher times parameterize the general function
$\Phi(u)$. Indeed, how it can be easily obtained from (\ref{whian}),
the solution to the Whitham equations with general $dS$ (\ref{dSexp})
is given by formula (\ref{soli}) with

\be
\Phi'(u)=\sum_{k=1}t_{2k+1}(-)^k{(2k-1)!!\over 2^k}u^k
\ee
Therefore, the differential $dS$ (\ref{dSexp}) actually leads to the general
solution to the Whitham hierarchy, the higher times being just constants
which parameterize the solution.

To complete our consideration of particular
system (\ref{hrenazh}), we discuss what happens when one
considers $u$ as
a function of {\it three} independent variables: $t_0$, $t_1$ and
$\alpha$. In this case,
one still gets the same solution depending on {\it two} independent
variables.

Indeed, since

\be\label{alpha1}
\alpha\equiv\oint_A dS=t_1a+t_0
\ee
(cf. with eq.(\ref{cf})) and using (\ref{asigma}),
(\ref{hrenazh})
turns into

\be\label{more}
t_1{\d u\over\d t_1}=-\alpha{\d u\over\d \alpha}+t_0{\d
u\over\d\alpha},\\
{\d u\over\d t_0}=-{\d u\over\d \alpha}
\ee
with the solution

\be\label{solu}
u=\Psi\left({\alpha-t_0\over t_1}\right)
\ee
where $\Psi(x)$ is an arbitrary function. Note that (\ref{cf})
now can be understood just as the condition of constant $\alpha$.

Let us now note that $a(u)$ for the given data is some known
function of $u$. Then, the inverse (not arbitrary!) function $u=\Psi(a)$ can
be rewritten in the form (\ref{solu}) using (\ref{alpha1}) with independent
variables $\alpha$, $t_0$ and $t_1$. It means that one can not
consider arbitrary function $\Psi(x)$ with the particular anzatz
of zero higher times in (\ref{dSexp}), i.e.  solutions to the equations
(\ref{more}) should be additionally constrained.

If now one forgets about the specific anzatz for $dS$
with zero higher times and considers
unconstrained $dS$ (\ref{dSexp}), one should
also consider the additional Whitham equation

\be
{\d d\Omega_n\over\d\alpha} = {\d d\omega\over\d t_n}
\ee
that exactly gives rise to the equations (\ref{more})

\be
{\d u\over\d t_1}=-a{\d u\over\d \alpha},\\
{\d u\over\d t_0}=-{\d u\over\d \alpha}
\ee
These equations are equivalent to (\ref{whisl2}) with $u$ being a function
of $t_0-\alpha$. It means that we again obtain the same solution
(\ref{soli}) that describes $u$ as a function of {\it two} independent
variables $t_1$ and $t_0-\alpha$. This perfectly matches the number $g+n-1=2$
of independent variables.

\subsection{WDVV equations}

We started this review with discussing integrability of \SW and
continued with Whitham hierarchies associated both with integrable and
topological properties of Seiberg-Witten theory. Now we are going to
discuss briefly a feature of the prepotential that has much to do more with
topological theories, while is not that straightforwardly (to our present
knowledge) related to integrability. This property briefly mentioned
in the Introduction is the WDVV equations (\ref{wdvv}),
which are solved by the prepotentials of theories with
the spherical bare spectral curve.

Note that we still do not understand any special reason for the WDVV
equations to emerge in topological and Seiberg-Witten theories,
although, first, the latter two are connected through the Whitham hierarchies
(see the previous subsection) and there are also tight connections of the
Whitham hierarchies with the WDVV equations \cite{typeB}.

To get some insight of general structure of the WDVV
equations, we start with examples not related to Seiberg-Witten theory but
covering the main fields where the WDVV equations have been
met so far (that is, type A and type B $2d$ topological theories).

The simplest non-trivial example of $N_c=3$ WDVV (the equations at
$N_c=1,2$ are trivially satisfied) comes from the type B
topological theory, namely, \N2 SUSY Ginzburg-Landau
theory with the superpotential $W'(\lambda)=\lambda^3-q$ \cite{typeB}.  In
this case, the prepotential reads as

\be
F=\2 a_1a_2^2+\2 a_1^2a_3+{q\over
2}a_2a_3^2
\ee
and the matrices $F_i$ (the third derivatives of the
prepotential) are

\be
F_1 = \left(\begin{array}{ccc} 0&0&1\\0&1&0\\1&0&0
\end{array}\right), \ \ \ \ F_2 = \left(\begin{array}{ccc}
0&1&0\\1&0&0\\0&0&q\end{array}\right),
\ \ \ \
F_3 = \left(\begin{array}{ccc}
1&0&0\\0&0&q\\0&q&0 \end{array}\right).
\label{Fexpl}
\ee
One can easily check that these matrices do really satisfy the WDVV equations
(\ref{wdvv}).

The second example is the quantum cohomologies of $\C\bP ^2$ (type A
topological $\sigma$-model).
In this case, the prepotential is given by the formula \cite{typeA}

\be
F=\2 a_1a_2^2+\2 a_1^2a_3+\sum_{k=1}^{\infty}{N_ka_3^{3k-1}\over
(3k-1)!} e^{ka_2}
\ee
where the coefficients $N_k$ (describing the rational Gromov-Witten classes)
counts the number of the rational curves in
$\C\bP ^2$ and are to be calculated. Since the matrices $F$ have the form

\be
F_1 = \left(\begin{array}{ccc}
0&0&1\\0&1&0\\1&0&0\end{array}\right), \ \ \ \
F_2 = \left(\begin{array}{ccc}
0&1&0\\1&F_{222}&F_{223}\\0&F_{223}&F_{233}\end{array}\right),
\ \ \ \
F_3 = \left(\begin{array}{ccc}
1&0&0\\0&F_{223}&F_{233}\\0&F_{233}&F_{333}\end{array}\right)
\ee
the WDVV equations are equivalent to the identity

\be
F_{333}=F_{223}^2-F_{222}F_{233}
\ee
which, in turn, results into the recurrent relation defining the coefficients
$N_k$:

\be
\frac{N_k}{(3k-4)!} = \sum_{a+b=k}
\frac{a^2b(3b-1)b(2a-b)}{(3a-1)!(3b-1)!}N_aN_b.
\ee
The crucial feature of the presented examples is that, in both cases, there
exists a constant matrix $F_1$. Following \cite{typeB},
one can consider it as a flat
metric on the moduli space. In fact, in its original version, the WDVV
equations have been written in a slightly different form, that is, as
the associativity condition of some algebra. We shall discuss this later, and
now just remark that, having distinguished (constant) metric $\eta\equiv F_1$,
one can naturally rewrite (\ref{wdvv}) as the equations

\be\label{Cas2}
C_iC_j=C_jC_i
\ee
for the matrices $C_i\equiv
\eta^{-1} F_i$, i.e. $\left(C_i\right)^j_k=\eta^{jl}F_{ilk}$. Formula
(\ref{Cas2}) is equivalent to (\ref{wdvv}) with $j=1$. Moreover, this
particular relation is already sufficient \cite{WDVVlong,WDVVr} to reproduce
the whole set of the WDVV equations (\ref{wdvv}). Indeed, since $F_i=F_1
C_i$, we obtain

\be
F_iF^{-1}_jF_k=F_1\left(C_iC^{-1}_kC_j\right)
\ee
which is obviously symmetric under the permutation $i\leftrightarrow j$.
Let us also note that, although the WDVV equations can be fulfilled only for
some specific choices of the coordinates $a_i$ on the moduli space, they
still admit any linear transformation. This defines the flat structures on
the moduli space, and we often call $a_i$ flat coordinates.

In fact, the existence of the flat metric is not necessary for
(\ref{wdvv}) to be true, how we explain below. Moreover, the
Seiberg-Witten theories give
exactly an example of such a case, where there is no distinguished constant
matrix. This matrix can be found in topological theories because of
existence their field theory interpretation where the unity operator is always
presented.

Now note that in the context of the two-dimensional Ginzburg-landau
topological theories, the
WDVV equations arose as associativity condition of some polynomial algebra.
We will prove below that the equations in \SW have the same
origin. Now we briefly remind the main ingredients of this approach in the
standard case of the Ginzburg-Landau theories.

In this case, one deals with the chiral ring formed by a set of polynomials
$\left\{\Phi_i(\lambda)\right\}$ and two co-prime (i.e. without common
zeroes) fixed polynomials $\CQ(\lambda)$ and $P(\lambda)$. The polynomials
$\Phi$ form the associative algebra with the structure constants $C_{ij}^k$
given with respect to the product defined by modulo $P'$:

\be
\Phi_i\Phi_j=C_{ij}^k\Phi_k\CQ'+(\ast)P'\longrightarrow C_{ij}^k\Phi_k\CQ'
\ee
the associativity condition being

\be
\left(\Phi_i\Phi_j\right)\Phi_k=\Phi_i\left(\Phi_j\Phi_k\right),
\ee
\be\label{ass}
\hbox{ i.e. }\ \ \
C_iC_j=C_jC_i,\ \ \ \left(C_i\right)_k^j=C_{ik}^j
\ee
Now, in order to get from these conditions the WDVV equations, one needs to
choose properly the flat moduli \cite{typeB}:

\be
a_i=-{n\over i(n-i)}\hbox{res}\left(P^{i/n}d\CQ\right),\ \ \ n=\hbox{ord} (P)
\ee
Then, there exists the prepotential whose third derivatives are given by
the residue formula

\be\label{resLG}
F_{ijk}=
\stackreb{P' = 0}{\hbox{res}}
\frac{\Phi_i\Phi_j\Phi_k}{P'}
\ee
On the other hand, from the associativity condition (\ref{ass}) and
the residue formula (\ref{resLG}), one obtains that

\be\label{im}
F_{ijk}=\left(C_i\right)_j^lF_{\CQ'lk},\ \ \hbox{ i.e. }\ \ C_i=
F_iF^{-1}_{\CQ'}
\ee
Substituting this formula for $C_i$ into (\ref{ass}), one finally reaches the
WDVV equations in the form

\be
F_i G^{-1} F_j = F_j G^{-1} F_i,\\
G \equiv F_{\CQ'}
\label{WDVVgen}
\ee
The choice $\CQ'=\Phi_l$ gives the standard
equations (\ref{wdvv}). As we already mentioned,
in two-dimensional topological theories, there is
always the unity operator that corresponds to $\CQ'=1$ and leads to the
constant metric $F_{\CQ'}$.

Thus, from this short study of the WDVV equations in the Ginzburg-Landau
theories, we can
get three main ingredients necessary for these equations to hold. These are:
\begin{itemize}
\item
associative algebra
\item
flat moduli (coordinates)
\item
residue formula
\end{itemize}
As we already know, in \SW only the first ingredient requires
a non-trivial check, while the other two are automatically presented due to
the proper integrable structures.

\subsection{WDVV equations in \SW}

Now we are ready to study the most interesting for us example of \SW
and determine when the
prepotential satisfies the WDVV equations.
First, let us note that if the prepotential satisfies the WDVV equations,
its leading perturbative part satisfies them too (since the classical
quadratic piece does not contribute into the third derivatives). In each
case it can be checked by straightforward calculations.

We already discussed the perturbative prepotentials in detail in
the previous section, see also \cite{WDVVlong} which also contains
a proof of the WDVV
equations for the perturbative prepotentials. Here we just list some results of
\cite{WDVVlong} adding some newer statements.

{\bf i)} The WDVV equations always hold for the pure gauge theories
$\CF_{pert}=\CF^V_{pert}$ (including the exceptional groups)
\footnote{The rank of
the group should be bigger than 2 for the WDVV equations not to be empty,
thus for example in the pure gauge $G_2$-model they are satisfied trivially.
It have been checked in \cite{WDVVlong,IY} (using MAPLE package)
that the WDVV
equations hold for the perturbative prepotentials of the pure gauge $F_4$,
$E_6$ and $E_7$ models.
Note that the corresponding non-perturbative Seiberg-Witten curves are not
obviously hyperelliptic.}. In fact, in \cite{WDVVlong} it has been proved
that, if one starts with the general prepotential of the form

\be
\CF={1\over 4}\sum_{i,i} \left(\alpha_-\left(a_{i}-a_{j}
\right)^2\log\left(a_{i}-a_{j}\right)+\alpha_+\left(a_{i}+a_{j}
\right)^2\log\left(a_{i}+a_{j}\right)\right)+{\eta\over 2}\sum_i a_i^2\log a_i
\ee
the WDVV hold iff $\alpha_+=\alpha_-$ or $\alpha_+=0$, $\eta$ being
arbitrary. A new result due to A.Veselov \cite{Ves} shows that this form can
be further generalized by adding boundary terms. For this latter case,
however, we do not know the non-perturbative extension.

{\bf ii)} If one considers the gauge supermultiplets interacting with the
$n_f$
matter hypermultiplets in the first fundamental representation with masses
$m_{\alpha}$

\be\label{complete}
\CF_{pert}=\CF^V_{pert}+r\CF^F_{pert}+Kf_F(m)
\ee
(where $r$ and $K$
are some undetermined coefficients), the WDVV equations do not hold unless
$K=r^2/4$, the masses are regarded
as moduli (i.e. the equations (\ref{wdvv}) contain the derivatives with
respect to masses) and

\be
f_F(m)= {\f
4}
\sum_{\alpha,\beta}\left(\left(m_{\alpha}-m_{\beta}
\right)^2\log\left(m_{\alpha}-m_{\beta}\right)
\right)
\ee
for the $SU(N_c)$ gauge group and

\be
f_F(m)= {\f
4}
\sum_{\alpha,\beta}\left(\left(m_{\alpha}-m_{\beta}
\right)^2\log\left(m_{\alpha}-m_{\beta}\right)+\left(m_{\alpha}+m_{\beta}
\right)^2\log\left(m_{\alpha}+m_{\beta}\right)\right)+\\+
{r(r+s)\over 4}
\sum_{\alpha}m_{\alpha}^2\log m_{\alpha}
\ee
for other classical groups, $s=2$ for the orthogonal groups and
$s=-2$ for the symplectic ones.

Note that at value $r=-2$ the prepotential (\ref{complete}) can be considered
as that in the pure gauge theory with the gauge group of the higher rank,
$\hbox{rank}({\sf G})+n_f$. At the same time,
at value $r=2$, like $a_i$'s lying in irrep of ${\sf G}$, masses
$m_{\alpha}$'s can be regarded as lying in irrep of some $\widetilde {\sf G}$
so
that if ${\sf G}=A_n$, $C_n$, $D_n$, $\ \widetilde
{\sf G}=A_n$, $D_n$, $C_n$
accordingly (this is nothing but the notorious (gauge group
$\longleftrightarrow$ flavor group)
duality, see, e.g. \cite{AS}). These correspondences "explain" the form of
the mass term in the prepotential $f(m)$.

{\bf iii)} The set of the perturbative prepotentials satisfying the WDVV
equations can be further extended. Namely, one can consider higher
dimensional SUSY gauge theories, in particular, $5d$ theories.
Then, the perturbative prepotentials in the $SU(N_c)$ theory (without
adjoint matter) is given by formula (\ref{gp}) with $d=5$.
One can check that these perturbative prepotentials satisfy the WDVV equations.
Note that the presence of the cubic terms in (\ref{gp})
turns to be absolutely crucial for the WDVV equations to hold. This means that
the coefficient $c_c$ in (\ref{seith}) is fixed by the equations. This
is not surprising since that the WDVV equations are not necessarily
satisfied when $c_c\ne N_c-N_f/2$. Since it corresponds to degeneration of an
integrable system, one fixes some moduli (masses) which are the parameters
to vary in the WDVV equations (see \cite{WDVVlong} for details).

One can also consider other classical groups. Then, the perturbative
prepotentials acquire the form (\ref{adjvo}) with all $x^2\log x$ substituted
by ${i\over 3}x^3+{1\over 2} \hbox{Li}_3\left(e^{-2iRx}\right)$. One can
easily check, along the line of \cite{WDVVlong} that these prepotentials
satisfy the WDVV equations.

{\bf iv)} If in the $4d$ theory the adjoint matter hypermultiplets are
presented, i.e. $\CF_{pert}=\CF^V_{pert}+
\CF^A_{pert}+f_A(m)$, the WDVV equations
never hold. At the same time, the WDVV equations are fulfilled for the
theory with matter hypermultiplets in the symmetric/antisymmetric
square of the fundamental representation\footnote{These hypermultiplets
contribute to the prepotential
\be
\CF^{S}_{pert} =  -\frac{1}{4}\sum_{i\leq j} (a_i+a_{j}+m)^2
\log(a_i+a_{j} +m)\\
\CF^{AS}_{pert} =  -\frac{1}{4}\sum_{i < j} (a_i+a_{j}+m)^2
\log(a_i+a_{j} +m)
\ee
} iff the masses of these hypermultiplets are equal to zero.

{\bf v)} Our last example \cite{bmmm1,bmmm2}
has the most unclear status, at the moment. It
corresponds to the $5d$ theory with adjoint matter.
It is described by the perturbative prepotential (\ref{appR}).
This prepotential satisfies the WDVV equations iff $\epsilon=\pi/2$
\cite{bmmm1}. Moreover, it gives the most general solution
for a general class of perturbative prepotentials
${\CF}_{pert}$ assuming the functional form

\be\label{funform}
{\CF}=\sum_{\alpha\in \Phi}f(\alpha\cdot a),
\ee
where the sum is over the root system $\Phi$ of a Lie algebra.
The mystery about this prepotential is that, on one hand, it never satisfies
the WDVV equations unless $\epsilon=\pi$ and we do not know if the
corresponding complete non-perturbative prepotential satisfies the WDVV even
for $\epsilon=\pi/2$. On the other hand, in the limit $R_5 \rightarrow 0$ and
$\epsilon \sim MR$ for finite $M$, (when the mass spectrum (\ref{spectrum})
reduces to the two points $M_n = 0$ and $M_n = M$), the theory is the four
dimensional SYM model with ${\cal N}=4$ SUSY softly broken to \N2, i.e.
includes
the adjoint matter hypermultiplet. It corresponds to item {\bf iv)} when the
WDVV always do {\it not} hold. By all these reasons, this exceptional case
deserves further investigation.

 From the above consideration of the WDVV equations for the perturbative
prepotentials, one can learn the following lessons:
\begin{itemize}
\item
masses are to be regarded as moduli

\item
as an empiric rule, one may say that the WDVV
equations are satisfied by perturbative prepotentials which depend only on
the pairwise sums of the type $(a_i\pm b_j)$, where moduli $a_i$ and $b_j$ are
either periods or masses\footnote{This general rule can be
easily interpreted in D-brane terms, since the interaction of branes
is caused by strings between them. The pairwise structure $(a_i\pm
b_j)$ exactly reflects this fact, $a_i$ and $b_j$ should be identified with
the ends of string.}. This is the case for the models that contain either
massive matter hypermultiplets in
the first fundamental representation (or its dual), or massless
matter in the square product of those.
Troubles arise in all other situations because of the terms
with $a_i\pm b_j\pm c_k\pm\ldots$. (The inverse statement is wrong --
there are some exceptions when the WDVV equations hold despite the presence
of such terms -- e.g., for the exceptional groups.)
\end{itemize}

Note that, for the non-UV-finite theories with the perturbative prepotential
satisfying the WDVV equations, one can add one more parameter to the set of
moduli -- the parameter $\Lambda_{QCD}$ that enters all the logarithmic terms
as
$x^2\log x/\Lambda_{QCD}$. Then, some properly defined WDVV equations still
remain correct despite the matrices $\CF_i^{-1}$ no longer exist (one just
needs to consider instead of them the matrices of the proper minors)
\cite{BM}, see footnote \ref{matone}.

Now let us check the WDVV equations for full Seiberg-Witten prepotentials.
As we already discussed, in order to derive the WDVV equations along the line
used in the context of the Ginzburg-Landau theories,
we need three crucial ingredients:
flat moduli, residue formula and associative algebra. However, the first two
of these are always contained in the SW construction provided the
underlying integrable system is known.
Thus, the only point to be checked is the existence of the associative
algebra. The residue formula (\ref{res}) hints that this algebra is to be
the algebra $\Omega^1$ of the holomorphic differentials $d\omega_i$. In the
forthcoming discussion we restrict ourselves to the case of pure gauge
theory, the general case being treated in complete analogy.

Let us consider the algebra $\Omega^1$ and fix three differentials $dQ$,
$d\omega$, $d\lambda\ \in\Omega^1$. The product in this algebra is given
by the expansion

\be\label{prod}
d\omega_id\omega_j=C^k_{ij}d\omega_kd\CQ+(\ast)d\omega+(\ast)d\lambda
\ee
that should be factorized over the ideal spanned by the differentials
$d\omega$ and $d\lambda$.
This product belongs to the space of quadratic holomorphic differentials:

\be
\Omega^1\cdot\Omega^1\in\Omega^2\cong\Omega^1\cdot\left(d\CQ
\oplus d\omega\oplus d\lambda\right)
\ee
Since the dimension of the space of quadratic holomorphic differentials is
equal to $3g-3$, the l.h.s. of (\ref{prod}) with arbitrary $d\omega_i$'s is
the vector space of dimension $3g-3$. At the same time, at the
r.h.s. of (\ref{prod}) there are $g$ arbitrary coefficients $C_{ij}^k$ in the
first term (since there are exactly so many holomorphic 1-differentials that
span the arbitrary holomorphic 1-differential $C_{ij}^kd\omega_k$), $g-1$
arbitrary holomorphic differentials in the second term (one differential
should be subtracted to avoid the double counting) and $g-2$ holomorphic
1-differentials in the third one. Thus, totally we get that the r.h.s. of
(\ref{prod}) is spanned also by the basis of dimension $g+(g-1)+(g-2)=3g-3$.

This means that the algebra exists in the general case of the SW construction.
However, generally this algebra is not associative. This is because, unlike
the LG case, when it was the
algebra of polynomials and, therefore, the product
of the two belonged to the same space (of polynomials), product in the algebra
of holomorphic 1-differentials no longer belongs to the same space but to the
space of quadratic holomorphic differentials. Indeed, to check associativity,
one needs to consider the triple product of $\Omega^1$:

\be\label{assSW}
\Omega^1\cdot\Omega^1\cdot\Omega^1\in\Omega^3=
\Omega^1\!\cdot
\left(d\CQ\right)^2\oplus\Omega^2\!\cdot d\omega\oplus\Omega^2
\!\cdot d\lambda
\ee
Now let us repeat our calculation: the dimension of the l.h.s. of this
expression is $5g-5$ that is the dimension of the space of holomorphic
3-differentials. The dimension of the first space in expansion of the r.h.s.
is $g$, the second one is $3g-4$ and the third one is $2g-4$. Since
$g+(3g-4)+(2g-4)=6g-8$ is greater than $5g-5$ (unless $g\le 3$), formula
(\ref{assSW}) {\bf does not} define the unique expansion of the triple
product of $\Omega^1$ and, therefore, the associativity spoils.

The situation can be improved if one considers the curves with additional
involutions. As an example, let us consider the family of hyperelliptic
curves: $y^2=Pol_{2g+2}(\lambda)$. In this case, there is the involution,
$\sigma:\ y\to -y$ and $\Omega^1$ is spanned by the $\sigma$-odd holomorphic
1-differentials ${x^{i-1}dx\over y}$, $i=1,...,g$. Let us also note that both
$dQ$ and $d\omega$ are $\sigma$-odd, while $d\lambda$ is $\sigma$-even. This
means that $d\lambda$ can be only meromorphic on the surface without
punctures (which is, indeed, the case in the absence of mass
hypermultiplets). Thus,

$d\lambda$ omits from formula (\ref{prod}) that acquires the form
\be\label{prodhe}
\Omega^2_+=\Omega^1_-\cdot dQ\oplus\Omega^1_-\cdot d\omega
\ee
where we expanded the space of
holomorphic 2-differentials into the parts with definite $\sigma$-parity:
$\Omega^2=\Omega^2_+\oplus\Omega^2_-$, which are manifestly given by the
differentials ${x^{i-1}(dx)^2\over y^2}$, $i=1,...,2g-1$ and
${x^{i-1}(dx)^2\over y}$, $i=1,...,g-2$ respectively.  Now it is easy to
understand that the dimensions of the l.h.s. and r.h.s. of (\ref{prodhe})
coincide and are equal to $2g-1$.

Analogously, in this case, one can check the associativity. It is given by the
expansion

\be
\Omega^3_-=\Omega_-^1\cdot \left(dQ\right)^2\oplus\Omega_+^2\cdot d\omega
\ee
where both the l.h.s. and r.h.s. have the same dimension: $3g-2=g+(2g-2)$.
Thus, the algebra of holomorphic 1-differentials on hyperelliptic curve
is really associative. This completes the proof of the WDVV equations in this
case.

Now let us briefly look at cases when there exist the associative algebras
basing on the spectral curves discussed in the previous section. First of
all, it exists in the theories with the gauge group $A_n$, both in the pure
gauge $4d$ and $5d$ theories and in the theories with fundamental matter,
since, in accordance with the previous section, the corresponding spectral
curves are hyperelliptic ones of genus $n$.

The theories with the gauge groups $SO(n)$ or $Sp(n)$ are also described
by the hyperelliptic curves. The curves, however, are of higher genus
$2n-1$. This would naively destroy all the reasoning of this section.
The arguments,
however, can be restored by noting that the corresponding curves (see
(\ref{charpo})) have yet {\bf another} involution, $\rho:\
\lambda\to-\lambda$. This allows one to expand further the space of
holomorphic differentials into the pieces with definite $\rho$-parity:
$\Omega^1_-=\Omega^1_{--}\oplus\Omega^1_{-+}$ etc. so that the proper algebra
is generated by the differentials from $\Omega^1_{--}$. One can easily check
that it leads again to the associative algebra.

Consideration is even more tricky for the exceptional groups, when the
corresponding curves are looking non-hyperelliptic. However, additional
symmetries should allow one to get associative algebras in these cases too.

There are more cases when the associative algebra exists. First of all, these
are $5d$ theories, with and without fundamental matter \cite{WDVVlong}. One
can also consider the SYM theories with gauge groups being the product of
several factors, with matter in the bi-fundamental representation \cite{W}.
As we discussed earlier, these theories are described by
$sl(p)$ spin chains \cite{GGM1} and the
existence of the associative algebra in this case has been checked in
\cite{Isid}.

The situation is completely different in the adjoint matter case. In four
dimensions, the theory is described by the Calogero-Moser integrable
system. Since, in this case, the curve is non-hyperelliptic and has no
enough symmetries, one needs to include into consideration both the
differentials $d\omega$ and $d\lambda$ for algebra to exist. However, under
these circumstances, the algebra is no longer to be associative how it was
demonstrated above. This can be done also by direct calculation for several
first values of $N_c$ (see \cite{WDVVlong}). This also explains the lack of the
perturbative WDVV equations in this case.

\subsection{Covariance of the WDVV equations}

After we have discussed the role of the (generalized) WDVV equations in SYM
gauge theories of the Seiberg-Witten type, let us briefly describe the
general structure of the equations themselves. We look at
them now just as at some over-defined set of non-linear equations for a
function (prepotential) of $r$ variables\footnote{We deliberately
choose different notations for these variables, $t$ instead of $a$
in the gauge theories,
in order to point out more general status
of the discussion.} (times),
$F(t^i)$, $i=1,\ldots,r$, which can
 be written in the form (\ref{WDVVgen})

\be
F_i G^{-1} F_j = F_j G^{-1} F_i, \nn \\
G = \sum_{k=1}^r \eta^k F_k, \ \ \
\forall i,j = 1,\ldots,r \ \ {\rm and} \ \
\forall \eta^k(t)
\label{WDVV}
\ee
$F_i$ being $r\times r$ matrices
$\displaystyle{(F_i)_{jk} = F_{,ijk} = \frac{\partial^3 F}
{\partial t^i\partial t^j\partial t^k}}$
and the "metric" matrix $G$ is an arbitrary linear
combination of $F_k$'s, with coefficients
$\eta^k(t)$ that can be
time-dependent.\footnote{\label{matone}We already remarked
in the previous subsection
that one can add to the set of times (moduli) in the WDVV
equations the parameter $\Lambda_{QCD}$ \cite{BM}. In this case,
the prepotential that depends on one extra variable $t^0\equiv\Lambda$
can be naturally considered as a homogeneous function of degree $2$:
$$
{\cal F}(t^0,t^1,\ldots,t^r) = (t^0)^2F(t^i/t^0),
$$
see \cite{KriW2,RG} for the general theory.
As explained in \cite{BM}, the WDVV equations
(\ref{WDVV}) for $F(t^i)$ can be also rewritten
in terms of ${\cal F}(t^I)$:
$$
{\cal F}_I \hat{\cal G}^{-1} {\cal F}_J =
{\cal F}_J \hat{\cal G}^{-1} {\cal F}_I,  \ \ \
\forall I,J = 0,1,\ldots,r; \{ \eta^K(t)\}
$$
where this time ${\cal F}_I$ are $(r+1)\times(r+1)$ matrices
of the third derivatives of ${\cal F}$ and
$$
{\cal G} = \sum_{k=0}^r \eta^K {\cal F}_K, \ \ \
\hat{\cal G}^{-1} = (\det {\cal G}) {\cal G}^{-1}
$$
Note that the homogeneity of ${\cal F}$ implies that
$t^0$-derivatives are expressed through those w.r.t. $t^i$, e.g.
$$
t^0{\cal F}_{,0ij}=-{\cal F}_{,ijk}t^k,\ \ \ \
t^0{\cal F}_{,00i}={\cal F}_{,ikl}t^kt^l,\ \ \ \
t^0{\cal F}_{,000}=-{\cal F}_{,klm}t^kt^lt^m \ \ \ \hbox{etc.}
$$
Thus, all the "metrics" ${\cal G}$ are degenerate, but
$\hat{\cal G}^{-1}$ are non-degenerate.
One can easily reformulate the entire present section in terms of
${\cal F}$. Then, e.g., the Baker-Akhiezer vector-function
$\psi(t)$ should be just substituted by the manifestly
homogeneous (of degree 0) function $\psi(t^i/t^0)$.
The extra variable $t^0$ should not be mixed with the
distinguished "zero-time" associated with the constant metric
in the $2d$ topological theories which generically
does not exist (when it does, see comment 3 below,
we identify it with $t^r$).
\label{f1}
}

The WDVV equations imply consistency of the following system
of differential equations \cite{WDVVv1}:

\be
\left( F_{,ijk}\frac{\partial}{\partial t^l} -
       F_{,ijl}\frac{\partial}{\partial t^k} \right)\psi^j(t) = 0,
\ \ \ \forall i, j, k
\label{ls}
\ee
Contracting with the vector $\eta^l(t)$, one can also rewrite
it as

\be\label{*}
\frac{\partial \psi^i}{\partial t^k} = C^i_{jk} D\psi^j,
\ \ \ \forall i, j
\ee
where

\be\label{3}
C_k = G^{-1}F_k, \ \ \ G = \eta^lF_l,\ \ \ D = \eta^l\partial_l
\ee
(note that the matrices $C_k$ and the differential $D$ depend on
choice of $\{\eta^l(t)\}$, i.e. on choice of the metric $G$)
and (\ref{WDVV}) can be rewritten as

\be
\left[C_i,C_j\right]=0,
\ \ \ \forall i, j
\ee

As we already discussed,
the set of the WDVV equations (\ref{WDVV}) is invariant under
{\it linear} change of the time variables with the prepotential
unchanged \cite{WDVVlong}.
According to the second paper of \cite{typeB} and especially to \cite{L},
there can exist also {\it non-linear} transformations
which preserve the WDVV structure, but they generically change
the prepotential.
In \cite{WDVVv2}, it is shown that such transformations are naturally
induced by solutions of the linear system (\ref{ls}):

\be
t^i \ \longrightarrow \ \tilde t^i = \psi^i(t), \nn \\
F(t) \ \longrightarrow \ \tilde F(\tilde t),
\label{tr}
\ee
so that the period matrix remains intact:

\be
F_{,ij} = \frac{\partial^2 F}{\partial t^i\partial t^j}
= \frac{\partial^2 \tilde F}{\partial \tilde t^i\partial \tilde t^j}
\equiv \tilde F_{,\hat i\hat j}
\label{pm}
\ee

Now let us make some comments.

\noindent
{\bf 1.} As explained in \cite{WDVVv1}, the linear system (\ref{ls})
has infinitely many solutions. The "original" time-variables
are among them: $\psi^i(t) = t^i$.

\noindent
{\bf 2.} Condition (\ref{pm}) guarantees that the transformation
(\ref{tr}) changes the linear system (\ref{ls}) only by
a (matrix) multiplicative factor, i.e. the set of solutions
$\{\psi^i(t)\}$ is invariant of (\ref{tr}). Among other
things this implies that successively applying (\ref{tr})
one does not produce new sets of time-variables.

\noindent
{\bf 3.} We already discussed that, in the case of $2d$ topological
models \cite{typeA,typeB,L}, there is a distinguished time-variable,
say, $t^r$, such that all $F_{,ijk}$ are independent of $t^r$:

\be
\frac{\partial}{\partial t^r} F_{ijk} = 0 \ \ \
\forall i,j,k = 1,\ldots, r
\ee
(equivalently, $\frac{\partial}{\partial t^i} F_{rjk} = 0$
$\forall i,j,k$).
Then, one can make the Fourier transform of (\ref{ls})
with respect to $t^r$ and substitute it by the system

\be
\frac{\partial}{\partial t^j} \hat\psi^i_z = zC^i_{jk}
\hat\psi^k_z,
\ \ \ \forall i, j
\ee
where
$\hat\psi^k_z(t^1,\ldots,t^{r-1}) =
\int \psi^k_z(t^1,\ldots,t^{r-1},t^r) e^{zt^r}dt^r$.
In this case, the set of transformations (\ref{tr})
can be substituted by a family, labeled by a single variable $z$:

\be
t^i \ \longrightarrow \ \tilde t^i_z = \hat\psi^i_z(t)
\ee
In the limit $z \rightarrow 0$ and for the particular choice of the
metric, $\check G=F_r$, one obtains the particular
transformation

\be
{\partial \tilde t^i \over \partial t^j}
={\check C}^i_{jk}h^k, \ \ \ h^k = const,
\label{L}
\ee
discovered in \cite{L}. (Since ${\check C}_i=\partial_j \check C$, one can
also write ${\check C}_i={\check C}^i_kh^k$, ${\check
C}^i_k=\left(F_r^{-1}\right)^{il}F_{,lk}$.)

\noindent
{\bf 4.} Parameterization like (\ref{L}) can be used in
the generic situation
(\ref{tr}) as well (i.e. without distinguished $t^r$-variable and for
the whole family (\ref{tr})), the only change is that
$h^k$ is no longer a constant,
but a solution to

\be\label{10}
\left(\partial_j - DC_j\right)^i_kh^k = 0
\ee
($h^k = D\psi^k$ is always a solution, provided $\psi^k$
satisfies (\ref{*})).

Note also that, although we have described a set of non-trivial non-linear
transformations which preserve the structure of the WDVV
equations (\ref{WDVV}), the consideration above does not
{\it prove} that {\it all} such transformations are of the
form (\ref{tr}), (\ref{pm}).
Still, (\ref{tr}) is already unexpectedly large,
because (\ref{WDVV}) is an {\it over-defined} system and
it could seem to be very {\it restrictive}, if to have any solutions
at all.

To conclude this short survey of the prepotentials and WDVV equations,
let us emphasize that there are a lot of problems to be solved.
We already mentioned the problem of lack of the WDVV
equations for the Calogero-Moser system.
The way to resolve this problem might be to construct higher
associativity conditions like it has been done by E.Getzler in the elliptic
case \cite{Getzler}, that is to say, for the elliptic Gromov-Witten
clas\-ses. Even if this will succeed, it is not an {\it explanation} why
at all generalized WDVV equations should be satisfied for the
Seiberg-Witten prepotentials.

\section{Dualities in the many-body systems and gauge theories}

In this section, we are going to discuss duality, a property
of integrable systems which has much to do with the
possibility of exact solving SUSY gauge theories at low energies and with
the possibility of re-interpreting them in terms of strings/branes (see also
the Introduction). This property also will allow us to fill in the right bottom
corner of table 1, i.e. to construct a double elliptic system.

In fact, as in string theory there are several dualities (S-, T- and
U-dualities),
as in integrable systems there are several kinds of duality \cite{gnr}.
We briefly survey all
of them here, and then concentrate on the most interesting, important and still
remaining mysterious one which can be described via
canonical transformations \cite{fgnr,bmmm3}.
Note that all the dualities at the moment are attributed only with
the Hitchin type systems, i.e.
the Seiberg-Witten theories with adjoint matter, while
spin chains
are still not embedded into this treatment.

Let us point out that the dualities under
discussion interchanges different theories, while there is also a duality that
interchanges descriptions of {\it the same} integrable system. This is the
duality
between $N_c\times N_c$ and $2\times 2$ Lax representations, we discuss it in
the
next section.

\subsection{D-branes, gauge theory and separation of variables}

Thus, similarly to branes/string theories, the Hitchin type systems
are involved into three different dualities.
Let us start with considering a counterpart  of the T-duality
which can be identified with the separation of variables
in dynamical systems \cite{gnr}.
The separation of variables (SoV) in integrable system allows one to reduce
the system with $N$ degrees of freedom to $N$ identical systems with one
degree of freedom each.

Recently,
E.Sklyanin formulated a recipe for the  SoV in a
large
class of quantum integrable models \cite{sklyanin}.
The method is to study the
pole dynamics of the Baker-Akhiezer
function $\Psi (z, \lambda)$:

\beq
L(\xi) \Psi(\xi, \lambda) = \lambda (\xi) \Psi(\xi,
\lambda)
\eeq
with some clever choice of normalization which is the main, difficult
and artistic part of the method.
The poles $\xi_{i}$ of $\Psi(\xi, \lambda)$
together with the eigenvalues
$\lambda_{i} = \lambda(\xi_{i})$ are then identified with the
(canonically conjugated) separated variables.

In \cite{gnr} there was constructed a geometrical picture behind the
recipe (see also \cite{efr,takr}). Their construction of the separation of variables
in integrable systems on moduli spaces of holomorphic bundles
with some additional structures that can be described
as a symplectomorphism between the moduli spaces of
bundles (more precisely, torsion free sheaves)
having different Chern classes. In \cite{gnr}, there were discussed
three equivalent realizations of the Hitchin integrable systems.

The first one uses non-Abelian gauge fields on the base curve $\Sigma$
embedded into a symplectic surface $S$. The phase space of the integrable
system is the moduli space of the stable pairs $(\CE,\phi)$, where
$\CE$ is the rank $r$ vector bundle over $\Sigma$ of degree $l$, while
$\phi$ is the holomorphic section of $\omega^1_{_{\Sigma}}\otimes
{\rm End}(\CE)$.
In the second realization, one deals with the moduli space of
pairs $(\CE, \phi)$, where $C$ is the curve (divisor) in $S$ which realizes
the homology class $r[\Sigma]$ and $\CL$ is the line bundle on $C$.
The third realization is the Hilbert scheme of points on $S$
with the number of points determined by the rank of the gauge group\footnote{
The role of the Hilbert schemes in the context of Hitchin systems was
first pointed out in \cite{Hurt} (see also the original context of
\cite{ruijh}) for the surfaces without marked points and
then generalized to the Calogero systems in \cite{wilson,gnr}.}.

The equivalence of these three realizations can be interpreted in
string theory terms. To this end, let us specify the moduli space
$\CM_{\vec v}$ of stable torsion free coherent sheaves ${\CE}$
on $S$. Let ${\hat A}_{S} = 1 - [ {\rm pt} ] \in H^{*} (S,\Z)$
be the $A$-roof
genus of $S$. The vector
$\vec v = Ch ({\CE}) \sqrt{\hat A_{S}} =
(r; \vec w; d - r) \in {H}^{*}(S,\Z), \vec w \in \Gamma^{3,19}$
corresponds to the sheaves
with the Chern numbers:

\be
ch_{0} ({\CE})  = r \in {H}^{0}(S;\Z)\\
ch_{1}({\CE})  = \vec w \in {H}^{2} (S;\Z)\\
ch_{2}({\CE})  = d \in {H}^{4}(S;\Z)
\ee

Now let us describe geometrically string theory objects.
In type IIA string theory compactified onto $S$, there are
Dp-branes with even p which wrap different supersymmetric cycles in $S$
labelled by $\vec v\in H^*(S,\Z)$. The BPS states correspond to the
cohomology classes of the moduli spaces, $\CM_{\vec v}$
of brane configurations.
These moduli spaces can be identified with the moduli spaces of
appropriate sheaves, $\CM_{\vec v}$.

The string theory compactified on $S$, has the following
vacuum moduli space

$$
\CM_{A} = O\left( {\Gamma}^{4,20} \right) \backslash
O(4,20; R) / O(4;R) \times O(20;R)
$$
where the arithmetic group $O({\Gamma}^{4,20})$ is the group
of discrete automorphisms that
maps between the states with the different $\vec v$.
The ${\vec v}^{2}$ is the only invariant of the group action.

Now, the equivalence of the first and the second realizations
of Hitchin systems
corresponds to the physical statement that the bound
states of $N$ $D2$-branes wrapped around $\Sigma$ are represented
by a single $D2$-brane which wraps a holomorphic curve $C$
which
is an $N$-sheet covering of the base curve $\Sigma$.
The equivalence of the second and the third descriptions
is naturally attributed to T-duality.

Let us mention that the separation of variables
above provides some insights on the Langlands duality
which involves the spectrum of the Hitchin Hamiltonians.
An attempt to reformulate Langlands duality as a quantum
separation of variables has been presented in \cite{frenkel}, while
in \cite{fgnr} the authors proposed to identify the  proper
classical version of the Langlands correspondence
with the map of the phase space of the Hitchin system
to the Hilbert scheme of points on four-dimensional manifold.

\subsection{S-duality}

Now let us see that S-duality
well-established in field theory  also
has a clear counterpart
in the holomorphic dynamical system.
In field theory, S-duality just maps one theory onto another so that
the weak coupling regime of one theory maps onto the strong coupling
regime of the other one and vice versa. Therefore, a hint of what is
S-duality in integrable systems can come from \SW where this duality
just maps the weak coupling, perturbative limit of the gauge theory
(due to the asymptotic freedom at large energies,
where the gauge bosons are almost massless, with their masses
proportional to $a_i$) onto the strong coupling limit (where
the BPS monopole is almost massless, with the mass proportional to
$a_i^D$). This means that S-duality interchanges theories with small $a$
and small $a_D$. Note now that, in the integrable set-up, formula (\ref{aad})
it just corresponds to interchanging $A$- and $B$-cycles. Therefore, the
duality transform is just the SP$(2g,\Z)$ group acting on the $2g$ cycles
and the whole set of variables $\{a_i,a_i^D\}$ is just local coordinates
on the SP$(2g,\Z)$ bundle.

It is an easy exercise on the
properties of the period matrix of Abelian varieties
to check that the two form

\beq
da^{i} \wedge da^{D}_{i}
\eeq
vanishes. Therefore one can always locally  find a function
$\CF$, {\it prepotential} such that

\beq
a^{D}_{i} =\frac {\partial \CF}{\partial a^{i}}
\eeq

To illustrate this scheme with some formulas, let us look at the
two-body system, relevant for the
$SU(2)$ $\CN=2$ supersymmetric gauge theory. It is the periodic Toda chain
with the Hamiltonian (see s.3.2)

\beq
H =
{{p^{2}}\over{2}} + \cos (q)
\eeq
The action variable is
given by one of the periods of the differential $pdq$ (\ref{dSTC})
on the curve (\ref{sc2}). Then, the two periods $a$ and $a^D$
obey the  Picard-Fuchs equation

$$
\left( {{d^{2}}\over{du^{2}}} +
 {1\over{4(u^{2}-1)}} \right) \pmatrix{& a\cr& a^{D}\cr} = 0
$$
which can be used to obtain the asymptotic expansions of the action
variable and the prepotential in the vicinity of $u=\infty$ or
$u= \pm 1$. The S-duality manifests itself in the fact that
near $u =\infty$ (which
corresponds to high energies in the two-body problem
and also to the
perturbative regime of the $SU(2)$ gauge theory) the appropriate action
variable
is $a$, while near $u = 1$ (which corresponds to the dynamics of the
two-body
system near the top of the potential and to the strongly coupled
$SU(2)$ gauge theory)
it is $a^{D}$. The monodromy
invariant combination of these
periods \cite{Ma}

\beq
aa^{D} - 2\CF = u
\eeq
(which originates from the periods of
Calabi-Yau manifolds, on one hand,
and from the properties of anomaly in field theory, on the other) can be chosen
as a global coordinate on the space of integrals of motion.
At $u \to \infty$ the prepotential has an expansion of the form

$$
\CF \sim {\half}  u \log u + \ldots \sim a^{2}
\log a + \sum_{n}f_{n} a^{2-4n}
$$

Let us emphasize that S-duality maps the dynamical
system to itself, just at a different energy.

\subsection{``Mirror" symmetry in dynamical systems}

Now we come to the most interesting type of duality, a
duality between pairs of dynamical systems (which is self-duality for
some systems) \cite{rud,ruijh,fgnr,bmmm3}. This duality plays a role
similar to the mirror symmetry.

To start with, let us see how this symmetry is formulated
in field theory. The initial motivation comes
from the 3d theory example \cite{is}, where the mirror symmetry
interchanges the Coulomb and Higgs branches of the moduli space.
The specifics of three dimensions is that both the Coulomb and
Higgs branches are hyperkahler manifolds and the mirror
symmetry can be formulated as a kind of hyperkahler rotation.

Taking into account the relation between dynamical
systems and low-energy effective actions, one can formulate
a counterpart of this mirror symmetry for gauge
theories in different dimensions (see further details in \cite{fgnr}).

Let us consider a system with the phase space
being a hyperkahler four dimensional manifold. Let this manifold
at most involve two tori, i.e. be the
elliptically fibered K3 manifold.
One torus is where momenta live, and
the second one is where coordinates do. The duality at hands
actually interchanges momentum and coordinate tori. However
due to the nontrivial fibration this interchange is far from being
naive.
Note that typically one deals with
degenerations of these tori and only in the double elliptic case
two elliptic moduli survive.

One can also consider a theory which is self-dual. This integrable system
is called Dell (double elliptic)
system\footnote{In fact, this is an example of the Dell system with 2 degrees
of freedom. One may equally consider the Dell system with arbitrary many
degrees of freedom, see ss.6.3,6.4.} and
describes the Seiberg-Witten theory for the six-dimensional
theory with adjoint matter.

The other cases correspond to some degenerations of Dell system.
Say, the degeneration
of the momentum torus to $C/Z_{2}$ leads
to the five-dimensional theory, while the degeneration to $R^2$
leads to the four-dimensional theory. Since the modulus
of the coordinate torus has meaning of the
complexified bare coupling constant in field theory, interpretation
of the degenerations of the coordinate torus is different.
In particular, the degeneration to the cylinder corresponds to switching
off the instanton effects, i.e. to the perturbative regime.

We shall see below that the ``mirror" symmetry maps theories
in different dimensions to each other. Instanton effects
in one theory ``get mapped" into an additional compact dimension
in the dual counterpart. We discuss here the $pq$-duality mostly for
classical integrable systems only. Since
the wave functions in the Hitchin like systems can be
identified with some solutions to the KZ or qKZ equations,
the quantum duality would imply some relations between
solutions to the rational, trigonometric and elliptic KZ
equations. Recently the proper symmetries for the KZ equations
were discussed in \cite{etingof}.

Now let us formulate the same mirror symmetry in terms of dynamical
system.
The idea of duality here expresses a relationship between two completely
integrable systems $S_1, S_2$ on a fixed symplectic manifold with given
symplectic structure $(M,\omega)$ and goes back to
\cite{rud}.
We say the Hamiltonian systems are {\it dual} when the conserved quantities of
$S_1$ and $S_2$ together form a coordinate system for $M$. Consider for
example free particles, $H^{(1)}_k=\sum_i p^k_i/k$. For this system the free
particles momenta are identical to the conserved quantities or action
variables. Now consider the Hamiltonian $H^{(2)}_k=\sum_i q^k_i/k$ with
conserved quantities $q_i$. Together $\{p_i,q_i\}$ form a coordinate system
for phase space, and so the two sets of Hamiltonians are dual. Duality then
in this simplest example is a transformation which interchanges momenta and
coordinates. For more complicated interacting integrable systems finding dual
Hamiltonians is a nontrivial exercise. Note that this whole construction
manifestly depends on the particular choice of conserved quantities. A
clever choice may result in the dual system arising by simply interchanging
the momentum and coordinate dependence, as in the free system.

Some years ago Ruijsenaars \cite{rud} observed such
dualities between various members of the Calogero-Moser and Ruijsenaars
families: the rational Calogero and trigonometric Ruijsenaars models were
dual to themselves while trigonometric Calogero model was dual with the
rational Ruijsenaars system.
These dualities were shown by starting with a
Lax pair $L=L(p,q)$ and an auxiliary diagonal matrix $A=A(q)$. When $L$ was
diagonalized the matrix $A$ became the Lax matrix for the dual Hamiltonian,
while $L$ was a function of the coordinates of the dual system. Dual systems
for
a model possessing a Lax representation are then related to the eigenvalue
motion of the Lax matrix. We discuss this approach in more details below.

Another approach to finding a dual system is to make a canonical
transformation which
substitutes the original set of Poisson-commuting coordinates $q_i$,
$\{q_i,q_j\} = 0$, by another obvious
set of the Poisson-commuting
variables: the Hamiltonians $h_i(\vec p,\vec q)$ or, better, the
action variables $a_i(\vec h) = a_i(\vec p,\vec q)$.
It will be clear below that in practice really interesting
transformations are a little more sophisticated: $h_i$ are identified with
certain functions of the new coordinates (these functions determine the
Ruijsenaars matrix $A(q)$), which -- in the most interesting cases --
are just the same Hamiltonians with the interactions switched-off. Such free
Hamiltonians are functions of momenta alone, and
the dual coordinates substitute
these momenta, just as one had for the system of free particles.

The most interesting question for our purposes is: {\it what are the duals of
the elliptic Calogero and Ruijsenaars systems ?}
Since the elliptic Calogero (Ruijsenaars) is rational
(trigonometric) in momenta and elliptic in the coordinates, the dual
will be elliptic in momenta
and rational (trigonometric) in coordinates.
Having found such a model the final elliptization of the coordinate
dependence is straightforward, providing us with the wanted
double-elliptic systems.

\subsection{Duality as the canonical transformation: two-body system}

The way of constructing dual systems via a canonical transformation
is technically quite tedious for many-particle systems, but
perfectly fits the case of $SU(2)$
which, in the center-of-mass frame, has only one coordinate and
one momentum.
In this case the duality transformation can be described explicitly since
the equations of motion can be integrated in a straightforward way.
Technically, given two Hamiltonian systems, one with the momentum $p$,
coordinate $q$ and
Hamiltonian $h(p,q)$ and another with the momentum $P$,
coordinate $Q$ and
Hamiltonian $H(P,Q)$ we may describe duality by the relation

\be
h(p,q) = f(Q), \nn \\
H(P,Q) = F(q).
\ee
Here the functions $f(Q)$ and $F(q)$ are such that

\be
dP\wedge dQ = -dp\wedge dq,
\ee
which expresses the fact we have an (anti-) canonical transformation. This
relation entails that

\be
F'(q)\frac{\partial h(p,q)}{\partial p} =
f'(Q)\frac{\partial H(P,Q)}{\partial P}.
\label{pbconseq}
\ee

At this stage the functions $f(Q)$ and $F(q)$ are arbitrary.
However, when the Hamiltonians depend on a coupling constant $\nu^
2$
and are such that their ``free'' part can be separated and depends
only on the momenta,\footnote{
Note that this kind of duality relates the weak coupling
regime for $h(p,q)$ to the weak coupling regime for
$H(P,Q)$. For example, in the rational Calogero case
\be\label{ratCdual}
h(p,q) = \frac{p^2}{2} + \frac{\nu^2}{q^2} = \frac{Q^2}{2}\\
H(P,Q) = \frac{P^2}{2} + \frac{\nu^2}{Q^2} = \frac{q^2}{2}
\ee
We recall that, on the field theory side, the coupling constant
$\nu$ is related to the mass of adjoint hypermultiplet and
thus remains unchanged under duality transformations.
}
the free Hamiltonians provide a natural choice for these functions:
$F(q) = h_0(q)$ and $f(Q) = H_0(Q)$ where

\be
h(p,q)\big|_{\nu^2 = 0} = h_0(p), \\
H(P,Q)\big|_{\nu^2 = 0} = H_0(P).
\ee
With such choice the duality equations become

\be
h_0(Q) = h(p,q) \label{duality1},\\
H_0(q) = H(P,Q) \label{duality2},\\
\frac{\partial h(p,q)}{\partial p}H'_0(q) =
h'_0(Q)\frac{\partial H(P,Q)}{\partial P}.
\label{duality3}
\ee

Free rational,
trigonometric and elliptic Hamiltonians are
$h_0(p) = \frac{p^2}{2}$, $h_0(p) = \cosh p$ and $h_0(p) =\cn(p|k)$
respectively.

Note that from the main duality relation,

\be
H_0(q) = H(P,Q)
\label{duality}
\ee
it follows that

\be
\left.\frac{\partial q}{\partial P}\right|_Q = \frac{1}{H_0'(q)}
\frac{\partial H(P,Q)}{\partial P}
\ee
which together with (\ref{duality3}) implies:

\be
\left.\frac{\partial q}{\partial P}\right|_Q =
\frac{1}{h_0'(Q)}\frac{\partial h(p,
q)}{\partial p}
\ee
When compared with the Hamiltonian equation for the
original system,

\be
\frac{\partial q}{\partial t} = \frac{\partial h(p,q)}{\partial p},
\ee
we see that $P = h'_0(Q)t$ is proportional to the ordinary time-variable $t$,
while $h_0(Q)=h(p,q)=E$ expresses $Q$ as a function of the energy $E$.
This is a usual feature of classical integrable systems, exploited in
Seiberg-Witten theory \cite{GKMMM}:
in the $SU(2)$ case the spectral curve $q(t)$ can be described by

\be
h\left(p\left(\frac{\partial q}{\partial t},q\right),\ q\right) = E.
\label{times}
\ee
where $p$ is expressed through $\partial q/\partial t$ and $q$
from the Hamiltonian equation
$\partial q/\partial t = \partial H/\partial p$.
In other words, the spectral curve is
essentially the solution of the equation of motion of integrable system,
where the time $t$ plays the role of the spectral parameter
and the energy $E$ that of the modulus.

Let us consider several simple examples. The simplest one is
the rational Calogero. In this case, the duality transformation connects
two identical Hamiltonians, (\ref{ratCdual}). Therefore, this system is
self-dual.
Somewhat
less trivial example is the trigonometric Calogero-Sutherland model. It leads
to the following dual pair

\be
h(p,q) = \frac{p^2}{2} + \frac{\nu^2}{\sin^2 q} = \frac{Q^2}{2}\\
H(P,Q) = \sqrt{1-{2\nu^2\over Q^2}}\cos{P} = \cos{q}
\ee
One may easily recognize in the second Hamiltonian the rational Ruijsenaars
Hamiltonian. Thus, the Hamiltonian rational in coordinates and momenta
maps onto itself under the duality, while that with rational momentum
dependence and trigonometric coordinate dependence maps onto the Hamiltonian
with inverse, rational coordinate and trigonometric momentum dependencies.
Therefore, we can really see that the $pq$-duality exchanges
types of the coordinate and momentum dependence. This implies that the
Hamiltonian
trigonometrically dependent both on coordinates and momenta is to be
self-dual. Indeed, one can check that the trigonometric
\RS system is self-dual:

\be
h(p,q) = \sqrt{1-{2\nu^2\over q^2}}\cos{p} = \cos{Q}\\
H(P,Q) = \sqrt{1-{2\nu^2\over Q^2}}\cos{P} = \cos{q}
\ee
Now we make the next step and introduce into the game elliptic
dependencies.

\subsection{Two-body systems dual to Calogero and Ruijsenaars}

We begin with the elliptic Calogero
Hamiltonian\footnote{\label{sf}We use
below the Jacobi functions. Recall the standard relations:
\be
\sn (q) = \sqrt{\frac{e_{13}}{\wp(\check q) - e_3}} =
\frac{1}{\sqrt{k}}\frac{\vartheta_1(\hat q)}{\vartheta_4(\hat q)} =
\frac{1}{\sqrt{k}}\frac{\theta_{11}(\hat q|\tau)}
{\theta_{01}(\hat q|\tau)},
\nn \\
\cn (q) = \sqrt{\frac{\wp(\check q) - e_1}{\wp(\check q) - e_3}} =
\sqrt{\frac{k'}{k}}\frac{\vartheta_2(\hat q)}{\vartheta_4(\hat q)} =
\sqrt{\frac{k'}{k}}\frac{\theta_{10}(\hat q|\tau)}
{\theta_{01}(\hat q|\tau)}, \nn \\
\dn (q) = \sqrt{\frac{\wp(\check q) - e_2}{\wp(\check q) - e_3}} =
\sqrt{k'}\frac{\vartheta_3(\hat q)}{\vartheta_4(\hat q)} =
\sqrt{k'}\frac{\theta_{00}(\hat q|\tau)}
{\theta_{01}(\hat q|\tau)}.
\label{ellid}
\ee
Here the Jacobi moduli $k^2$ and $k'^2 = 1 - k^2$
are the cross-ratios of the ramification points of the (hyper-) elliptic
representation of the torus,
\be
y^2 = \prod_{a=1}^3(x - e_a(\tau)), \ \ \
\sum_{a=1}^3 e_a = 0, \ \ \ x = \wp(\check q), \ \ \
y = \frac{1}{2}\wp'(\check q).
\ee
Then
\be
k^2 = \frac{e_{23}}{e_{13}} =
\frac{\vartheta_2^4(0)}{\vartheta_3^4(0)} =
\frac{\theta_{10}^4(0|\tau)}{\theta_{00}^4(0|\tau)}, \ \ \
k'^2 = 1 - k^2 = \frac{e_{12}}{e_{13}} =
\frac{\vartheta_4^4(0)}{\vartheta_3^4(0)} =
\frac{\theta_{01}^4(0|\tau)}{\theta_{00}^4(0|\tau)}
\ee
and
\be
e_{ij} = e_i - e_j, \ \ \
q = 2K(k)\hat q,\ \ \ \check q = 2\omega\hat q, \ \ \
e_{13} = \frac{K^2(k)}{\omega^2}
%\ \ \ \omega = \frac{1}{2}.
\ee
where $K(k)$ is the complete elliptic integral of the first kind.
}

\be
h(p,q) = \frac{p^2}{2} + \frac{\nu^2}{\sn^2(q|k)},
\ee
and seek a dual Hamiltonian elliptic in the momentum. Thus
$h_0(p)=\frac{p^2}{2}$ and we seek $H(P,Q)=H_0(q)
$ such that $H_0(q) =
\cn(q|k)$. Eqs.(\ref{duality}) become

\be\label{QeCH}
\frac{Q^2}{2} = \frac{p^2}{2} + \frac{\nu^2}{\sn^2(q|k)},\ \ \
\cn(q|k) = H(P,Q), \ \ \
p \cdot \cn'(q|k) = Q\frac{\partial H(P,Q)}{\partial P}.
\label{Calduality}
\ee
Upon substituting

\be
\cn'(q|k) = -\sn(q|k)\dn(q|k) =
%-\sqrt{(1 - \\cn^2(q|k))(k'^2 + k^2\cdot \cn^2(q|k))} = \nn \\ =
-\sqrt{(1 - H^2)(k'^2 + k^2 H^2)},
\label{cnid}
\ee
(this is because $\sn^2 q = 1 - \cn^2 q$,
$\dn^2 q = k'^2 + k^2\cn^2 q$, $k'^2 + k^2 = 1$ and
$\cn q = H$)
we get for (\ref{Calduality}):

\be
\left(\frac{\partial H}{\partial P}\right)^2 = \frac{p^2}{Q^2}
(1-H^2)(k'^2 + k^2H^2).
\ee
Now from the first eqn.(\ref{Calduality}) $p^2$ can be expressed
through $Q$ and $\sn^2(q|k) = 1 - \cn^2(q|k) = 1 - H^2$ as

\be
\frac{p^2}{Q^2} = 1 - \frac{2\nu^2}{Q^2(1- H^2)},
\ee
so that

\be
\left(\frac{\partial H}{\partial P}\right)^2
= \left( 1 - \frac{2\nu^2}{Q^2} - H^2\right)
\left( k'^2 + k^2H^2\right).
\ee
Therefore $H$ is an elliptic function of $P$, namely \cite{fgnr,bmmm3}

\be
H(P,Q) = \cn(q|k) = \alpha(Q) \cdot
\cn\left(P\sqrt{k'^2 + k^2\alpha^2(Q)}\ \bigg| \
\frac{k\alpha(Q)}{\sqrt{k'^2 + k^2\alpha^2(Q)}}\right)
\label{dualCal}
\ee
with

\be\label{alpha}
\alpha\sp2(Q) = \alpha\sp2_{rat}(Q) = 1 - \frac{2\nu^2}{Q^2}.
\ee

In the limit $\nu^2 = 0$, when the interaction is switched off,
$\alpha(q) = 1$ and $H(P,Q)$ reduces to $H_0(P) =\ \cn(P|k)$,
as assumed in (\ref{Calduality}).

We have therefore obtained a dual formulation of the elliptic Calogero
model (in the simplest $SU(2)$ case). At first glance our dual Hamiltonian
looks somewhat unusual. In particular, the relevant elliptic curve is
``dressed'': it is described by an effective modulus

\be\label{keff}
k_{eff} = \frac{k\alpha(Q)}{\sqrt{k'^2 + k^2\alpha^2(Q)}}
= \frac{k\alpha(Q)}{\sqrt{1 - k^2(1- \alpha^2(Q))}},
\ee
which differs from the ``bare'' one $k$ in a $Q$-dependent way.
In fact $k_{eff}$ is nothing but the
modulus of the ``reduced''
Calogero spectral curve \cite{IM1,IM2}.

Now one can rewrite (\ref{dualCal}) in many different forms, in particular,
in terms of $\theta$-functions, in hyperelliptic parameterization etc.
This latter is rather illuminating and allows one to picture
mapping between the bare and dressed tori. Therefore,
we refer the reader for the details to
the paper \cite{bmmm3}.

Let us now construct the action-angle variables for the
2-particle Calogero system
and see how they related with the dual variables $P$ and $Q$.

Since this system possesses one degree of freedom,
we proceed similarly to the 2-particle Toda case of s.3.2 and construct
the action variable, $a$ as the integral of differential $dS=pdq$ on the
level submanifold of the phase space. Then, using formula (\ref{QeCH}),
one obtains

\be
a=\oint\sqrt{Q^2-{2\nu^2\over\sn ^2(q)}}dq
\ee
which expresses $a$ through $Q$ quite transcendentally. Using this relation,
one obtains

\be
da=dQ\oint{Q\over\sqrt{Q^2-{2\nu^2\over\sn ^2(q)}}}dq
\ee
We already met this integral in s.5.6, formula (\ref{intAdvtorus}). It can be
manifestly calculated giving

\be\label{da}
da={1\over K(k)}{K\left(k_{eff}(Q)\right)
\over \sqrt{k'^2 + k^2\alpha^2(Q)}}dQ
\ee
where we used the formulas of footnote \ref{sf} for the both bare and dressed
tori as well as the definitions (\ref{alpha}) and (\ref{keff}). Note that
$da=dQ$ at large $Q=\sqrt{2E}$.

Since the coordinate on the Jacobian differs from the argument of the Jacobi
function by a factor $2K(k)$ (see footnote \ref{sf}), one expects for the
angle variable $p^{Jac}$ (looking at the Hamiltonian (\ref{dualCal}))

\be\label{dpJac}
p^{Jac}={P\sqrt{k'^2 + k^2\alpha^2(Q)}\over 2K\left(k_{eff}(Q)\right)}
\ee
This is, indeed, the case, up to inessential
constant normalization factor $2K(k)$
(emerging here since the original Calogero symplectic structure is
defined here as $dp\wedge dq$ with $q$ being the argument of the Jacobi
function, but not $\theta$-function), since it follows from
(\ref{da}) and (\ref{dpJac}) that

\be
dP\wedge dQ={1\over 2K(k)}dp^{Jac}\wedge da
\ee

To conclude our discussion of the Calogero model, we make some
comments on its solution.
According to the argument of the previous subsection, our Hamiltonian
(\ref{dualCal}) should be simply related to the solution $q(t)$
of the equations of motion of the Calogero Hamiltonian,
which in the case of $SU(2)$ are immediately integrated to give

\be
H^{Cal}\left(\frac{\partial q}{\partial P},q\right) = E
\ee

More explicitly, the equation

\be
\frac{dq}{dt} = \sqrt{2E - \frac{2\nu^2}{\sn^2(q|k)}},
\label{caleq}
\ee
has a solution \cite{Per}:

\be
\cn(q|k) = \sqrt{1 - \frac{\nu^2}{E}}\cdot
\cn\left(t\sqrt{2E - 2\nu^2k^2}\left|
k\sqrt{\frac{E-\nu^2}{E-\nu^2k^2}}\right.\right).
\label{calsol}
\label{elcalt}
\ee
This may be derived straightforwardly by
differentiating both sides and applying (\ref{cnid}).
Note that the Calogero equation (\ref{caleq}) and the {\it family}
of Calogero spectral curves are essentially independent of the
value of coupling constant $\nu^2$: it can be absorbed into
rescaling of moduli (like $E$) and the time-variables (like $t$).

In order to see that (\ref{calsol}) is identical to
(\ref{dualCal}) one needs
to put $E = Q^2/2$ and make the rescaling $P = h'_0(Q)t = Qt$.
With these substitutions we find that

\be
\sqrt{1 - \frac{\nu^2}{E}} = \sqrt{1 - \frac{\nu^2}{Q^2}} = \alpha_{rat}(Q),
\ee
and

\be
t\sqrt{2E - 2\nu^2k^2} = P\sqrt{1 - \frac{2\nu^2k^2}{Q^2}} =
P\sqrt{k'^2 + k^2\alpha_{rat}^2(Q)}, \nn \\
k\sqrt{\frac{E-\nu^2}{E-\nu^2k^2}} =
k\sqrt{\frac{1-2\nu^2/Q^2}{1-2\nu^2k^2/Q^2}} =
\frac{k\alpha_{rat}(Q)}{\sqrt{k'^2 + k^2\alpha_{rat}^2(Q)}}.
\ee
We then see that (\ref{calsol}) is identical to (\ref{dualCal}).

We remark that the relevant symplectic structure here is\footnote{
In what follows we prove that not only the
elliptic-rational (the dual of the elliptic Calogero model) but also
the elliptic-trigonometric (the dual of the elliptic Ruijsenaars model)
and the elliptic-elliptic (our new double-elliptic) Hamiltonians
have the same form (\ref{elcalt}), but the latter with the identifications
$E = {1\over 2}\sinh^2Q$ and $E ={1\over 2} \sn^2(Q|\tilde k)$.
Thus they are also related to {\it Calogero} equation
(\ref{caleq}). However, the relevant symplectic structures
-- which are always given by $dP\wedge dQ = h'_0(Q)dt\wedge
dQ = dh_0(Q)\wedge dt$ -- are no
longer equivalent to $dE\wedge dt$ (since $E \neq h_0(Q)$, i.e. $E$ is no
longer associated with the proper Hamiltonian). }

\be
dE\wedge dt =
QdQ\wedge dt = -dP\wedge dQ.
\ee

All of the above formulae are straightforwardly generalized
from the Calogero (rational-elliptic) system to the Ruijsenaars
(tri\-go\-no\-met\-ric-elliptic) system. The only difference ensuing is that
the $q$-dependence of the dual (elliptic-trigonometric) Hamiltonian
is now trigonometric rather than rational \cite{bmmm3} (see also \cite{fgnr}
for
a more involved formula in the action-angle variables; this, however, makes
formula not
convenient for the double elliptic generalization):

\be
\alpha\sp2(q) = \alpha\sp2_{trig}(q) = 1 - \frac{2\nu^2}{\sinh^2 q}
\ee
Rather than giving further details we will proceed directly to
a consideration of the double-elliptic model.

\subsection{Double-elliptic two-body system}

In order to get a double-elliptic system one needs to
exchange the rational $Q$-dependence
in (\ref{Calduality}) for an elliptic one, and so we substitute
$\alpha\sp2_{rat}(Q)$ by the obvious elliptic
analogue $\alpha\sp2_{ell}(Q) = 1 - \frac{2\nu^2}{\sn^2(Q|\tilde k)}$.
Moreover, now the elliptic curves for $q$ and $Q$ need not in general
be the same, i.e. $\tilde k \neq k$.

Instead of (\ref{Calduality}) the duality equations now become

\be
\cn(q|k) = H(P,Q|k,\tilde k), \nn \\
\cn(Q|\tilde k) = H(p,q|\tilde k,k), \nn \\
\cn'(Q|\tilde k)\frac{\partial H(P,Q|k,\tilde k)}{\partial P} =
\cn'(q|k)\frac{\partial H(p,q|\tilde k,k)}{\partial p},
\label{dellduality}
\ee
and the natural anzatz for the Hamiltonian (suggested
by (\ref{dualCal})) is

\be
H(p,q|\tilde k,k) = \alpha(q|\tilde k,k)\cdot
\cn\left(p\;\beta(q|\tilde k,k)\ |\ \gamma(q|\tilde k,k)\right)
= \alpha\cn(p\beta|\gamma), \nn \\
H(P,Q|k,\tilde k) = \alpha(Q|k,\tilde k)\cdot
\cn\left(P\;\beta(Q|k,\tilde k)\ |\ \gamma(Q|k,\tilde k)\right)
= \tilde\alpha \cn(P\tilde\beta|\tilde\gamma).
\label{dellans}
\ee

Substituting these anzatz into (\ref{dellduality}) and making use
of (\ref{cnid}), one can arrive, after some calculations \cite{bmmm3},
at\footnote{For ease of expression hereafter we suppress the dependence of
$\alpha,\beta,\gamma$ on $k$ and $\tilde k$ in what follows
using $\alpha(q)$ for $\alpha(q|\tilde k,k)$
and $\tilde\alpha(Q)$ for $\alpha(Q|k,\tilde k)$ etc.}

\be\label{ab}
\alpha^2(q|\tilde k,k) = \alpha^2(q|k) =
1 - \frac{2\nu^2}{\sn^2(q|k)}, \ \ \
\beta^2(q|\tilde k,k) = \tilde k'^2 + \tilde k^2\alpha^2(q|k), \ \ \
\gamma^2(q|\tilde k,k) = \frac{\tilde k^2\alpha^2(q|k)}
{\tilde k'^2 + \tilde k^2\alpha^2(q|k)},
\ee
with some constant $\nu$ and finally the double-elliptic duality becomes
\cite{bmmm3}

\be
H(P,Q|k,\tilde k)= \cn(q|k) =
\alpha(Q|\tilde k) \cn\left(
P\sqrt{k'^2 + k^2\alpha^2(Q|\tilde k)}\ \bigg|\
\frac{k\alpha(Q|\tilde k)}{\sqrt{k'^2 + k^2\alpha^2(Q|\tilde k)}}
\right),
\ee
\be\label{dellHsu2}
H(p,q|\tilde k,k) = \cn(Q|\tilde k) =
\alpha(q|k) \cn\left(p\sqrt{\tilde
k'^2 + \tilde k^2\alpha^2(q|k)}\ \bigg|\
\frac{\tilde k\alpha(q|k)}
{\sqrt{\tilde k'^2 + \tilde k^2\alpha^2(q|k)}}
\right).
\ee

We shall now consider various limiting cases arising from these
and  show that the double-elliptic Hamiltonian (\ref{dellHsu2})
contains the entire Ruijsenaars-Calogero and Toda family
as its limiting cases, as desired. (Of course we have restricted ourselves
to the $SU(2)$ members of this family.)

In order to convert the elliptic dependence of the momentum $p$
into the trigonometric one, the corresponding ``bare''
modulus $\tilde k$ should vanish: $\tilde k \rightarrow 0, \
\tilde k'^2 = 1 - \tilde k^2 \rightarrow 1$ (while $k$
can be kept finite). Then, since
$\cn(x|\tilde k = 0) = \cosh x$,

\be
H^{dell}(p,q) \longrightarrow \alpha(q)\cosh p =
H^{R}(p,q)
\ee
with the same

\be
\alpha^2(q|k) = 1 - \frac{2\nu^2}{\sn^2(q|k)}.
\ee
Thus we obtain the $SU(2)$ elliptic Ruijsenaars Hamiltonian.\footnote{
Indeed,
$$
F^2(q) = c^2(\check\epsilon|k)
\left(\wp(\check\epsilon) - \wp(\check q)\right)
= c^2(\check\epsilon|k)\left(\frac{e_{13}}
{\sn^2(\sqrt{e_{13}}\check \epsilon|k)} -
\frac{e_{13}}{\sn^2(\sqrt{e_{13}}\check q|k)}\right)
=
\frac{c^2(\check\epsilon|k) e_{13}}{\sn^2(q|k)}
\left(1 - \frac{\sn^2(\epsilon|k)}{\sn^2(q|k)}\right)
$$
where $q = 2\omega\hat q\sqrt{e_{13}}$
and $2\nu^2 = \sn^2(\epsilon|k)$.
}
The trigonometric and rational Ruijsenaars
as well as all of the Calogero and
Toda systems are obtained through further limiting procedures
in the standard way.

The other limit $k \rightarrow 0$ (with $\tilde k$ finite) gives
$\alpha(q|k) \rightarrow \alpha_{trig}(q) = 1 - \frac{2\nu^2}{\cosh q}$
and

\begin{equation}
\begin{array}{rl}
H^{dell}(p,q) \longrightarrow &
\alpha_{trig}(q)\cdot
\cn\left(
p\sqrt{\tilde k'^2 + \tilde k^2\alpha_{trig}^2(q)}\left|
\frac{\tilde k\alpha_{trig}(q)}
{\sqrt{\tilde k'^2 + \tilde k^2\alpha_{trig}^2(q)}}
\right.\right) = \tilde H^{R}(p,q).\\
\end{array}
\end{equation}
This is the elliptic-trigonometric model, dual to the
conventional elliptic Ruijsenaars (i.e. the tri\-go\-no\-met\-ric-elliptic)
system.
In the further limit of small $q$
this degenerates into the elliptic-rational
model with $\alpha_{trig}(q) \rightarrow \alpha_{rat}(q) =
1 - \frac{2\nu^2}{q^2}$, which is dual to the conventional elliptic
Calogero (i.e. the rational-elliptic) system, analyzed in some detail
above.

Our approach has been based on choosing appropriate functions
$f(q)$ and $F(Q)$ and implementing duality. Other choices of
functions associated with alternative free Hamiltonians may be possible.
Instead of the duality relations (\ref{dellduality}) one could consider
those based on $h_0(p) = \sn(p|\tilde k)$ instead of $\cn(p|\tilde k)$.
With this choice one gets somewhat simpler expressions for $\beta_s$
and $\gamma_s$:

\be
\beta_s = 1, \ \ \
\gamma_s(q|\tilde k,k) = \tilde k\alpha_s(q|k), \ \ \
\alpha_s(q|k) = 1 - \frac{2\nu^2}{\cn^2(q|k)}
\ee
and the final Hamiltonian is now

\be
H_s(p,q|\tilde k,k) = \alpha_s(q|k) \cdot
\sn(p|\tilde k\alpha_s(q|k)).
\label{dellHamsin}
\ee
Although this Hamiltonian is somewhat simpler than our earlier choice,
the limits involved in obtaining the Ruijsenaars-Calogero-Toda
reductions are somewhat more involved, and that is why we
chose to present the Hamiltonian (\ref{dellHsu2}) first.

One might further try other elliptic functions for $h_0(p)$.
Every solution we have obtained by making a different anzatz
has been related to our solution (\ref{dellHsu2}) via modular transformations
of the four moduli $\tilde k$, $k$, $\tilde k_{eff} =
\tilde\gamma$ and $k_{eff} = \gamma$.

Another approach to the 2-particle double elliptic systems was proposed
in \cite{fgnr}. It is suggested there to start with the four dimensional
manifold, namely,
the elliptically fibered K3 manifold. This manifold
is the space where the separated variables
live. Since it is a phase space manifold, it is provided with nontrivial
Poisson brackets. They follow from
the natural Poisson structure on the K3 manifold. The Hamiltonian
can be chosen as a linear function on the base manifold while
the dual Hamiltonian as a linear function on the fiber.
Using explicit formulas for the Poisson brackets of coordinates on the
elliptically fibered
K3 manifold, in principle, one may obtain explicit Hamiltonians. We are
planning to report on these calculations as well as on equivalence of the
two approaches in a separate publication \cite{neCh}.

\subsection{Duality, $\tau$-functions and Hamiltonian reduction}

Now one should extend the results of the two-particle case to the generic
number of particles. Unfortunately, it is still unknown how to get very
manifest formulas. Here we describe a method that ultimately allows one to
construct
commuting Hamiltonians of the $N$-particle Dell system, but very explicit
expressions will be obtained only for first several terms in a perturbation
theory.

However, in order to explain the idea, we need some preliminary work.
In fact, we start from the old observation \cite{KriF,zeroes,krispin2,KriZab}
that, if one
study the elliptic (trigonometric, rational) solution to the KP hierarchy,
the dynamics (dependence on higher times of the hierarchy)
of zeroes of the corresponding $\tau$-function w.r.t. the first time
is governed by the elliptic (trigonometric, rational) Calogero system.
Moreover, if one starts with the two-dimensional Toda lattice hierarchy
and studies the zeroes of the $\tau$-function
w.r.t. the zero (discrete) time, the corresponding
system governing zeroes is the Ruijsenaars system \cite{KriZab}.

Let us illustrate this with the simplest example of the trigonometric solution
of the KP hierarchy. This is, in fact, the $N$-solitonic solution, the
corresponding $\tau$-function being

\be\label{soltau}
\tau(\{t_k\}|\left\{\nu_i,\mu_i,X_i\right\})=
\det_{N\times N}\left(\delta_{ij}-L_{ij}
e^{\sum_k t_k(\mu_i^k-\nu_i^k)} \right),\ \ \
L_{ij}\equiv {X_i\over \nu_i-\mu_j}
\ee
Here $\nu_i$, $\mu_i$ and $X_i$ are the soliton parameters. Let us impose
the constraint $\nu_i=\mu_i+\epsilon$. Then, the standard Hamiltonian
structure
of the hierarchy on the soliton solution gives rise to the Poisson brackets
(in \cite{FT,solitons3} similar calculation was done for the sine-Gordon case)

\be
\left\{X_i,X_j\right\}=-{4X_iX_j\epsilon^2\over (\mu_i-\mu_j)(\mu_i-\mu_j
-\epsilon)(\mu_i-\mu_j+\epsilon)},\ \ \
\left\{\mu_i,\mu_j\right\}=0,\ \ \
\left\{X_i,\mu_j\right\}=X_i\delta_{ij}
\ee
Note that, upon identification $\mu_i=q_i$, $X_i\equiv e^{p_i}
\prod_{l\ne i}\left({1\over (q_i-q_l)^2} -{1\over\epsilon^2}
\right)$, the matrix
$L_{ij}$ in (\ref{soltau}) becomes the rational Ruijsenaars Lax operator,
with the proper symplectic structure $\{p_i,q_j\}=\delta_{ij}$.

At the same time, (\ref{soltau}) can be rewritten in the form

\be
\tau(\{t_k\}|\left\{\nu_i,\mu_i,X_i\right\})\sim
\det_{N\times N}\left(e^{\epsilon x}\delta_{ij}-L_{ij}
e^{\sum_{k\ne 1} t_k(\mu_i^k-\nu_i^k)} \right)
\ee
where $x\equiv t_1$. This determinant
is the generating function of the rational Ruijsenaars Hamiltonians.
On the other hand, the $N$ zeroes of
the determinant as the function of $\epsilon x$ are just logarithms of the
eigenvalues of the Lax operator. These zeroes are governed,
as we know from \cite{KriF,zeroes} by the trigonometric Calogero
system, and are nothing but the $N$ particle coordinates in the Calogero
system:

\be
\tau(\{t_k\}|\left\{\nu_i,\mu_i,X_i\right\})\sim
\sum_k e^{k\epsilon x}H_k^R\sim \prod_i^N\sin ({\epsilon x}-q_i)
\ee
Therefore, the eigenvalues of the (rational Ruijsenaars) Lax operator are the
exponentials $e^{q_i}$ of coordinates in the (trigonometric Calogero) dual
system,
while the $\tau$-function simultaneously is the generating function of
the Hamiltonians in one integrable systems and the function of coordinates
in the dual system. This is exactly the form of duality, as it
was first realized by S.Ruijsenaars \cite{rud}. Its relation with
$\tau$-functions was first observed by D.Bernard and O.Babelon \cite{solitons3}
(see also \cite{Khar2}).

Similarly, one can consider the $N$-solitonic solution in the two-dimensional
Toda lattice hierarchy. Then,

\be\label{soltaut}
\tau_n(\{t_k\}|\left\{\nu_i,\mu_i,X_i\right\})=
\det_{N\times N}\left(\delta_{ij}-L_{ij}
\left({\mu_i\over\nu_i}\right)^n
e^{\sum_k t_k(\mu_i^k-\nu_i^k)+\bar t_k(\mu_i^{-k}-\nu_i^{-k})}\right),\ \ \
L_{ij}\equiv {X_i\over \nu_i-\mu_j}
\ee
This time one should make a reduction $\nu_i=e^{\epsilon}\mu_i$.
Then, $L_{ij}$ becomes
the Lax operator of the trigonometric Ruijsenaars, and the zeroes w.r.t. to
$\epsilon n$ are governed by the same, trigonometric Ruijsenaars system.
This is another check that this system is really self-dual.

Thus constructed duality can be also interpreted
in terms of the
Hamiltonian reduction \cite{ruijh}. In doing so, one starts
with the moment map which typically is a constraint for two matrices or
two matrix-valued functions. In order to solve it,
one diagonalizes one of the matrices, its diagonal elements ultimately
being functions (exponentials) of the coordinates, while the other matrix
gives the Lax operator (its traces are the Hamiltonians).
Now what one needs in order to construct the dual
system is just to diagonalize the second matrix in order to get
coordinates in the dual system, while the first matrix will provide one with
the new
Lax operator (its traces are the Hamiltonians in the dual system).

As an illustrative example, let us consider a Hamiltonian reduction
of the trigonometric
Calogero-Sutherland
system which differs from that discussed in s.2.1 \cite{olper}.
Now one starts from the free Hamiltonian system
$H=\tr A^2$ given on the phase space $(A,B)$ of two $N\times N$ matrices, with
the symplectic form $\tr (\delta A\wedge B^{-1}\delta B)$ (i.e.
$A$ lies in the $gl(N)$ algebra, while $B$ lies in the $GL(N)$ group).
Then, one
can impose the constraint

\be\label{mm}
A-BAB^{-1}=\nu ({\bf I}-P)
\ee
where $\nu$ is a constant, ${\bf I}$ is the unit matrix and
$P$ is a matrix of rank 1. Now one has to make the Hamiltonian reduction
with this
constraint.
First of all, one solves the constraint diagonalizing matrix $B$ and using
the gauge freedom of the system to transform $P$ into the matrix with all
entries unit. It gives\footnote{Since (\ref{mm}) does not fix the
diagonal elements of $A$, they are parametrized by arbitrary numbers $p_i$.}
$A_{ij}=p_i\delta_{ij}+(1-\delta_{ij}){\nu b_j\over b_i-b_j}$, where $b_i$ are
the diagonal elements of $B$.
This is nothing but the Lax operator of the trigonometric Calogero-Sutherland
model.
Traces of powers of this Lax operators give the Calogero Hamiltonians.

One can also diagonalize the other matrix, $A$. Then, (\ref{mm}) can be
rewritten
in the form

\be\label{mm2}
AB-BA=\nu (B-\bar P)
\ee
where $\bar P$ is another rank 1 matrix which is can be gauged out
to the matrix with all entries depending only on the number of row (let denote
them
$X_i/\nu$). Solving then this constraint, one obtains
$B={X_i\over a_i-a_j-\nu}$, where $a_i$ are the diagonal elements of the
matrix $A$.
This is nothing but the Lax operator of the rational \RS system, traces of its
powers
give a set of the dual Hamiltonians.

Certainly, this picture \cite{ruijh} nicely explains the Ruijsenaars
observation
of duality \cite{rud} and is just the $\tau$-function approach told in
different words.

The same scheme, certainly, works in for the trigonometric Ruijsenaars too,
where one can use a similar procedure in order to show that diagonalizing
either of the matrices $A$ and $B$,
one arrives at the same system, trigonometric Ruijsenaars
model, checking again that it is self-dual.

In fact, similarly to the trigonometric Calogero system,
there are several equivalent ways to obtain the trigonometric Ruijsenaars
model. Say, instead of the Hamiltonian reduction described in
section 2, one can make the Hamiltonian reduction of the gauge theory given on
a two-torus ${\bf T}^2=S^1\times S^1$ \cite{fgnr}. The other way of doing is to
consider the Poisson reduction \cite{g13}.
Namely, one introduces a set of commuting
functions on the space of graph connection on a graph corresponding to a moduli
space of flat connections on a torus with one hole. Being reduced to a
particular
symplectic leaf of the moduli space of flat connections on the torus
in the symplectic structure \cite{fr}, this set of
functions turns out to be a full set of commuting Hamiltonians. This picture
is far more geometrical and the self-duality of trigonometric Ruijsenaars
system gets within this framework rather
transparent geometrical meaning. That is, this Poisson reduction physically is
nothing but the three dimensional Chern-Simons theory on
${\bf T}^{2} \times {\bf R}^{1}$ (the line gives the temporal variable)
with the insertion of an appropriate
temporal Wilson line and spatial Wilson loop. It is the freedom
to choose the place of the latter one that is in charge of the duality.
The group of (self-) dualities of this model is therefore generated
by the (modular) group acting on the non-trivial torus cycles.

\subsection{Dual Hamiltonians for many-body systems}

Unfortunately, the scheme described in the previous subsection does not
that immediately applied to
elliptic models with arbitrary number of particles. The reason is
that even having the dual Lax operator calculated, one typically meets
in elliptic cases a problem of constructing the dual Hamiltonians, i.e.
invariant combinations involving the Lax operator.

Instead, in order to obtain dual Hamiltonians, here
we use the other approach discussed above, that dealing with
zeroes of the $\tau$-functions \cite{bmmm3,MM1}. This approach gives no
very explicit form of the dual Hamiltonians. Therefore, we
construct later a kind of perturbative procedure for the Hamiltonians
which allows one to get them absolutely explicitly term by term \cite{MM1}.

Consider $SU(N)$ ($N=g+1$) system.
The whole construction is based on the fact that
the spectral curve of the original integrable system
(Toda chain, Calogero, Ruijsenaars or the most interesting double
elliptic system) has a period matrix $T_{ij}(\vec a)$,
$i,j = 1,\ldots,N$ with the special property:

\be
\sum_{j=1}^N T_{ij}(a) = \tau, \ \ \forall i
\label{sumT}
\ee
where $\tau$ does not depend on $a$.
As a corollary, the genus-$N$ theta-function is naturally decomposed
into a linear combination of genus-$g$ theta-functions:

\be
\Theta^{(N)}(p_i|T_{ij}) =
\sum_{n_i \in Z} \exp\left(i\pi\sum_{i,j=1}^NT_{ij}n_in_j +
2\pi i\sum_{i=1}^N n_ip_i\right)
=\sum_{k=0}^{N-1}
\theta_{\left[{k\over N},{0}\right]}(N\zeta|N\tau)
\cdot \check\Theta_k^{(g)}(\check p_i|\check T_{ij})
\ee
where
\be
\Theta_k \equiv
\check\Theta_k^{(g)}(\check p_i|\check T_{ij}) \equiv
\sum_{{n_i \in Z}\atop{\sum n_i = k}}
\exp\left(-i\pi\sum_{i,j=1}^N\check T_{ij}n_{ij}^2 +
2\pi i\sum_{i=1}^N n_i\check p_i\right)
\ee
and $p_i = \zeta + \check p_i$, $\sum_{i=1}^N \check p_i = 0$;
$T_{ij} = \check T_{ij} + \frac{\tau}{N}$,
$\sum_{j=1}^N \check T_{ij} = 0$, $\forall i$.

Now we again use the argument that zeroes of the KP (Toda)
$\tau$-function (i.e. essentially the Riemannian theta-function), associated
with the spectral curve (\ref{sc}) are nothing but the coordinates $q_i$ of
the original (Calogero, Ruijsenaars) integrable system.
In more detail, due to the property (\ref{sumT}),
$\Theta^{(N)}(p|T)$ as a function
of $\zeta = \frac{1}{N} \sum_{i=1}^N p_i$ is an elliptic function on the
torus $(1,\tau)$ and, therefore, can be decomposed into an $N$-fold product
of the genus-one theta-functions. Remarkably, their arguments are just
$\zeta-q_i$:

\be
\Theta^{(N)}(p|T) =
c(p,T,\tau)\prod_{i=1}^N \theta(\zeta - q_i(p,T)|\tau)
\ee
(In the case of the Toda chain when $\tau\to
i\infty$ this ``sum rule" is implied by the standard expression
for the individual $e^{q_i}$ through the KP $\tau$-function.) Since one
can prove that $q_i$ form a Poisson-commuting set of variables with respect to
the symplectic structure (\ref{ss}) \cite{KriF},
this observation indirectly justifies
the claim \cite{bmmm3,MM1} that all the ratios
$\Theta_k/\Theta_l$ (in order to cancel the non-elliptic factor $c(p,T,\tau)$)
are Poisson-commuting with respect to the
Seiberg-Witten symplectic structure

\be
\sum_{i=1}^N dp_i^{Jac}\wedge da_i
\label{ss}
\ee
(where $a_N \equiv -a_1 - \ldots - a_{N-1}$, i.e.
$\sum_{i=1}^N a_i = 0$):

\be
\left\{\frac{\Theta_k}{\Theta_l}, \frac{\Theta_m}{\Theta_n}\right\} = 0
\ \ \forall k,l,m,n
\label{comm}
\ee
It can be equivalently written as

$$
\Theta_i\{\Theta_j, \Theta_k\} +
\Theta_j\{\Theta_k, \Theta_i\} +
\Theta_k\{\Theta_i, \Theta_j\} = 0  \ \ \forall i,j,k
$$
or

$$
\{\log\Theta_i, \log\Theta_j\} =
\left\{\log\frac{\Theta_i}{\Theta_j}, \log\Theta_k\right\}
\ \ \forall i,j,k
$$
The Hamiltonians of the dual integrable system can be chosen in
the form $H_k = \Theta_k/\Theta_0$, $k = 1,\ldots, g$.
However, these dual Hamiltonians are not quite manifest,
since they depend on the period matrix of the original system
expressed in terms of its action variables.
Nevertheless, what is important, one can work with the $\theta$-functions
as with
series and, using the instantonic expansion for the period matrix
$T_{ij}$, construct Hamiltonians term by term in the series.

\subsection{Dual Hamiltonians perturbatively}

Let us see how it really works. We start with
the simplest case of the Seiberg-Witten family, the periodic Toda chain.

\paragraph{Perturbative approximation.}

First we consider the perturbative approximation which is nothing but the
open Toda chain, see (\ref{pertcurv}) and (\ref{pertdS}).
This is the case of the perturbative $4d$ pure $N=2$ SYM theory with
the prepotential

\be
{\cal F}_{pert}(a) = \frac{1}{2i\pi}\sum_{i<j}^N
a_{ij}^2\log a_{ij}
\ee
Then, the period matrix
is singular and only finite number of terms survives in the
series for the theta-function:

\be
\Theta^{(N)}(p|T) = \sum_{k=0}^{N-1} e^{2\pi ik\zeta}H^{(0)}_k(p,a),
\label{pertTodatheta}
\ee
\be
H^{(0)}_k(p,a) = \sum_{I, [I]=k} \prod_{i\in I}e^{2\pi ip_i}
\prod_{j\in \bar I} {\cal Z}^{(0)}_{ij}(a)
\label{pertTodaHam}
\ee
Here
\be
{\cal Z}^{(0)}_{ij}(a)=e^{-i\pi T^{(0)}_{ij}} = \frac{\Lambda_{QCD}}{a_{ij}}
\label{pertTodaF}
\ee
and $I$ are all possible partitions of $N$ indices into the sets
of $k = [I]$ and $N-k = [{\bar I}]$ elements. Parameter $\Lambda_{QCD}$
becomes significant only when the system is deformed:
either non-perturbatively or to more complex systems of the
Calogero--Ruijsenaars--double-elliptic  family.
The corresponding $\tau$-function $\Theta^{(N)}(p|T)$,
eq.(\ref{pertTodatheta}), describes an $N$-soliton solution
to the KP hierarchy and is equal to the determinant (\ref{soltau}).
The Hamiltonians $H^{(0)}_k$,
eq.(\ref{pertTodaHam}) dual to the open Toda chain,
are those of the degenerated rational Ruijsenaars
system, and they are well-known to Poisson-commute with respect
to the relevant Seiberg-Witten symplectic structure (\ref{ss}).
This is not surprising since the open Toda chain is obtained by a
degeneration from
the trigonometric Calogero system (s.3.3)
which is dual to the rational Ruijsenaars.

The same construction for other perturbative
Seiberg-Witten systems ends up with the Hamiltonians
of more sophisticated systems.

For the spectral curves (\ref{scfC}) of the trigonometric Calogero system
(perturbative $4d$ $\CN=4$ SYM with SUSY softly broken down
to $\CN=2$ by the adjoint mass $M$) the Poisson-commuting (with respect to
the same
(\ref{ss})), whose perturbative prepotential is given by
(\ref{appC}), Hamiltonians $H^{(0)}_k$ are given by
(\ref{pertTodaHam}) with

\be
{\cal Z}^{(0)}_{ij}(a) = \frac{\sqrt{a_{ij}^2 - M^2}}{a_{ij}}
\label{pertCalF}
\ee
i.e. are the Hamiltonians of the rational Ruijsenaars system which are really
dual to the trigonometric Calogero model.

For the spectral curve (\ref{scf})
of the trigonometric Ruijsenaars system
(perturbative $5d$ $\CN=2$ SYM compactified on a circle
with an $\epsilon$ twist as the boundary conditions) whose perturbative
prepotential is given by (\ref{appC}),
the Hamiltonians are given by (\ref{pertTodaHam}) with

\be
{\cal Z}^{(0)}_{ij}(a) =
\frac{\sqrt{\sinh(a_{ij}+\epsilon)\sinh(a_{ij}-\epsilon)}}
{\sinh a_{ij}}
\label{pertRuF}
\ee
i.e. are the Hamiltonians of the trigonometric Ruijsenaars system, which
is, indeed, self-dual.

Finally, for the perturbative limit of the
most interesting self-dual double-elliptic system
(the explicit form of its spectral curves is yet unknown except
the case of two particles, see the next subsection)
the relevant Hamiltonians are those of the elliptic Ruijsenaars
system, given by the same (\ref{pertTodaHam}) with

\be
{\cal Z}^{(0)}_{ij}(a) =\sqrt{1-{2g^2\over \hbox{sn}^2_{\tilde\tau}(a_{ij})}}
\sim\frac{\sqrt{\theta(\hat a_{ij}+\varepsilon|\tilde\tau)
\theta(\hat a_{ij}-\varepsilon|\tilde\tau)}}
{\theta(\hat a_{ij}|\tilde\tau)}
\label{pertdellF}
\ee
where $\tilde\tau$ is the modulus of the second torus associated with the
double elliptic system and $\hat a_{ij}$ are $a_{ij}$ rescaled, \cite{BEWW}.

In all these examples $H_0^{(0)} = 1$, the theta-functions
$\Theta^{(N)}$ are singular and given by determinant (solitonic)
formulas with finite number of items (only terms with $n_i =0, 1$
survive in the series expansion of the theta-function),
and Poisson-commutativity of arising Hamiltonians is analytically
checked within the theory of Ruijsenaars integrable systems.

\paragraph{Beyond the perturbative limit.}

Beyond the perturbative limit, the analytical evaluation of
$\Theta^{(N)}$ becomes less straightforward.

The non-perturbative deformation (\ref{fsc-Toda}) of the curve
(\ref{pertcurv}),
is associated with somewhat sophisticated prepotential
of the periodic Toda chain,

\be
{\cal F}(a) = \frac{1}{4\pi i}\sum_{i<j}^N a_{ij}^2\log\frac{a_{ij}}{\Lambda}
+ \sum_{k=1}^\infty \Lambda^2{\cal F}^{(k)}(a)
\ee
The period matrix is

\be
\pi iT_{ij} = \frac{\partial^2{\cal F}}{\partial a_i\partial a_j} =
-2\pi i\tau\delta_{ij}\ + \log\frac{a_{ij}}{\Lambda} +
\sum_{k=1}^\infty \Lambda^2
\frac{\partial^2{\cal F}^{(k)}(a)}{\partial a_i\partial a_j},
\ \  i\neq j, \\ T_{ii} =-2 \tau - \sum_{j\neq i} T_{ij}
\ee
and $\tau$ in this case can be removed by the rescaling of $\Lambda$.
Then,

\be
\Theta^{(N)}(p|T) = \sum_{k=0}^{N-1} e^{2\pi ik\zeta}
\Theta_k(p,a)
= \sum_{k=0}^{N-1} e^{2\pi i k\zeta}
\sum_{n_i,\ \sum_i n_i = k}
e^{-i\pi\sum_{i<j} T_{ij}n_{ij}^2} e^{2\pi i \sum_i n_ip_i} = \\
= \left( 1 + \sum_{i\neq j}^N e^{2\pi i(p_i - p_j)}
{\cal Z}_{ij}^4\prod_{k\neq i,j} {\cal Z}_{ik}{\cal Z}_{jk} + \right. \\
\left. +
\sum_{i\neq j\neq k\neq l}^N  e^{2\pi i(p_i + p_j - p_k - p_l)}
{\cal Z}_{ik}^4 {\cal Z}_{il}^4 {\cal Z}_{jk}^4  {\cal Z}_{jl}^4
\prod_{m\neq i,j,k,l} {\cal Z}_{im}{\cal Z}_{jm}{\cal Z}_{km}{\cal Z}_{lm}
+ \ldots\ \right) + \\
+ e^{2\pi i \zeta}
\left(\sum_{i=1}^N e^{2\pi i p_i} \prod_{j\neq i} {\cal Z}_{ij} +
\sum_{i\neq j\neq k} e^{2\pi i (p_i + p_j - p_k)}
{\cal Z}_{ik}^4{\cal Z}_{jk}^4\prod_{l\neq i,j,k} {\cal Z}_{il}{\cal Z}_{jl}
{\cal Z}_{kl} + \right.\\
\left. + \sum_{i\neq j} e^{2\pi i (2p_i - p_j)}
{\cal Z}_{ij}^9\prod_{k\neq i,j} {\cal Z}_{ik}^4 {\cal Z}_{jk} + \ldots \
\right) + \\
+ e^{4\pi i\zeta}
\left(\sum_{i\neq j} e^{2\pi i(p_i+p_j)}\prod_{k\neq i,j}{\cal Z}_{ik}
{\cal Z}_{jk} +
\sum_i e^{4\pi ip_i} \prod_{k\neq i}{\cal Z}_{ik}^4 + \ldots \ \right) + \\
+ \ldots
\label{Thetaexp}
\ee
with

\be
{\cal Z}_{ij} = e^{-i\pi T_{ij}} = {\cal Z}^{(0)}_{ij}
\left( 1 -
i\pi\frac{\partial^2{\cal Z}^{(1)}(a)}{\partial a_i\partial a_j}
+ \ldots\ \right), \ \ \      i\neq j
\label{F}
\ee
The first few corrections  ${\cal Z}^{(k)}$ to the prepotential
are explicitly known in the Toda-chain case \cite{IC},
for example,

\be
{\cal Z}^{(1)} = -\frac{1}{2i\pi}\sum_{i=1}^N\prod_{k\neq i}
\left({\cal Z}_{ik}^{(0)}\right)^2
\label{F1}
\ee

The coefficients $\Theta_k$ in (\ref{Thetaexp})
are expanded into powers of $(\Lambda/a)^{2N}$ and the leading (zeroth-order)
terms are exactly the perturbative expressions (\ref{pertTodaHam}).
Thus, the degenerated Ruijsenaars Hamiltonians (\ref{pertTodaHam}) are just
the
perturbative approximations to the $H_k = \Theta_k/\Theta_0$ --
the full Hamiltonians of the integrable system dual to the Toda
chain.

The same expansion can be constructed for the other, elliptic systems.
Note that the period matrix in these cases is expanded into powers of
$q=e^{2\pi i\tau}$:

\be
\pi iT_{ij} = \frac{\partial^2{\cal F}}{\partial a_i\partial a_j} =
-2\pi i\tau\delta_{ij}\ + \log\frac{a_{ij}}{\Lambda} +\mu^2
\sum_{k=1}^\infty q^k
\frac{\partial^2{\cal F}^{(k)}(a)}{\partial a_i\partial a_j},
\ \  i\neq j
\ee
and the dimensional constant $\mu$ depends on the system. E.g., in the
Calogero
model $\mu=iM$ etc. The limit to the Toda system corresponds to $q\to 0$, $q
\mu^{2N}$=fixed.

Note that the first (instanton-gas) correction to the prepotential
is indeed known to be of the same universal form (\ref{F1}) not only for the
Toda
chain, but also  for the Calogero system \cite{DPhongCP}. For the
Ruijsenaars and double-elliptic systems eq.(\ref{F1})
is not yet available in the literature.

In \cite{MM1} the perturbative expansions for the Hamiltonians were
tested for Poisson commutativity, up to few first terms and small number of
particles $N$. In particular, it was checked that
(\ref{F1}) is true for both the Ruijsenaars and Dell systems.

The main lesson we could get from this consideration is that
if the {\it universal} expressions like (\ref{F1})
in terms of perturbative ${\cal Z}^{(0)}_{ij}$, the same   for all
the systems, will be found for higher corrections
to the prepotentials\footnote{
To avoid possible confusion, the recurrent relation
\cite{LNS,RG,Jose1,Jose4} for the Toda-chain prepotential,
$$
\frac{\partial^2{\cal F}}{\partial\log\Lambda^2} \sim
\frac{\partial^2{\cal F}}{\partial\log\Lambda\partial a_i}
\frac{\partial^2{\cal F}}{\partial\log\Lambda\partial a_j}
\left.\frac{\partial^2}{\partial p_i\partial p_j}
\log\Theta_0\right|_{p=0}
$$
does {\it not} immediately provide such  {\it universal}
expressions. Already for ${\cal Z}^{(1)}$ this relation gives
$$
{\cal Z}^{(1)} \sim\sum_{i<j}^N \left({\cal Z}^{(0)}_{ij}\right)^2
\prod_{k\neq i,j}^N  {\cal Z}^{(0)}_{ik}{\cal Z}^{(0)}_{jk}
$$
which coincides with (\ref{F1}) for the Toda-chain
${\cal Z}^{(0)}_{ij}$, eq.(\ref{pertTodaF}), but is not true
(in variance with (\ref{F1})) for Calogero ${\cal Z}^{(0)}_{ij}$,
eq.(\ref{pertCalF}). Meanwhile, the recurrent relations of ref.\cite{MNW} for
the Calogero system provide more promising expansion.

All this emphasizes the need to study extension of the Whitham theory and
the WDVV equations from the Toda chains to the Calogero and Ruijsenaars
systems.},  this will immediately give an explicit
(although not the most appealing) construction of the
Hamiltonians dual to the Calogero and Ruijsenaars models and, especially
important, the self-dual double-elliptic system which
was explicitly constructed above only for $N=2$.

Let us note now that the
$SU(2)$ Hamiltonian (\ref{dellHsu2}) has a rather nice form in terms
of the $p,q$ variables or their duals $P,Q$.
However, this our general $SU(N)$ construction of the Hamiltonians as ratios
of
genus $g$ $\theta$-functions uses another kind of
canonical variables -- angle-action variables $p^{Jac}_i, a_i$.
The {\it flat} moduli $a_i$ play the central role in
\SW theory, while $Q_i$ are rather generalizations
of the algebraic moduli.

It is an interesting open problem to express the Hamiltonians
in terms of $P_i$ and $Q_i$, perhaps, they can acquire a more
transparent form, like it happens for $SU(2)$.
Another option is to switch to the
``separated" variables ${\cal P}_i$, ${\cal Q}_i$ \cite{sklyanin} such that
$\oint_{A_j}{\cal P}_id{\cal Q}_i=a_i\delta_{ij}$ (while generically
$a_i=\oint_{A_j}\sum_i P_idQ_i$). Generically, they are different from $P_i$,
$Q_i$ but coincide with $P$, $Q$ in the case of $SU(2)$.

\subsection{Dell systems: 6d adjoint matter}

Thus, we have constructed the Dell system and a set of dualities that
acts on the Toda-Calogero-Ruijsenaars-Dell family. They are all
put together in Fig.2.

\begin{figure}[t]
%TexCad Options
%\grade{\on}
%\emlines{\on}
%\beziermacro{\off}
%\reduce{\on}
%\snapping{\off}
%\quality{2.00}
%\graddiff{0.01}
%\snapasp{1}
%\zoom{1.00}
\special{em:linewidth 0.4pt}
\unitlength 1mm
\linethickness{0.4pt}
\begin{picture}(141.33,120.33)
\emline{18.33}{110.00}{1}{141.33}{110.00}{2}
\emline{141.33}{110.00}{3}{141.33}{110.00}{4}
\emline{18.33}{95.00}{5}{141.33}{95.00}{6}
\emline{18.33}{80.00}{7}{141.33}{80.00}{8}
\emline{18.33}{65.00}{9}{141.00}{65.00}{10}
\emline{141.00}{120.00}{11}{141.00}{65.00}{12}
\emline{110.33}{120.00}{13}{110.33}{65.00}{14}
\emline{80.00}{120.33}{15}{80.00}{65.00}{16}
\emline{50.00}{120.00}{17}{50.00}{65.00}{18}
\emline{50.00}{110.00}{19}{18.33}{120.00}{20}
%\vector(70.67,91.67)(84.00,101.33)
\put(84.00,101.33){\vector(4,3){0.2}}
\emline{70.67}{91.67}{21}{84.00}{101.33}{22}
%\end
%\vector(85.67,99.33)(73.33,90.33)
\put(73.33,90.33){\vector(-4,-3){0.2}}
\emline{85.67}{99.33}{23}{73.33}{90.33}{24}
%\end
%\vector(102.67,76.33)(114.00,85.33)
\put(114.00,85.33){\vector(4,3){0.2}}
\emline{102.67}{76.33}{25}{114.00}{85.33}{26}
%\end
%\vector(115.33,83.33)(105.67,75.67)
\put(105.67,75.67){\vector(-4,-3){0.2}}
\emline{115.33}{83.33}{27}{105.67}{75.67}{28}
%\end
%\bezvec{96}(54.67,102.67)(47.67,112.00)(58.33,105.33)
\put(58.33,105.33){\vector(2,-1){0.2}}
\emline{54.67}{102.67}{29}{53.40}{104.44}{30}
\emline{53.40}{104.44}{31}{52.52}{105.86}{32}
\emline{52.52}{105.86}{33}{52.02}{106.94}{34}
\emline{52.02}{106.94}{35}{51.90}{107.67}{36}
\emline{51.90}{107.67}{37}{52.17}{108.05}{38}
\emline{52.17}{108.05}{39}{52.82}{108.08}{40}
\emline{52.82}{108.08}{41}{53.85}{107.77}{42}
\emline{53.85}{107.77}{43}{55.27}{107.11}{44}
\emline{55.27}{107.11}{45}{58.33}{105.33}{46}
%\end
%\bezvec{108}(84.33,87.67)(76.33,97.67)(88.67,91.33)
\put(88.67,91.33){\vector(2,-1){0.2}}
\emline{84.33}{87.67}{47}{83.03}{89.38}{48}
\emline{83.03}{89.38}{49}{82.07}{90.81}{50}
\emline{82.07}{90.81}{51}{81.46}{91.96}{52}
\emline{81.46}{91.96}{53}{81.20}{92.83}{54}
\emline{81.20}{92.83}{55}{81.28}{93.43}{56}
\emline{81.28}{93.43}{57}{81.72}{93.74}{58}
\emline{81.72}{93.74}{59}{82.50}{93.77}{60}
\emline{82.50}{93.77}{61}{83.64}{93.52}{62}
\emline{83.64}{93.52}{63}{85.12}{92.99}{64}
\emline{85.12}{92.99}{65}{88.67}{91.33}{66}
%\end
%\bezvec{80}(114.00,73.33)(108.33,81.00)(117.33,76.67)
\put(117.33,76.67){\vector(2,-1){0.2}}
\emline{114.00}{73.33}{67}{112.81}{75.06}{68}
\emline{112.81}{75.06}{69}{112.08}{76.42}{70}
\emline{112.08}{76.42}{71}{111.81}{77.40}{72}
\emline{111.81}{77.40}{73}{112.00}{78.00}{74}
\emline{112.00}{78.00}{75}{112.65}{78.23}{76}
\emline{112.65}{78.23}{77}{113.75}{78.08}{78}
\emline{113.75}{78.08}{79}{115.31}{77.56}{80}
\emline{115.31}{77.56}{81}{117.33}{76.67}{82}
%\end
%\bezvec{224}(74.00,76.67)(108.00,77.67)(114.67,98.67)
\put(114.67,98.67){\vector(1,4){0.2}}
\emline{74.00}{76.67}{83}{77.17}{76.81}{84}
\emline{77.17}{76.81}{85}{80.21}{77.04}{86}
\emline{80.21}{77.04}{87}{83.13}{77.36}{88}
\emline{83.13}{77.36}{89}{85.92}{77.77}{90}
\emline{85.92}{77.77}{91}{88.60}{78.27}{92}
\emline{88.60}{78.27}{93}{91.15}{78.86}{94}
\emline{91.15}{78.86}{95}{93.57}{79.54}{96}
\emline{93.57}{79.54}{97}{95.88}{80.31}{98}
\emline{95.88}{80.31}{99}{98.06}{81.17}{100}
\emline{98.06}{81.17}{101}{100.11}{82.12}{102}
\emline{100.11}{82.12}{103}{102.05}{83.16}{104}
\emline{102.05}{83.16}{105}{103.86}{84.29}{106}
\emline{103.86}{84.29}{107}{105.55}{85.51}{108}
\emline{105.55}{85.51}{109}{107.11}{86.82}{110}
\emline{107.11}{86.82}{111}{108.55}{88.22}{112}
\emline{108.55}{88.22}{113}{109.87}{89.72}{114}
\emline{109.87}{89.72}{115}{111.07}{91.30}{116}
\emline{111.07}{91.30}{117}{112.14}{92.97}{118}
\emline{112.14}{92.97}{119}{113.09}{94.73}{120}
\emline{113.09}{94.73}{121}{113.92}{96.58}{122}
\emline{113.92}{96.58}{123}{114.67}{98.67}{124}
%\end
%\bezvec{220}(117.00,96.67)(107.67,75.00)(76.00,73.33)
\put(76.00,73.33){\vector(-1,0){0.2}}
\emline{117.00}{96.67}{125}{116.05}{94.64}{126}
\emline{116.05}{94.64}{127}{115.01}{92.70}{128}
\emline{115.01}{92.70}{129}{113.86}{90.85}{130}
\emline{113.86}{90.85}{131}{112.60}{89.10}{132}
\emline{112.60}{89.10}{133}{111.25}{87.44}{134}
\emline{111.25}{87.44}{135}{109.79}{85.87}{136}
\emline{109.79}{85.87}{137}{108.23}{84.39}{138}
\emline{108.23}{84.39}{139}{106.57}{83.00}{140}
\emline{106.57}{83.00}{141}{104.81}{81.71}{142}
\emline{104.81}{81.71}{143}{102.94}{80.50}{144}
\emline{102.94}{80.50}{145}{100.97}{79.39}{146}
\emline{100.97}{79.39}{147}{98.90}{78.37}{148}
\emline{98.90}{78.37}{149}{96.72}{77.44}{150}
\emline{96.72}{77.44}{151}{94.45}{76.60}{152}
\emline{94.45}{76.60}{153}{92.07}{75.86}{154}
\emline{92.07}{75.86}{155}{89.58}{75.21}{156}
\emline{89.58}{75.21}{157}{87.00}{74.64}{158}
\emline{87.00}{74.64}{159}{84.31}{74.17}{160}
\emline{84.31}{74.17}{161}{81.53}{73.80}{162}
\emline{81.53}{73.80}{163}{78.63}{73.51}{164}
\emline{78.63}{73.51}{165}{76.00}{73.33}{166}
%\end
\put(39.33,119.00){\makebox(0,0)[cc]{{coordinate}}}
\put(26.67,113.33){\makebox(0,0)[cc]{{momentum}}}
\put(64.67,114.33){\makebox(0,0)[cc]{rational}}
\put(95.00,113.67){\makebox(0,0)[cc]{trigonometric}}
\put(125.00,114.00){\makebox(0,0)[cc]{elliptic}}
\put(33.67,102.00){\makebox(0,0)[cc]{rational}}
\put(33.33,86.67){\makebox(0,0)[cc]{trigonometric}}
\put(33.33,72.67){\makebox(0,0)[cc]{elliptic}}
\put(73.00,102.00){\makebox(0,0)[cc]{\parbox{.27\linewidth}{rational
Calogero}}}
\put(98.67,102.00){\makebox(0,0)[cc]{\parbox{.19\linewidth}
{trigonometric Calogero}}}
\put(130.67,102.33){\makebox(0,0)[cc]{\parbox{.15\linewidth}{elliptic
Calogero}}}
\put(65.67,86.67){\makebox(0,0)[cc]{\parbox{.2\linewidth}{rational
Ruijsenaars}}}
\put(95.67,86.33){\makebox(0,0)[cc]{\parbox{.15\linewidth}
{trigonometric Ruijsenaars}}}
\put(129.67,86.33){\makebox(0,0)[cc]{\parbox{.18\linewidth}
{elliptic Ruijsenaars}}}
\put(64.67,72.67){\makebox(0,0)[cc]{\parbox{.17\linewidth}
{dual Calogero}}}
\put(98.67,71.00){\makebox(0,0)[cc]{\parbox{.15\linewidth}
{dual Ruijsenaars}}}
\put(129.67,73.00){\makebox(0,0)[cc]{\parbox{.2\linewidth}{Dell system}}}
\end{picture}
\vspace{-6.8cm}
\caption{Action of the coordinate-momentum duality on the
Calogero-Ruijsenaars-Dell family. Hooked arrows mark self-dual
systems. The duality leaves the coupling constant $\nu$ intact.}\label{dual}
\end{figure}
\vspace{10pt}

Since the theories of this family describes low energy limits of different
SUSY gauge theories, this table describes the dualities between different
gauge theories too. Say, perturbative limit of the $4d$ theory (trigonometric
Calogero) is dual to a special degeneration \cite{bm}
of the perturbative limit of the $5d$ theory (rational Ruijsenaars), while
full, {\it non-perturbative} $5d$ theory (elliptic Ruijsenaars) is dual to
the perturbative limit of the $6d$ theory (Dell system).

To conclude our discussion of dualities and Dell systems, we calculate the
perturbative prepotential of the 2-particle Dell system (\ref{dellHsu2})
and show that it coincides with what is to be expected for the $6d$ theory
with adjoint matter, supporting the identification of the Dell system with the
$6d$ gauge theory.

Let us start from the full Dell model for 2-particles. Then, it is
given by the Hamiltonian (\ref{dellHsu2})
parametrized by two independent (momentum and coordinate)
elliptic curves with elliptic moduli $k$ and $\tilde k$.
As usual, for two particles we can construct the {\it full}
spectral curve from the Hamiltonian (\cite{GKMMM}, see also ss.3.2,
s.3.7 and
(\ref{times})),

\be\label{sc}
H(\zeta,\xi|k,\tilde k) = u \ \ \ \ \left(\ = \cn(Q|k)\ \right)
\ee
It is characterized by the effective elliptic moduli

\be
k_{eff} = \frac{k \alpha(q|\tilde k)} {\beta(q|k,\tilde k)},\ \ \ \
\tilde k_{eff} = \frac{\tilde k \alpha(q|k)}
{\beta(q|\tilde k,k)}
\ee
where the functions $\alpha$ and $\beta$ are manifestly given in (\ref{ab}).
Coordinate-momentum duality interchanges
$k \leftrightarrow \tilde k$, $k_{eff} \leftrightarrow \tilde k_{eff}$
(and $q,p \leftrightarrow Q,P$), while
in general $SU(N)$ case, the model describes an interplay between
the four tori: the two bare elliptic curves and two
effective Jacobians of complex dimension $g = N-1$.

 From our previous discussions it should be clear that the
generating differential in $6d$ theories is of the form
$dS=\zeta d\xi$, with $\xi$ living on the coordinate torus and
$\zeta$ -- on the momentum torus.

Now  we calculate the leading order of the prepotential expansion in
powers of $\tilde k$ when the bare spectral torus degenerates into
sphere. In the
forthcoming calculation we closely follow the line of s.5.6.

When $\tilde k\to 0$, $\sn(q|\tilde k)$ degenerates into the ordinary sine.
For further convenience, we shall
parameterize the coupling constant $2\nu^2\equiv\sn^2(\epsilon|k)$. Now the
spectral curve (\ref{sc}) acquires the form

\be\label{psc}
\alpha(\xi)\equiv\sqrt{1-{\sn^2(\epsilon|k)\over\sin^2\xi}}=
{u\over \cn
\left(\zeta\beta\bigg|k_{eff}\right)}
\ee
Here the variable $\xi$ lives in the cylinder produced after degenerating
the bare coordinate torus. So does the variable $x=1/\sin^2\xi$. Note
that the $A$-period of the dressed torus shrinks on the sphere to a contour
around $x=0$. Similarly, $B$-period can be taken as a contour passing from
$x=0$ to $x=1$ and back.

The next step is to calculate variation of the generating differential
$dS=\eta d\xi$ w.r.t. the modulus
$u$ in order to obtain a holomorphic differential:

\be
dv=\left(-i\sn(\epsilon|k)\sqrt{k'^2+k^2u^2}\right)^{-1}
{dx\over x\sqrt{(x-1)(U^2-x)}}
\ee
where $U^2\equiv{1-u^2\over\sn^2(\epsilon|k)}$.
Since

\be
{\partial a\over\partial u}=\oint_{x=0}{\partial dS\over\partial u}
=\oint_{x=0}dv=
-{1\over \sqrt{(1-u^2)(k'^2+k^2u^2)}}
\ee
we deduce that $u=\cn (a|k)$ and $U={\sn(a|k)\over\sn(\epsilon|k)}$. The
ratio of the $B$- and $A$-periods of $dv$
gives the period matrix

\be
T={U\over\pi}\int_0^1 {dx\over x\sqrt{(x-1)(U^2-x)}}=
-{1\over i\pi}\lim_{\kappa\to 0}\left(\log{\kappa\over 4}\right)+
{1\over i\pi}\log{U^2\over 1-U^2}
\ee
where $\kappa$ is a small-$x$ cut-off. The $U$ dependent part of this
integral is finite and can be
considered as the \lq\lq true" perturbative correction, while
the divergent part just renormalizes the bare  \lq\lq classical"
coupling constant $\tau$, {\it i.e.} classical part of the prepotential
(see s.5.6 for further details).
Therefore, the perturbative period matrix is finally

\be
T_{finite}={i\over\pi}\log{\sn^2(\epsilon|k)-
\sn^2 (a|k)\over\sn^2 (a|k)}+\ const\
\longrightarrow
{i\over\pi}\log{\theta_1(a+\epsilon)\theta_1(a-\epsilon)\over\theta_1^2(a)}
\ee
and the perturbative prepotential is the elliptic tri-logarithm.
Note that it is again of the form
(\ref{appC}) and (\ref{appR}) but with the function $f^{(6)}$ instead
of $f^{(4)}$ and $f^{(5)}$.
Remarkably, it lives on
the {\it bare} momentum torus, while the modulus of the perturbative curve
(\ref{psc}) is the dressed one.

\subsection{Quantum duality}

Our discussion so far concerned the classical systems. However
duality is expected to be a powerful tool to deal with quantum
problems too. This point has been first recognized by Ruijsenaars in his
early papers. The very idea is that Hamiltonians and their
duals have common wave-functions and this fact could
be exploited to consider one or another set of eigenvalue problems,
with respect to different variables.
Since the wave functions in the systems under consideration
can be realized in group terms, one could expect
that quantum duality can be formulated purely in
group terms too. Recent activity in this direction supports
such expectations \cite{qdual}.

Here we work out a couple of simple quantum 2-particle
examples of dual systems in order to illustrate the general picture.
Let us start with the most standard oscillator example with the Hamiltonian

\beq
{\hat H} = {-1/2}{{{\partial}^{2}}\over{{\partial} q^{2}}} +
{{{\omega}^{2} q^{2}}\over{2}}
\eeq
Its normalized eigenfunctions are

\be
{\hat H}
{\psi}_{n}  =  {\omega} (n + {1/2}) \psi_{n}
\\
{\psi}_{n} (q) =  \bigl(
{{\omega}\over{\pi}} \bigr)^{1/4}  {{
e^{-{{\omega
q^{2}}\over{2}}}
}\over{2^{n/2}\sqrt{n!}}} H_{n}(q \sqrt{\omega})
\ee
where
$H_{n}(\xi)$ are the Hermite polynomials
$H_{n}(\xi) = e^{\xi^{2}} (-
\partial_{\xi})^{n} e^{-{\xi}^{2}}$.
Using this representation for the
wave-function one can easily obtain a recurrence
relation

\beq
\sqrt{n+1} {\psi}_{n+1}(q) + \sqrt{n} {\psi}_{n-1}
(x) = \sqrt{2\omega} q {\psi}_{n}(x)
\eeq
It means that $\psi_{n}(q)$ is the
eigenfunction of the difference operator

\beq
{\hat H}_{D}
= T_{+} \sqrt{n} + \sqrt{n} T_{-}, \quad T_{\pm} = e^{\pm {{\partial}
\over{\partial
n}}}
\eeq
acting on the  subscript $n$, which is nothing but the
quantized dual Hamiltonian.

Less trivial example whose classical version we discussed throughout
this whole section is the quantum (2-particle) Hamiltonian of the
trigonometric Calogero model

\beq
{\hat H} = {-\half }{{{\partial}^{2}}\over{{\partial}
q^{2}}} + {{{\nu}({\nu} -1)}\over{2 \sin^{2}(q)}}
\eeq
Its normalized
eigenfunctions are

\be
{\hat H} \psi_{n}  =
{{n^{2}}\over{2}} \psi_{n}
\\
\psi_{n}(q) = \sin^{\nu}(q)   \sqrt{n
{{(n-{\nu})!}\over{(n+{\nu}-1)!}}} \Pi_{n-\half}^{\nu -
\half}\bigl(\cos(q)\bigr)
\\
\Pi_{l}^{m}(x) = {1\over{l!}} {\partial}_{x}^{l+m}
\bigl({{x^{2}-1}\over{2}}\bigr)^{l}
\ee
For the sake of simplicity, we take half-integer $\nu$ and
$n$.
One can make the change $\nu \to - \nu - 1$ and get
another eigenfunction with the same eigenvalue.
Using the fact that the
generating function for $\Pi_{l}^{0}$'s

\beq
Z(y,x) =
\sum_{l=0}^{\infty} y^{l} \Pi_{l}^{0} = {1\over{\sqrt{1 - 2xy + y^2}}}
\eeq
satisfies the two obvious equations

\be
{(x-y)\partial_{x} Z = y\partial_{y} Z}
\\
{(1 - 2xy + y^{2})\partial_{y} Z = (x-y)Z}
\ee
which implies

\beq
(y\partial_{y} - m) \partial_{x}^{m} Z = (x- y) \partial^{m+1}_{x} Z
\eeq
and, hence, yields

\beq
{ \bigl( (1 - 2xy + y^{2}) \partial_{y}  + y - x \bigr) \partial^{m}_{x}Z
= m ( 1 + 2y\partial_{y}) \partial^{m-1}_{x} Z}
\eeq
one derives the
recurrence relations

\be
x \Pi_{l}^{m} = {{l+1-m}\over{2l+1}} \Pi_{l+1}^{m} + {{l+m}\over{2l+1}}
\Pi_{l-1}^{m}
\\
\cos(q) \psi_{n} = {\half} \Bigl( \sqrt{1-
{{\nu}({\nu}-1)}\over{n(n+1)}}\psi_{n+1}
+\sqrt{1-
{{\nu}({\nu}-1)}\over{n(n-1)}}\psi_{n-1} \Bigr)
\ee
i.e. $\psi_{n}$ is an
eigenfunction of the finite-difference operator
acting on the subscript $n$

\beq
{\hat H}_{D} \psi(q) = \cos(q) \psi(q)
\eeq
\beq
{\hat H}_{D} = T_{+} \sqrt{1- {{{\nu}({\nu}-1)}\over{n(n-1)}} }+
\sqrt{1- {{{\nu}({\nu}-1)}\over{n(n-1)}} }T_{-}
\eeq
which is  a quantum
version of the rational Ruijsenaars model.

Summarizing,
when the system with
trigonometric dependence on momentum
is quantized, its Hamiltonian becomes
a finite-difference operator. The
wave-functions become the functions
of discrete variables. The origin of
this is in the Bohr-Sommerfeld quantization
condition. Indeed, since the
trigonometric dependence on momenta
implies that the leaves of
polarization are compact and, moreover,
non-simply connected,  covariantly
constant
sections of the prequantization connection along the polarization
fiber generically ceases
to exist. It is only for special ``quantized''
values of the action variables that the section exists.
In the elliptic case,
the quantum
dual Hamiltonian is going to be a difference operator of
infinite order.

\section{Branes as low energy degrees of freedom
in supersymmetric gauge theories}
\setcounter{equation}{0}
\subsection{Field theory on D-brane worldvolume}

We begin this section with the description of nonperturbative states
in string theory
which play the key role for constructing the low-energy degrees of freedom
in SUSY gauge theories. In particular,
the phase space of the integrable systems, considered in the previous sections
is associated with the collective coordinates of branes of different
dimensions. Besides it, nonperturbative dualities in string theory
suggest several different formulations for the same single field theory.

The central objects introduced into string theory by exact solving
the \N2 SYM theory are the duality (between the strong and week coupling
regimes) and the exact spectrum of BPS
solitons. It appeared that in order to consistently
describe duality in string theory,
one has to introduce solitonic objects of different
dimensions, ``p-branes" with $p$ space coordinates
\cite{polchinski1}. In quasi-classical approximation, their masses behave as
$m=g^{-2}$, where $g$ is the string coupling constant, and the mass formulae
for the
BPS states, being related to the central extension of the SUSY algebra remains
exact
at quantum level. At the strong coupling regime, the solitons become the
lightest objects in the theory and define the dynamics in that region.
Technically, it is possible to introduce dual variables which yield a
week coupling theory in terms of the solitonic fields.

All this picture is literally inherited from Seiberg-Witten theory.
On the other hand, \SW itself gets a new description in string/brane terms.
Indeed, the appearance of Riemann surfaces in \SW can be explained
in terms of branes.
It is this surface which is the part of the worldvolume of the larger brane.
The structure of the brane intersections is fixed by the SUSY and the
set of configurations yielding the field theory become finite or
even single.

As we already explained in the Introduction,
the key point which admits the use of branes for obtaining field theories
is the $U(1)$ gauge theory on the brane worldvolume. It is known that
branes are charged with respect to the RR sector of the theory
\cite{polchinski1} and can serve as a space of endpoints of
strings. Using this property Witten formulated the new view on the
gauge theory as the theory on the worldvolume of coinciding branes which
carry $U(1)$ factor each. The open strings connecting branes give rise to
non-Abelian theory, rank of the gauge group being defined by the number of
branes.
In what follows we shall follow this line: SUSY field theories are defined on
the brane worldvolume and the parameters are introduced via the geometry of the
brane configuration.

Some parameters of \N2 SUSY theories (condensates of scalars)
coordinatize the Coulomb branch of the corresponding vacuum moduli space,
while other parameters (bare masses and coupling constants) remain external
parameters. All these parameters are defined in the brane
terms (see the Introduction):
scalar condensates yield the distance between branes (or determine embedding
of the single M-theory brane into the target space),
coupling constant is just the distance between
additional, background branes (or also determines geometry of the M-theory
brane),
while the hypermultiplet mass is the coordinate
of another brane (often of different dimension)
in the proper direction. In what follows we shall
present examples of particular brane configurations.
Coordinates of branes appear to be associated with the degrees of freedom of
the
integrable system governing the low energy action in the SUSY gauge theory.

Let us emphasize that the gauge theories of
the Hitchin type responsible for the
integrable many-body systems arise in the ``momentum" space
or ``the space of fields" and should not be thought of as those
obtained by the
dimensional reduction from the $4d$ SUSY gauge theory. In fact, these theories
emerge from different degenerations
of the toric geometry.

It is useful to start with the softly broken $\CN=4$ theory. In \n=1
terms, there are 3 adjoint hypermultiplets whose complex scalar
components parameterize the embedding of branes into six dimensions.
The potential of scalars is

\be\label{pot}
V=\frac{1}{g^{2}}\Tr\sum_{i=1,2,3}[\phi_{i},\phi^{+}_{i}]^{2}
\ee
and
and is induced
by the kinetic term (as a result of integration over auxiliary fields).
There is also a non-trivial superpotential, $W=\Tr\Phi_{1}[\Phi_{2},\Phi_{3}]
+M^{2}\Tr(\Phi_{2}^{2}+\Phi_{3}^{2})$, where $\Phi_{i}$ are $\CN=1$
hypermultiplets with lowest components $\phi_i$ respectively
and M is the mass of the hypermultiplet. The minimum
of the potential energy amounts to the vacuum configuration with non-trivial
commutation relations between the 3 scalars. In the large mass
limit, the fluctuations orthogonal to the algebraic surface
in the two dimensional complex space

\be
\Tr(\Phi_{2}^{2}+\Phi_{3}^{2})=const ,
\ee
are frozen. Note that, since one of the scalars remains massless, it jointly
with the gauge superfield forms one \N2 supermultiplet, while the two massive
scalars are the components of the \N2 adjoint massive hypermultiplet, i.e.
the $\CN=4$ SUSY is broken down to the \N2 SUSY. In this \N2 theory, there are
valleys
for the massless scalar $\phi_1$ due to the potential term (\ref{pot}), i.e.
this scalar may have a condensate which parameterizes the Coloumb branch.

There are two equivalent approaches to \N2 theories in the string
theory framework which start from the theory in ten dimensions. In the
first, geometrical engineering
approach, one considers the compact Calabi-Yau (CY) threefold near the
singularity
providing the desired gauge group  \cite{eng,V1}.
In the second approach proposed
in \cite{W}, the theory on the worldvolume
of D4-branes stretched between two NS 5-branes
is considered.
The location of the branes is
summarized in the following table

\begin{equation} \ \ \ \left\{
\begin{array}{c|cccccccccc}
      & 0& 1& 2& 3& 4& 5& 6& 7& 8& 9\cr
NS5   & +& +& +& +& +& +& -& -& -& -\cr
D4    & +& +& +& +& -& -& +& -& -& -\cr
D6    & +& +& +& +& -& -& -& +& +& +
\end{array}\right.
\label{tabl}
\end{equation}
where plus stands for the extended direction
of each brane and minus for the transverse direction.
The fourbranes are finite in the $x_6$ direction stretching between the two NS
branes,
therefore, the long range theory is a
four-dimensional theory as we  need. While in this type IIA picture
the original SUSY (32 real supercharges) is broken down to $N=2$ in $4d$
(8 real supercharges) due to the concrete brane configuration\footnote{The
presence of
D4-branes breaks symmetry down to 16 supercharges and two NS branes breaks
it down further to 8 supercharges.}, in the IIB picture (related to CY),
the original SUSY is broken via geometry of the CY manifold. D-branes wrapped
around shrinking cycles provide the field theory and break half of the
supersymmetries indeed. Note that the IIA picture can be lifted
to M-theory in $11d$ which has the $11d$ SUGRA as its low-energy limit.

\subsection {Branes and degrees of freedom in many-body systems}

Let us start with the description of degrees of freedom for the
$4d$ $SU(N_c)$ theory without matter in IIB/F picture \cite{g19}.
It emerges as the worldvolume
theory on  D7-branes wrapped around an elliptically fibered K3
\cite{V1}. The bare coupling of the gauge theory yields the complex
structure of the elliptic fiber, therefore, the spectral curve of the
Calogero model enters the picture automatically.

To identify the degrees of freedom using the approach of section 2.3,
one has to define the holomorphic
one-form $\phi$ as well as the anti-holomorphic vector potential $\bar {A}$
on the spectral curve. The holomorphic form can be identified if one takes
into account that the torus is embedded into the
K3 space as a supersymmetric
cycle. Therefore the embedding of D7 branes is described in terms of
the one-form
\cite{bsv}
which after all turns out to be the Lax operator for the many-body system.
The anti-holomorphic vector field $\bar{A}$ whose eigenvalues are
the coordinates
of particles in the many-body system comes from the $N_c$ branes
wrapped around the spectral curve with the worldvolume gauge fields.
Finally the marked point on the spectral curve with nontrivial
monodromy is introduced via additional
``background" branes localized on the spectral curve.  Such ``background"
branes play the role of regulators in field theory. Let us emphasize
that in F-theory the 11-th and 12-th dimensions just provide the bare coupling
constant.

Since the background brane does not change the interpretation of  degrees
of freedom let us consider the simplest case of the free many-body system.
The trigonometric Ruijsenaars system, i.e. the $5d$ SUSY theory
is the most illustrative, since
in this case the brane geometry is in one-to-one correspondence with the
Hamiltonian reduction procedure of section 2.
Let us remind that the system originates from the Chern-Simons theory
given on the product of torus and real (time) line. There is the relation
for the monodromies of the gauge field  around the torus cycles
$g_A, g_B$

$$
g_{A}g_{B}g_{A}^{-1}g_{B}^{-1}=1
$$
In the free case the matrices can be diagonalized simultaneously
and their diagonal elements represent the coordinates and momenta of the
dynamical system
$g_{A}=diag(e^{iq_{i}})$, $g_{B}=diag(e^{ip_{i}})$. Note that from the F-theory
point of view we have degenerated the fiber torus to a cylinder.

Let us compare the geometrical descriptions following from the
Chern-Simons Lagrangian and the brane configuration. Two coordinates
$x_5, x_6$ are essential for the description of the five dimensional theory.
The typical scale along $x_5$ is $R_{5}^{-1}$, while
that along $x_6$ is $g^{-2}$ in the string units \cite{ah}.
We shall assume that both coordinates are compactified onto circles
of the corresponding radii.

The (coordinate and momentum) circles
in the Ruijsenaars model are related to the circles in the brane
configuration via the T-duality transformation along the both
directions. Therefore, the radii of circles in the Ruijsenaars model are
$R_{5}$, $g^{2}$,
and after the T-duality transformation the eigenvalues of the monodromy matrix
transform into the brane positions along the dual circle.
There are precisely $N_c$ 5-branes in the IIB theory extended along
$x_6$ and localized at $p_i$ along $x_5$. The coordinates in the
dynamical system are provided by $N_c$ 1-branes localized at $q_i$
along $x_6$. The SUSY gauge theory is defined on
$(x_{0},x_{1},x_{2},x_{3},x_{4},x_{6})$,
and the coordinates of 5-branes along $x_5$ define the vacuum expectation
values of the real scalar field.

To switch on the interaction in the dynamical system,
one has to add a massive hypermultiplet in adjoint in the field theory
language, to introduce
a marked point with non-trivial monodromy around in the Chern-Simons
description or to add the background branes in the string theory
approach.

The new relation for the monodromy matrix is

$$
g_{A}g_{B}g_{A}^{-1}g_{B}^{-1}=g_{c},
$$
where $g_{c}$ corresponds to inserting the $\C\bP^{n-1}$
orbit at the marked point. After
the T-duality transformation, a brane configuration with
non-trivial  flux arises. The color orientation of the fluxes
follows from the structure of the monodromy matrix $g_c$.

Another approach to description of field theories, for definiteness let them be
$4d$ theories,
is based on the consideration of field theory on the worldvolume
of M5-brane, the solitonic state in M-theory \cite{W}.
We look for an M5-brane configuration that provides a theory
with 8 real supercharges, i.e.
in $\CN=2$ theory. This can be achieved  by considering the
product of $R^4$ and a fixed genus compact surface $\Sigma$
as the worldvolume of M5-brane. After compactification onto the circle (11-th
dimension), the
M5-brane in the IIA theory looks as a collection of D4-branes located
between two NS-branes.

To get the $SU(N_c)$ theory, one has to introduce $N_c$ D4-branes of finite
size
along one of
the coordinates, say, $x_6$.
Then, the plane $x_4+ix_5$ in the perturbative regime
is where the momentum variables in the many-body system live. Moreover, there
are D0-branes on the D4-worldvolume, one per each D4-brane, which are ordered
along
the $x_6$ coordinate. In the low-energy limit, the D0-brane coordinates
are defined

\be
q_{j}=j\Delta + \phi_{j},
\ee
where $\Delta$
fixes their equilibrium locations,
while $\phi_{i}$ correspond to small fluctuations and coincide with the
variables in the Toda system. Let us note that $q_i$ can be also
considered as the coordinates in the Calogero system, without any
ordering. Comparing with the field theory variables
one immediately arrives at
$N_c\Delta=\frac{1}{g^{2}(M)}$ which perfectly fits the
suggested ordering of the D0-branes along the $x_6$ direction.

A qualitative interpretation of the $2\times 2$ Toda Lax operator
is then as follows. The
element $L_{11}$
corresponds to the D0-brane confining with the D4-brane
so that they have the same $(x_4+ix_5)$-coordinate. Since
Lax operators are related with fermionic zero modes in a nontrivial
background field, the size of the Lax matrix indicates the
presence of the two dimensional space of zero modes. Elements
$L_{12},L_{21}$ fix the matrix elements of the spectral parameter
between the fermionic zero modes
localized on the neighbour D0-branes.

Let us note once again that qualitatively in any brane
configuration, there are
three essential elements presented. First, there are the branes
that provide the worldvolume for the field theory,
$N_c$ D7-branes in IIB/F or one M5-brane in IIA/M picture. Second,
it is necessary to introduce the branes responsible for the variables
in the dynamical system, which are zero size instantons
in the field theory language \cite{barbon}. Finally,
the account of the regulators in field theory corresponds to an
additional Wilson line in the IIB/F picture or fluxes
in the IIA/M description.

Recently the correspondence between integrable systems and
branes has been pushed forward even further. Namely, one can
start with the $4d$ theory with one compact dimension
\cite{kapustin} and map the system to the Hitchin phase space by a
chain of dualities. On one hand, such a moduli  space
coincides with the moduli space of the non-commutative instantons
on $T^4$ or K3 manifolds \cite{Ganor}. On the other hand
non-commutative instantons
introduced in \cite{neksw} allow a deformed ADHM description,
moreover, the explicit map between the moduli space of non-commutative
instantons and the trigonometric Calogero model has been
elaborated \cite{Braden}.

\subsection{Branes in five dimensional theory}

While considering $5d$ gauge theories the role of the brane configuration
becomes the most transparent. It appears that in five dimensions
the brane configuration fixes the structure of the spectral curve of the
integrable system. We shall consider below the configuration of branes
in the IIB theory which is related to the toric manifold data providing
the compactification of M-theory
\cite{V1,eng,nikita3}.

We start from the geometrical engineering picture \cite{GGM2}.
As a compact manifold we take the CY manifold in the vicinity of singularities.
Being interested
only in this region of parameters, we discuss exclusively a local
model implying that its embedding into a compact Calabi-Yau is
always possible.
To this end, we consider $A_{n-1}$ ALE spaces
fibered over a set of the base spheres, each sphere corresponding
to an $SU(n)$ gauge theory\footnote{Throughout this subsection we consider
the gauge theory whose gauge group is a product of $k-1$ simple factors,
typically $SU(n_i)$.}.
The size of the base sphere
$V \propto {1 \over g^2}$ is governed by the coupling constant of
the corresponding $SU(n)$ factor. Kahler moduli of
$\bP^1$'s associated with blowing up the ALE space define the Cartan
scalars $a_i$ on the Coulomb branch of the moduli space.
This non-compact Calabi-Yau space is a hypersurface in
the holomorphic quotient and
can be represented as a set of Kahler cones, or, in other words,
by the toric polyhedron $\De$.

For example, one can start from the $\CN=(2,2)$ linear sigma model
construction of Calabi-Yau \cite{AGM} and define $U(1)$
charge vectors of matter fields

\be
{\bf Q^a}=(q^a_1, \ldots , q^a_k)
\nn
\ee
 From the $\sig$-model point of view these vectors define
gauge-invariant variables, while from the Calabi-Yau standing point they
act on the homogeneous coordinates in projective space and
fulfill $r$ linear relations

\be
\sum_J Q^a_J {\bf \nu_J} =0
\label{rel}
\ee
between vectors ${\bf \nu_J} \in {\bf N} = {\bf Z}^d$ which define
coordinate patches in each Kahler cone.
Vertices ${\bf \nu_J}$ in lattice ${\bf N}$ form a toric polyhedron
$\De = \{ {\bf \nu_J} \}$.
For the toric variety to be smooth, $\De$ must be convex.
Thus, having defined the original Calabi-Yau in terms of
holomorphic quotients, we now turn to the solution
of the model via mirror symmetry.

As it was explained in \cite{eng,V1}, Kahler moduli of the
original Calabi-Yau specify Coulomb moduli of the low-energy
effective theory. Under mirror transform they get mapped to the complex
structure of the mirror manifold, which is just Riemann surface
$E$ having the same moduli space. There is a profound reason in
this symmetry, because the Riemann surface is the spectral curve
of the associated integrable system, \XXZ spin chain, section 3.5.

In terms of charge vectors this mirror manifold is

\be
\prod_J y_J^{q^a_J} =1
\nonumber
\ee
For the models we consider here it is just a spectral
curve of underlying integrable system.

Let us now turn to the description of $5d$ theories in terms of
brane configurations. One can derive the  brane configuration
using T-duality along $x_4$ from the IIA picture
\cite{W}.
Thus, one obtains the IIB picture \cite{ah}
with the brane configuration \cite{ah}

\begin{equation} \ \ \ \left\{
\begin{array}{c|cccccccccc}
      & 0& 1& 2& 3& 4& 5& 6& 7& 8& 9\cr
NS5   & +& +& +& +& +& +& -& -& -& -\cr
D5    & +& +& +& +& +& -& +& -& -& -
\end{array}\right.
\label{tabl5}
\end{equation}
where the field theory is realized on the D5-brane
worldvolume.
The holomorphic solution to the $2d$ Laplace equation
in the space $(x_4,x_5)$, $\Delta s=0$ ($s\equiv x_6+ix_{10}$):
$s=\log (x_4+ix_5)$
transforms into the solution to the $1d$ Laplace equation
$s(\equiv x^6)= {1 \over 2} | x_5 | =
{x_5 \over 2} {\rm sign} (x_5)$. If one demands
for SUSY  to be broken up to 1/4, a restriction
on the brane intersection arises. In particular,
when charged $n$ D5-brane meet charged $k$ NS5-branes they must merge into
$(k,n)$ bound state with the slope $\tan \phi = {k \over n}$ on
$(x_5, x_6)$ plane \cite{ah}.

As usual we divide moduli of brane configurations into two groups.
The first group of parameters, external parameters
(e.g. coupling constants, bare quark
masses) define the five-dimensional theory, while the rest are
dynamical moduli.
In brane terms, changing the parameters of the first group one
changes the asymptotic (i.e. infinite in all their non-zero dimensions)
branes\footnote{In our case, NS5 branes.},
while changing the parameters of the second group does not.

This is a good point to make a contact
with the geometrical engineering approach \cite{eng,V1}. In that
language each link in the toric polyhedron $\Delta$ is associated
with a 1-cycle on the corresponding Riemann surface $E$. Periods
along 1-cycles in the compact part of $E$ (i.e. internal links
in $\Delta$) define the dynamical moduli whereas bare parameters
arise from 1-cycles that wrap the non-compact part of $E$ (i.e.
related to the links on the boundary of $\Delta$). One can go
further and note that the number of such external links in $\Delta$
corresponding to some low-energy effective field theory coincide
with the number of infinite fivebranes in the brane construction
of the same theory.

The reason for this is that the brane picture and the toric polyhedron
corresponding to a five-dimensional field theory are intimately
related to each other. To establish other connections let us
visualize brane diagrams in other terms. Any such configuration
depicts the kinematics of two-dimensional ($x_5$, $x_6$) free
particles. The charge conservation in the vertices where branes merge
together is nothing but the momentum conservation law for free
particles. The integer-valued momentum $p^{(5,6)} =(k,n)$ of
particle is just the $SL(2,\Z)$ charge of the brane. The
vertical branes in terms of \cite{ah}
carry the $(1,0)$ charge (momentum) vector,
and horizontal ones -- $(0,1)$. Notice that the number of vertical (NS5)
branes specifies the number of simple gauge factors whereas the number of
horizontal (D5) branes defines the rank of each $SU(n_i)$ factor.

For instance, for $SU(n)^k$ gauge theory there are $k+1$
NS-branes and $n$ D-branes in each stack.
And, similarly,
the number of vertical links in the toric polyhedron,
i.e. the height of $\Delta$ equals $k+1$ and the number
of horizontal links, i.e. its horizontal size is $n$.
At generic point on the moduli space, the vertices of
$\Delta$ lie on a two-dimensional integer lattice. In these
conventions we can endow each link in $\Delta$ with an integer-valued
two-dimensional vector $(k_{\delta},n_{\delta})$.

One can exploit \cite{GGM2} the fact that type IIB brane configurations
for five-dimensional field theories are in one-to-one correspondence
with toric polyhedrons of Calabi-Yau M-theory compactifications
describing these field theories. The exact relation between
the two is provided by identification of the $(k_{\delta},n_{\delta})$
charge (momentum) vector with a link vector in $\Delta$.
This powerful tool allows one to solve any five-dimensional
$\prod SU(n)$ gauge theory with arbitrary matter.

The toric polyhedron $\Delta$ for the
case of pure gauge $SU(n)$ theory can be found in \cite{V1}.
Adding a massive hypermultiplet,
we break one link on the boundary of $\Delta$ into
two pieces and add one more internal link. Therefore, the
number of internal and external links representing massive hypermultiplets
increases by one. Within the brane approach, the picture is exactly
the same: adding a fundamental matter hypermultiplet to the theory
one attaches a new semi-infinite brane.
It can be done in four
different ways: between the pairs of the four semi-infinite branes
already presented in the configuration..
The same
possibilities can be found in breaking one of the four
external links of the polyhedron into two parts.
For the $\prod_{i=1}^k SU(n_i)$
case, one has to take a fibration
of $n_i$ spheres over each of $k$ base spheres.
The total number of external links is $2(k+1)$
as the number of semi-infinite branes in the brane configuration.
The number of finite branes is
$\sum_{i=1}^k (3 n_i -1) -1$ which matches exactly the number of
internal links in $\Delta$. The relation between toric diagrams
and brane configurations has been also considered in \cite{Leung}.

\subsection{Branes and equations of motion in many-body systems}

Our next step in clarification of the role of
brane moduli as degrees of freedom in integrable systems
is to interpret the equations of motion
in the brane terms  \cite{g19}. To this end, let us relate the brane
configuration with the monopole moduli space and identify
the equations of motion with the Nahm equations defining
the structure of the moduli space.

D0-branes localized on D4-branes can be considered as the
degrees of freedom corresponding to the BPS monopoles. They can be related
to other D4-branes by open strings. Following \cite{Dia},
the theory on the string worldvolume can be obtained by the
dimensional reduction from the $10d$ SYM theory. Moreover, there
is one-to-one correspondence between the equation defining the
supersymmetric
vacuum state in $\sigma$-model and the Nahm equation for monopoles.
Now, in order to get the equations of motion in the Toda system,
one should take into account the connection between the spectral curve
in the Toda system and the spectral curve amounted from the
Nahm equations for the charge $N_c$ cyclic monopole \cite{sat}.
The spectral curve for the cyclic configuration can be derived
from the generic monopole spectral curve

\be
{\lambda}^{n}+{\lambda}^{n-1}a_{1}({w})+....+a_{n}(w)=0
\ee
where $a_i(w)$ is a polynomial of degree $2i$. Then,
assuming the center of mass is fixed at the origin, the
overall $U(1)$ phase is unit and imposing the (order $N_c$) cyclic group
condition, one arrives at the Toda curve.

One can also connect the Lax operator for the Toda
system
with the Nahm matrices $T_i$ known in the monopole theory \cite{Nahm}

\be
T(w)=T_{+}+2T_{0}w+T_{-}w^{2}
\ee
where $T_{\pm}$ are respectively the (strictly)
upper and low triangular parts of the Toda Lax
operator (\ref{LaxTC}), while $T_0$ is its diagonal part\footnote{In monopole
terms, the variable $w$ is the coordinate on the twistor manifold
$\C\bP^{1}$.}.
Adopting the notation
$T_{\pm}\equiv T_1\pm iT_2$,  one can identify the
Toda Lax equation (i.e. the equations of motion)

\be
\frac{dT(w)}{dt}=\epsilon_{ijk}[T(w),M]
\ee
where $M\equiv T_+w^{-1}-T_-w$, with the Nahm equations

\be
\frac{dT_i}{dt}=2i\epsilon_{ijk}[T_j,T_k]
\ee

To develop a more geometrical picture, it is useful to
consider fermions in the background  monopole field

\be
({\sigma}D+{\phi}-t){\Psi}=0 .
\ee
where $t$ is an external parameter.
The equation has $k$ linear independent solutions for
the magnetic charge $k$. It is the variable $t$ which
becomes the time variable in the Toda system and
simultaneously plays the role of the space variable
in the brane interpretation. The Nahm matrices
in terms of the fermionic zero modes are

\be
T_{j}=\int x_{j}{\Psi}^{+}{\Psi} d^{3}x;\ \ \ j=1,2,3
\ee
These expressions have the structure of the Berry connection
if the momentum space is considered as the parameter space.
There exists also an inverse construction which allows one
to derive the multi-monopole fields starting from the auxiliary
one-dimensional problem for the fermions in the external Nahm connection

\be
({i\frac{d}{dt}} +i{\vec \sigma}{\vec x}+{\vec\sigma}{\vec T}^{+})v=0\\
A_{j}=\int v^{+}{\partial}_{j}vdt \ \ \
\phi=\int tv^{+}vdt
\ee
where the superscript plus of ${\vec T}$ means Hermitian conjugation.
The limits of the integration are fixed by the asymptotics of
the scalar field $\phi$. Note that the fermions at $x=0$ are
identified with the spectral fermions of
the Toda (or, more generally Nahm) system.

One may also consider the generalization of the Nahm construction for
the $SU(n)$ case \cite{Dia}. Then, one needs to fix asymptotics
of the $n$ components of the
scalar field to be some (ordered) numbers $\mu_i$, $i=1,...,n$,
$\sum_i\mu_i=0$. It naturally divides the domain where the Nahm
system is given onto the $n-1$ intervals
$(\mu_1,\mu_2)$, ..., $(\mu_{n-1},\mu_n)$. On each $a$-th interval
the Nahm system for the three
$p_k\times p_k$ matrices ${\cal T}_{i,a}(t)$ is

\be
\label{Nahmp1}
{d{\cal T}_{i,a}\over dt}=
2\epsilon_{ijk}\left[{\cal T}_{j,a},{\cal T}_{k,a}\right]
\ee
Thus, the $SU(n)$ monopole is characterized by the set of topological
charges $p_1,...,p_{n-1}$.

The correspondence between solutions to the Nahm
equations and the spectral curves of integrable systems
is generalized to the $SU(N_c)$ case \cite{Dia}, where the connection
with the higher spin magnets emerges \cite{GGM1}. The solution to the
Nahm equations is a product of
$n-1$ components, each of them being a spectral curve for the $SU(2)$
monopole of charge $p_a$, while the
whole monopole spectral curve is a special case
of the curve for the $sl(p)$ spin magnet with $p=\sum_a p_a$.
The configuration of the magnet is specified by a peculiar
pair clasterization, each of them describing the $sl(p_a)$
magnet. There are also additional matching conditions for the
correlations  between the two-site groups.

The peculiarity of the spin chain configuration has a clear explanation in the
brane language. The monopole moduli space is described via the
configuration of D1- and D3-branes \cite{Dia}. The clasterization in the brane
language means that the D1-branes have their endpoints only on the neighbor
D3-branes. Thus, appealing to the spin chain interpretation,
one can identify the D3-branes with the sites of the spin chain,
while the D1-branes give rise to the $SU(p)$ group, $p$ being the full number
of
the D1-branes. Let us note that the twisting procedure used in section 3 is
equivalent to adding semi-infinite D1-branes to the very left
and the very right D3-branes.

\subsection{Branes and two Lax representations}

Now let us consider two distinct Lax representations of the Toda chain that we
discussed
in section 3.2
and explain their interpretation
in terms of brane diagrams \cite{GGM1}. Remind that
in the $2\times 2$ representation the
system is formulated in terms of the transfer matrix, while for
the $N_c\times N_c$ representation the notion of the transfer matrix is missed.

{}From the brane pictures point of view the two Lax representations
are nothing but the choice of parametri\-za\-ti\-on of the fivebrane
worldvolume $\Sigma$. Of two holomorphic coordinates (which are
Cartan elements of scalars) $\la = x_4 + i x_5$ and $s = x_6 + i x_{10}$,
\footnote{Equivalently, $s= x_6 + i x_9$ in the Type IIB language.} we can
regard any we wish as a spectral parameter. In the ``standard" $2 \times 2$
representation, $w=\exp{-s \over R}$ is a spectral parameter.
Another possibility is to choose
$\la$ as a spectral parameter. It leads to the $N_c \times N_c$ representation.

Geometrical image of this phenomenon is the base-fiber
symmetry of the Calabi-Yau threefold \cite{V1}. In the
``standard" $2 \times 2$ language $w$ parameterizes
the base $A_{m}$ singularity and counts the number $m$ of gauge factors.
$\la$ is a coordinate in the fiber. If each coefficient at $w^i$ is
a polynomial of degree $n$, we deal with the $SU(n)^m$ gauge theory with
$n$ extra fundamentals at each end of the chain of the gauge factors.
The duality
$SU(n)^m \leftrightarrow SU(m+1)^{n-1}$ proposed in \cite{V1} is
nothing but the $m+1 \times m+1 \leftrightarrow n \times n$ symmetry
of the Lax representations.
Under this symmetry
the role of coordinates on the Newton polygon
describing the manifold is changed.

Note that the generating differential

\be
dS = \la {dw \over w} = \la ds \approx - s d \la
\ee
also suggests this symmetry.

Let us extend this argumentation onto description of the
fundamental matter in $4d$ gauge theory. There are two different
realizations of the matter in the brane language: via
semi-infinite D4-branes and via D6-branes.
On the integrable side, the first possibility corresponds to the
additionally twisted
Toda chain \cite{KPbc}, while the second one to the \XXX magnet.

Indeed, the curve  for the theory with the fundamental matter can be
also obtained from the $N_c \times N_c$ representation by deforming
the Lax operator. To this end, one has to add non-trivial
entries to the first and the last columns which corresponds
to attaching the semi-infinite branes to the end branes of the configuration.
On the contrary, in $2\times 2$ representation D6-branes
correspond to the fundamentals which are localized at the points
corresponding to the masses. It is in agreement with the relation
between the masses and inhomogeneities in the spin chain. The both
representations can be related if one takes into account the phenomenon of
the brane creation when branes pass through each other \cite{HW}.

In \cite{HW} the authors considered the
three dimensional theory in the common longitudinal directions of
branes (in the type IIB theory):

\begin{equation} \ \ \ \left\{
\begin{array}{c|cccccccccc}
      & 0& 1& 2& 3& 4& 5& 6& 7& 8& 9\cr
NS5   & +& +& +& +& +& +& -& -& -& -\cr
D3    & +& +& +& -& -& -& +& -& -& -\cr
D5    & +& +& +& -& -& -& -& +& +& +
\end{array}\right.
\end{equation}

By T-duality in the third direction, this setup gets
mapped onto the Type IIA brane picture \cite{W}
so that the statement of
creating a new D3-brane when the D5-brane passes through the NS5-brane
transforms into the claim that, moving the D6-brane through the NS5-brane
creates a new D4-brane stretched between them.

\subsection{Whitham dynamics via branes}

So far we discussed the
brane interpretation of the Toda/spin chains. Now we are going
to say some words about the other, Whitham integrability of \SW
and start with the brane configuration in F-theory, i.e.
on the elliptically fibered K3 manifold or, equivalently,
on orientifold in IIB theory.

According to \cite{sen1}, \N2 theory in $4d$ can be considered
as a theory on the worldvolume of 3-branes in the background of four 7-branes
located at the points
$\pm \Lambda$ on the complex plane of $u=\Tr\phi^{2}$
for the $SU(2)$ gauge group. We assume that masses of all fundamentals
are large and, therefore, we can work with the pure gauge theory.

Consider now the dynamics of the 3-branes along the directions transverse
to the background 7-branes. We shall restrict ourselves to the
$SU(2)$ gauge group. However, since the whole picture is very transparent, the
generalization to higher rank groups is straightforward.

We are going to study the Whitham dynamics of
moduli parametrized by branching points. In the $SU(2)$ case,
one can write the Seiberg-Witten curve in the hyperelliptic form
(\ref{2})

\be\label{3w}
Y^2=(\lambda^2 -u)^2-\Lambda_{QCD}^4, \ \ \ dS=2\lambda^2{d\lambda\over Y}
\ee
i.e. there are 4 branching points $\pm\sqrt{u\pm\Lambda_{QCD}^2}$. Since the
Whitham
dynamics
governs the $u$-behaviour, all the four points move under the
Whitham flows. Therefore, it is more convenient to use another parameterization
that goes back to the original paper \cite{SW1}

\be\label{4w}
Y^2=(x+u)(x-\Lambda^2_{QCD})(x+\Lambda^2_{QCD}),\ \ \ dS=(x+u){dx\over Y}
\ee
This parameterization is obtained from (\ref{3w}) by change of variables
$x=\lambda^2-u$.
This is a different curve but the new differential $dS$ has on this curve
the same periods as the differential (\ref{3w}).

The parameterization (\ref{4w}) has the evident advantage that only one of its
branching
points is moved by the Whitham flows. This Whitham dynamics is nothing but the
Gurevich-Pitaevskii solution first emerged in the problem of modulation
of the one-zone solution of the KdV equation \cite{GP}. In the brane language
this solution looks as follows.

Initially, the 3-branes coincide with a pair of 7-branes and then
they scatter during the evolution process. After all,
they are absorbed by the remaining pair of 7-branes. Therefore, the exact
metric on the moduli space is provided by the 3-brane exchange
process which generalizes the closed string exchange picture
suggested in \cite{douglasli}.

Returning back to the Whitham dynamics,
the coordinate $u$ defines positions of the two 3-branes located at
the points $\pm \sqrt{u}$  on the $a$-plane\footnote{Since
the whole our consideration here is done at the quasi-classical
(perturbative) level, $u=a^2$.}, while the
other branching  points fix locations of the background branes.
The branching point corresponding to the dynamical
3-branes moves in accordance with the Whitham flows for the one-zone
solution from $u=\Lambda^2_{QCD}$ to $u=-\Lambda_{QCD}^2$,
while 7 branes remain at rest.

Let us emphasize the close analogy between the dynamics of branes
and the low-energy monopole scattering. Indeed, there
is a geodesics which corresponds to the transformation of the
monopoles into dyons \cite{athi}. The brane evolution above is equivalent
to this process in the following sense. At the initial moment,
when branes are located at the points $\pm \Lambda^2_{QCD}$, there are
massless monopoles in the theory in accordance with the exact
formula (\ref{BPS}) for the BPS spectrum.
After the scattering process at the right angle at $u=0$, the
branes acquire an electric charge which can be explained as follows.
At the endpoint of the evolution when branes are at $u=-\Lambda^2_{QCD}$,
there are massless dyons in the theory. However, the dyons
can be treated as (1,1) strings stretched between the dynamical
and background branes \cite{sen2}. Therefore, the dynamical branes have to
acquire the electric charge necessarily.

\section{Conclusion}

We have tried to clarify the role and interpretation of classical
integrability in the vacuum sector of supersymmetric gauge
theories which we basically understand as obtained via string
compactifications.
Briefly this can be formulated as follows.
The vacuum sector is described in terms of  moduli spaces which
describe the phase spaces of two different integrable systems.
An integrable many-body system
defines a dynamics on a moduli space which
can be formulated in terms of the geometry of the compactification manifold.
On the other hand, the second integrable dynamics, the Whitham dynamics
can be interpreted as governing the
renormalization group flows in the theory and develops on the Coulomb
branch of the moduli space. Both integrable systems can be interpreted
in terms of the dynamics of collective coordinates of branes in different
dimensions.

The effectiveness and the validity of the approach is out of doubts
at the moment, despite the lack of its derivation from the first principles.
The most important problem to be solved seems to be to sum up
the infinite number of instantons into the finite number
of degrees of freedom that provide the phase space for the
integrable system.
One more question essential from the conceptual point of view is to
clarify the meaning of quantization of the integrable system.
Role of the corresponding ``Planck constant" is  uncertain at the
moment.

It is absolutely clear that the branes are extremely important for
dealing with nonperturbative contributions
and play the role analogous to the Feynman diagrams in the
perturbation theory. We have shown that, in simplest situations,
integrability can be described in brane terms, however, the
extensive formulation of these relations is still missed and,
therefore, further
work in this direction is desired.

The authors are grateful to H.W.Braden, V.Fock, S.Gukov, A.Marshakov,
A.Morozov, I.Krichever, N.Nekrasov and B.Rubtsov for
discussions and collaboration
and to A.Gerasimov, S.Kharchev, A.Losev, M.Olshanetsky, A.Rosly, K.Selivanov
and A.Zabrodin for useful
discussions. A.G. thanks University of Angers where the work
was partially done for the hospitality.
A.M. thanks T.Takebe for discussions on duality and kind hospitality at the
Ochanomizu University, Tokyo, where this work was partially done.
The work was supported in part by grants INTAS-99-1705, RFBR-98-01-00327,
CRDF-RP1-2108 and CNRS grant for research 2000  (A.G.) and by grants INTAS
99-0590, CRDF \#6531,
RFBR 00-02-16101a and the JSPS fellowship for research in Japan (A.M.).

\newpage
\renewcommand{\refname}{References}

\end{document}